\documentclass{aa}
\usepackage{graphicx}
\usepackage{txfonts}
\usepackage{lscape}
\begin{document}
   \title{HST/WFPC2  morphologies and color maps of distant luminous infrared galaxies}

   \author{X. Z. Zheng \inst{1}
          \and F. Hammer \inst{1}
          \and H. Flores \inst{1}
          \and F. Ass$\acute{\rm e}$mat \inst{1}
          \and D. Pelat \inst{2}
          }

   \offprints{X. Z. Zheng \& F. Hammer, E-mails: Xianzhong.Zheng@obspm.fr \& Francois.Hammer@obspm.fr}

   \institute{GEPI, Observatoire de Paris-Meudon, 92195 Meudon, France  
          \and
              LUTH, Observatoire de Paris-Meudon, 92195 Meudon, France
              }

   \date{Received December 03, 2003; accepted March 17, 2004}

   \abstract{ Using  HST/WFPC2 imaging in  F606W (or F450W)  and F814W
filters,   we  obtained   the  color   maps in observed frame  for  36   distant
(0.4$\,<\,z\,<\,$1.2) luminous infrared galaxies 
(LIRGs, L$_{\rm IR}(8-1000\,\mu$m) $\geq$\,$10^{11}$\,L$_\odot$), 
with average star  formation  rates  of $\sim$100\,M$_\odot$\,yr$^{-1}$.
Stars  and compact  sources are  taken as  references to  align images
after correction of geometric distortion. { This} leads  to an alignment
accuracy  of  0.15\,pixel,  which  is  a prerequisite for studying the
detailed color properties of galaxies with complex morphologies. A new
method is  developed to  quantify the reliability  of each pixel  in the
color map without any bias against very red or blue color regions.

Based on analyses of  two-dimensional structure and spatially resolved
color  distribution, we carried out morphological classification for LIRGs.
About 36\% of  the LIRGs were classified as disk galaxies
and  22\% as  irregulars.  Only 6  (17\%) systems  are
obvious ongoing  major mergers. An upper  limit of  58\%   was found for
the  fraction of  mergers in  LIRGs with all  the possible merging/interacting
systems included.  Strikingly,  the  fraction  of  compact
sources  is as high as 25\%, similar to  that found  in
optically selected samples.  From their K band luminosities, LIRGs are
relatively  massive  systems, with  an average stellar  mass of about 1.1$\times$10$^{11}$\,M$_\odot$. They  are related to the  formation of massive
and large disks,  from their morphologies and also  from the fact that
they  represent a significant  fraction of  distant disks  selected by
their sizes.  If sustained at  such large rates, their  star formation
can double their stellar masses in less than 1 Gyr.  The compact LIRGs
show blue cores,  which could be associated with  the formation of the
central  region  of  these  galaxies.   We find  that  all  LIRGs  are
distributed along a sequence which relate their central color to their
concentration  index. This  sequence links  compact objects  with blue
central  color to  extended ones  with relatively  red  central color,
which are  closer to  the local disks.  We suggest that there are many
massive disks which are  still forming a large fraction of their
stellar mass since $z$\,=\,1. For most of them, their central 
parts (bulge?) were formed prior to the  formation of  their disks.
  }  

\titlerunning{Distant LIRGs  --  morphologies and  color maps} 

\maketitle

%________________________________________________________________
%%%%%%%%%%%%%%%%%%%%%%
\section{Introduction}

The evolution of the cosmic star formation density (CSFD) exhibits the
history of  the stellar mass  assembly averaged over all  galaxies.  A
sharp decline of the CSFD  since $z\,\sim$\,1 has been found, whereas large
uncertainties still remain at higher redshifts, particularly due to
the uncertainties and biases  regarding dust extinction (e.g. Madau et
al.~\cite{Madau};  Hammer et al.~\cite{Hammer97}).   Investigations of
global stellar mass  density as a function of  redshift indicate that
more than one quarter, probably  up to half of the present day stars were
formed  since  $z\,\sim$\,1  (Dickinson  et al.  \cite{Dickinson}  and
references therein).  This is in  agreement with an integration of the
CSFD if the latter accounts for all the light re-radiated at IR wavelengths
(Flores  et  al.~\cite{Flores}).   Hence the  star-forming  activities
since $z\,\sim$\,1  still play an  important role in the  formation of
the galaxy Hubble sequence seen in the local universe.

{\it Hubble Space Telescope} (HST) observations show that  the merger 
rate increases significantly at
$z\,\sim$\,1,  compared  with   that  in   the  local   universe  (Le
F$\grave{\rm e}$vre     et      al.     \cite{Fevre};     Conselice     et
al. \cite{Conselice}).  Such events were claimed to be related to dwarf
galaxies while massive systems have  almost formed before  redshift 1
(Brinchmann     \&      Ellis~\cite{BrinchmannEllis};     Lilly     et
al. \cite{Lilly98}; Schade et  al. \cite{Schade}).  However, with {\it
Infrared  Space  Observatory} (ISO)  mid-infrared  imaging, Flores  et
al.  (\cite{Flores})  inferred that  a  substantial  fraction of  star
formation since $z\,\sim$\,1 is associated with the LIRGs.
These objects are luminous star-forming
galaxies  at intermediate  redshifts ($z\,\sim\,$0.5  to  1), different
from the faint  blue galaxy population (Genzel  \& Cesarsky \cite{Genzel};
Franceschini  et al.~\cite{Franceschini03}).   It  is widely  accepted
that merger/interaction is very  efficient in pushing gas into nuclear
region and triggering violent star formation. Therefore LIRGs are suspected
to  be merging systems  and the evolution  of these  galaxies is
linked to the decline of  the merger rate (Elbaz et al. \cite{Elbaz}).
Although HST imaging  showed that
most of  the LIRGs are  luminous disk/interacting galaxies  (Flores et
al.~\cite{Flores}),  systematic investigation  of their  properties is
still required  to understand their  formation and evolution, as well as link
them to the counterparts in the local universe.

Morphological classification  is essential to revealing  the nature of
the distant LIRGs. However, at  high redshifts, it becomes difficult to
classify galaxy  morphology securely because,
the images  of the high-z  galaxies suffer from  reduced resolution,
band-shifting  and cosmological  surface  brightness dimming  effects,
compared with the local objects.  With  HST {\it Wide Field Planet 
Camera 2} (WFPC2)  high resolution
imaging in  two or more  bands, spatially resolved  color distribution
can be used to investigate the distribution of the stellar population, which
is  complementary  to  addressing   the  appearance  in  single  band.
Furthermore, the star-forming regions  and dusty regions can stand out
in the color map. This is  very important to the study of LIRGs, in which
these regions  are  expected to be numerous.

Canada-France Redshift Survey (CFRS) fields are among the most studied
fields at various wavelengths.   Two CFRS fields 0300+00 and 1415+52
had  been   observed  deeply  by  ISOCAM  at   15\,$\mu$m  (Flores  et
al.  ~\cite{Flores}; 2004 in  preparation) and  by HST  (Brinchmann et
al.~\cite{Brinchmann}).   Aimed   at  performing  detailed   analyses  of
morphology,  photometry  and  color  distribution for  distant  LIRGs,
additional HST images through blue  and red filters have been taken to
complement the color information for  the two CFRS fields (PI: Hammer,
Prop. 9149).  In this work,  we present the preliminary results of the
color distribution of  the distant LIRGs.  We correct additional
effects in HST images and  recenter them accurately, which allow us to
access the color maps of complex galaxies. We also implement a method
to quantify the signal-to-noise  (S/N) ratio of  the color
image in order  to give a reasonable cut for the  target area in color
maps.

This paper is organized as  follows.  Sect. 2 describes the HST imaging
observations and  the archive data we  adopt. In Sect.  3, we describe
the various effects which have  to be corrected for aligning images in
different WFPC2 filters.  In Sect. 4, we describe a method we use to 
generate the color  maps.   In Sect.  5, we  summarize  the
morphological properties of the distant LIRGs. The results we obtained
of the LIRGs are discussed in Sect. 6.  Brief conclusions are given in
Sect.      7.       Throughout      this      paper      we      adopt
H$_0$\,=\,70\,km\,s$^{-1}$\,Mpc$^{-1}$,  $\Omega_{\rm  M}$\,=\,0.3 and
$\Omega_\Lambda$\,=\,0.7. Unless specified, we exclude PC chip and the
unit  of pixel  refers  to that  in  WF chips.   The bands  B$_{450}$,
V$_{606}$ and  I$_{814}$ refer to  HST filters F450W, F606W  and F814W,
respectively.  Vega system is adopted for our photometry.
%__________________________________________________________________

\begin{table*}
\centering 
   \caption[]{HST imaging with two bands observations in CFRS fields 0300+00 and 1415+52}
   \label{hstlist}
   \begin{tabular}{cccccccccc}
   \hline
   \noalign{\smallskip}
Field &  BlueFilter & Total Exp. & N$^{\mathrm{a}}$ & Dither$^{\mathrm{b}}$ & RedFilter & Total Exp. & N$^{\mathrm{a}}$ & Dither$^{\mathrm{b}}$ & Prop.ID$^{\mathrm{c}}$\\
   \noalign{\smallskip}
   \hline
   \noalign{\smallskip}
030226+001348 & F606W & 6400 & 5 & 20& F814W & 6000 & 5 & 20& 9149 \\
030227+000704 & F450W & 7000 & 5 & 20& F814W & 6700 & 5 & 20& 6556,5996 \\
030233+001255 & F450W & 6600 & 6 & 0 & F814W & 6400 & 6 & 0 & 5449 \\
030237+001414 & F606W & 6400 & 5 & 20& F814W & 6400 & 5 & 20& 9149 \\
030240+000940 & F606W & 6400 & 5 & 20& F814W & 7000 & 5 & 12.5& 9149,8162 \\
030243+001324 & F450W & 6600 & 6 & 0 & F814W & 6400 & 6 & 0 & 5449 \\
030250+001000 & F606W & 6400 & 5 & 20& F814W & 7000 & 5 & 12.5& 9149,8162 \\
141743+523025 & F450W & 7800 & 6 & 0 & F814W & 7400 & 6 & 0 & 5449 \\
141803+522755 & F606W & 6400 & 5 & 20& F814W & 6800 & 5 & 20& 9149 \\
141809+523015 & F450W & 7800 & 6 & 0 & F814W & 7400 & 6 & 0 & 5449 \\
% 221755+001715 & F450W & 7000 & 5 & 20& F814W & 6700 & 5 & 20& 6556,5996 \\
% \noalign{\smallskip}
% \multicolumn{10}{c}{Groth Strip Fields}\\
%\multicolumn{10}{c}{See Simard et al. (\cite{Simard}) for details of Groth Strip fields} \\
%141527+520410 & F606W & 2800 & 4 & 0 & F814W & 4400  & 4 & 0 & 5090 \\
%141534+520520 & F606W & 2800 & 4 & 0 & F814W & 4400  & 4 & 0 & 5090 \\
%141540+520631 & F606W & 2800 & 4 & 0 & F814W & 4355.5& 4 & 0 & 5090 \\
%141547+520741 & F606W & 2800 & 4 & 0 & F814W & 4400  & 4 & 0 & 5090 \\
%141553+520851 & F606W & 2800 & 4 & 0 & F814W & 4400  & 4 & 0 & 5090 \\
%141600+521001 & F606W & 2800 & 4 & 0 & F814W & 4400  & 4 & 0 & 5090 \\
%141606+521111 & F606W & 2800 & 4 & 0 & F814W & 4400  & 4 & 0 & 5090 \\
%141613+521222 & F606W & 2800 & 4 & 0 & F814W & 4400  & 4 & 0 & 5090 \\
%141619+521332 & F606W & 2800 & 4 & 0 & F814W & 4400  & 4 & 0 & 5090 \\
%141626+521442 & F606W & 2800 & 4 & 0 & F814W & 4400  & 4 & 0 & 5090 \\
%141632+521552 & F606W & 2800 & 4 & 0 & F814W & 4400  & 4 & 0 & 5090 \\
%141638+521702 & F606W & 2800 & 4 & 0 & F814W & 4400  & 4 & 0 & 5090 \\
%141645+521812 & F606W & 2800 & 4 & 0 & F814W & 4400  & 4 & 0 & 5090 \\
%141651+521922 & F606W & 2800 & 4 & 0 & F814W & 4400  & 4 & 0 & 5090 \\
%141658+522032 & F606W & 2800 & 4 & 0 & F814W & 4400  & 4 & 0 & 5090 \\
%141704+522142 & F606W & 2800 & 4 & 0 & F814W & 4400  & 4 & 0 & 5090 \\
%141711+522252 & F606W & 2800 & 4 & 0 & F814W & 4400  & 4 & 0 & 5090 \\
%141717+522402 & F606W & 2800 & 4 & 0 & F814W & 4400  & 4 & 0 & 5090 \\
141724+522512 & F606W & 2800 & 4 & 0 & F814W & 4400  & 4 & 0 & 5090 \\
141731+522622 & F606W & 2800 & 4 & 0 & F814W & 4400  & 4 & 0 & 5090 \\
141737+522731 & F606W & 2800 & 4 & 0 & F814W & 4400  & 4 & 0 & 5090 \\
141750+522951 & F606W & 2800 & 4 & 0 & F814W & 4400  & 4 & 0 & 5090 \\
141743+522841 & F606W &24400 &12 & 0 & F814W &25200  &12 & 0 & 5109 \\ 
141757+523101 & F606W & 2800 & 4 & 0 & F814W & 4400  & 4 & 0 & 5090 \\
141803+523211 & F606W & 2800 & 4 & 0 & F814W & 4400  & 4 & 0 & 5090 \\
%141810+523320 & F606W & 2800 & 4 & 0 & F814W & 4400  & 4 & 0 & 5090 \\
%141816+523430 & F606W & 2800 & 4 & 0 & F814W & 4400  & 4 & 0 & 5090 \\
%141823+523540 & F606W & 2800 & 4 & 0 & F814W & 4400  & 4 & 0 & 5090 \\
   \noalign{\smallskip}
   \hline
   \end{tabular}

  \begin{list}{}{}
\item[$^{\mathrm{a}}$] Number of exposures.
\item[$^{\mathrm{b}}$] Dither offset among the consecutive exposures, aimed 
         to remove cosmic-rays and hot/bad pixels.  Here the largest offset 
         in unit of pixel along x axis is present.
\item[$^{\mathrm{c}}$] HST Proposal ID. One ID means that both the blue and 
         the red observations were carried out in the same proposal. Two IDs 
         refer to the proposals that the blue and the red observations were observed  respectively.
  \end{list}
\end{table*}

%%%%%%%%%%%%%%%%%%%%%%
\section{Observations}

In  the CFRS, ground-based  spectroscopic redshift  identification was
carried out with CFHT telescope  for objects brighter than 22.5 mag 
(I$_{\rm AB}$) in five 10$\arcmin \times 10\arcmin$ fields (see Crampton 
et al.~\cite{Crampton} for details). With improved data reduction,
Flores et  al. (\cite{Flores04}) present  the updated
catalogs of  the deep  ISOCAM observations at  15\,$\mu$m for  the two
CFRS fields 0300+00 and 1415+52 (see also Flores et al. \cite{Flores}).

Using high resolution HST/WFPC2 imaging, the distant galaxies up to
z$\sim$1 can be spatially resolved.  Three HST fields were observed in
F606W  and  F814W   filters  in  the  two CFRS fields  with  ISOCAM
observations (in Cycle  10, PI: Hammer, Prop. 9149),  and two fields in
F606W  filter were observed  to complement  the observations  in F814W
during  Cycle 8  (PI: Lilly,  Prop. 8162).   F606W and  F814W filters
respectively correspond to rest-frame U (3634\,\AA) and V (4856\,\AA) at a  redshift 0.65.  Those
fields were chosen so that to  maximize the number of LIRGs contained in
each field.   We also  collected the HST  imaging data with  two bands
observations in  the CFRS fields.  A detailed description  of the previous
CFRS  field  HST  imaging   survey  was  presented  in  Brinchmann  et
al.  (\cite{Brinchmann}).  The CFRS  1415+52 field  partially overlaps
the Groth  Strip Survey (GSS, Groth et  al.~\cite{Groth}). We included
the GSS imaging data into our analysis.

Table~\ref{hstlist}  summarizes  the HST  imaging  data  used in  this
analysis.  The total exposure time  is usually more than 6000 seconds,
in  which the  surface  brightness corresponding to 1$\sigma$ above  the
background  is $\sim$25.5\,mag\,arcsec$^{-2}$.  Note  that the  GSS is
relatively shallow except for one very deep field.  Here we list the seven
of 28 GSS fields  covering the CFRS  1415+52 field.  The 28 GSS fields 
are composed of 27 fields observed in Prop. 5090 (PI: Groth) and one very 
deep field in Prop. 5109 (PI: West Phal).
Simard  et  al. (\cite{Simard})  carried  out a  detailed
morphological analysis  on the  GSS observations.  They  also provided
the   physical  scale   and  absolute   magnitude  for   objects  with
spectroscopic redshift  identification using the Keck  telescope.  We 
quoted these results directly and  further details can be found in
their paper.  The  observation was divided into N  exposures (col. 4
and 8  in Table~\ref{hstlist}) aimed at removing  cosmic-rays and correcting
the bad/hot pixels.

%%%%%%%%%%%%%%%%%%%%%%%%%
\section{Image alignment}

   \begin{figure*}[]
   \centering
   \includegraphics[width=0.28\textwidth]{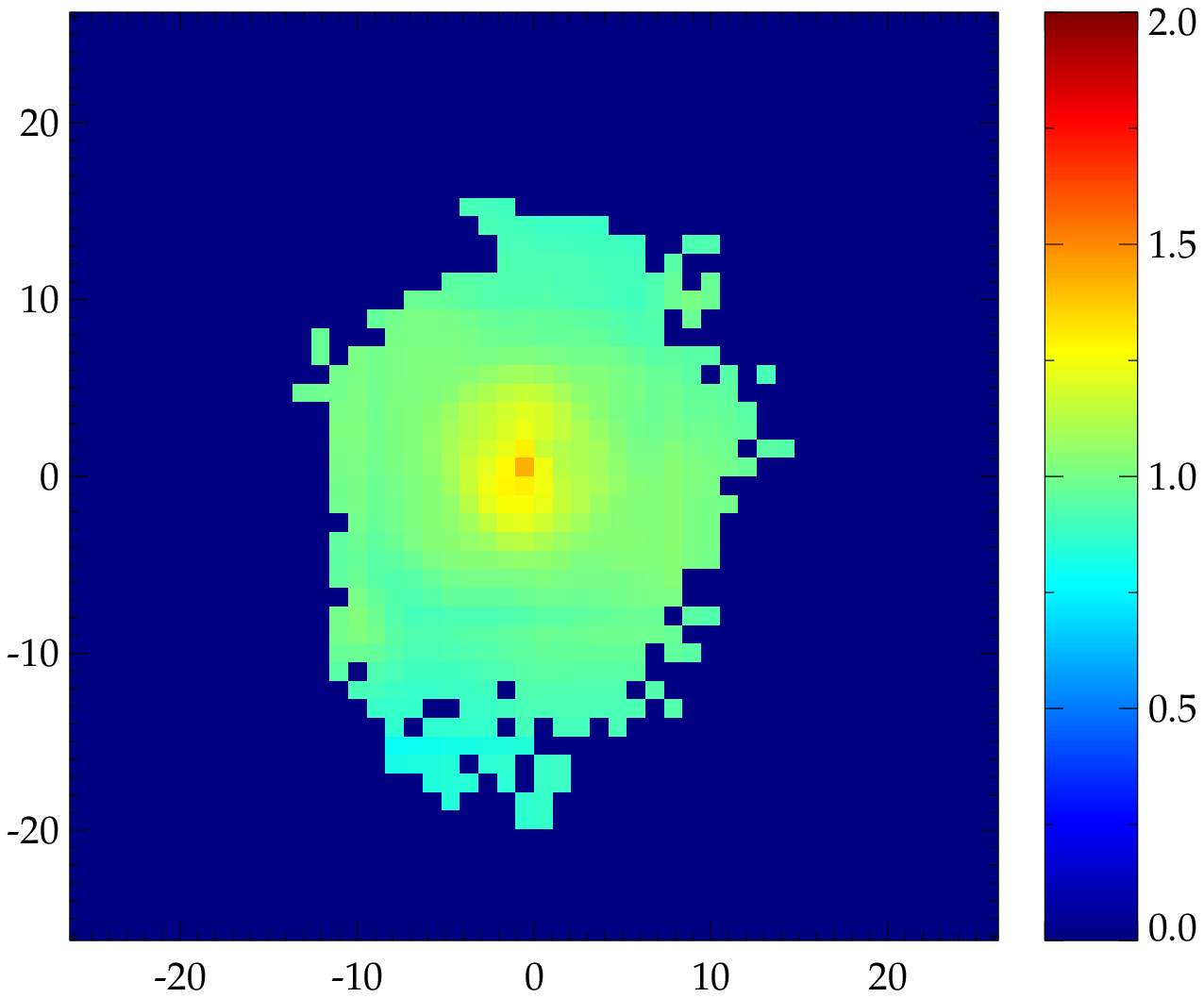}
   \includegraphics[width=0.28\textwidth]{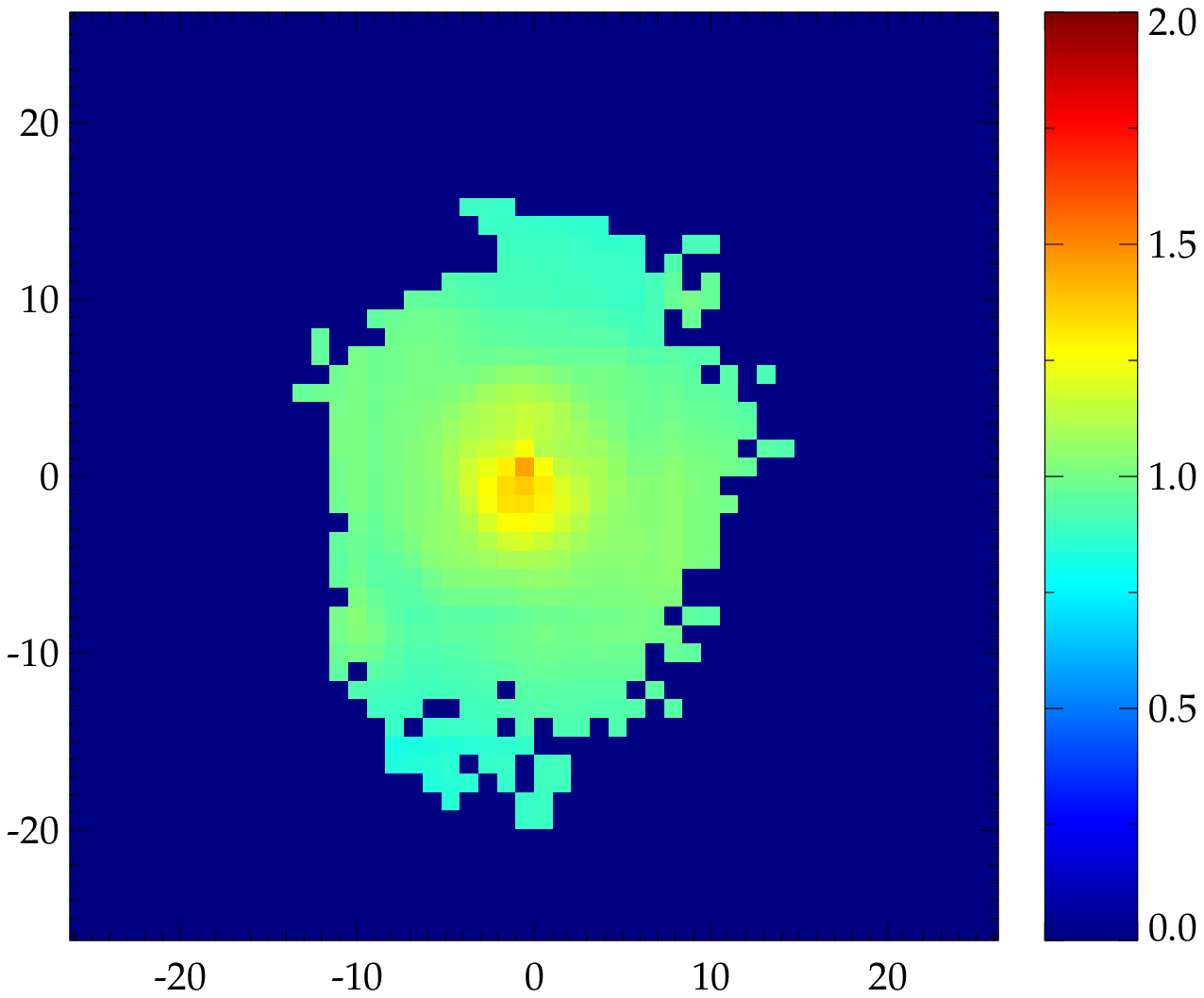}
   \includegraphics[width=0.28\textwidth]{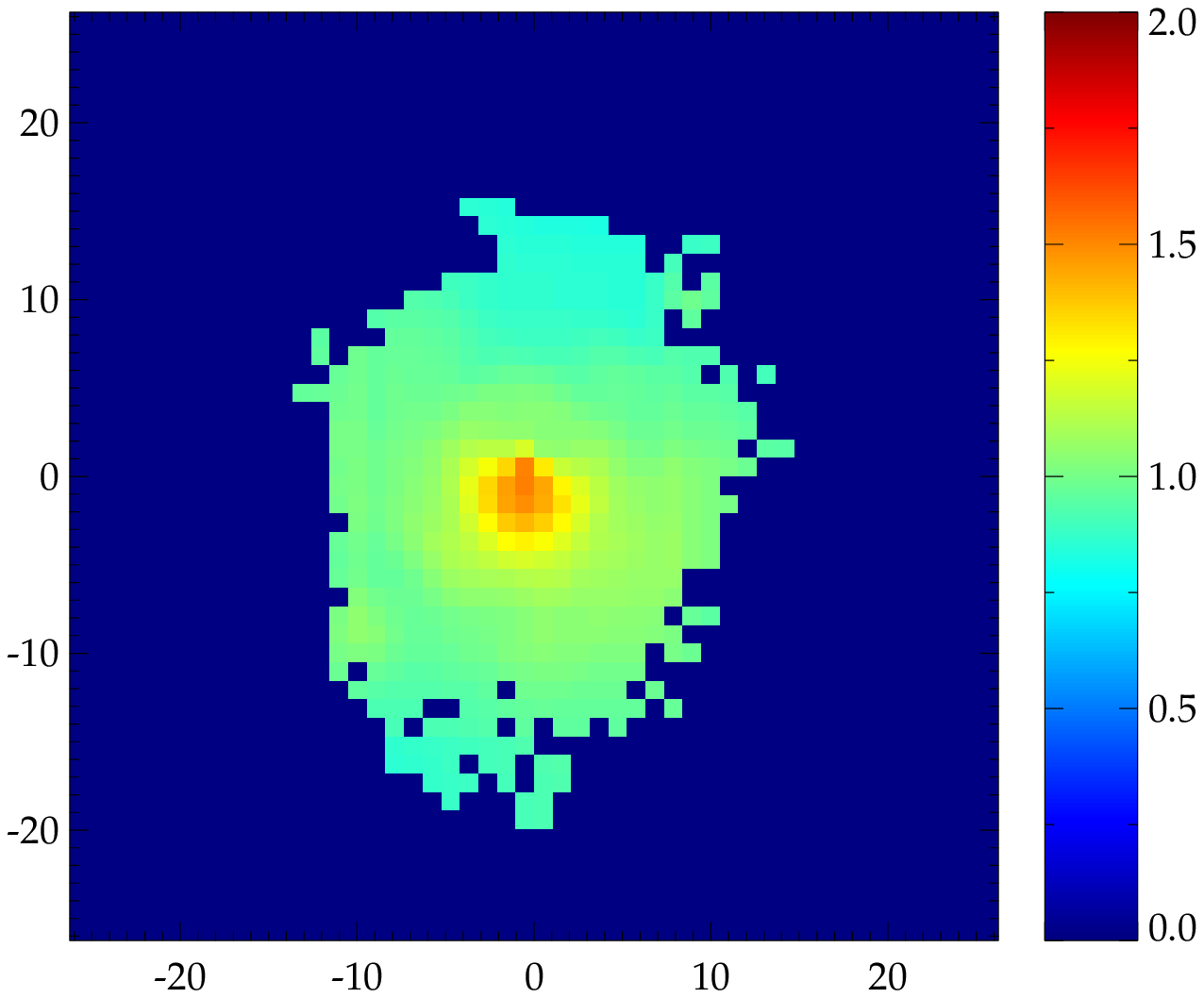}
   \caption{Imitated HST/WFPC2 V$_{606}-$I$_{814}$ color maps for an elliptical galaxy at redshift 0.299 when the offsets between blue and red images are 0.0 pixel (left panel), 0.15 pixel (middle panel) and 0.30 pixel (right panel).  The color maps are in a size of 50$\times$50\,pixel.
[{\it See the electronic edition for a color version of this figure.}]}
   \label{falsemap}%
   \end{figure*}
Studies  have  been  reported  on  investigating color  maps  for the
spheroidal  galaxies   at  intermediate  redshifts  (e.g.  Abraham  et
al.~\cite{Abraham};  Ellis et  al.~\cite{Ellis}).   In generating  the
color map,  a difficulty is  to align two  images accurately in  order to
keep each pixel of the same object at the same position in two images.  For
the spheroidal galaxies, generally the brightness peaks at 
the  galaxy center in different bands.  It  would be  technically easy  to
correct relative  shifts between  two images if  the galaxy  center is
used to be a reference.
For the galaxies with complex morphologies, however, caution should be
taken  in  determining the  relative  shifts  because their  irregular
morphologies as well as contamination from star-forming regions
could  easily blank the  efforts of finding a  reliable reference
point  (e.g. the galaxy  center  in  spheroidal  galaxies).  Such  a
difficulty  becomes more  serious  for  galaxies at  intermediate
redshifts  when  band-shifting   effect   becomes  significant.    The
uncertainty in  aligning two images is  required to be  much smaller than
one pixel  in addressing color distribution  pixel-by-pixel.  We model
the  color  map  V$_{606}-$I$_{814}$  to  imitate the  effect  of  the
alignment  offset on  the color  map. We  use the  modeled HST/WFPC2
images of  an elliptical galaxy at  redshift 0.299 with a de Vaucouleurs bulge+exponential disk structure (B/T\,=\,0.84) given by GIM2D (Simard et al. \cite{Simard}), in
order   to   avoid   the   contamination  by   the   intrinsic   color
fluctuation. Fig.~\ref{falsemap} illustrates that an offset
as small as $\sim$0.3 pixel will cause false structure ``half-blue and
half-red'' in the color map (right panel), compared with the color map
with zero offset (left panel).   Such effect becomes marginal when the
offset decreases  to 0.15 pixel  (middle panel), which is  the typical
uncertainty in our image alignment.
Instead  of operating on  individual objects,  we dealt with  the whole
images in data  processing.  The main  efforts in data  reduction to
align images are summarized below.

The raw  HST images were  processed using the standard  STScI pipeline.
For HST/WFPC2 observations in integer-pixel dither mode with telescope
relative offsets larger than a few pixels, camera geometric distortion
will cause an additional shift (increasing toward the CCD corners, see
HST/WFPC2  handbook  for  more  details).  We  correct  the  geometric
distortion for each individual exposure in a data set before combining
them.   Table~\ref{hstlist}  tabulates  the  dither  offsets,  usually
twenty pixels (at which, an additional shift of about half pixel due to
the geometric distortion is present at the WF chip corners) except for
those in Proposal 5449 and GSS, in
which consecutive  exposures were taken  at the same position  and the
distortion correction was not applied.
The shifts  between the exposures  were obtained using  two approaches,
cross-correlation  and point source  reference. The  cross correlation
technique is  to shift and/or rotate  one image relative  the other to
maximize the cross-correlation between the two images, i.e. the best
match.  In the later approach, the shifts are derived from a comparison of
the  locations  of  a  number  of  point  and  point-like  sources  in
individual exposures.   In our work, the point  and point-like sources
refer to  the objects satisfying  Full Width at Half  Magnitude (FWHM)
$<$\,2.5 pixels,  17\,$<\,m_{814}\,<$\,22, 17\,$<\,m_{450}\,<$\,23 and
17\,$<\,m_{606}\,<$\,23. These criteria exclude  the extended objects
and  those saturated  or faint.  For each  WFPC2 field, at  least three 
reference  sources are  used to
derive the  relative shifts between the different  exposures.  Both of
the two  approaches use real  images/objects to derive the  shifts and
hence are free from  guide star acquisition uncertainties.  In general,
the derived shifts are remarkably consistent with each other within 0.08
pixel and even better for  crowded fields.  For sparse fields and some
fields  with large  dither offsets,  the cross-correlation  is  not the  best
approach and the measurement using  the point source reference will be
adopted.   The cross-correlation  is  also not  suitable  to find  the
rotation  angle  and the  shifts  between  images  in different  bands.
The  morphologies and  brightness of the  astronomical sources
may  differ in  one  wavelength  window from  those  in another.   The
cross-correlation would be biased by  the sources with a real center
offset between  different bands,  which is often  seen in  spiral and
irregular galaxies.
Geometric   distortion   correction,   cosmic-ray  removement,   image
combination are accomplished using the STSDAS/DITHER package (version 2.0,
Koekemoer et al.~\cite{Trauger}).

   \begin{figure}[]
   \centering
   \includegraphics[width=0.40\textwidth]{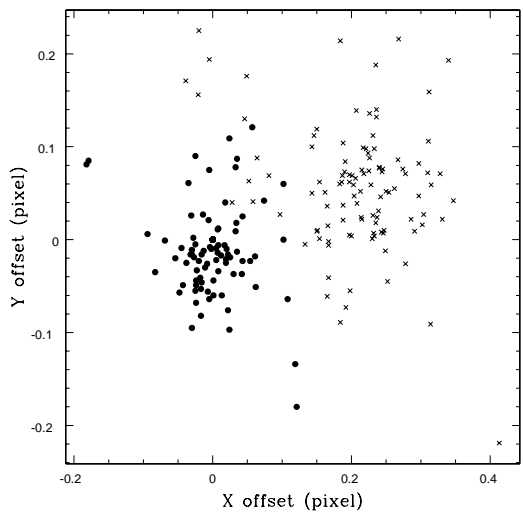}
   \caption{Systematic offsets between V$_{606}$ and I$_{814}$ images in 27 GSS fields. Solid circle is I$_{814}$ image and cross is V$_{606}$ image.}
   \label{grothoffsets}%
    \end{figure}
   \begin{figure}[]
   \centering
   \includegraphics[width=0.40\textwidth]{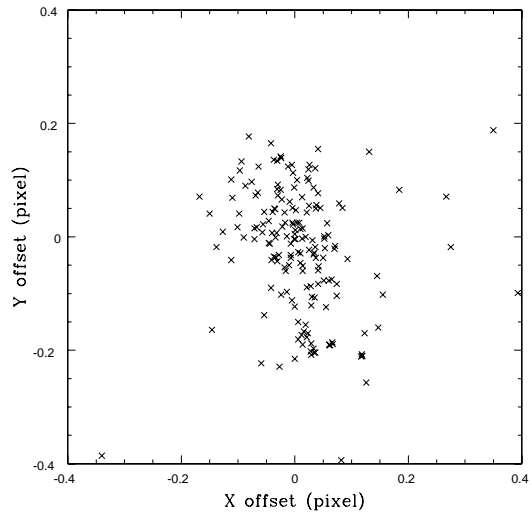}
   \caption{Position offset for 178 point/point-like sources between 
two aligned images. The offsets are in the coordinates of WF3 chip. The median
value in X is 0.006 pixel and the median value in Y is 0.004 pixel.}
   \label{composite}%
    \end{figure}

In  some of  the fields,  blue  images and  red images  were taken  in
different cycles, i.e.  HST telescope pointings in two  bands were not
the  same, which results in different relative  rotations and  shifts between
them.  We have used point-like  sources as references to determine the
rotations  and the shifts.   In practice,  we obtain  the  relative rotation
angle from  astrometric information recorded in the image header. The keyword
 ORIENTAT  provides the position angle of the telescope pointing.  
Normally the uncertainty of the position angle determination is 0.003
degree,  compared to  an rotation  deviation 0.01  degree  causing 0.1
pixel offset at the corners of a WF chip.  Table~\ref{align} lists the
shift and the rotation angle of the  blue image relative to the red one
for the WF3 chip.  Note that in  image combination, the first image of the
data set is always taken as reference to stack images.

To exhibit the advantages of  the point source reference method, it is
worth showing  that a systematic shift exists  between V$_{606}$ image
and I$_{814}$ image in the observations of the 27 GSS fields in Prop. 5090. 
The  observation of each field  was splitted
into 8  exposures in V$_{606}$  and I$_{814}$ bands alternately  at the
same  location, i.e.  no  offset between  consecutive exposures  and 4
exposures for each  filter.  For the 8 exposures of  each field, we derive
the  shifts  relative  to   the  first  I$_{814}$  exposure  by 
comparing  the  positions  of  point and  point-like  sources  in
individual  exposures.   It  is  assumed  that there  is  no  relative
shift/rotation among individual chips  during each exposure. At least
3 reference sources (usually 7 , even more than 10 for some fields) in
three  WF chips are  used to  give median  shifts in  X and  Y axes.
Fig.~\ref{grothoffsets} illustrates  the distribution of  the relative
shifts of  the rest 7 exposures  to the first  one for all the  27 GSS
fields.  It reveals  the  systematic offset  between  V$_{606}$  and
I$_{814}$ images.  The shifts for each field are tabulated in 
Table~\ref{align}.  Combining the  27 fields, we  get  median shift
of $\Delta$x\,=\,0.20 pixel and $\Delta$y\,=\,0.08 pixel.

Fig.~\ref{composite} shows the distribution of the position offsets of
178 point/point-like sources in two  aligned images for 11 CFRS fields
and  28 GSS  fields.   The median  of  X offset  is  0.006 pixel  with a
corresponding  semi-inter-quartile range  (SIQR) of 0.036  pixel  and the
median of Y offset is 0.004 pixel with an SIQR of 0.070 pixel. The figure 
shows that the systematic offsets have  been  corrected. The center
offset, $\sqrt{\Delta  x^2+\Delta y^2}$, can be used  to measure  the
uncertainty of an alignment.  For the 178  point/point-like sources,
the  mean  value  of  the  center  offset  is  0.117  pixel.  
This denotes that the images are well aligned and
can be used for generating color maps.
   \begin{figure*}[]
   \centering
   \includegraphics[width=0.32\textwidth]{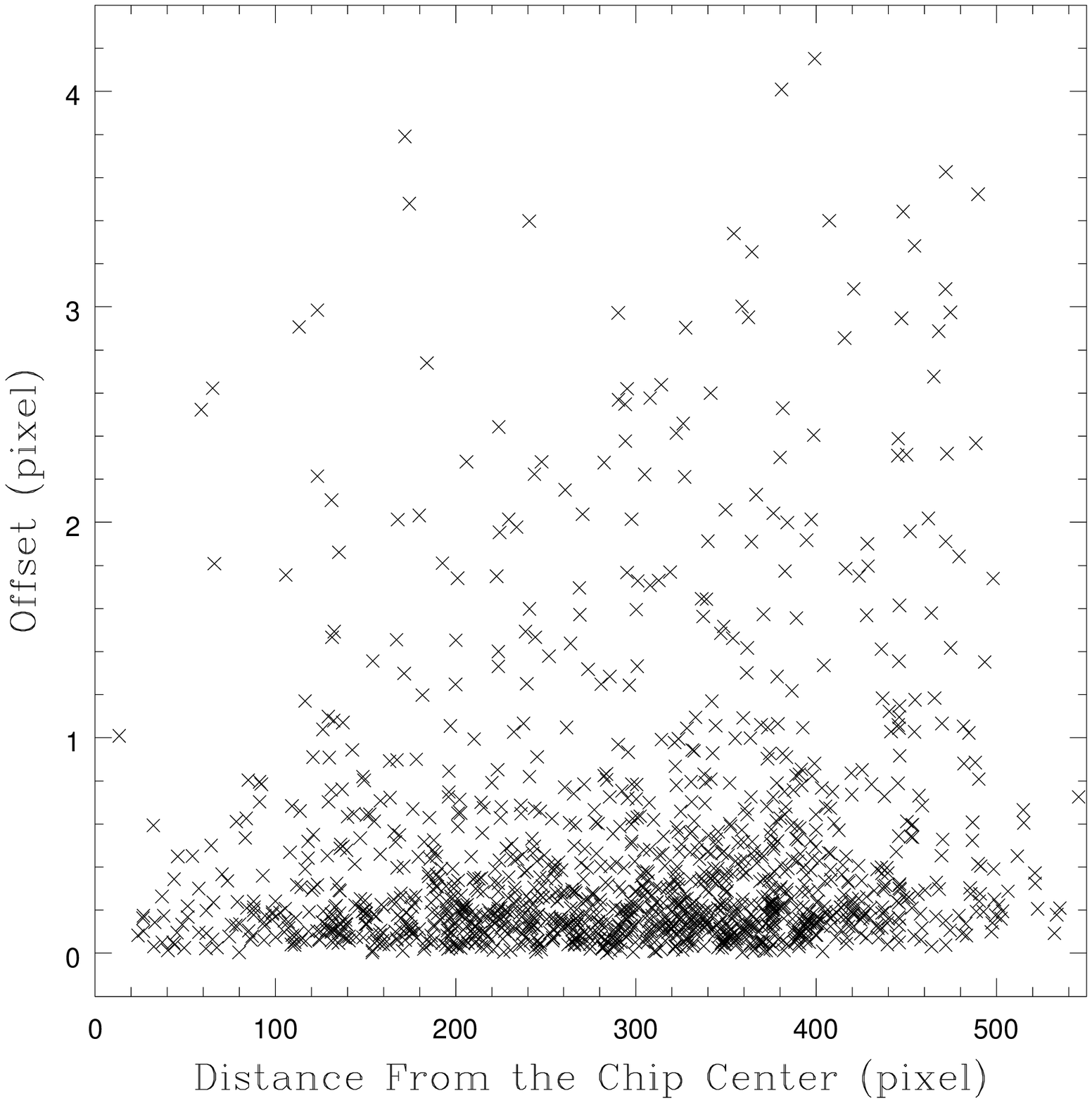}
   \includegraphics[width=0.32\textwidth]{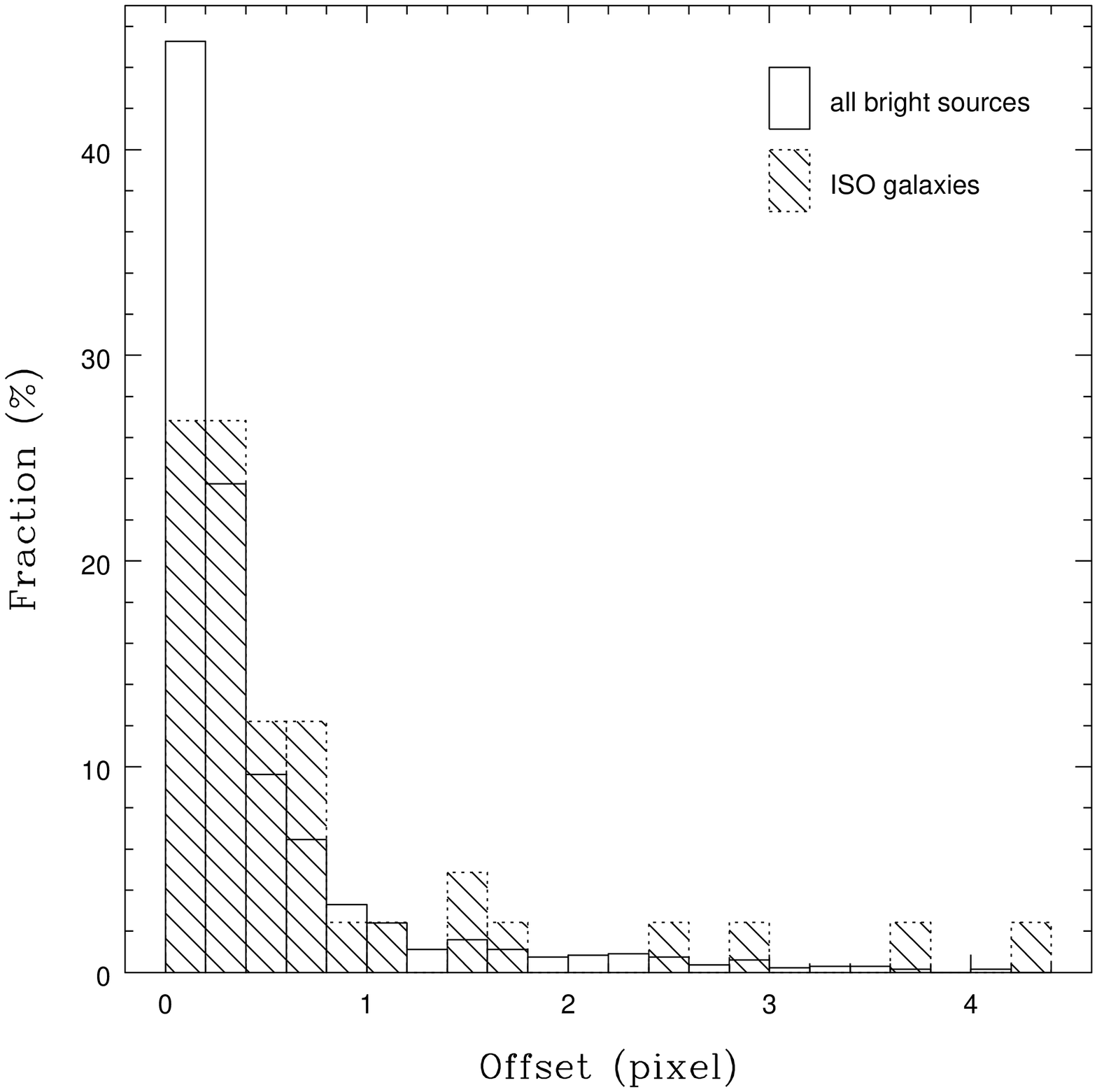}
   \includegraphics[width=0.32\textwidth]{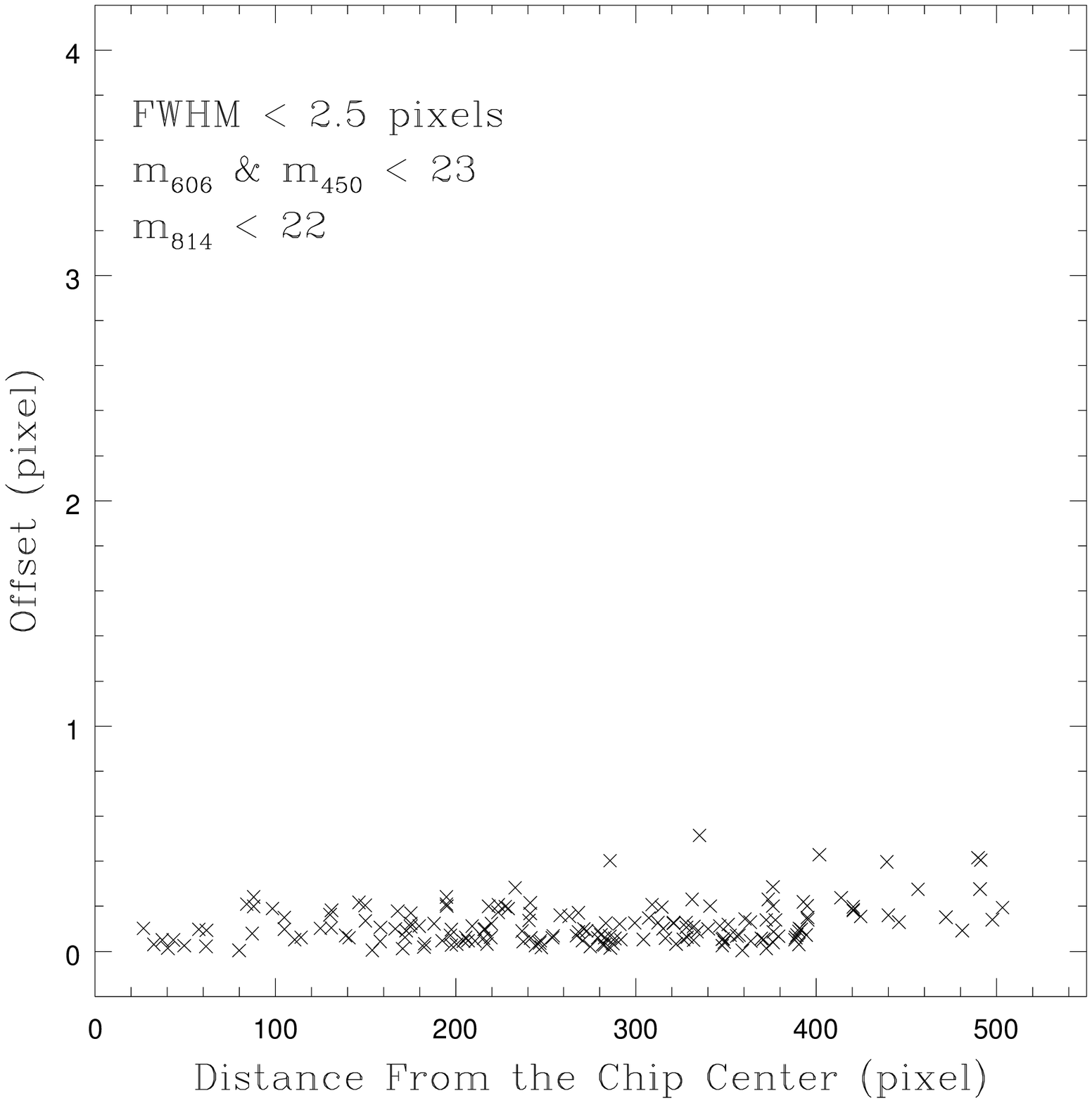}
   \caption{Left   panel:  center   offset  of   the   bright  sources
($m_{814}\,\leq\,22$)  between blue (V$_{606}$  or B$_{450}$)  and red
(I$_{814}$) image  as a  function of their  positions in  WFPC2 WF
chip, expressed by  the distance from the chip  center.  Middle panel:
histograms of  the center  offset for all  bright sources and  for our
ISOCAM  detected  galaxies.   Right   panel:  center  offset  for  the
point/point-like sources.  In this panel, the mean value of the offset
is 0.117 pixel with $\sigma$\,=\,0.088 pixel.}
   \label{offsets}%
    \end{figure*}

It  is worth  noting that  many  sources do  show a  center offset  in
different  band  images. Fig.~\ref{offsets}  shows  the center  offset
against the distance from the chip center for all sources (left panel)
with I$_{814}$ brighter than 22\,mag, compared to the distribution for
point/point-like sources (right panel). It is clear that a) the offset
is free from the position-dependent effect (e.g. geometric distortion); b) a
substantial fraction of sources show a large center offset (up to a few
pixels),  as shown  in  the  histogram of  the  center offset  (middle
panel). In addition,  we also plot the center  offset distribution for
our ISOCAM detected sample (see Sect. 5.1).
On average, ISOCAM galaxies present larger offsets than other sources,
which can be related  to their intrinsic morphological properties.  We
use  the  software  Sextractor  (Bertin \&  Arnouts~\cite{Bertin})  to
extract source  catalogs, including central  positions and integrated
fluxes. For galaxies with complex  morphologies, the central positions  
given in {Sextractor} are not always their brightness peaks.
An aperture of  3$\arcsec$ is adopted in our  photometry.  The updated
Charge Transfer Efficiency (CTE) correction (Dolphin~\cite{Dolphin02})
and  updated photometric  zeropoints are  adopted in  WFPC2 photometry
calibration in Vega system (Dolphin~\cite{Dolphin00}).
\begin{table}
  \footnotesize
  \centering
  \caption[]{Offsets and rotation between blue and red images}
  \label{align}
  \begin{tabular}{crrr}
  \hline
  \noalign{\smallskip}
Field & $\Delta$x(pixel) & $\Delta$y(pixel) & $\Delta\theta$($\degr$) \\
  \noalign{\smallskip}
  \hline
  \noalign{\smallskip}
\multicolumn{4}{c}{CFRS Fields} \\
  \noalign{\smallskip}
030226+001348   &$-$2.28&    2.50 &  0.550 \\  
030227+000704   &  4.31 &$-$14.00 &  0.188 \\  
030233+001255   &  0.00 & $-$0.15 &  0.000 \\  
030237+001414   &  5.54 & $-$5.37 &$-$1.200 \\  
030240+000940   &  9.76 &    8.90 &$-$0.185 \\ 
030243+001324   &$-$0.05&$-$0.36  &  0.000 \\  
030250+001000   & 20.65 &    7.04 &  0.070 \\  
141743+523025   &  0.35 & $-$0.32 &  0.000 \\  
141803+522755   &$-$6.47&    7.15 &  1.600 \\  
141809+523015   &  0.03 & $-$0.36 &  0.000 \\  
221755+001715   &$-$8.59& $-$3.99 &  0.188 \\  
  \noalign{\smallskip}
\multicolumn{4}{c}{Groth Strip Fields} \\
  \noalign{\smallskip}
141527+520410  &   0.25 &   0.05 & 0.000 \\
141534+520520  &   0.19 &   0.08 & 0.000 \\
141540+520631  &   0.19 &   0.13 & 0.000 \\
141547+520741  &   0.18 &   0.05 & 0.000 \\
141553+520851  &   0.18 &   0.08 & 0.000 \\
141600+521001  &   0.29 &   0.08 & 0.000 \\
141606+521111  &   0.17 &   0.07 & 0.000 \\
141613+521222  &   0.19 &   0.05 & 0.000 \\
141619+521332  &   0.21 &   0.04 & 0.000 \\
141626+521442  &   0.23 &   0.17 & 0.000 \\
141632+521552  &   0.26 &   0.13 & 0.000 \\
141638+521702  &   0.22 &   0.10 & 0.000 \\
141645+521812  &   0.31 &   0.10 & 0.000 \\
141651+521922  &   0.31 &   0.03 & 0.000 \\
141658+522032  &   0.19 &   0.07 & 0.000 \\
141704+522142  &   0.10 &   0.10 & 0.000 \\
141711+522252  &   0.20 &   0.11 & 0.000 \\
141717+522402  &   0.15 &   0.02 & 0.000 \\
141724+522512  &   0.21 &   0.08 & 0.000 \\
141731+522622  &   0.20 &$-$0.01 & 0.000 \\
141737+522731  &   0.20 &   0.03 & 0.000 \\
141750+522951  &$-$0.02 &   0.11 & 0.000 \\
141757+523101  &   0.21 &   0.09 & 0.000 \\
141803+523211  &   0.25 &   0.02 & 0.000 \\
141810+523320  &   0.27 &   0.08 & 0.000 \\
141816+523430  &   0.26 &   0.05 & 0.000 \\
141823+523540  &   0.09 &   0.06 & 0.000 \\
  \noalign{\smallskip}
  \hline
  \end{tabular}
\end{table}

%%%%%%%%%%%%%%%%%%%%
\section{Color map}

To obtain the color map image of an extended source, a key point is to
quantitatively  select the  pixels in  the color  map  with reliable
color  determination.  Instead  of adopting  a  semi-empirical noise
model  (Williams et al.   \cite{Williams}), a  method is  developed to
obtain the signal-to-noise (S/N) ratio of the color map image pixel-by-pixel.
Using  this  S/N  ratio  image,  the color  map area can be
constrained for  the extended source. Adopting  an approximation that
the Poisson  noise distribution  function in an HST  image is close  to a
Log-normal  law,  we can  obtain  that  for  two images  with  signals
$\mu_{\mathrm{F}_\mathrm{B}}$,    $\mu_{\mathrm{F}_\mathrm{R}}$    and
noises                              $\sigma^2_{\mathrm{F}_\mathrm{B}}$,
$\sigma^2_{\mathrm{F}_\mathrm{R}}$,  the  noise  of
the their color image satisfies
\begin{equation}
  \sigma^2 = \log ({\sigma^2_{\mathrm{F}_\mathrm{B}} \over \mu^2_{\mathrm{F}_\mathrm{B}}} + 1) + \log ({\sigma^2_{\mathrm{F}_\mathrm{R}} \over \mu^2_{\mathrm{F}_\mathrm{R}}} + 1). 
\end{equation}
Here, the signals and the noises include the ones from both source and 
sky background. A detailed explanation can be found in Appendix.  
The inverse of the noise is proportional to the S/N ratio.  Applying 
this formula to blue and red images pixel-by-pixel, an S/N ratio image 
associated with the color image can be obtained.  This S/N ratio image 
is indeed for the ``color image'' of source+sky background. In the pixels
that the source's signal is weak, the noise is dominated 
by that from the sky background. In the S/N ratio image, the fluctuation
of the background reflects the uncertainty in the color map image caused 
by the sky background noise. The color map pixels with S/N ratio value 
much higher than the mean S/N ratio background should be much less 
affected by the noise from the sky background and have reliable color
determination.  A criterion of 4$\sigma$ above the mean background of 
the S/N ratio image is adopted as the detection threshold in the color 
map image and conjunct pixels consist of the color map for a target.  
The color map is given in the HST Vega system and zero 
background is set in the area below the threshold.

\begin{figure}[]
%\centering
\includegraphics[height=4cm,clip]{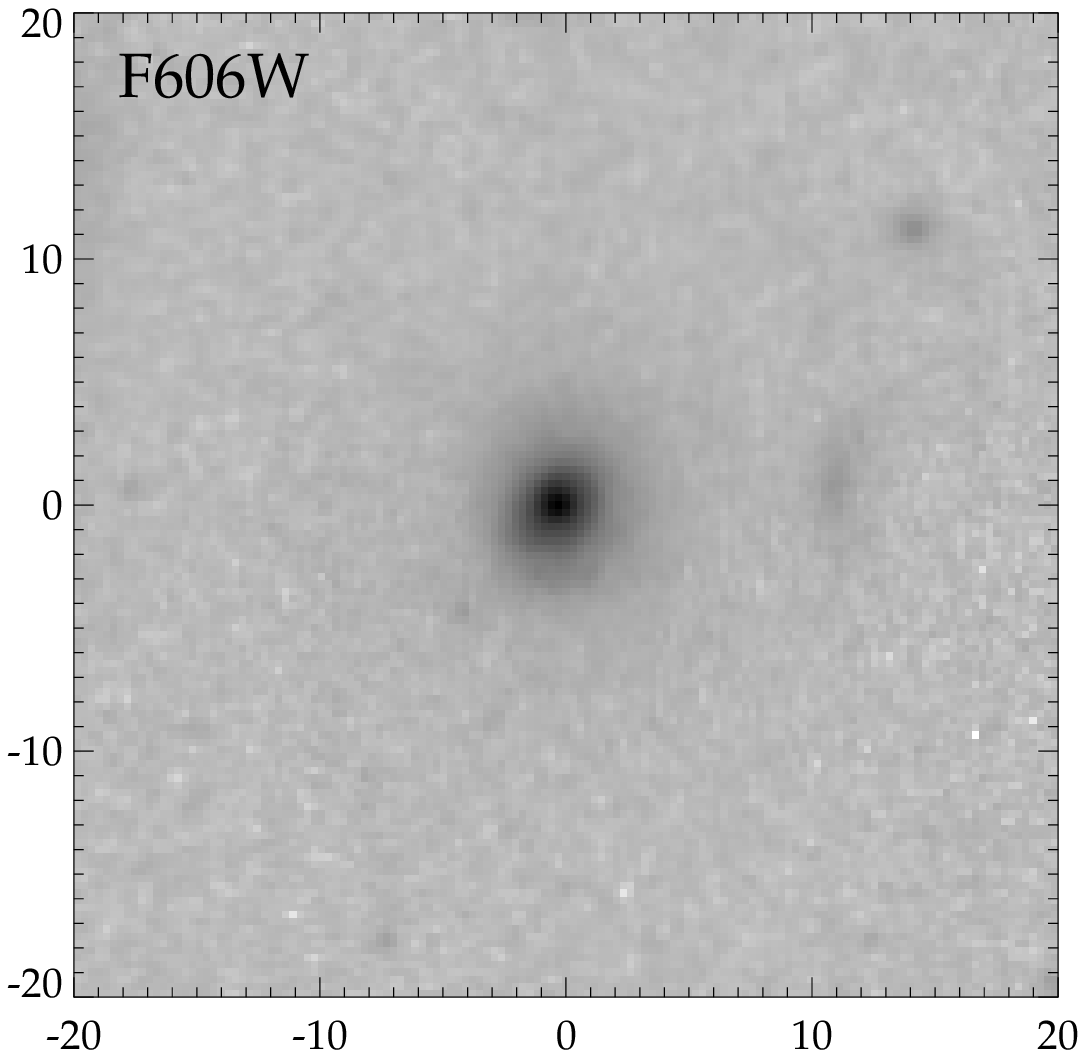}%
\includegraphics[height=4cm,clip]{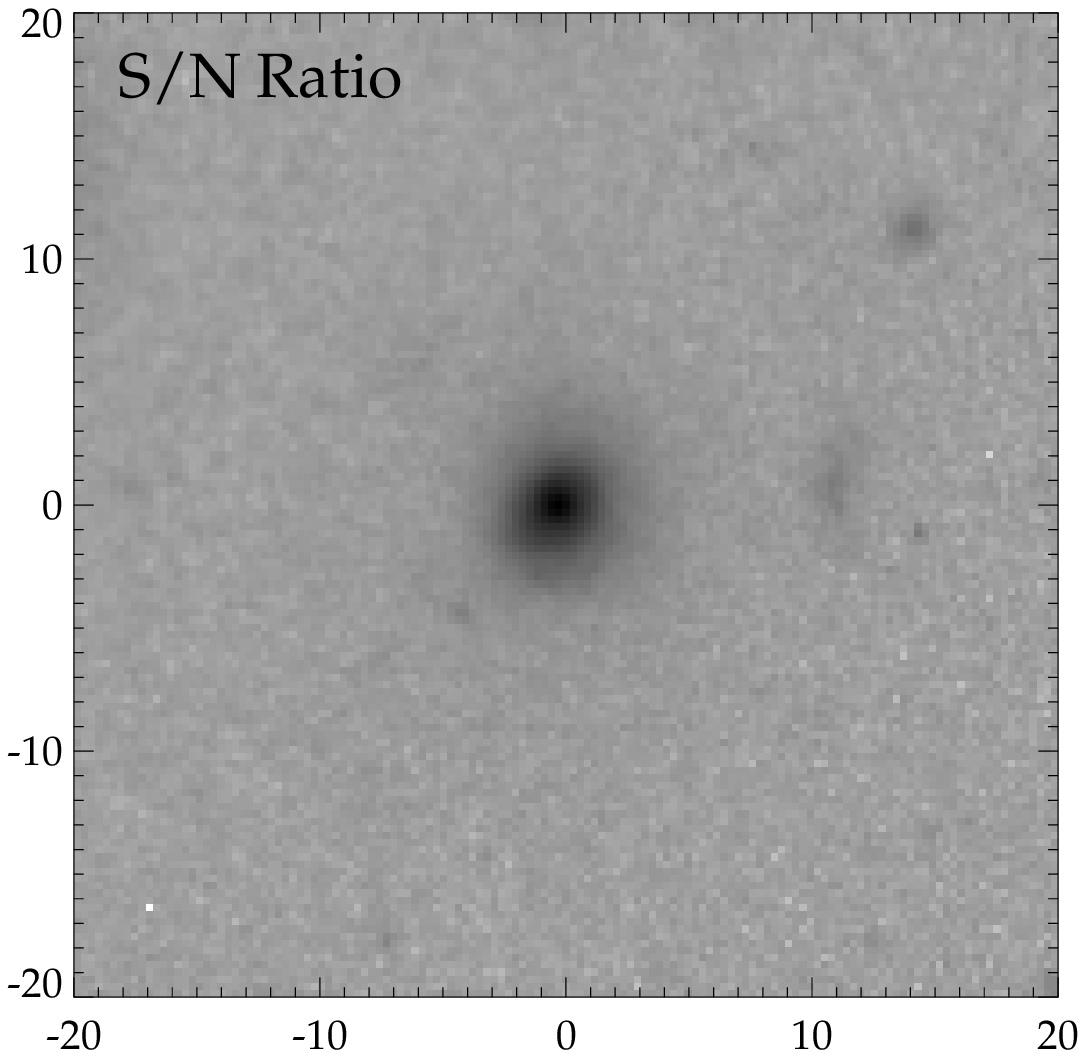}
\includegraphics[height=4cm,clip]{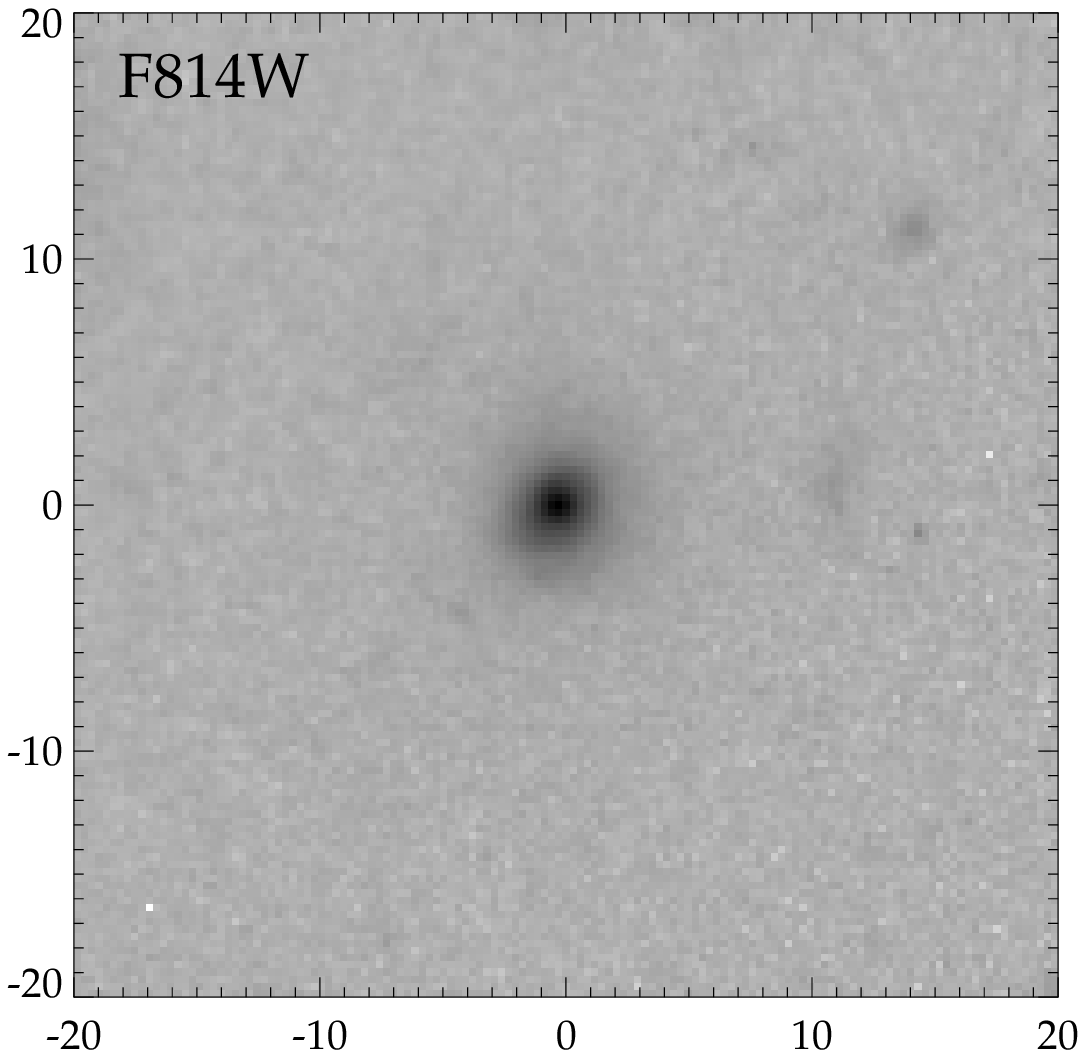}%
\includegraphics[height=4cm,clip]{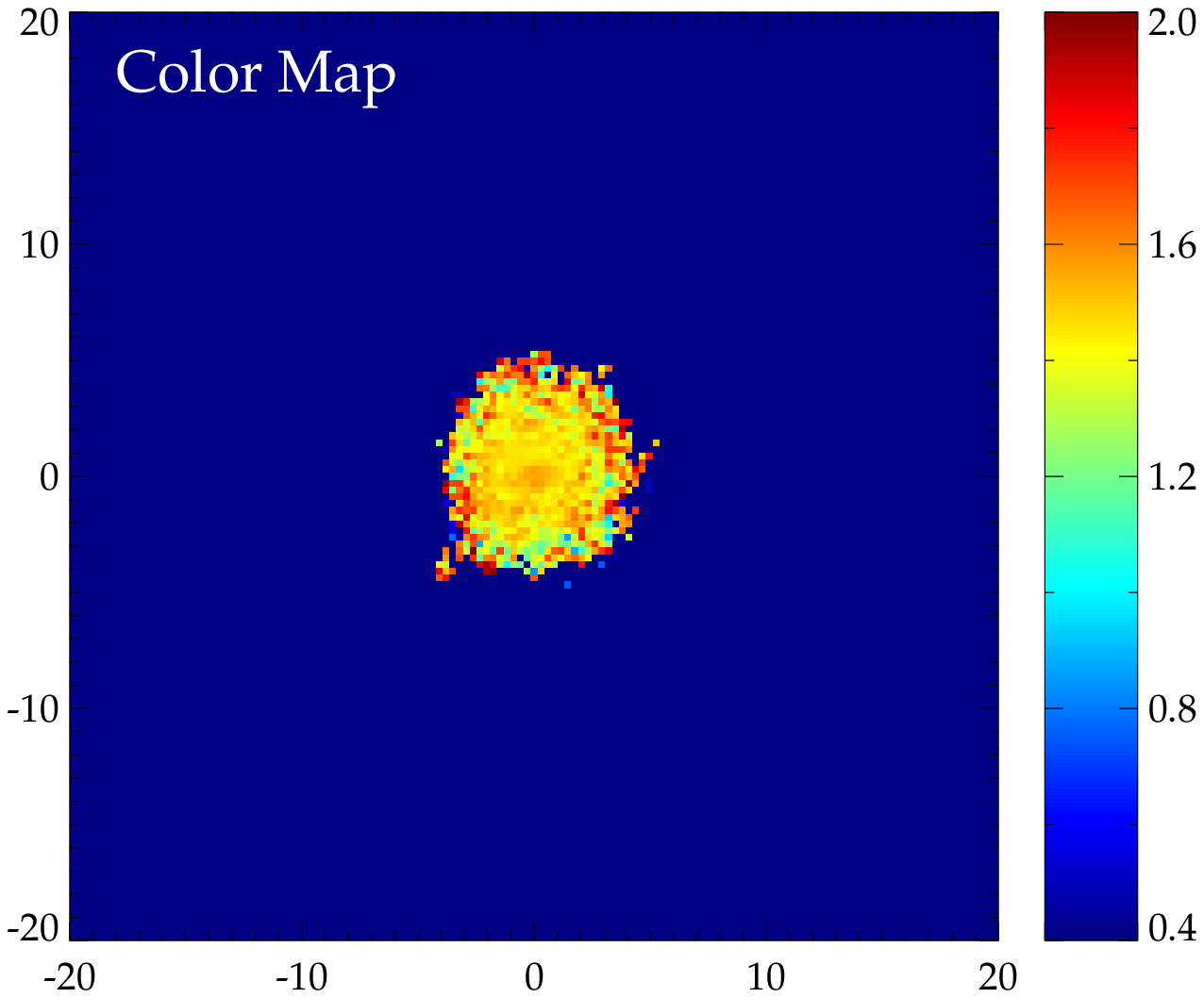}
\caption{V$_{606}$, I$_{814}$, the color map S/N ratio, as well as color map image stamps of elliptical galaxy 03.0037 at redshift 0.1730. 
[{\it See the electronic edition for a color version of this figure.}]}
\label{elliptical}
\end{figure}

\begin{figure}[]
%\centering
\includegraphics[height=4cm,clip]{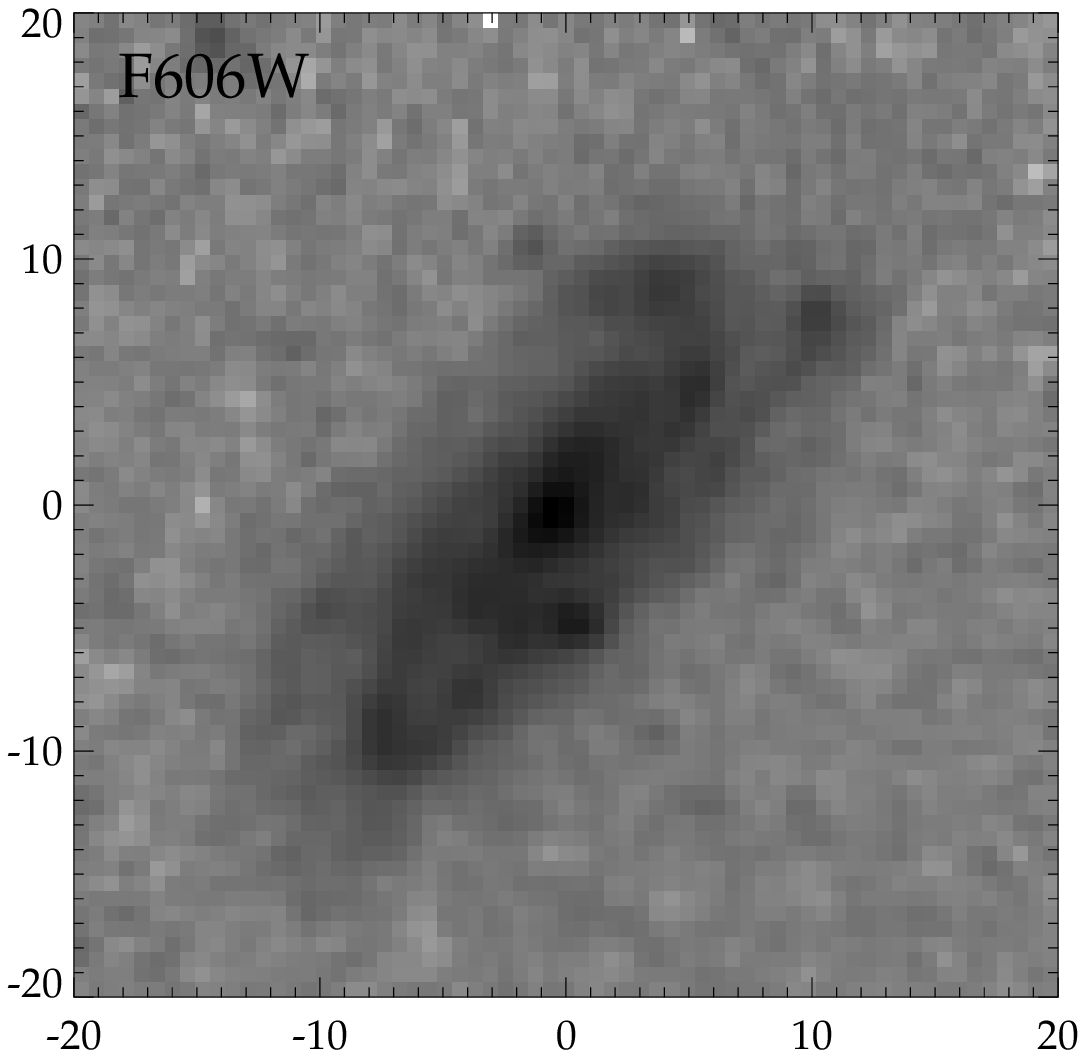}%
\includegraphics[height=4cm,clip]{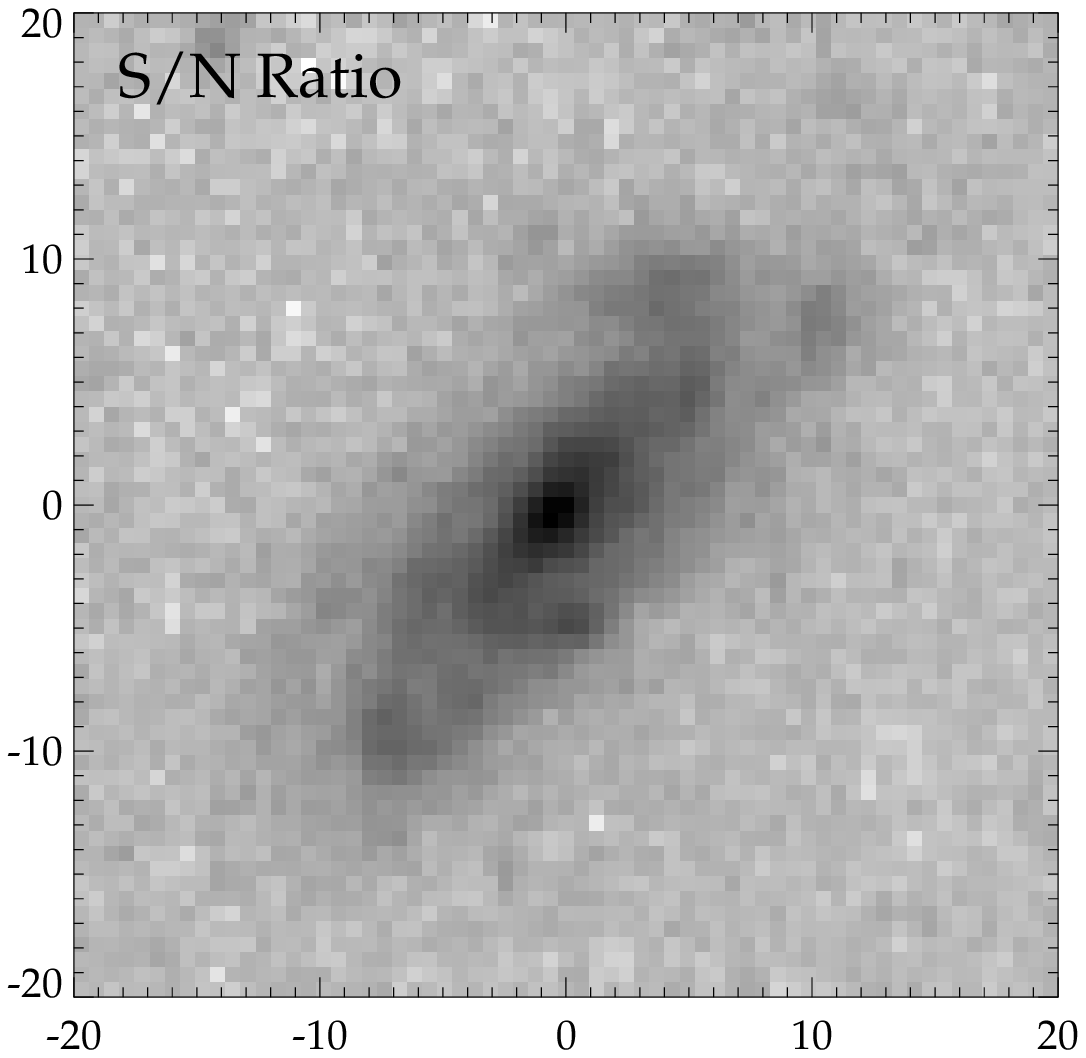}
\includegraphics[height=4cm,clip]{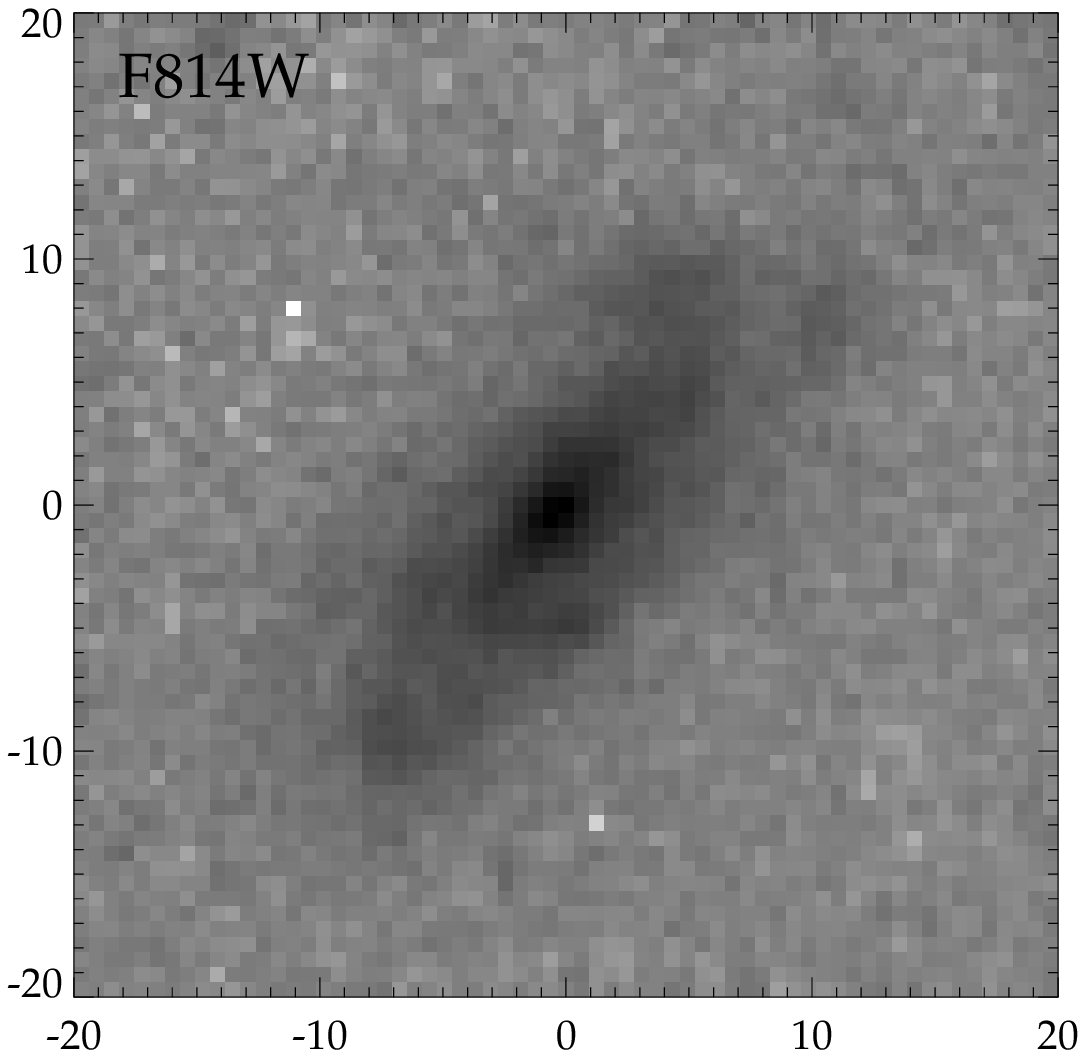}%
\includegraphics[height=4cm,clip]{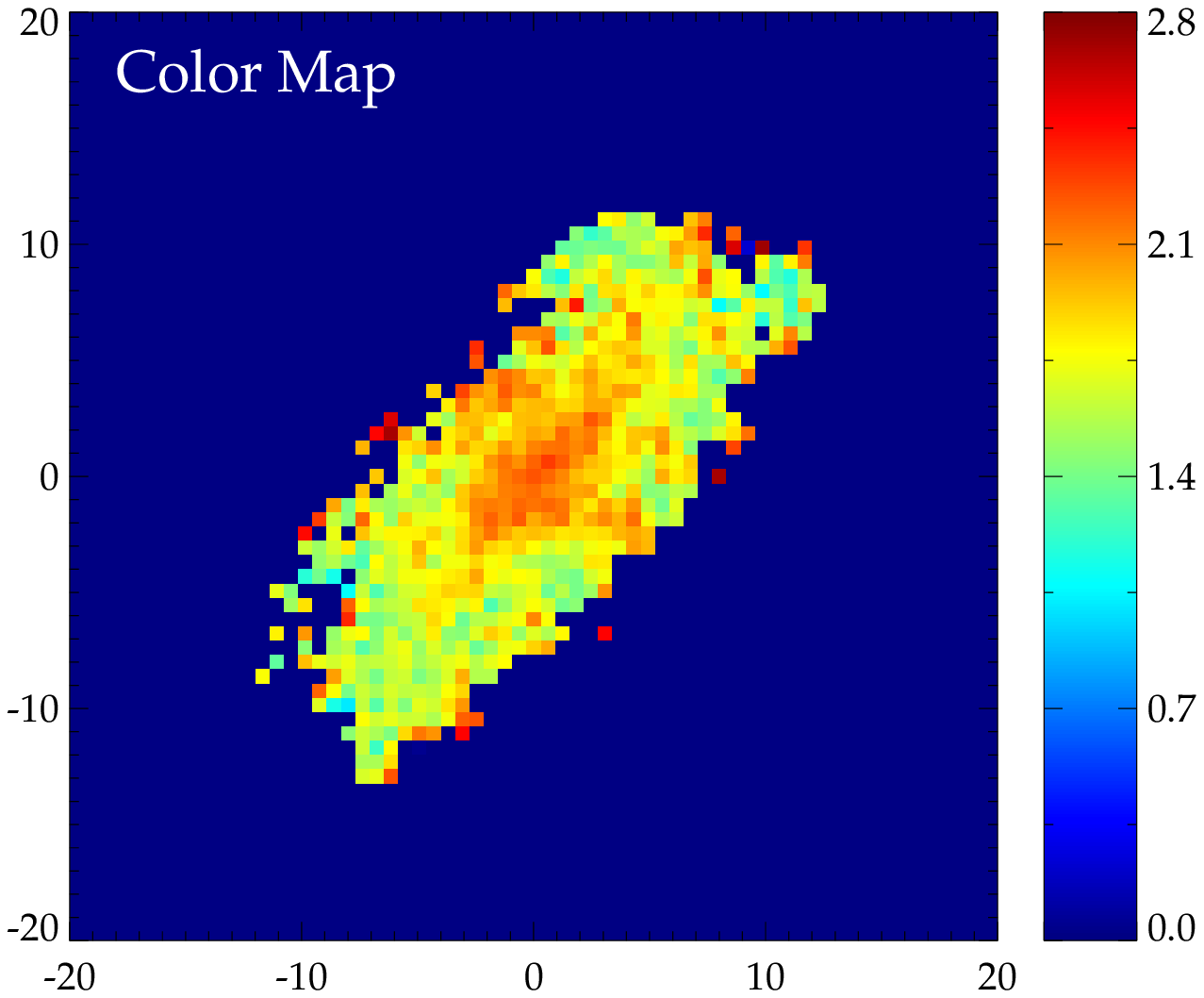}
\caption{Same as in Fig.~\ref{elliptical} for disk galaxy 03.0046 at redshift 0.5120.
[{\it See the electronic edition for a color version of this figure.}]}
\label{spiral}
\end{figure}

\begin{figure}[]
%\centering
\includegraphics[height=4cm,clip]{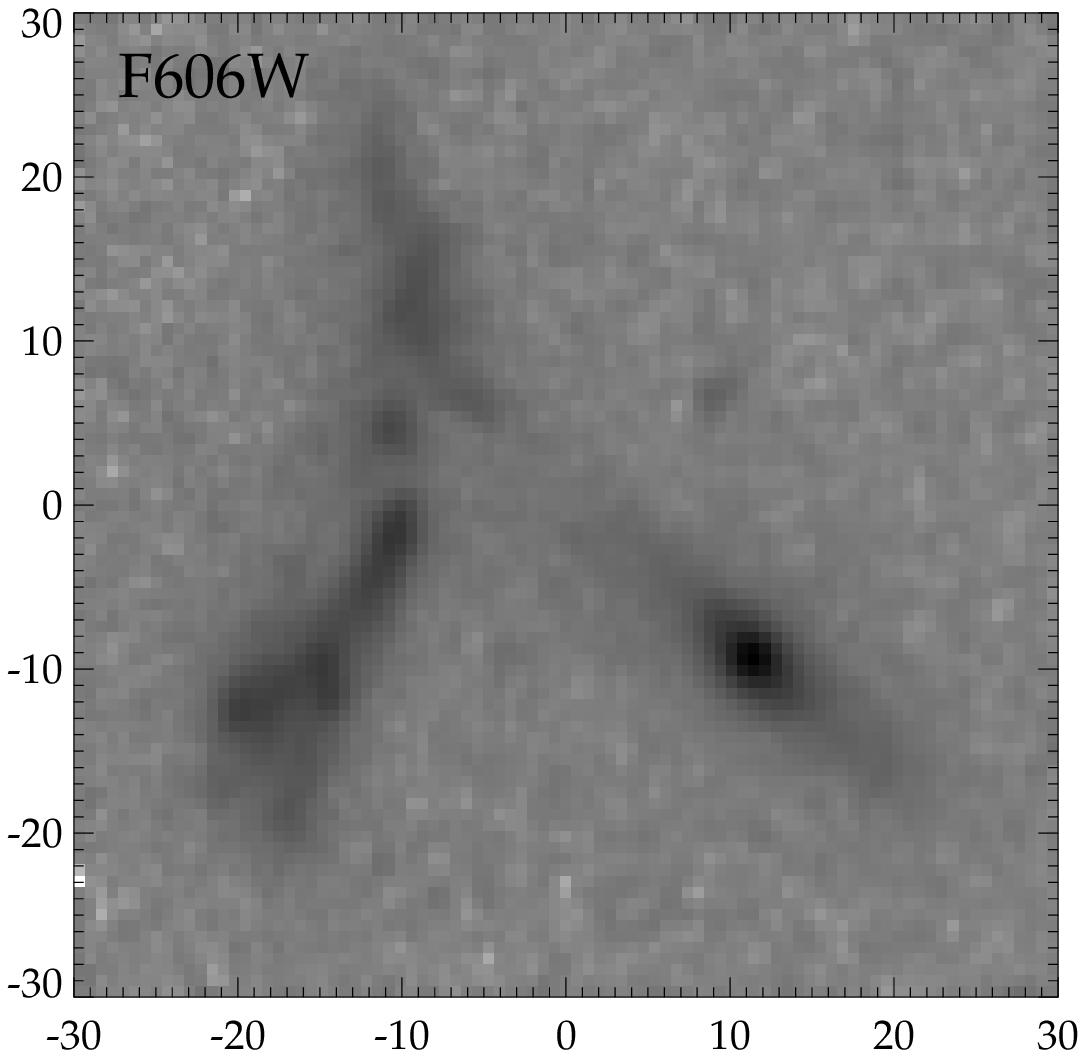}%
\includegraphics[height=4cm,clip]{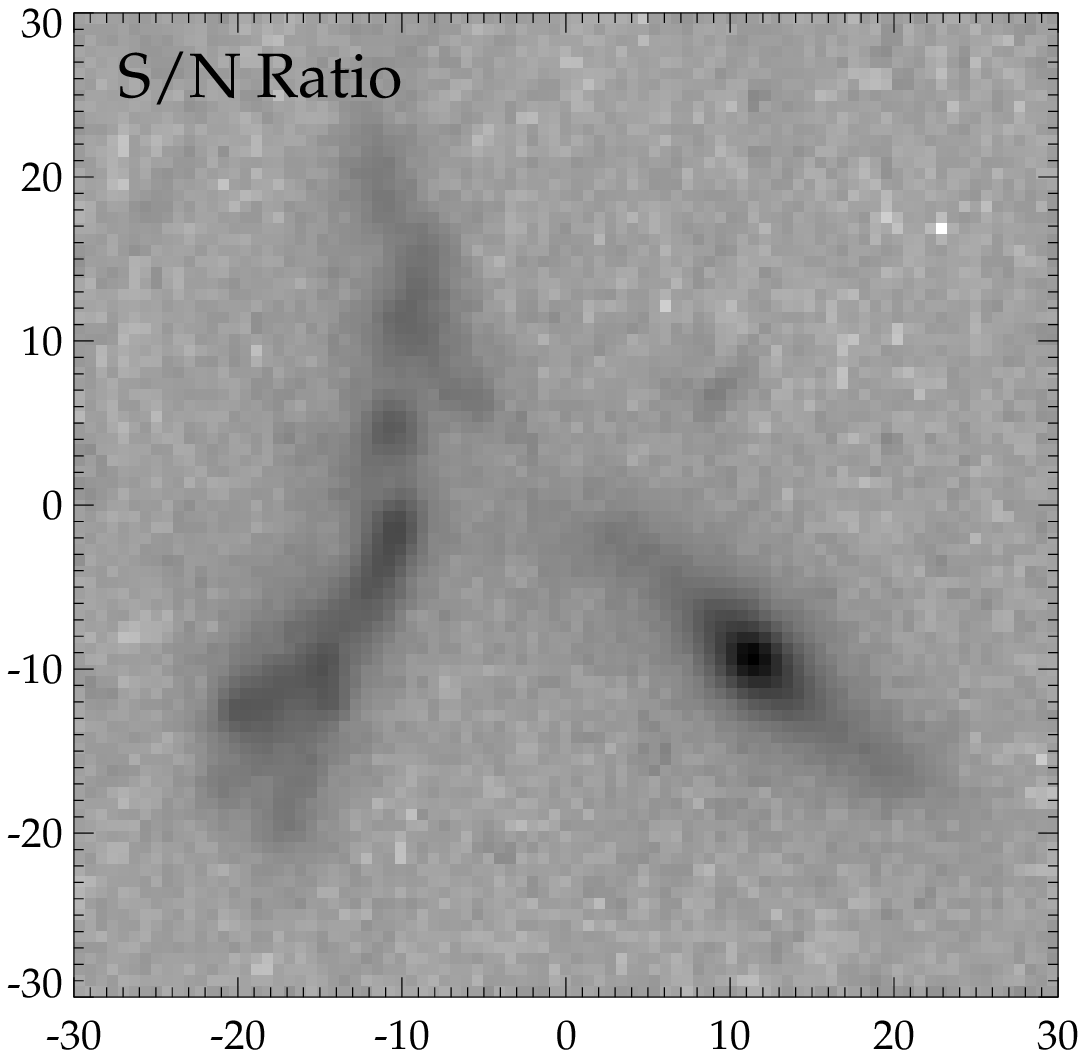}
\includegraphics[height=4cm,clip]{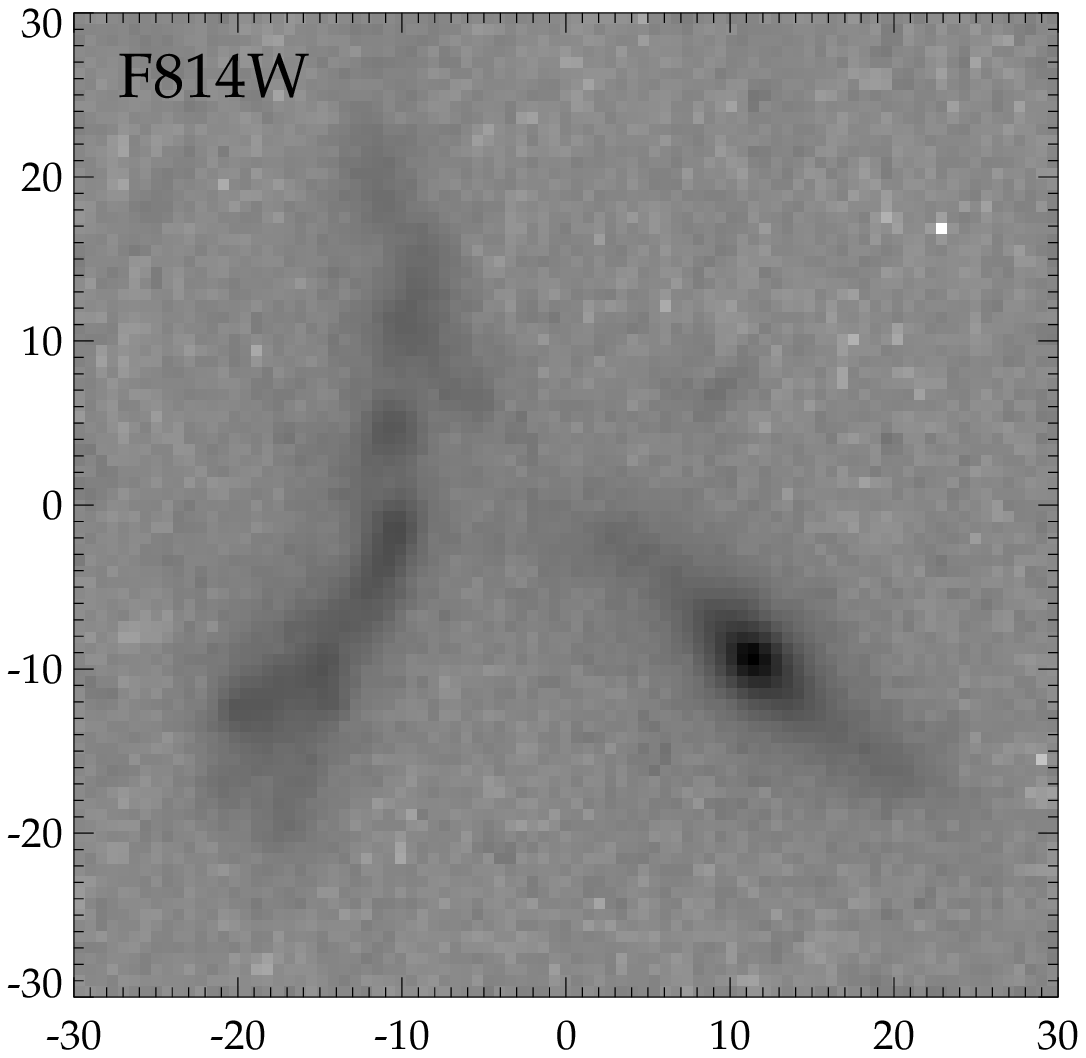}%
\includegraphics[height=4cm,clip]{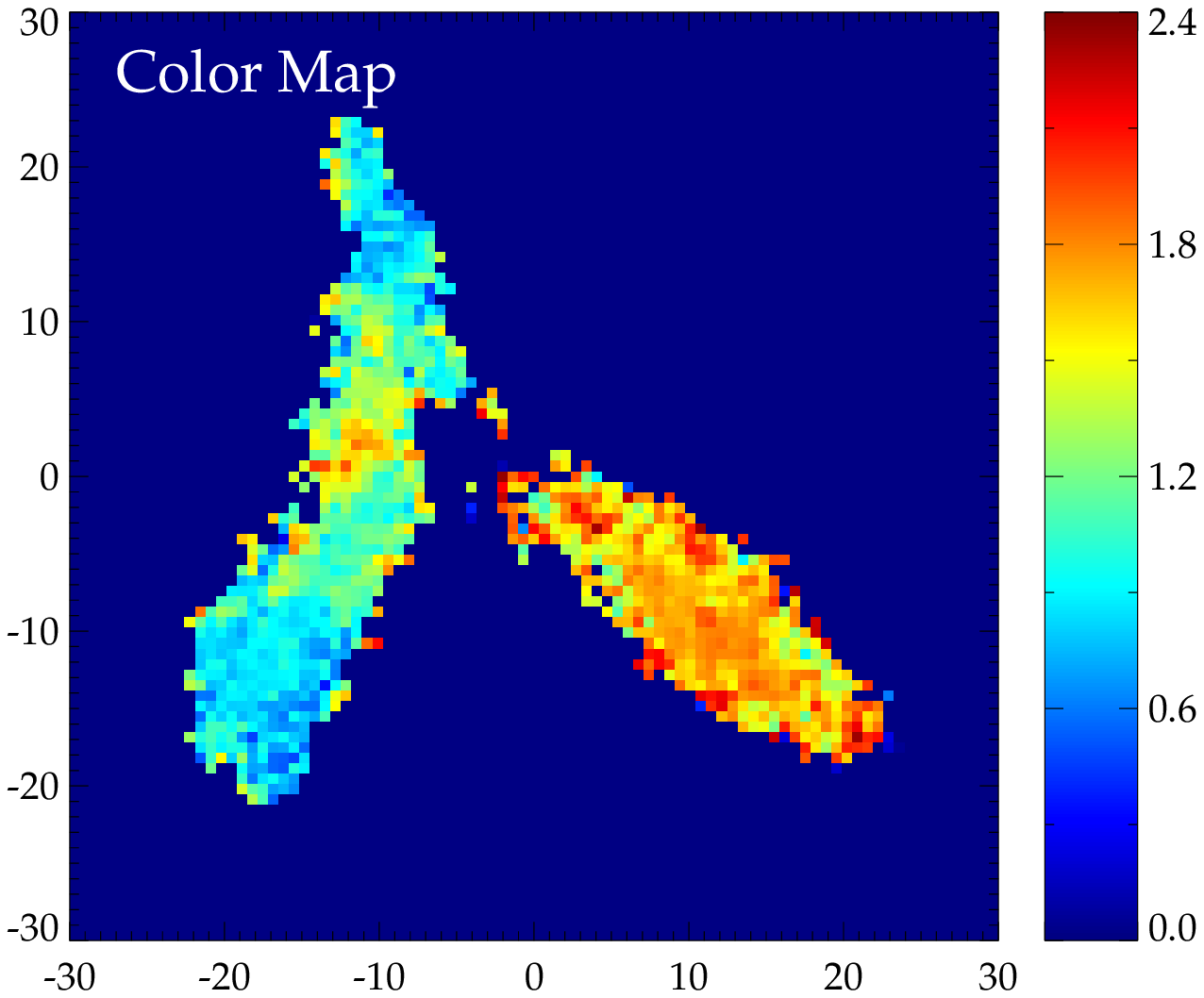}
\caption{Same as in Fig.~\ref{elliptical} for merging system 03.1309 at redshift 0.6170.
[{\it See the electronic edition for a color version of this figure.}]}
\label{merger}
\end{figure}

The S/N ratio image can give a quantitatively measurement of the 
reliability of each pixel in the color map image. The S/N ratio image does not introduce any bias against very red or very blue color regions because it accounts for both the S/N of the two WFPC2 images.
The S/N ratio image is shown with V$_{606}$ and I$_{814}$ images in three typical cases,  elliptical galaxy (Fig.~\ref{elliptical}), spiral galaxy (Fig.~\ref{spiral}) and merging system (Fig.~\ref{merger}). The resulted color map image with the S/N ratio image is displayed as well. The color bar next the color map stamp shows the color scheme in the observed frame. The color range is adjusted for a best visualization. Each stamp is labeled at its top-left corner. In Fig.~\ref{elliptical} and Fig.~\ref{spiral}, each stamp has a size of 40$\times$40\,kpc while in Fig.~ \ref{merger} the size of each stamp is 60$\times$60\,kpc.

%%%%%%%%%%%%%%%%%%%%
\section{Morphologies of LIRGs}

\subsection{LIRG sample}

The 17 HST imaging fields cover about 87  of 200 square arcminute
area of the CFRS fields 0300+00 and 1415+52. We obtained either
B$_{450}-$I$_{814}$ or V$_{606}-$I$_{814}$ color maps for 265 galaxies
(I$_{\rm AB}\,<\,22.5$). Among  them, 169 have spectroscopic redshifts
given  in the  CFRS redshift  catalog or  in the  literature.  ISOCAM
observations  detected  60  objects   and  77  objects  brighter  than
300\,$\mu$Jy at 15$\mu$m with S/N\,$>$\,5 in the CFRS fields 0300+00 and 1415+52,
respectively.   Of the  137 objects,  color  map is  available for  33
objects in the 0300+00 field and  26 objects in the 1415+52 field.  In
the   137  ISOCAM   detected  objects,   82.5\%  (113)   have  optical
counterparts  brighter than 22.5  mag and  94.2\% (129)  brighter than
23.5  mag in  I$_{\rm AB}$  band. For  objects fainter than
I$_{\rm  AB}$\,=\,22.5  mag, the  imaging sample  points are  limited for
deriving     reliable    morphological     parameters     and    color
distribution. Therefore they are not included  in this analysis.  Since the vast
majority of ISOCAM detected  objects are indeed optically bright, this
would  not cause  a significant  bias.   Note that  the fraction  of
ISOCAM  sources imaged  by  HST (59  among  113 I$_{\rm  AB}\,<\,22.5$
galaxies, 52.2\%) is  larger than the area fraction  of HST imaging of
the two CFRS fields  (87 against 200  square arcminute, 43.5\%),  because our
survey includes  complementary data aimed at  investigating ISO source
morphologies.

In the 59 ISOCAM detected galaxies, 53 objects are with known redshifts,
including 2 objects at z$>$ 1.2 which have been removed from our
sample.

\begin{figure}[]
   \centering
   \includegraphics[width=0.40\textwidth]{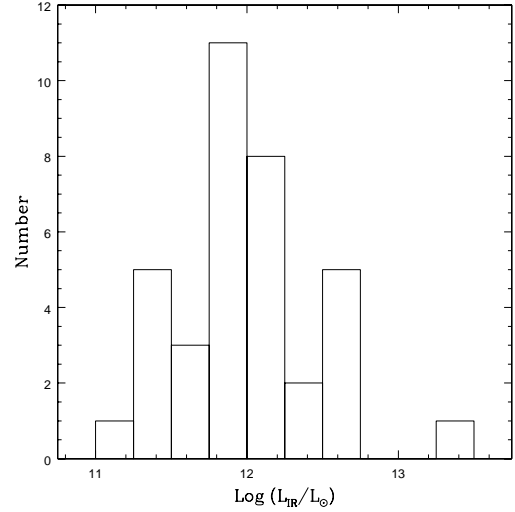}
   \caption{Histogram of the IR luminosity for 36 distant LIRGs.}
   \label{histogramLir}%
    \end{figure}

Table~\ref{morpho} lists the complete sample of ISOCAM galaxies which
have been imaged by HST in  two bands.  The objects are organized into
three groups, 6 with unknown redshift, 15 in nearby ($z\,\leq\,0.4$) and
36 in distant ($0.4\,<\,z\,<\,1.2$) universe.  Nearby galaxies are not IR
luminous (L$_{\rm IR}\sim 10^{10}$\,L$_{\odot}$)  and can not be taken
as  the local  counterparts  of  the distant  LIRGs.   The 36  distant
objects are used to reveal the morphological properties of the distant
LIRGs. In each group, the objects  are tabulated in the order of their
CFRS identifications.  In  Table~\ref{morpho}, the apparent magnitudes
$m_{450}$  (Col. 3),  $m_{606}$ (Col.  4) and  $m_{814}$ (Col.  5) are
given in  HST Vega system.  An  aperture of 3$\arcsec$  is adopted for
the HST image  photometry. Absolute B band magnitude  (Col. 6) and K
band magnitude  (Col. 7)  are provided in  AB system,  using isophotal
magnitudes    from    the    ground-based   imaging    (Lilly    et
al.~\cite{Lilly95}).   The  K-correction   is   calculated  based   on
ground-based  B,  V, I  and  K band  CFRS  photometry  (see Hammer  et
al.  \cite{Hammer01}  for details).   IR  luminosity  is tabulated  in
Col.  8.  Three models  are used  to calculate  the IR  luminosity and
derive proper uncertainties (see Flores et al.~\cite{Flores04} for
details).     Fig.~\ref{histogramLir}   shows   the    IR   luminosity
distribution for the 36 distant LIRGs.

\subsection{Morphological classification}

Based on HST imaging, two-dimensional  fitting has been carried out to
derive  structural   parameters  which   are  used  to   quantify  the
morphological   features.    With   the  structural   parameters   and
information derived  from color maps,  morphological classification is
performed by two researchers independently.

\subsubsection{Structural parameters}

The two-dimensional fitting is performed using the software GIM2D (see
Simard et al.~\cite{Simard} for  more details).  Two components, bulge
and disk,  are used  to fit the  surface brightness  distribution. The
fractions of bulge luminosity in total B/T and $\chi^2$  are listed  in  
Table~\ref{morpho} (from Col. 9 to Col.  14) .  The  parameter B/T  is
correlated with  the Hubble type  increasing for early  type galaxies.
In addition  to the ``quality'' parameter $\chi^2$,  the residual image,
which is a difference between the observed image and the modeled image, is
also used to estimate the quality  of the fitting. 
Good fitting is characterized by a $\chi^2$ close to unity and a residual image with random-distributed little residual emission. However, in the case of spiral galaxy with visible arms, the residual image exhibiting regular arms will refer to the fitting as good one even though $\chi^2$ value is biased to be different from unity due to the presence of arms.  Inclination  angle 
of the disk is also derived from the two-dimensional fitting.

\subsubsection{Color distribution properties}

With  information  provided by color  distributions,  morphological
classification can be improved dramatically. The appearance in a color
map  is free  from the  arbitrary  adjustment in  visualization of  an
image.   Physical  properties  can  be  derived  from  the  color  map
including whether  some regions are dusty or  star-forming. Apart from
the  usual  way  of  the  morphological classification  based  on  the
brightness distribution,  the color information provides  a new window
to  compare  distant  galaxies  with  local  galaxies  in  the  Hubble
sequence.

\begin{figure}[]
   \centering
   \includegraphics[width=0.45\textwidth]{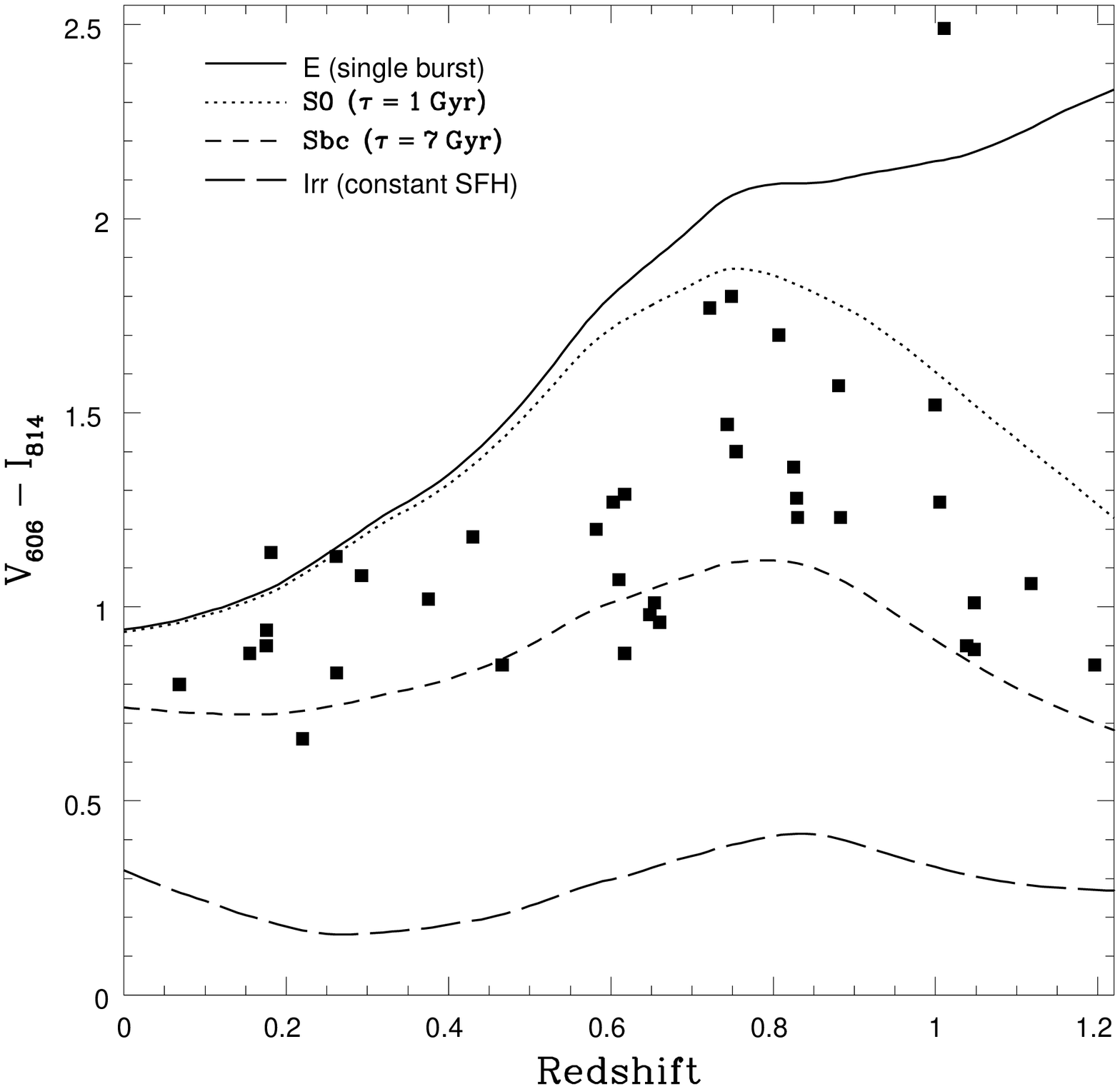}
   \includegraphics[width=0.45\textwidth]{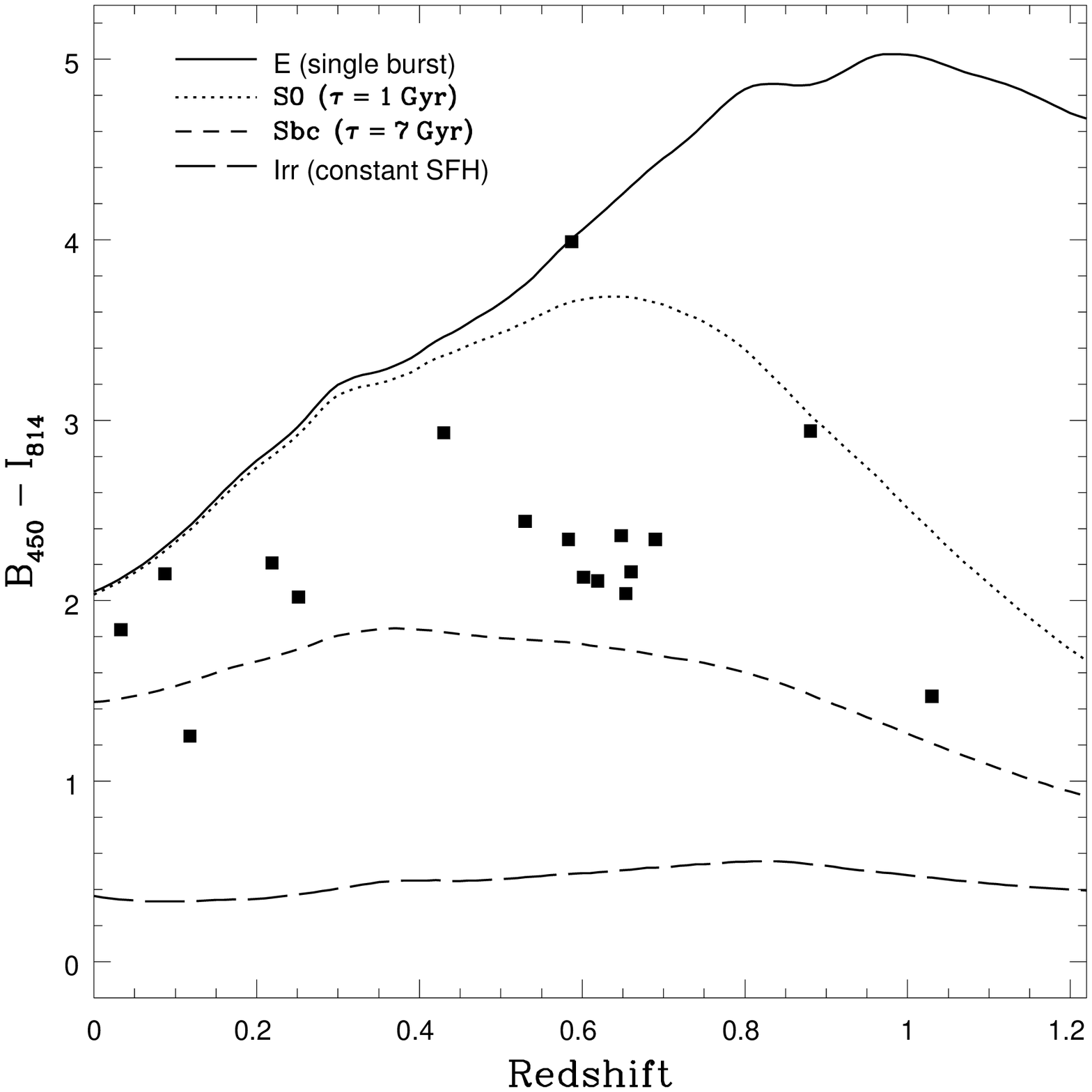}
   \caption{V$_{606}-$I$_{814}$ (top) and B$_{450}-$I$_{814}$ (bottom)
observed colors for LIRGs as  a function of redshift. Similar to fig.6
in  Menanteau  et al.  (\cite{Menanteau01}),  four solar  metallicity,
formed at $z$\,=\,5 galaxy models, including Elliptical galaxy (single
burst), S0 galaxy ($\tau\,=\,1\,$Gyr), Sbc galaxy ($\tau\,=\,7\,$Gyr),
irregular galaxy  (constant SFR) are present for  a comparison. Almost
all LIRGs are redder than Sbc galaxy.}
   \label{modelling}%
    \end{figure}

We obtain  color map for each  LIRG in the observed  frame. Instead of
applying K-correction and deriving the color in rest frame, we compare
the  observed color  with the  modeling  color.  Fig.~\ref{modelling}
illustrates the modeled color-redshift curves for V$_{606}-$I$_{814}$
and   B$_{450}-$I$_{814}$.   Using   GISSEL98   (Bruzual  \&   Charlot
\cite{Bruzual}), the observed  colors in HST Vega system  are given at
different   redshifts  for  four   galaxy  models,   corresponding  to
elliptical (single burst), S0 ($\tau=1\,$Gyr), Sbc ($\tau=7\,$Gyr) and
irregular galaxy  (constant star  formation rate with  a fixed  age of
0.06\,Gyr).  We assume  a  formation  epoch at  a  redshift of  $z=5$.
Distant  LIRGs are  compared with  the models  using  their integrated
colors.  Almost all LIRGs  are redder  than an Sbc galaxy. Using  the modeled
color-redshift  relations,  we  investigate  the  colors  of  specified
regions.  We refer to the region in outskirts as red as or redder than
an elliptical galaxy  as dusty region and the  region off the central
area bluer than an Sbc galaxy as star-forming region.

\subsubsection{Morphology labels}

By visually examining galaxy morphologies, we have tried to label each
target  with `Hubble  type'.   For  the  galaxies well  fitted  by
bulge+disk two-dimensional structure, we  divide them mainly into five
types  in terms of  the fraction  of bulge  luminosity in  total: E/S0
(0.8\,$<$\,B/T\,$\leq$\,1), S0   (0.5\,$<$\,B/T\,$\leq$\,0.8),     Sab
(0.15\,$<$B/T$\leq$\,0.5),  Sbc (0\,$<$\,B/T\,$\leq$\,0.15)   and   Sd
(B/T\,=\,0).  We also introduce three additional types to describe the
compact (C)  galaxy which is  too concentrated to  be decompositioned,
irregular  (Irr) galaxy and  `tadpole' (T)  galaxy. Quality  factor is
provided  to represent  our  confidence of  our  classification: 1  --
secure, 2 -- merely secure, 3 -- insecure and 4 -- undetermined.
We  also provide  a classification  for galaxies  which show  signs of
interacting or  merging: M1 -- obvious merging,  M2 -- possible merging,
I1 -- obvious  interaction,  I2 -- possible  interaction  and R -- 
relics  of merger/interaction.

Machine  is not  flexible  and  the experienced  eye  is essential  to
classify the  objects differing significantly from  the adopted models
in  machine.   With  information  from the  two-dimensional  structure
fitting and  the color map, the  visual examination is  carried out by
two of the authors F.H. and X.Z.Z. independently, aimed to reduce some
arbitrariness   from  different   viewers.  The   classifications  are
consistent with each other for most objects. After fully discussing on
a  few objects,  a  final morphological  classification  for the  LIRG
sample is  available.  In  Table~\ref{morpho}, galaxy type  (Col. 15),
quality factor  (Col. 16) and  Interaction/merging type (Col.  17) are
listed.

\subsection{Individual description}

In Fig.~\ref{colormap}, color  map stamps (right) of the  36 LIRGs are
shown, along with I$_{814}$  imaging negative greylevel stamps (left).
The stamps  are given  in an order of  CFRS identification (from  left to
right, top  to bottom). Color  bar in each  color map stamp  shows the
color range.   The same color  ranges over  $-$1 to 3  and 0 to  4 are
applied to all objects in V$_{606}-$I$_{814}$ and B$_{450}-$I$_{814}$,
respectively.  For the 36 LIRGs, the morphologies as well as the color
distributions  are remarkably  different. A description of each target
is present in turn.

\begin{description}

\item[{\bf  03.0035}] This  galaxy has  a morphology  of a  bright bulge
surrounded by diffuse emission.   Spiral arm-like structures, or tidal
tails are perceptible in the  diffuse region. Color map shows that the
bulge has a  color comparable to that of an  S0 galaxy.  Dusty regions
and star-forming regions are revealed in the color map.
\item[{\bf  03.0062}]  Its clear  and  regular  spiral  arms  surrounding  a
spheroid  core as well as its round appearance  suggest that this is a face-on
spiral galaxy. Color map confirms the red central region to be a bulge
with color  close to that of  an S0 galaxy.   Star-forming regions are
distributed along the arms.
\item[{\bf    03.0085}]    This   is    an    edge-on   spiral    galaxy
($i\sim$75$\degr$).  This  galaxy is imaged  close to the  chip border
and the  light distribution is not completely  recorded.  However, the
light distribution can still reveal  that the central bright region is
not a  bulge. Color map  exhibits symmetric arc structures  with color
close to that of an irregular galaxy, suggesting star formation in the disk.
\item[{\bf 03.0445}]  WFPC2 image and  color map show a spheroidal
component  in the  center surrounded  by blue  diffuse  emission.  Arm
structures  are visible  in the  surrounding regions.  This  galaxy is
classified as an Sbc galaxy.  The arms are asymmetric and it is not clear
whether  such asymmetric  arms  are related  to an  interaction/merger
event.
\item[{\bf 03.0507}]  This is a face-on  spiral galaxy with  a red bulge
surrounded by  blue diffuse emission. Arm  structures and star-forming
knots  can be seen  around the  red bulge.  A striking  long filament,
different from a spiral arm, could be associated with a stripe remnant
of an infalling dwarf galaxy.
\item[{\bf 03.0523}] This is a ``tadpole'' galaxy. It is widely believed
that  such ``tadpole''  feature is  an  indicator of  major merger  in
advanced phase.  The tidal tail  is clearly seen in WFPC2 image. Color
map exhibits a  conspicuous color gradient that the  central region is
bluer  than the  outer  region.  The  concentrated light  distribution
suggests this object to be a compact galaxy.
\item[{\bf 03.0533}] The morphology of  this galaxy is complex and shows
multiple  components. Clumpy structure  suggests an  irregular classification.
It  is however possible  that this  galaxy would  be a  nearly edge-on
spiral galaxy, with a  long tidal tail  roughly parallel to
the disk.  If this is the case,  the tidal tail is bluer than
the disk.  The central region of  this spiral galaxy is dusty and some
star-forming regions  are detected along  the arms and the  long tidal
tail.
\item[{\bf  03.0570}] This  is a  compact galaxy  with  integrated color
close to that of an Sbc  galaxy. 
The light distribution is  dominated by a bright
and compact nuclear component.
\item[{\bf 03.0603}]  An extremely blue  core dominates this  galaxy and
red  diffuse emission  extends about  10\,kpc. The  blue core  is even
bluer   than   an    irregular   galaxy.    Broad   Mg\,{\footnotesize
II}\,$\lambda$2799  line is detected with CFHT  spectroscopy. This  galaxy is
classified as a compact galaxy.  However, the  faint surrounding region
appears  round. This  object could  be  a galaxy  with very  faint
face-on disk.
\item[{\bf 03.0615}]  This galaxy is  compact and with round  and smooth
morphology. The global color is  comparable to that of a spiral galaxy
earlier than Sbc type.  A  star-forming region is visible in color map
5\,kpc off from the nucleus.
\item[{\bf  03.0776}] The  morphology of  this galaxy  is  elongated and
compact.  Distorted  fuzzy emission is  detected. Color map  reveals a
relatively red core surrounded  by blue regions.  The integrated color
of this galaxy is close to that of a Sbc galaxy.
\item[{\bf  03.0916}] This  is a  compact galaxy. The light distribution 
is too compact to obtain the structural parameters.  Broad
emission lines detected in CFHT spectroscopy suggests that this galaxy as a
type I AGN.
\item[{\bf 03.0932}] From WFPC2 I$_{814}$ image, we can see that this is
an extremely edge-on  galaxy with a bulge in  the center. However, the
bulge is not detected in B$_{450}$ band image.
\item[{\bf  03.1309}] This  chain galaxy  is a  merging system.  In this
system, the edge-on disk galaxy  has a dominant bulge with color
close  to that of an elliptical galaxy.  Along  the  chain,  star-formation
regions are visible.
\item[{\bf 03.1349}]  Regular appearance and a dominant  bulge suggest a
Sab  classification  for  this   galaxy.  The  bulge  is  compact  and
relatively  blue,  comparable to  the  disk.   Our spectroscopy confirms
that the companion galaxy 20\,kpc off is an interacting system.
.
\item[{\bf 03.1522}]  This is an edge-on  spiral galaxy. It  is not
detected in deep WFPC2 B$_{450}$ band (6600s exposure time) implying a
heavy extinction (dust screen).   CFRS ground-based photometry reveals
that this object is 0.3 mag redder than an elliptical galaxy at the same 
redshift.
\item[{\bf  03.1540}] This is  a compact  galaxy. Luminous  component is
surrounded by  fuzzy extended region (a disk?).  The compact component
looks like  a giant bar. Color map reveals a distinct  blue core
 surrounded by  dusty regions.   Off  the blue  core, a  blue
region appears at  one end of the bar. 
The  strong IR emission should relate to  the central star-forming regions.  
While  the starburst can
be triggered due to the instability of a bar, it is also conceivable
to ascribe the distinctive morphology to merger/interaction.
\item[{\bf 03.9003}] The brightness distribution of this galaxy is quite
complex and irregular.  From color  map, it can be seen that blue
star-forming regions surround a red central region (a bulge?).
\item[{\bf 14.0302}] This  galaxy is a face-on spiral  galaxy with clear
arms  and bulge/bar  detected  in WFPC2  image. In color map, 
star-forming  regions are revealed surrounding the  center and  the  arms.
\item[{\bf  14.0393}] This  is a  face-on spiral  galaxy with  wide star
formation spread over the arms.  This galaxy is very extended and with
a small bulge  which is distinctive in color map with  a color 1.5 mag
redder than that of the star-forming regions on the arms.
\item[{\bf  14.0446}] This galaxy  is a nearly edge-on ($i\sim  80\degr$) 
pure disk  galaxy. Color  map confirms  that  no  such  color  red central
region  similar  to a  bulge is detected. This disk galaxy is rather giant
characterized by a disk scale length of 6.8\,kpc in I$_{814}$ band. The mean
V$_{606}-$I$_{814}$  color is  close to  that  of an Sbc  galaxy at  that
redshift.
\item[{\bf  14.0547}]  Heavily  distorted  morphology showing  two  main
components and tidal tail features suggests that this is a pair system
undergoing a major merger episode. One component is elongated and blue,
while the other is relatively red.
\item[{\bf 14.0600}] This galaxy is a system with complex morphology and
clumpy light  distribution.  The clumpy  knots are bluer than  an Sbc
galaxy, and are associated with star formation. While it is classified
as an irregular galaxy, this galaxy is probably undergoing a minor merger
event.
\item[{\bf  14.0663}] This galaxy shows a round appearance.  The
central region exhibits a peanut-like structure, which could be linked
to a giant bar. The  color distribution shows blue regions surrounding
both ends of the ``peanut''.
\item[{\bf 14.0711}] The complex,  clumpy structure seen in WFPC2 image
suggests a classification of irregular galaxy.  Dust regions and  
star-forming regions are revealed in color map.
\item[{\bf  14.0725}] This  galaxy has elongated  morphology  and shows a
color  gradient. Color map shows two conjunctive regions. A  large  
region has  a  color close  to  that of  an elliptical  galaxy  
while  a small  region  is as  blue  as  an Sbc galaxy. It is not clear 
whether the elongated morphology is due to two
merging galactic  nuclei since  no tidal tail  features appear  in the
surrounded diffuse emission.
\item[{\bf 14.0814}] This  is a system with S-shape structure. This
galaxy could be a barred spiral  rather than a major merger because it
shows  no  evidence for  two  components.   While  it is  labeled  as
an irregular  galaxy, it  could  be  associated with  the  relics of
merger/interaction.
\item[{\bf 14.0998}] This system  has a peculiar morphology, composed of
two  very close  components surrounded  by diffuse  emission. Peculiar
substructures  like two  tidal  tails  can still  be  seen. The  minor
component is  bluer than  the major one by about 1 magnitude.  The color
distribution  of this  galaxy  is quite  complex. Broad dusty regions  and
star-forming regions can be viewed in color map.
\item[{\bf 14.1042}] Round, smooth  and elongated morphology can be seen
from WFPC2  image.  This galaxy is as red  as an elliptical galaxy.
while its brightness  distribution is  more concentrated. 
\item[{\bf 14.1129}]  This system consists of  two comparable components
separated  by  12\,kpc.   Strikingly,  a  distinct tidal  tail  and  a
bridge connecting the two components suggest that this system
is undergoing  a major merger episode.   The tidal tail  is bluer than
the two galactic nuclei.
\item[{\bf  14.1139}] From WFPC2  image, a  peculiar appearance  showing
the merging of two galaxies can be  seen.  The two galactic nuclei are as
  close   as   3.5\,kpc,    surrounded   by   diffuse   and   extended
  emission.  Substructures  like  tidal   tails  are  visible  in  the
  surrounding regions.  The major component has a blue core.
\item[{\bf 14.1157}]  This is a system composed  of multiple components.
Material streams connecting the components are detected. The brightest
component is very red.  This object  is detected by PC chip in the GSS
very  deep field.   The higher  resolution color  map reveals  a ring
structure  associated  with a  point  source  in  a very  red  region,
suggesting a heavily obscured AGN inside it.
\item[{\bf  14.1190}]  The  morphology  of  this  galaxy  is  round  and
clumpy. Spiral arms are visible  around the center with several bright
blue knots on them.  There is no dominant central components and this
galaxy is classified as an Sd galaxy.
\item[{\bf 14.1305}]  This system  is a merging system showing a dominant 
component and a conjoint one.
Seen in the color  map, the  dominant  component is  even
redder  than  an  elliptical galaxy,  suggesting  that  it  is  very
dusty.  Within 20\,kpc,  there  are two  other  components with  color
comparable to that of an Sbc galaxy. It is unclear whether they belong
to the same physical system.
\item[{\bf 14.1350}]  This galaxy is classified as an irregular galaxy due
to a  complex morphology and irregular light  distribution. The bright
components/knots are as  blue as an Sbc galaxy and  the central region is
very red (dust?).
\item[{\bf  14.1415}] Dominated  by a  major component,  this  galaxy is
classified  as a compact  galaxy.  The  central dominant  component is
surrounded by diffuse emission. Four minor components reside in the
surrounding emission,  which suggests that  this galaxy is  possibly the
relics of a merger event.  The main component is dusty.
\end{description}

\begin{figure*}[] \centering
\includegraphics[height=0.21\textwidth,clip]{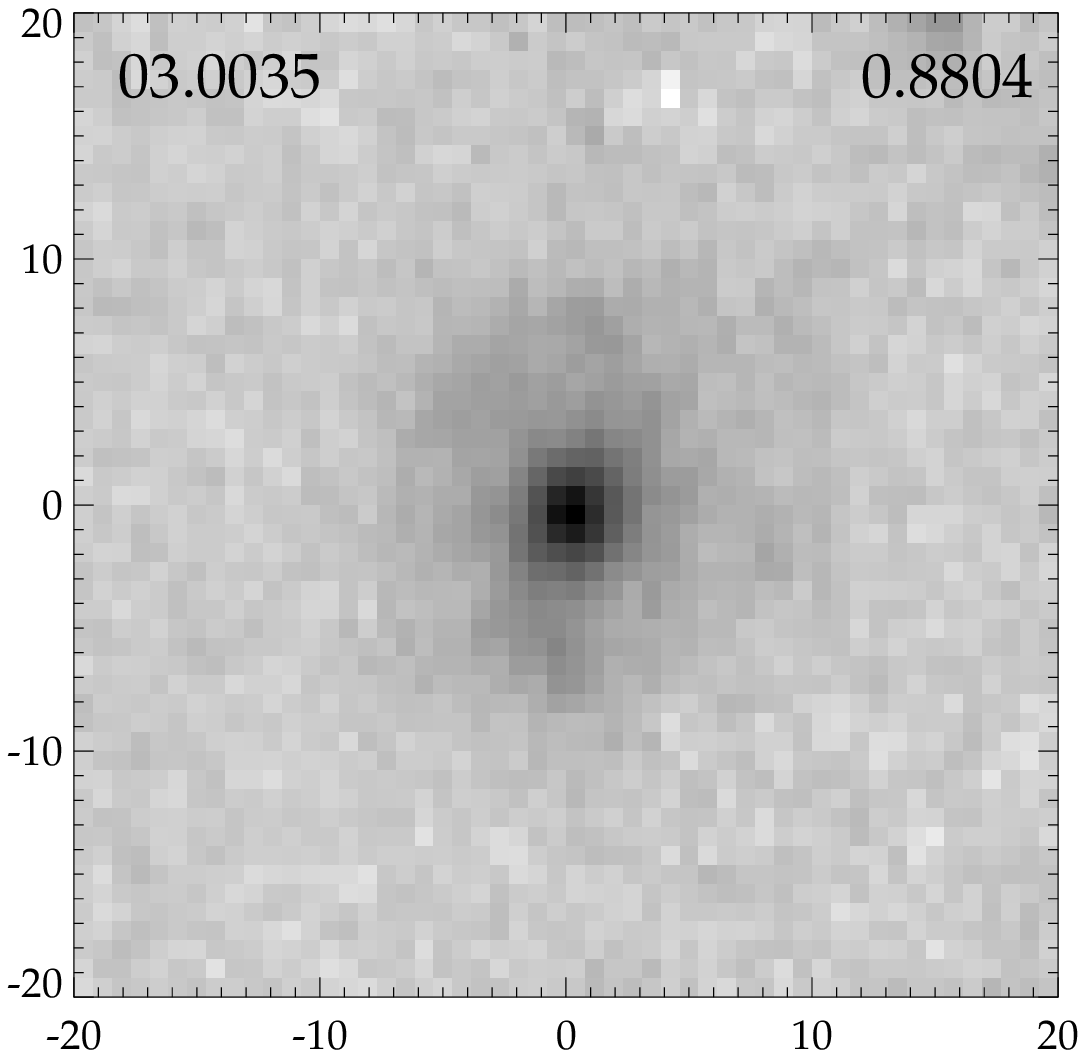}    \includegraphics[height=0.21\textwidth,clip]{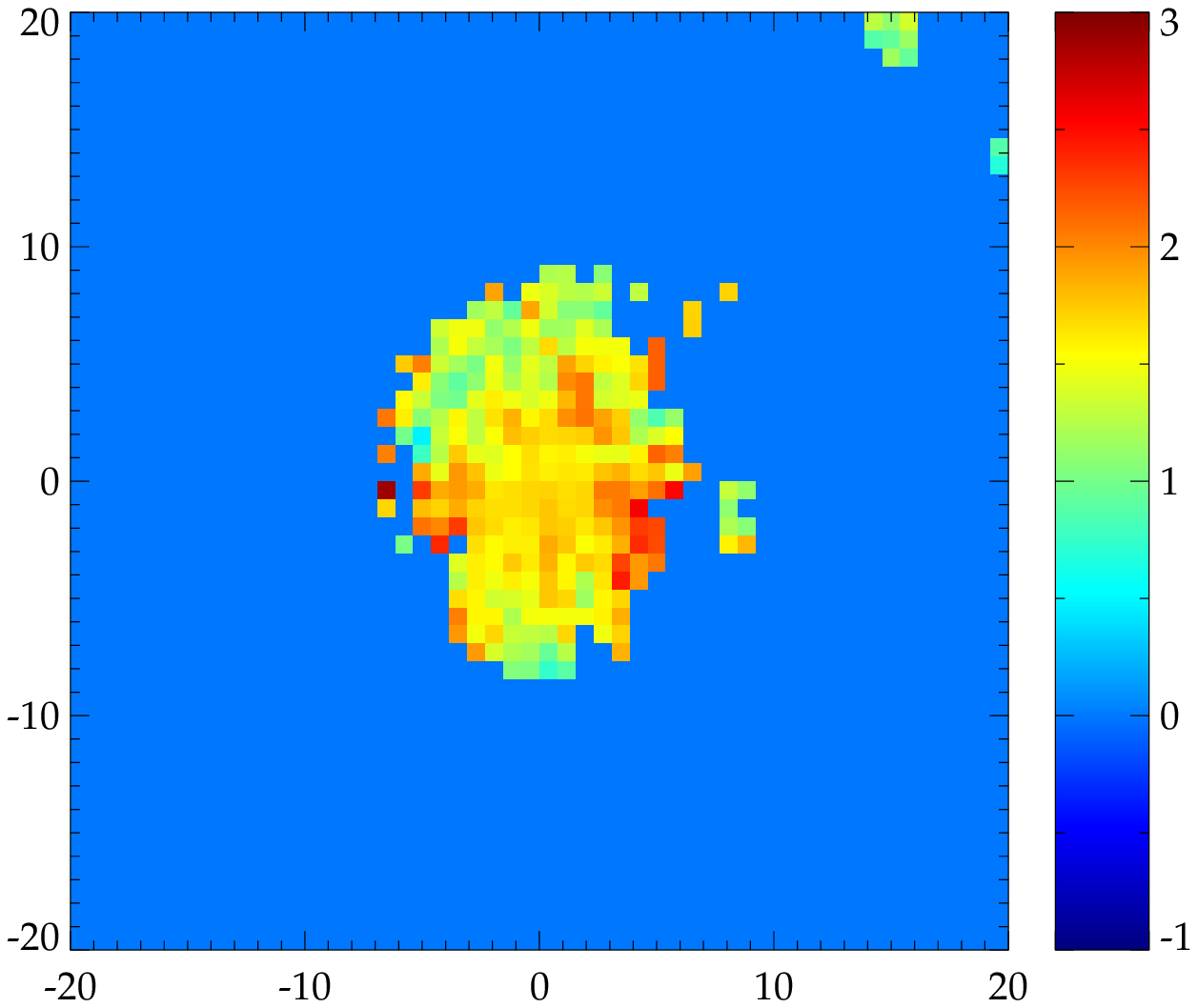} 
\includegraphics[height=0.21\textwidth,clip]{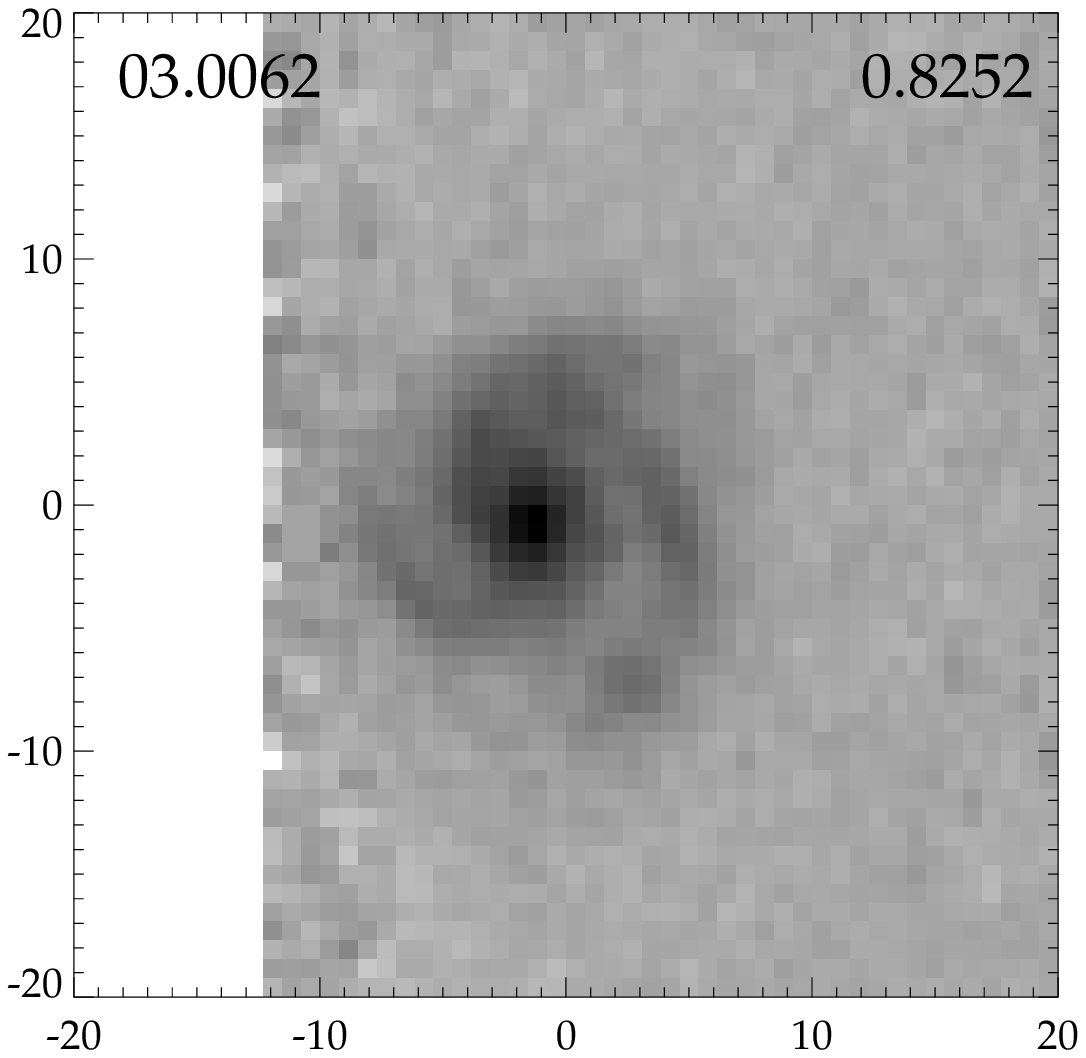}    \includegraphics[height=0.21\textwidth,clip]{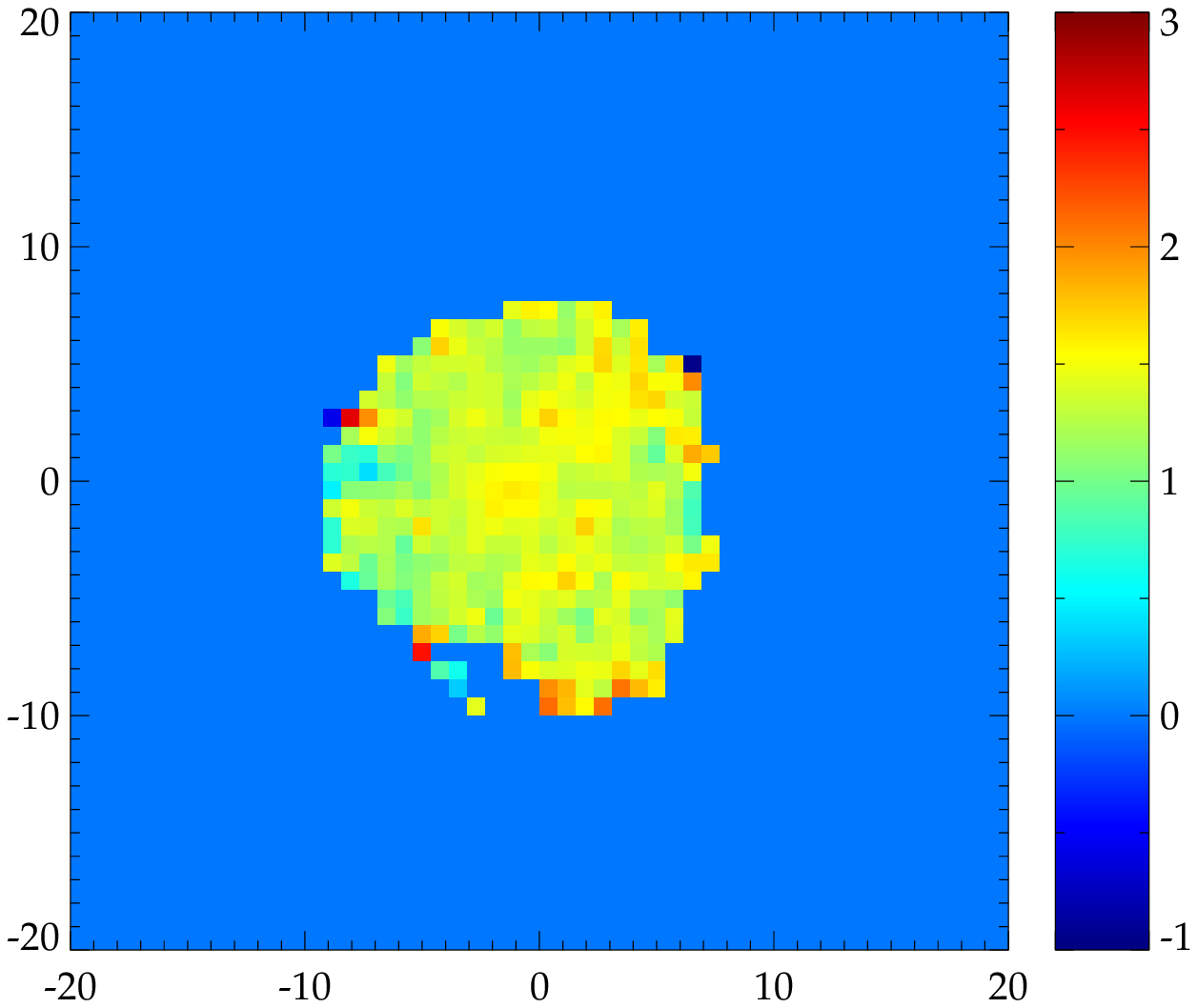}
\includegraphics[height=0.21\textwidth,clip]{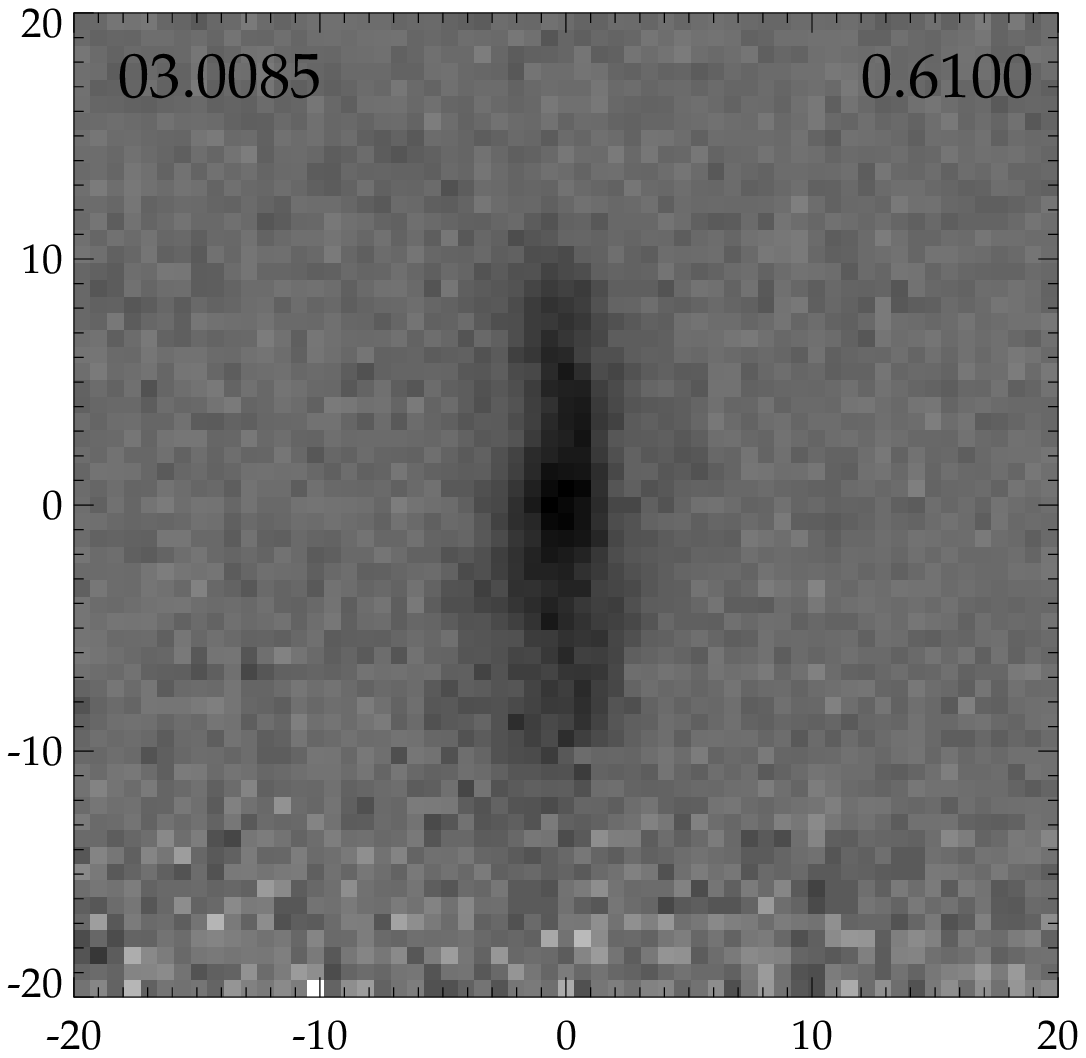}    \includegraphics[height=0.21\textwidth,clip]{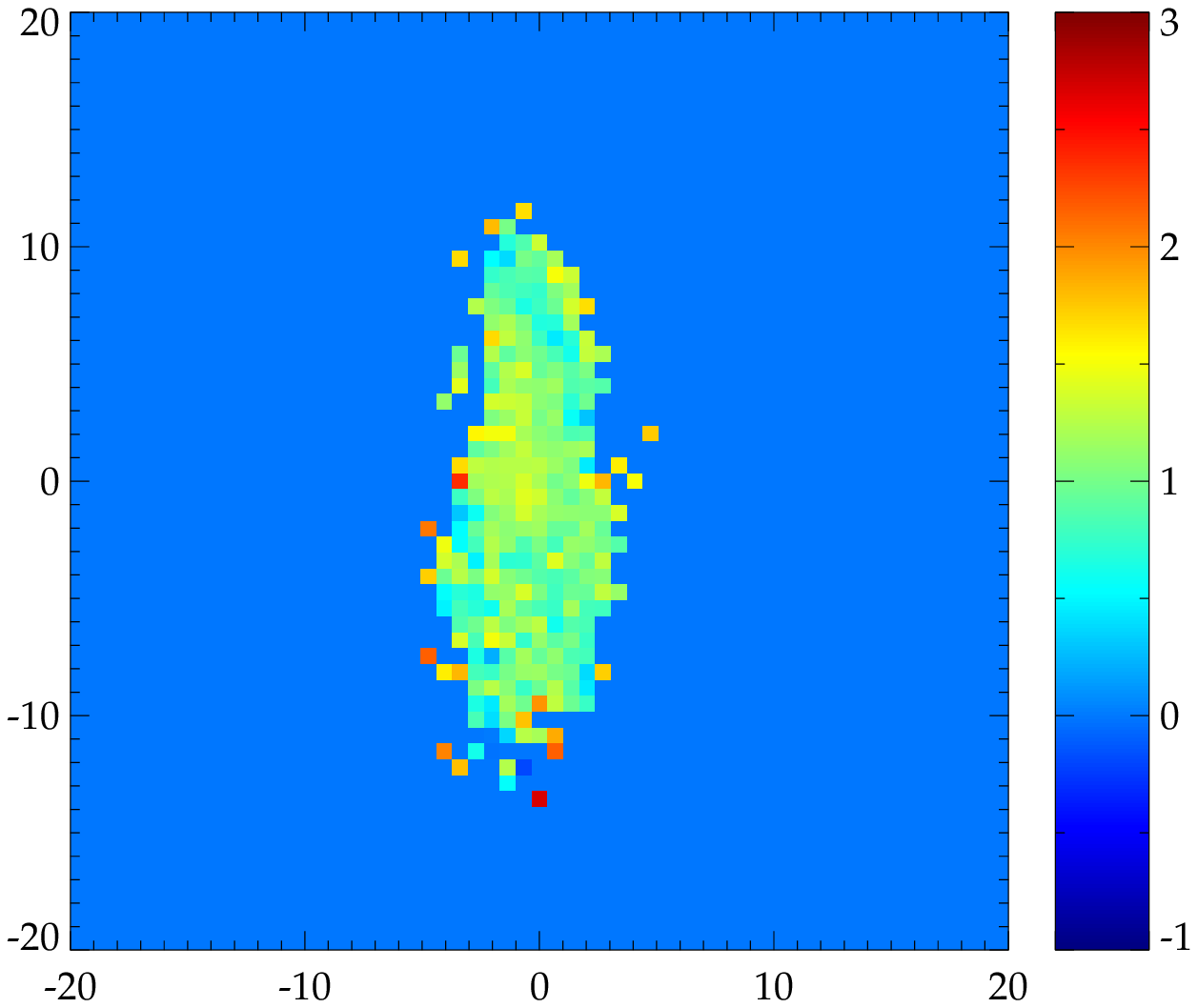}
\includegraphics[height=0.21\textwidth,clip]{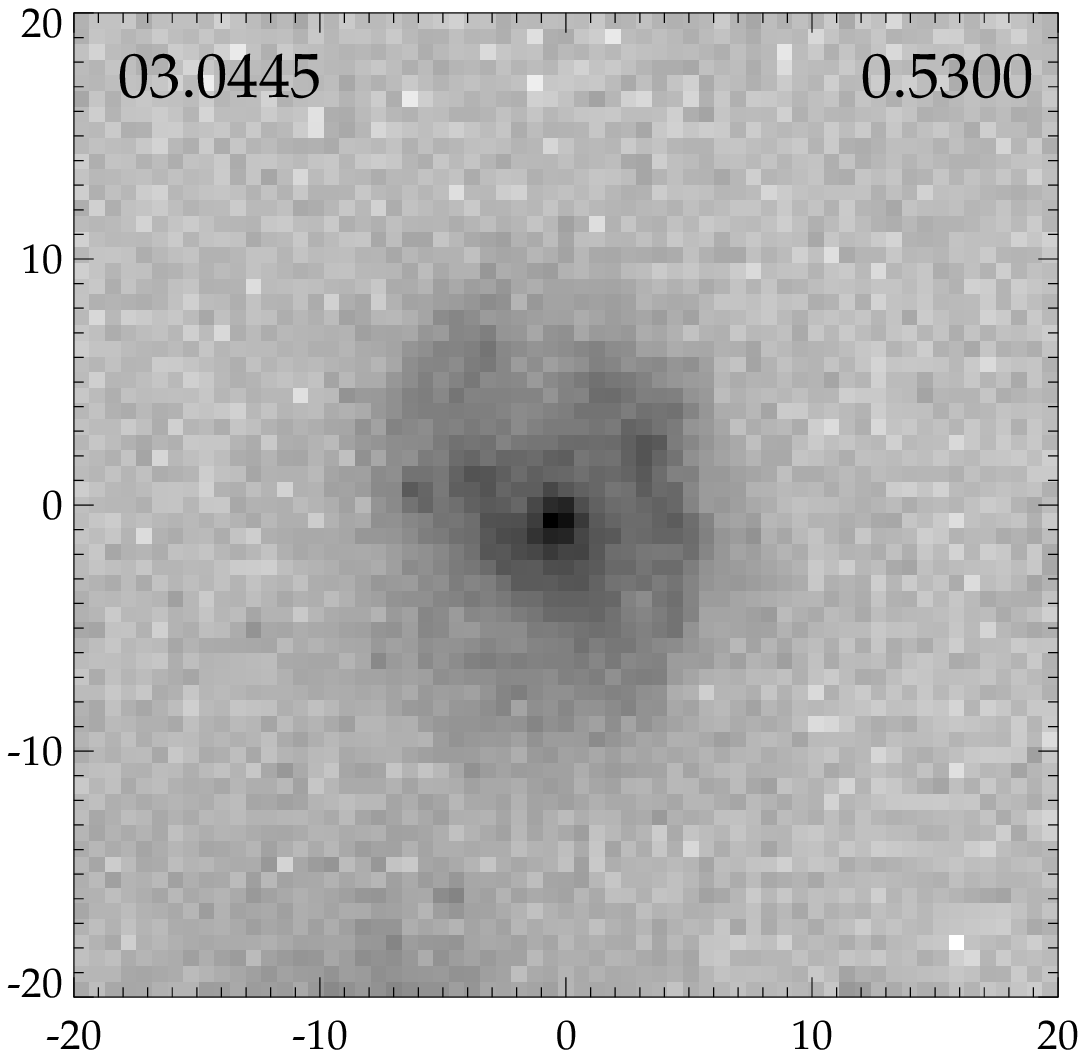}    \includegraphics[height=0.21\textwidth,clip]{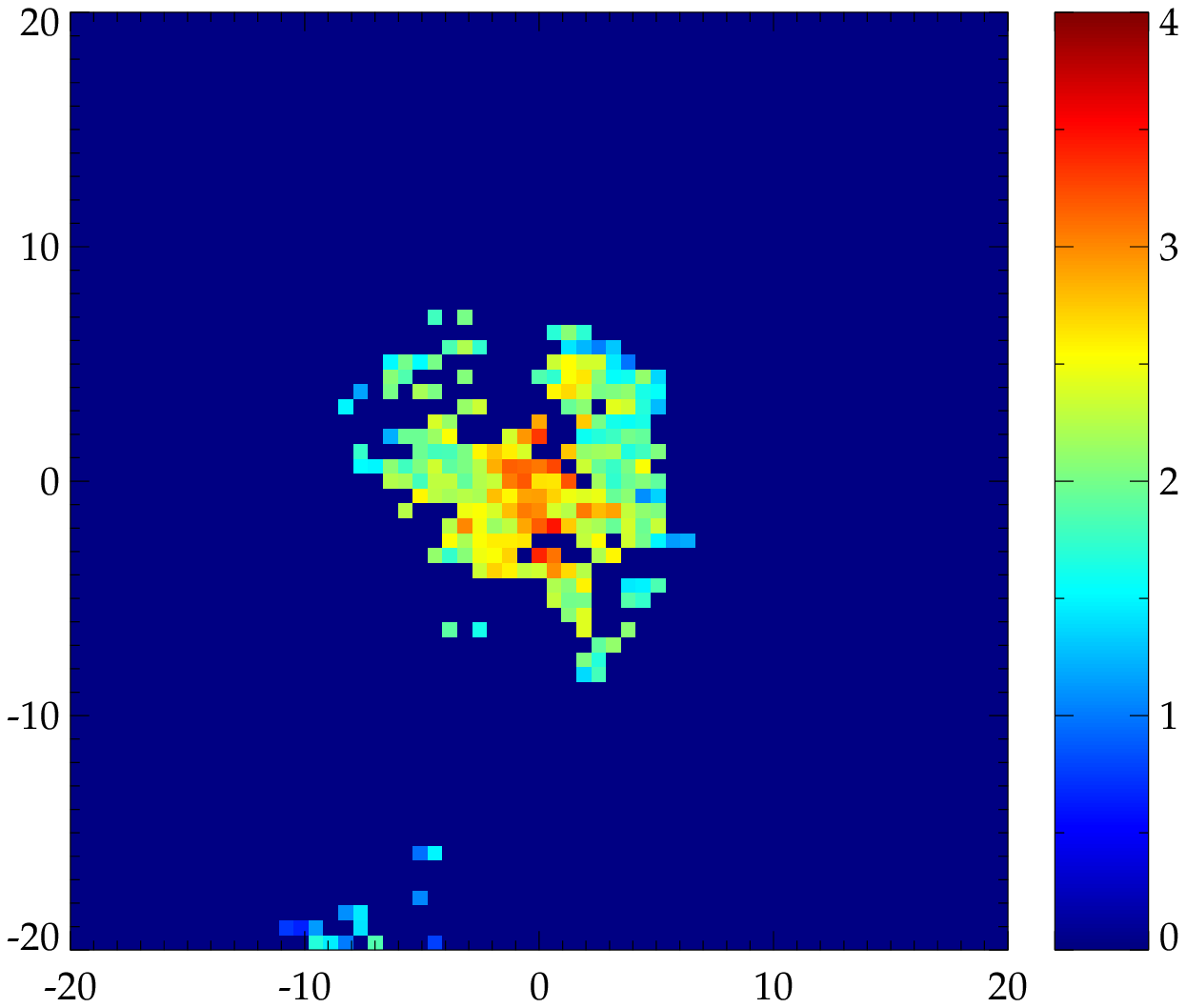}
\includegraphics[height=0.21\textwidth,clip]{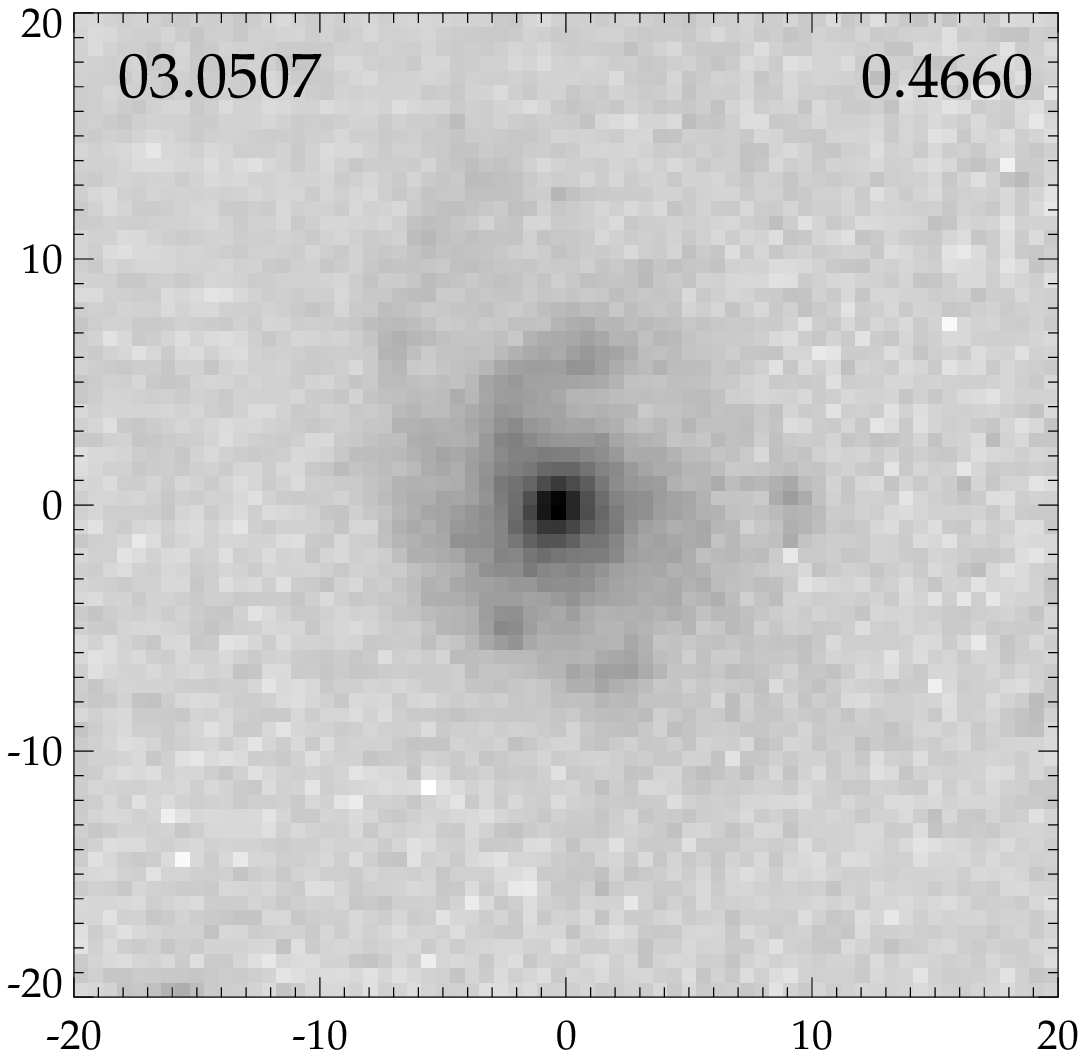}    \includegraphics[height=0.21\textwidth,clip]{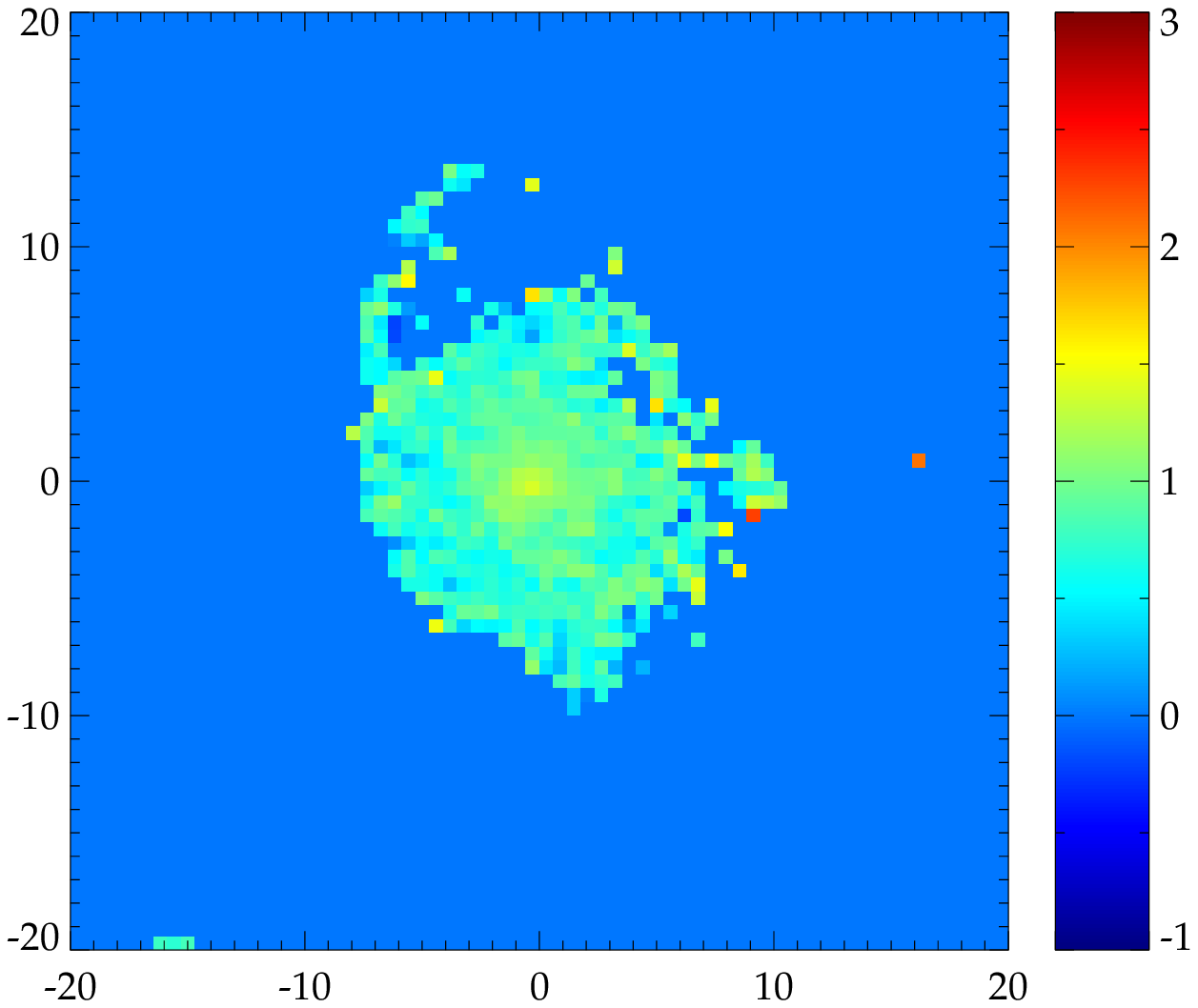}
\includegraphics[height=0.21\textwidth,clip]{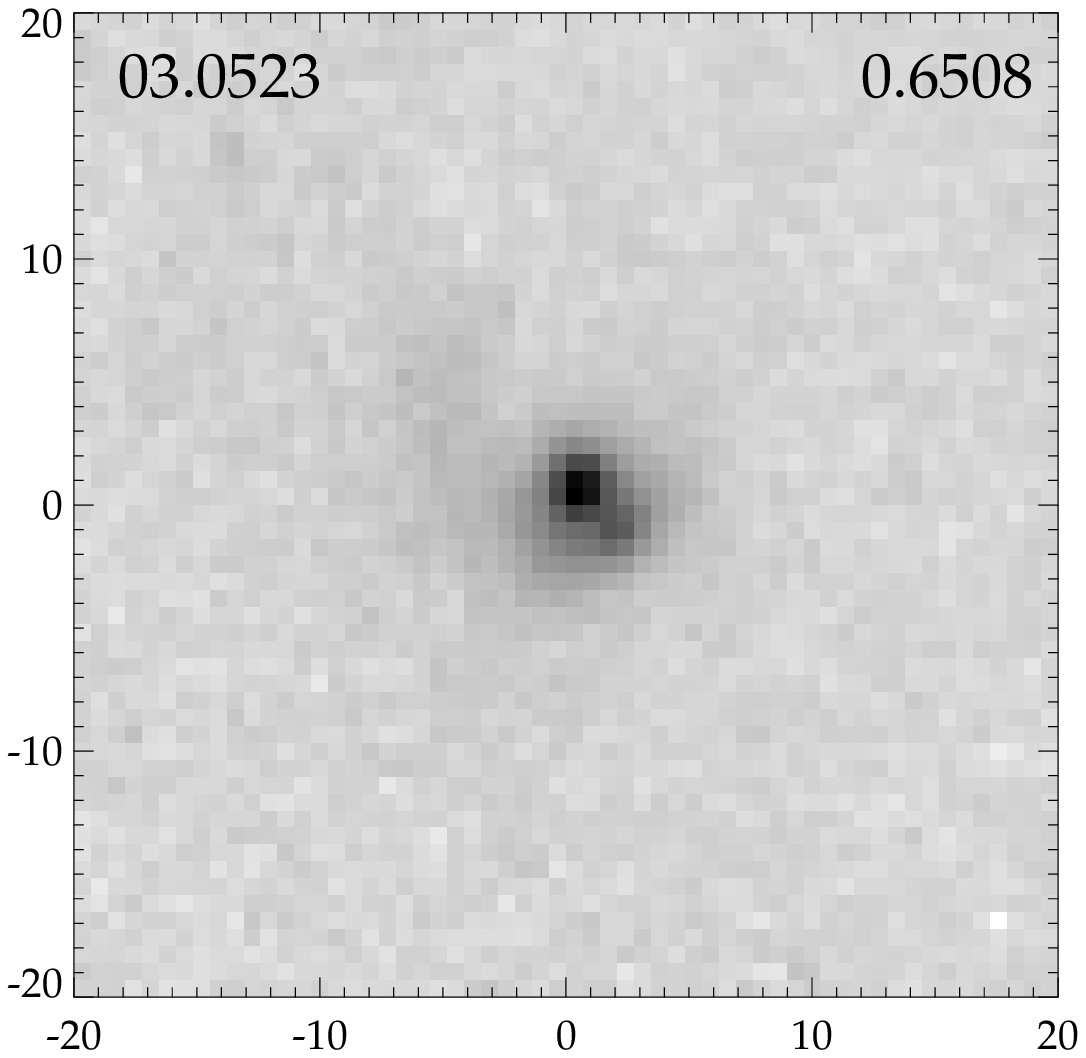}    \includegraphics[height=0.21\textwidth,clip]{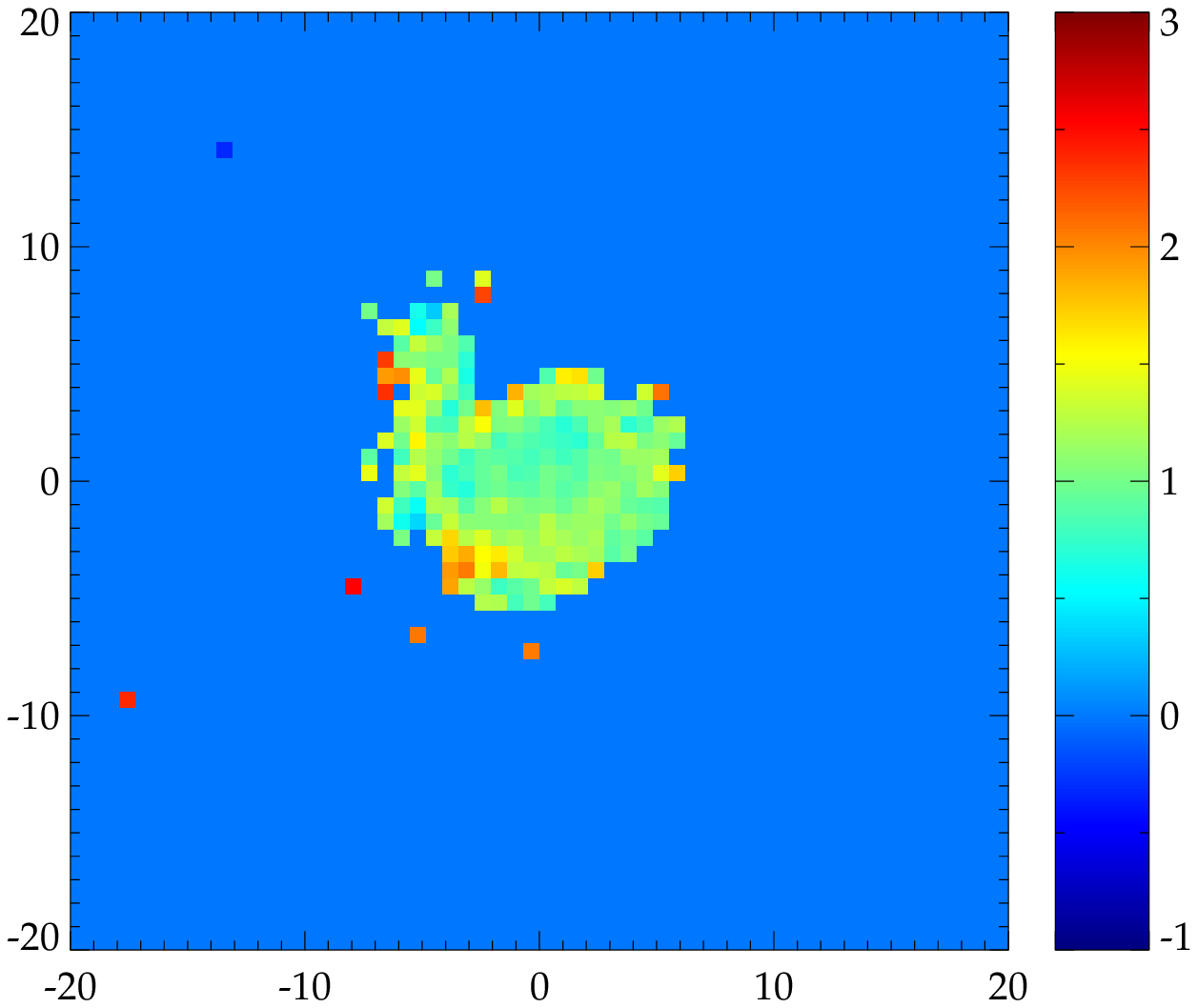}
\includegraphics[height=0.21\textwidth,clip]{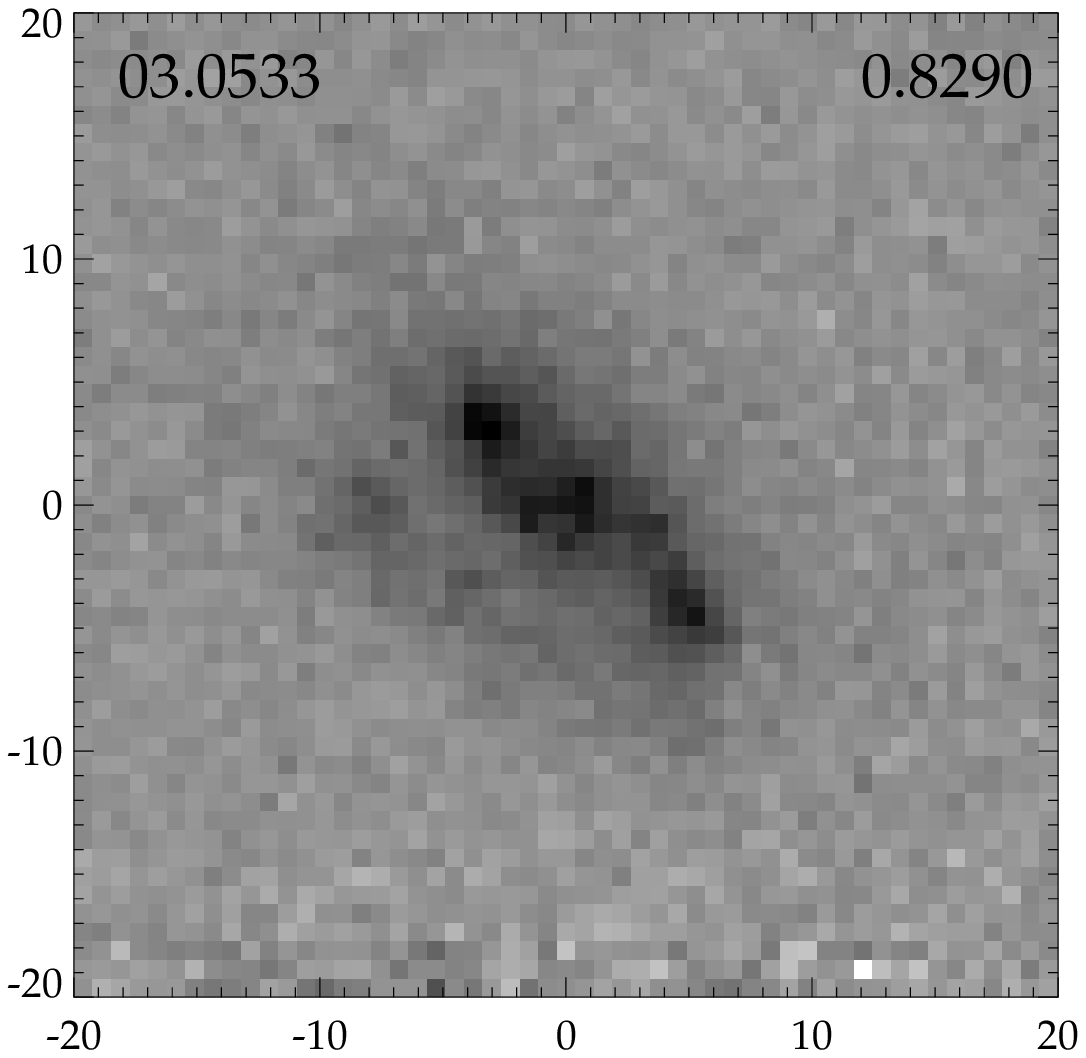}    \includegraphics[height=0.21\textwidth,clip]{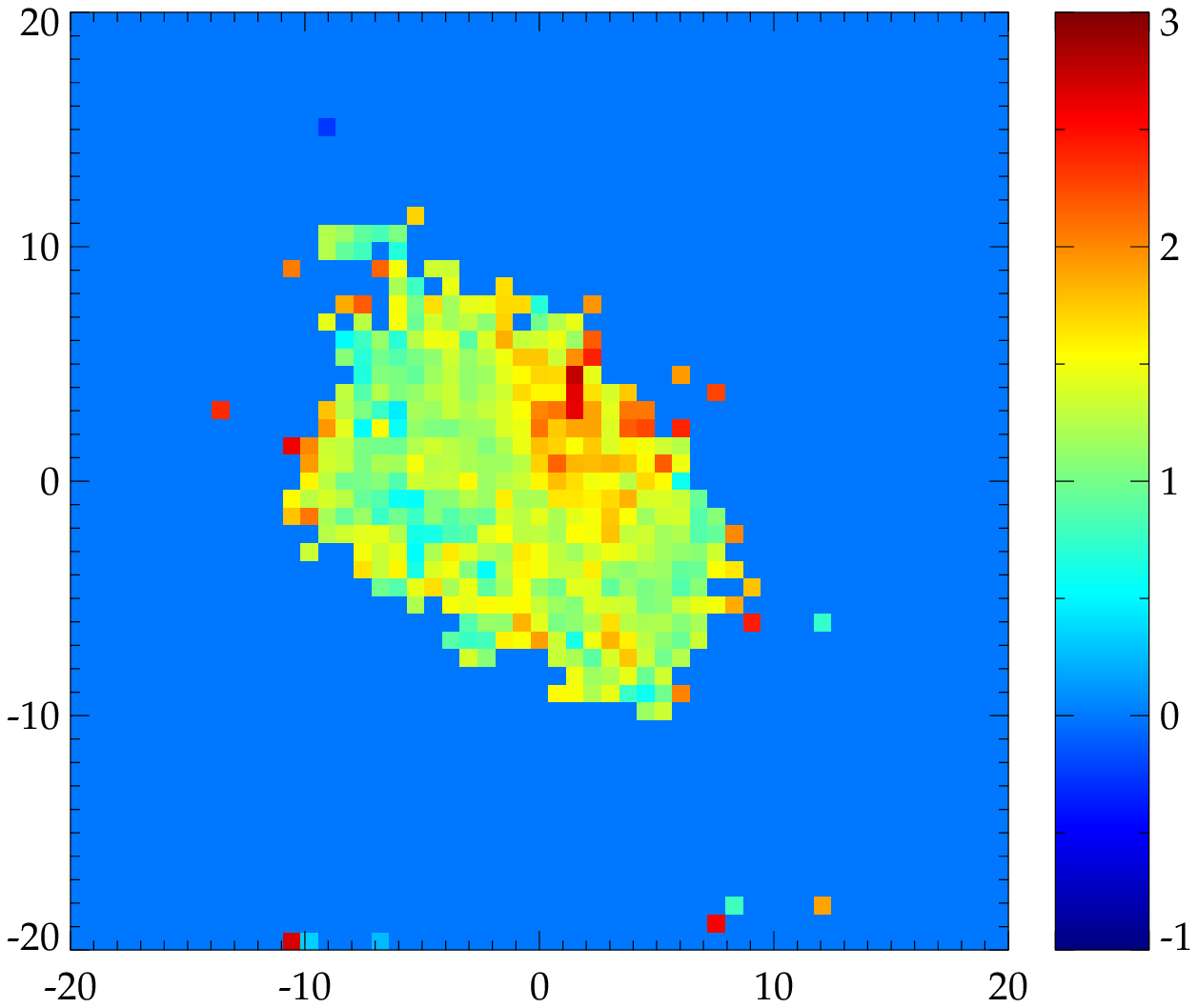}
\includegraphics[height=0.21\textwidth,clip]{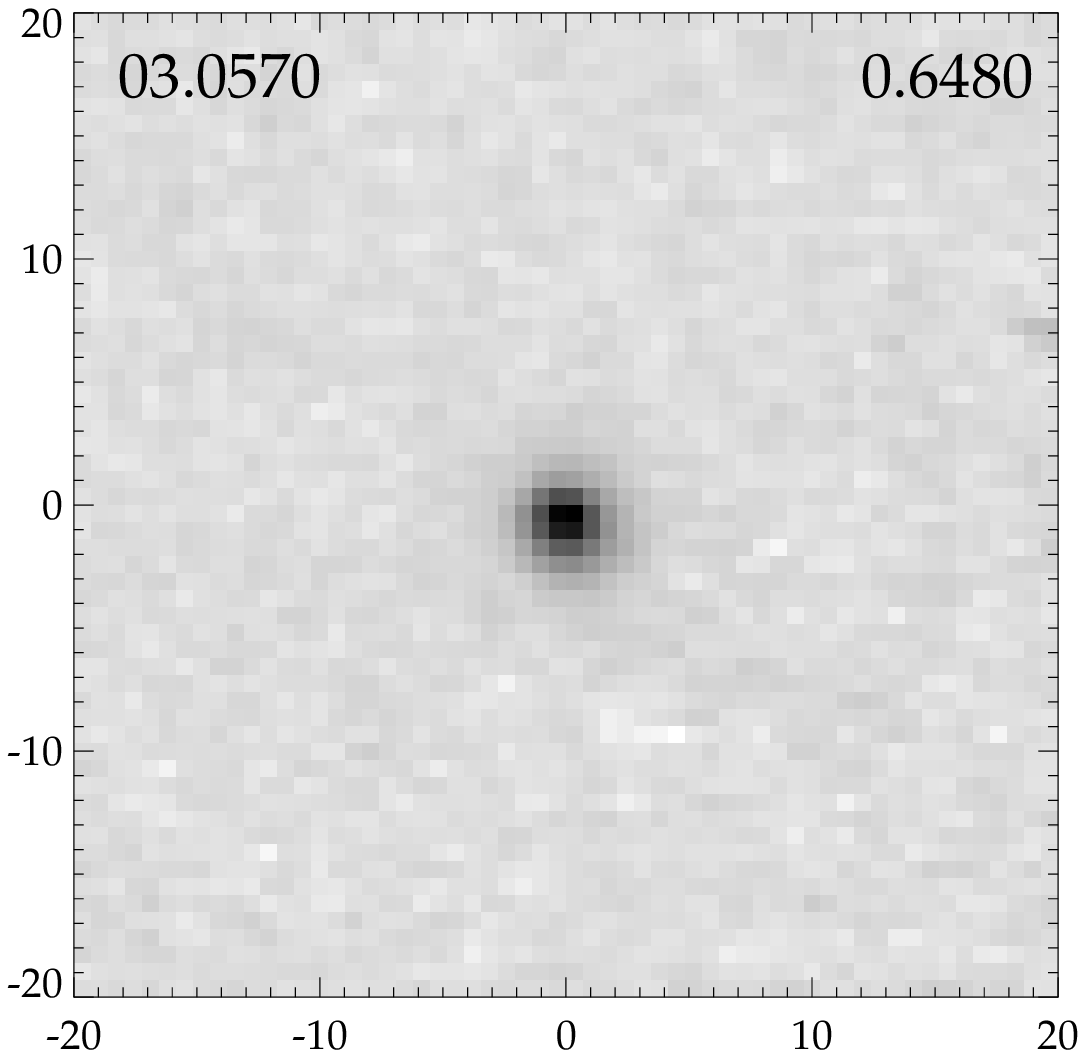}    \includegraphics[height=0.21\textwidth,clip]{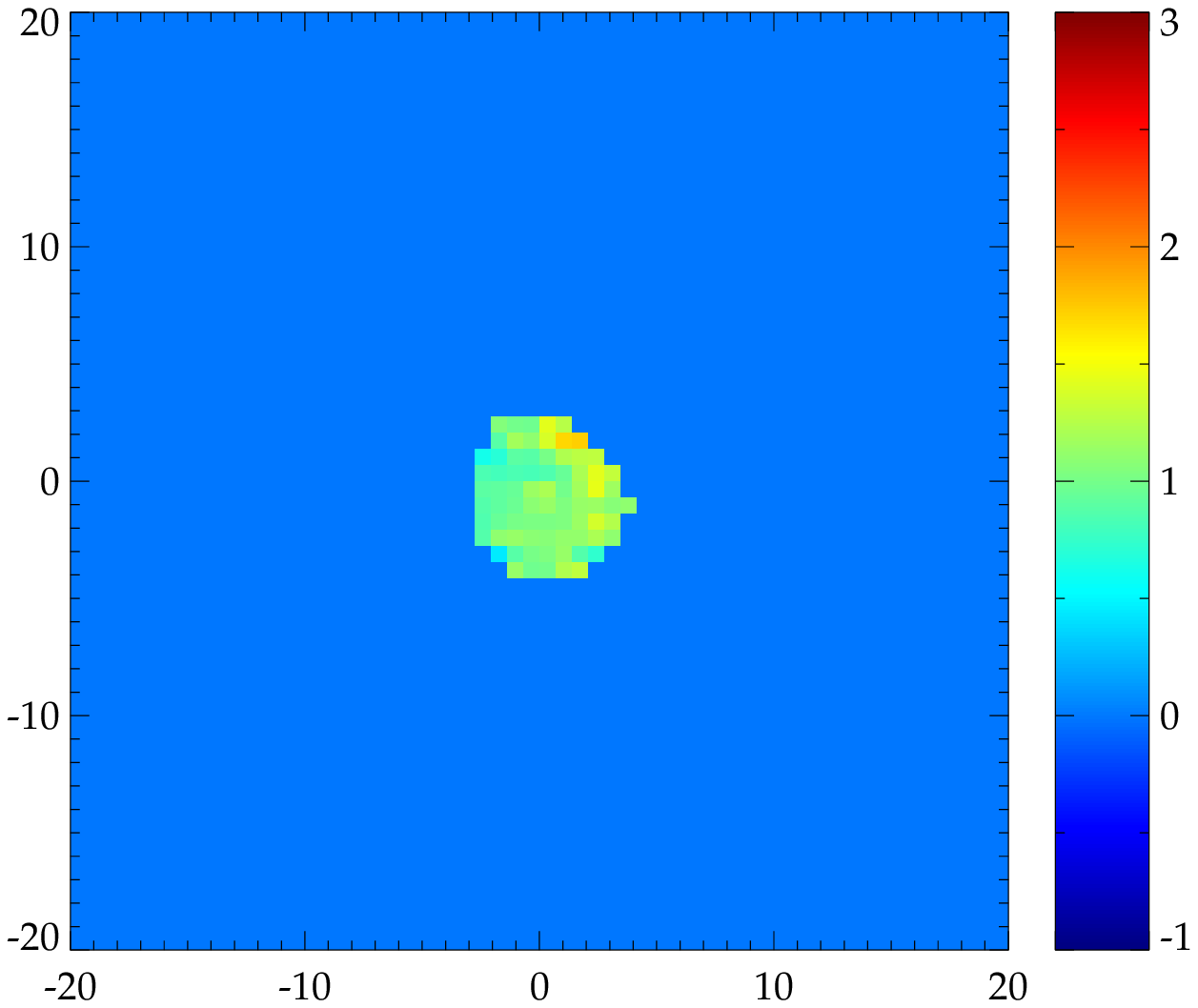}
\includegraphics[height=0.21\textwidth,clip]{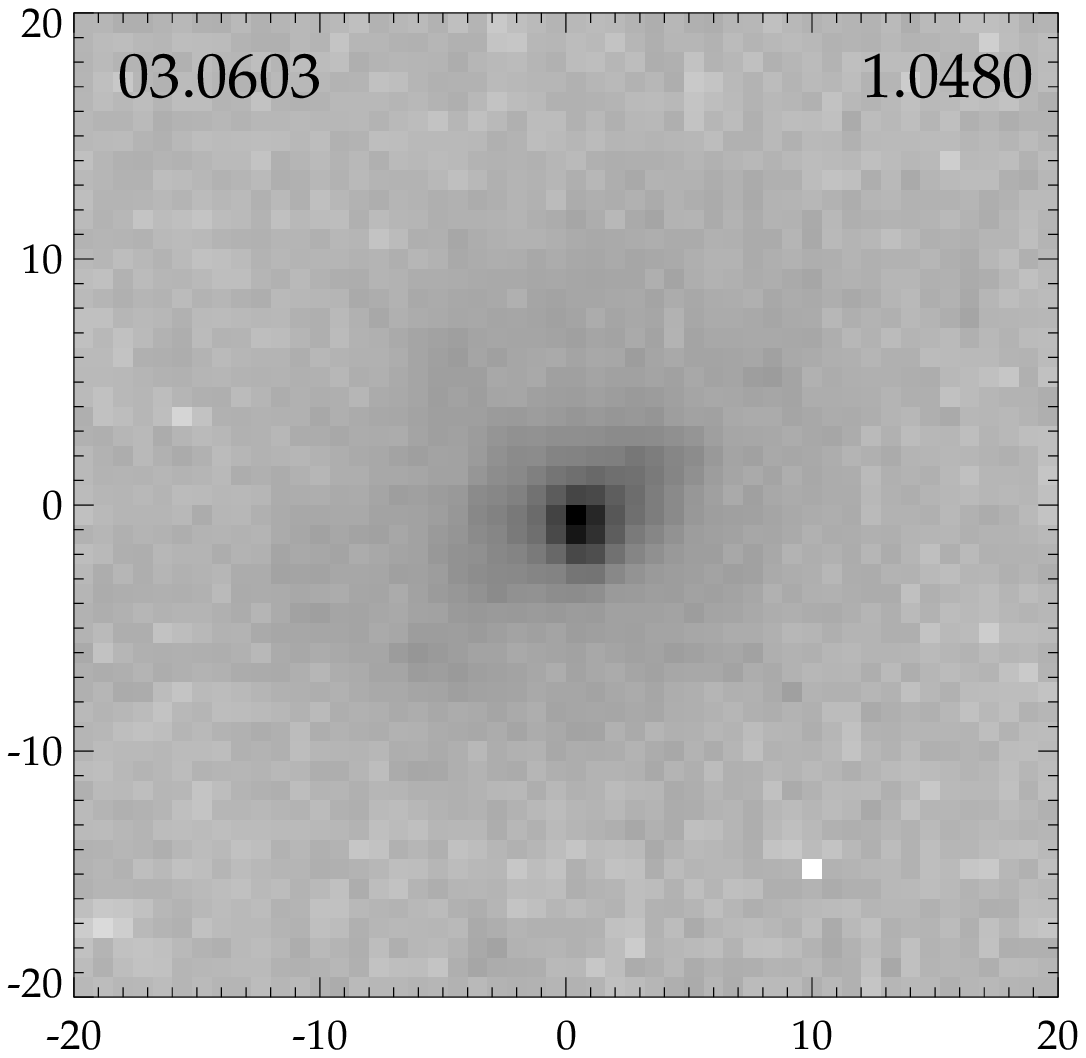}    \includegraphics[height=0.21\textwidth,clip]{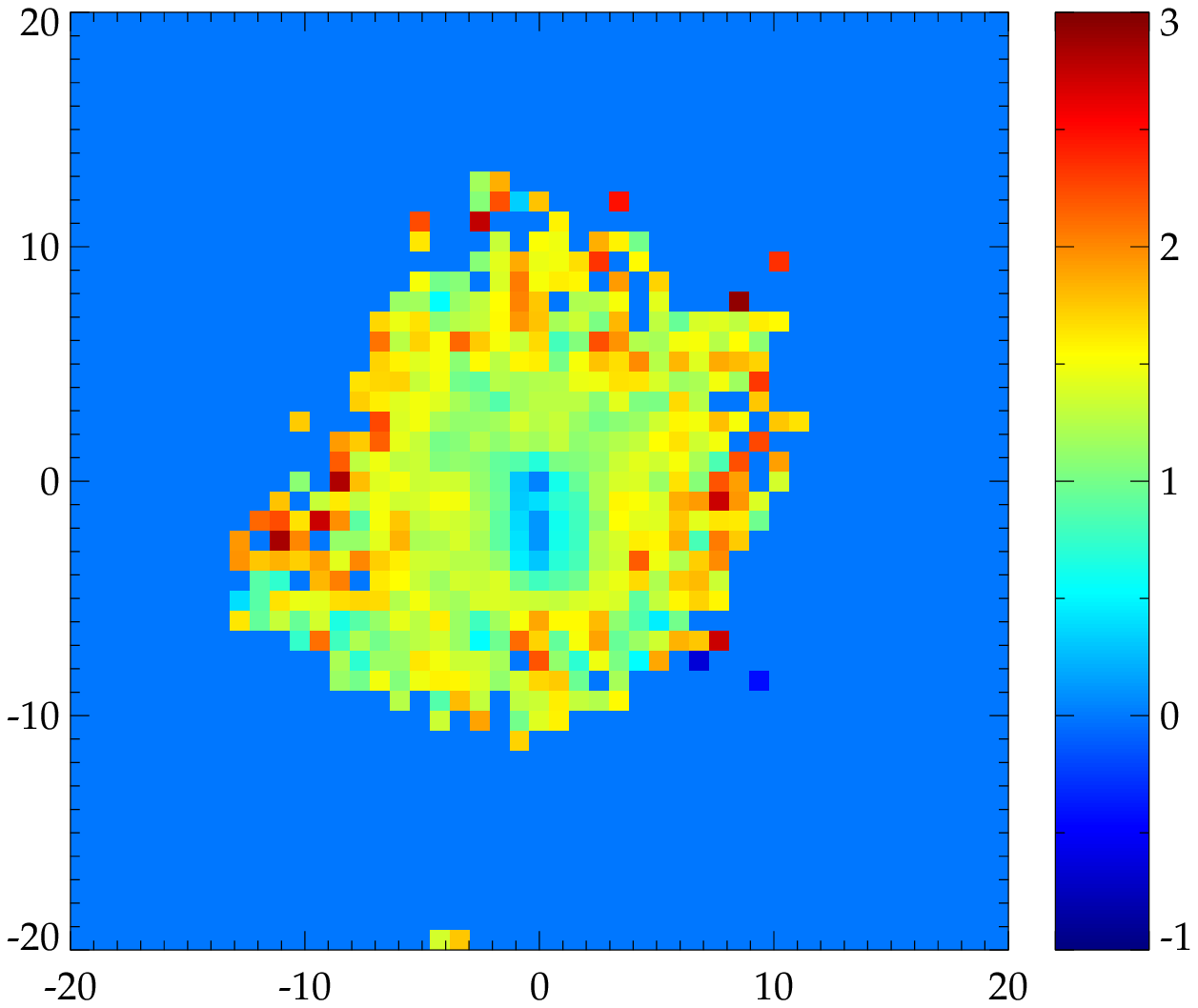}
\includegraphics[height=0.21\textwidth,clip]{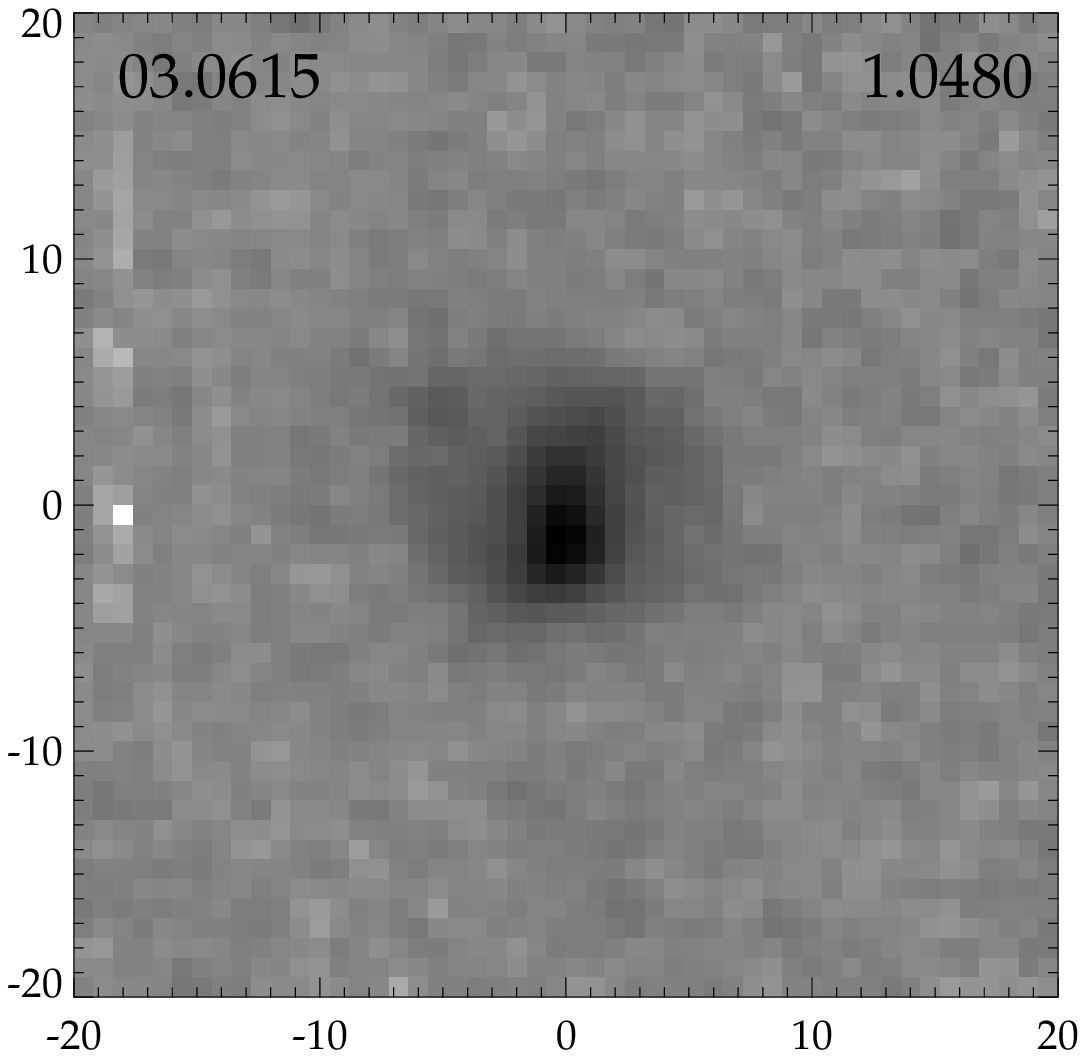}    \includegraphics[height=0.21\textwidth,clip]{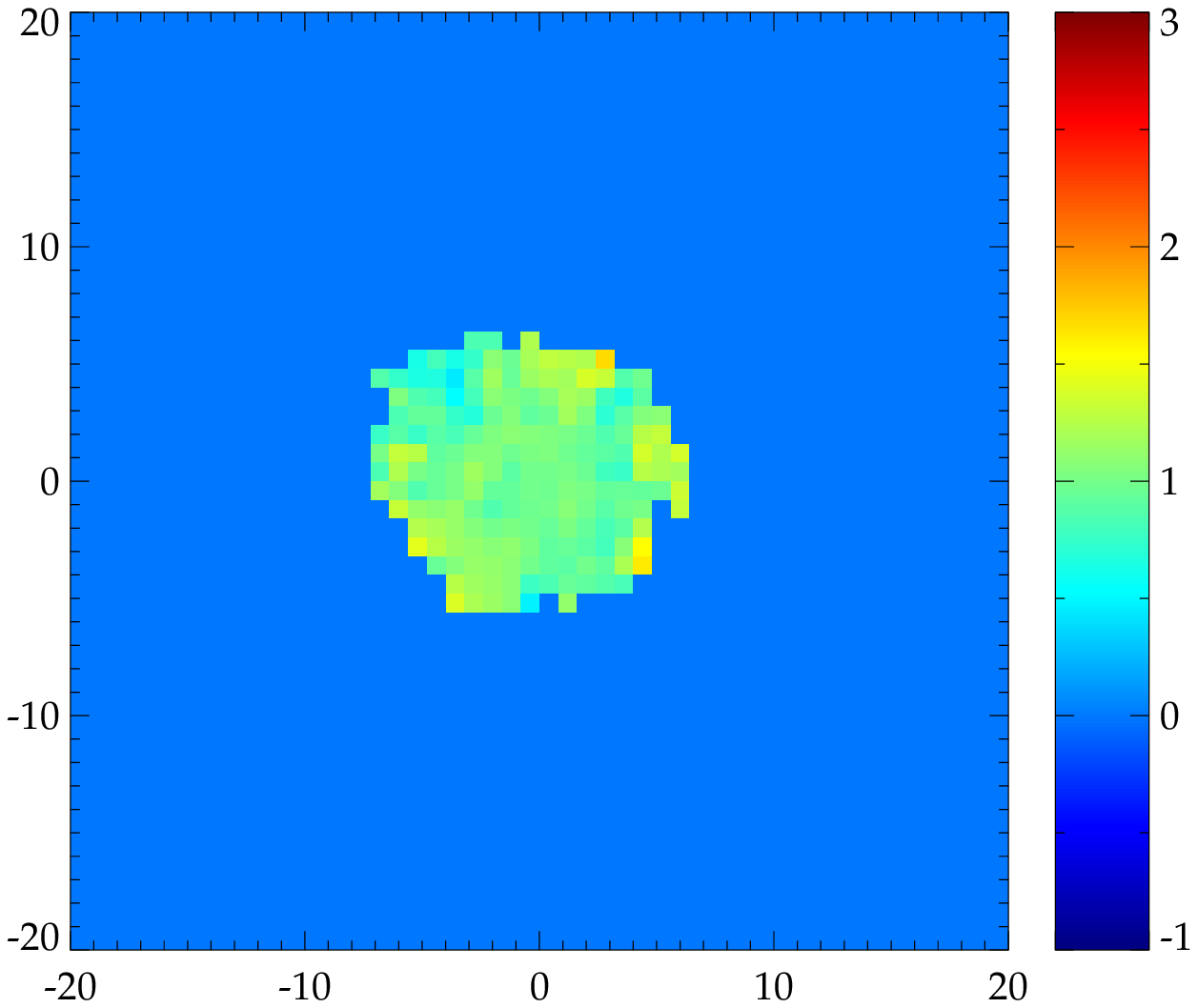}
\includegraphics[height=0.21\textwidth,clip]{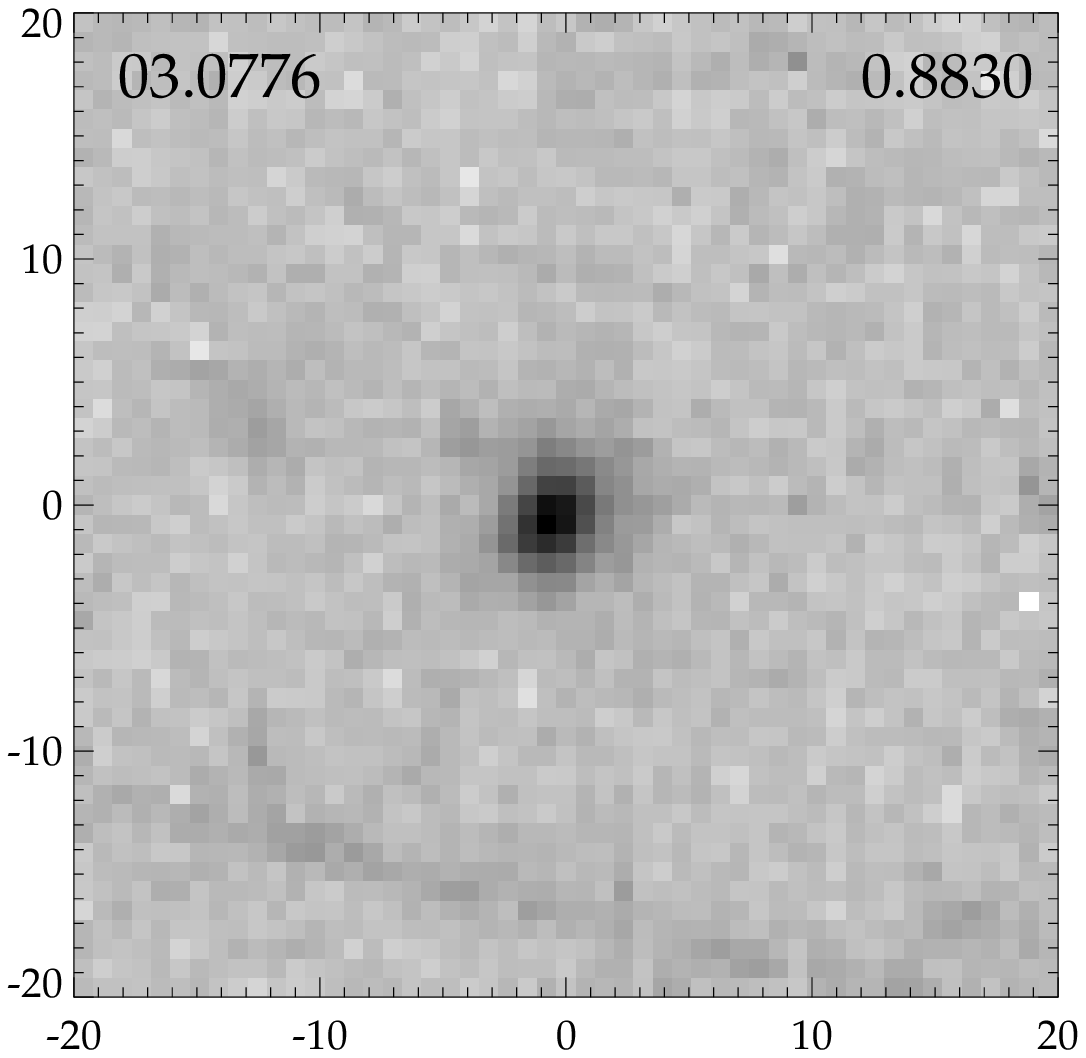}    \includegraphics[height=0.21\textwidth,clip]{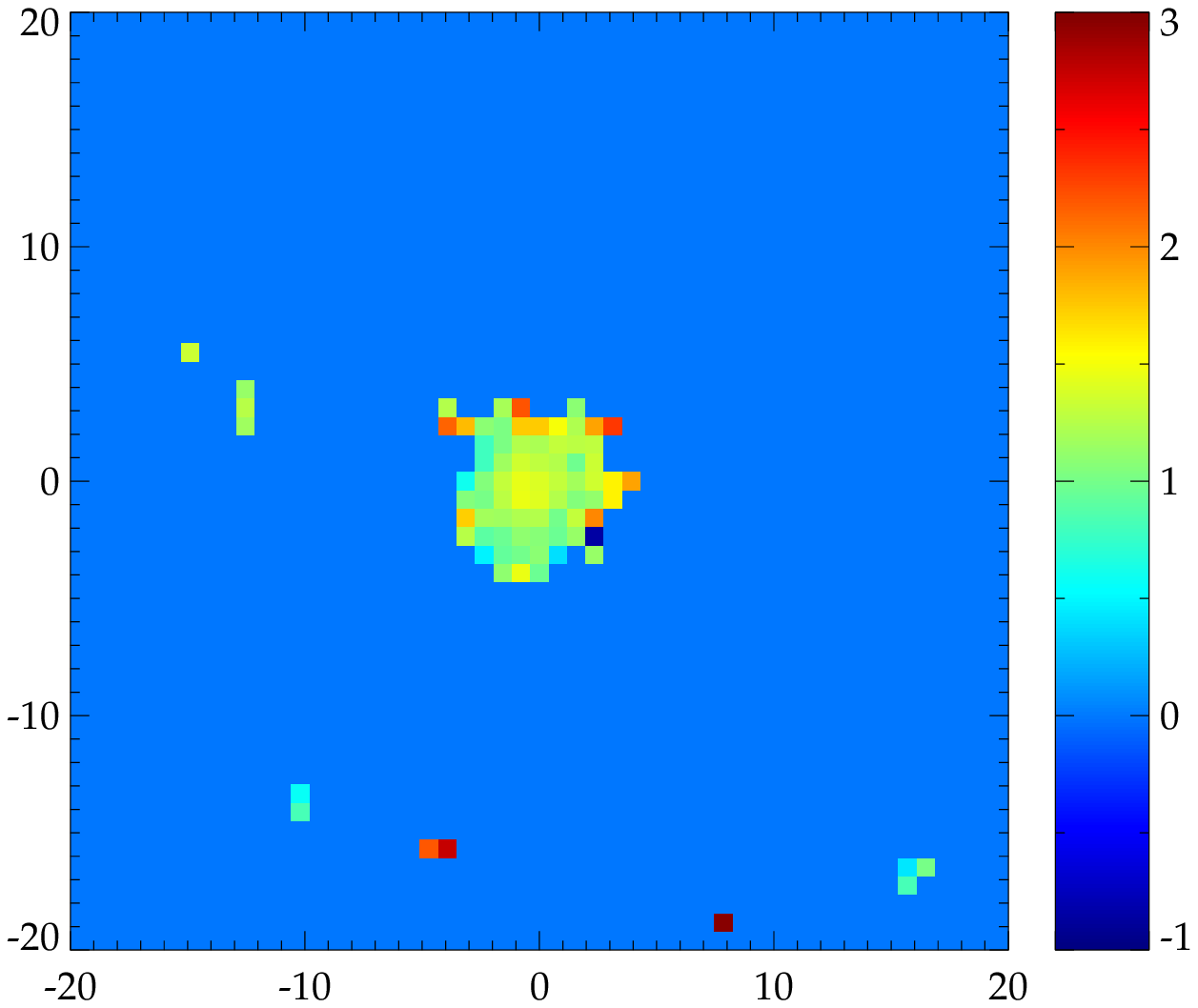}
\includegraphics[height=0.21\textwidth,clip]{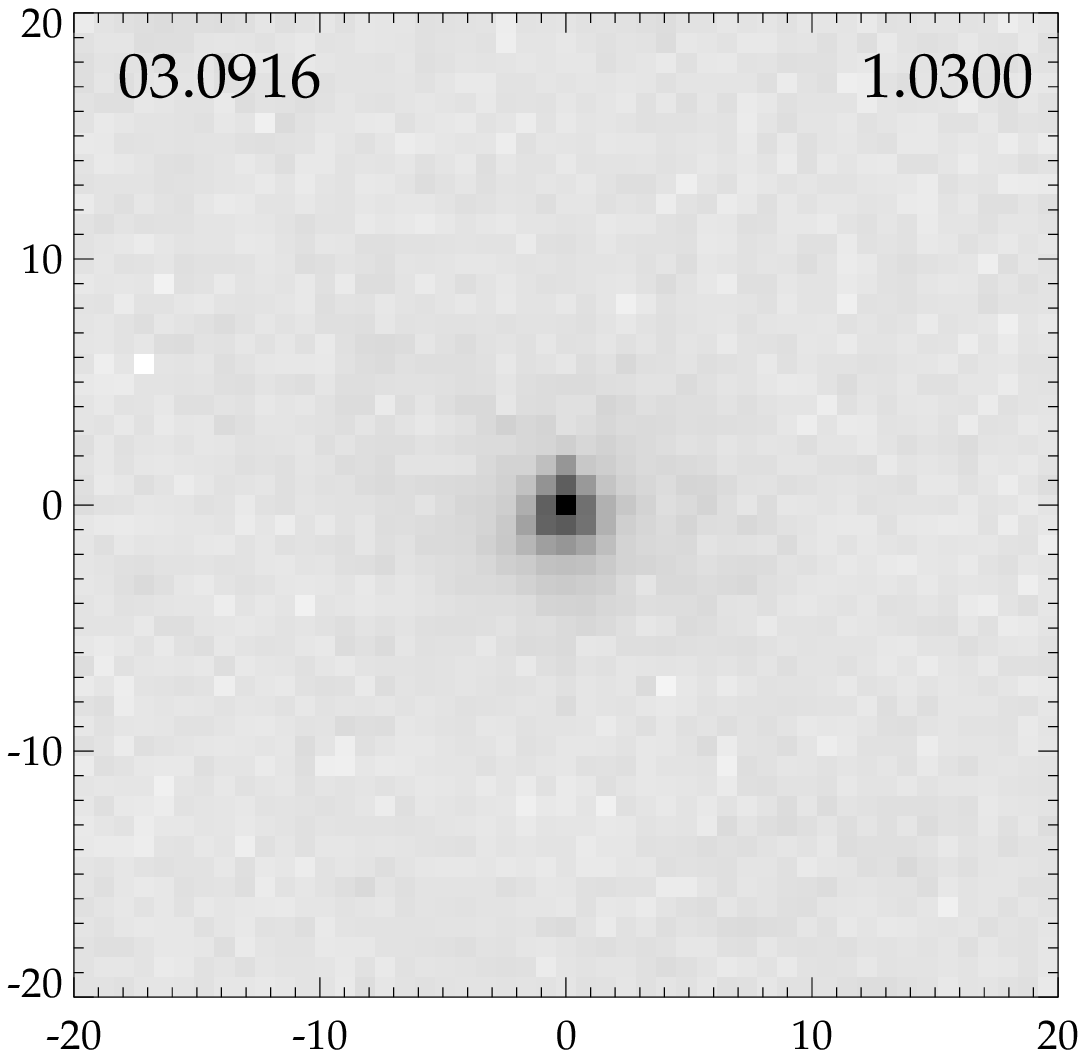}    \includegraphics[height=0.21\textwidth,clip]{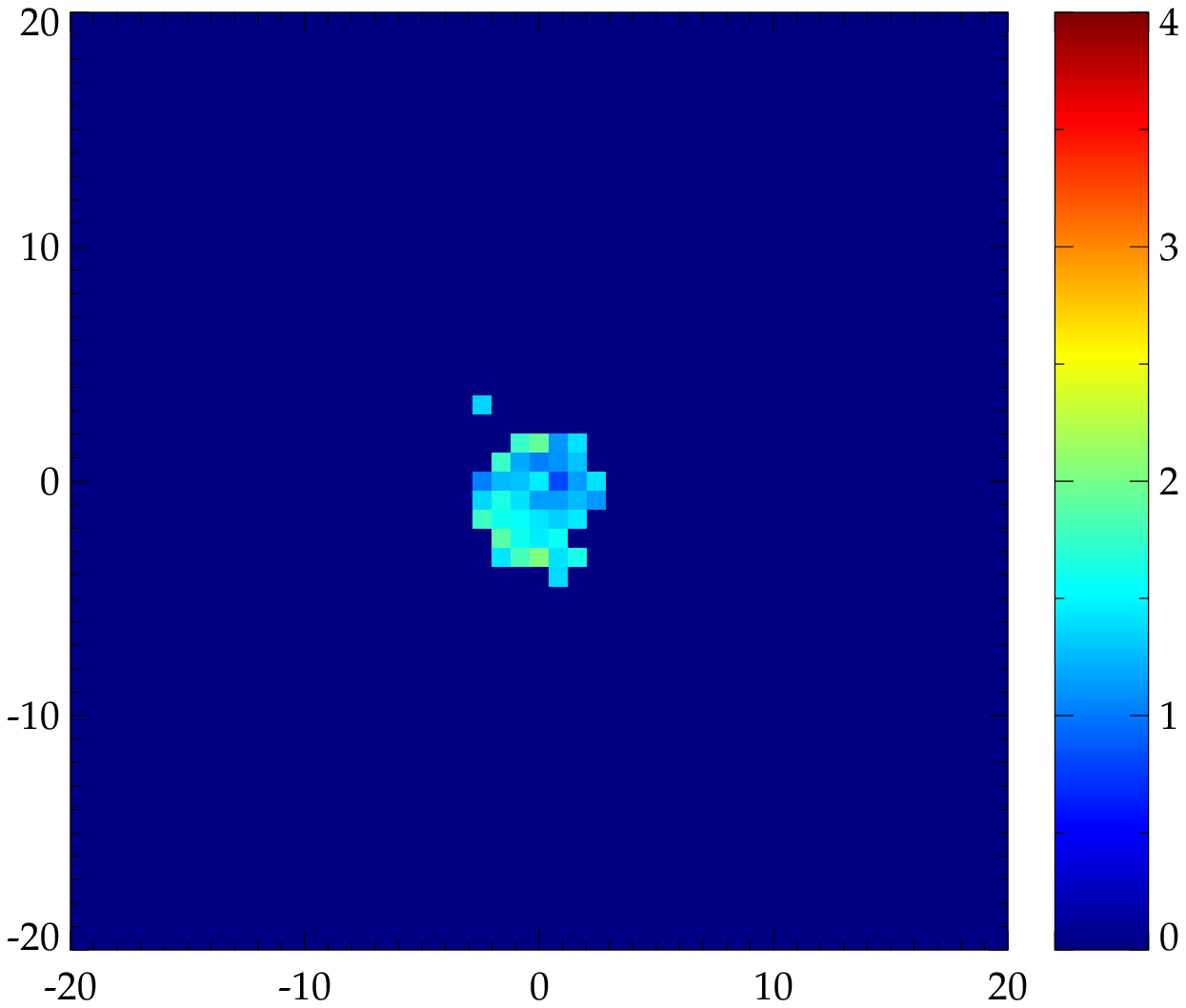}

\caption{I$_{814}$ imaging and color map stamps of distant LIRGs. 
For each target, the name and redshift are labeled top-left and top-right in the I$_{814}$ imaging stamp. Color bar ranges from $-$1 to 3 for V$_{606}-$I$_{814}$ color map and 0 to 4 for B$_{450}-$I$_{814}$ color map.  Blank in I$_{814}$ imaging is due to the target imaged close to the chip border. The size of each stamp is 40$\times$40\,kpc  except for the object 03.1309, which is given in 60$\times$60\,kpc in order to display the whole merging system.
[{\it See the electronic edition for a color version of this figure.}]
} 
\label{colormap} \end{figure*}

\addtocounter{figure}{-1}
\begin{figure*}[] \centering

\includegraphics[height=0.21\textwidth,clip]{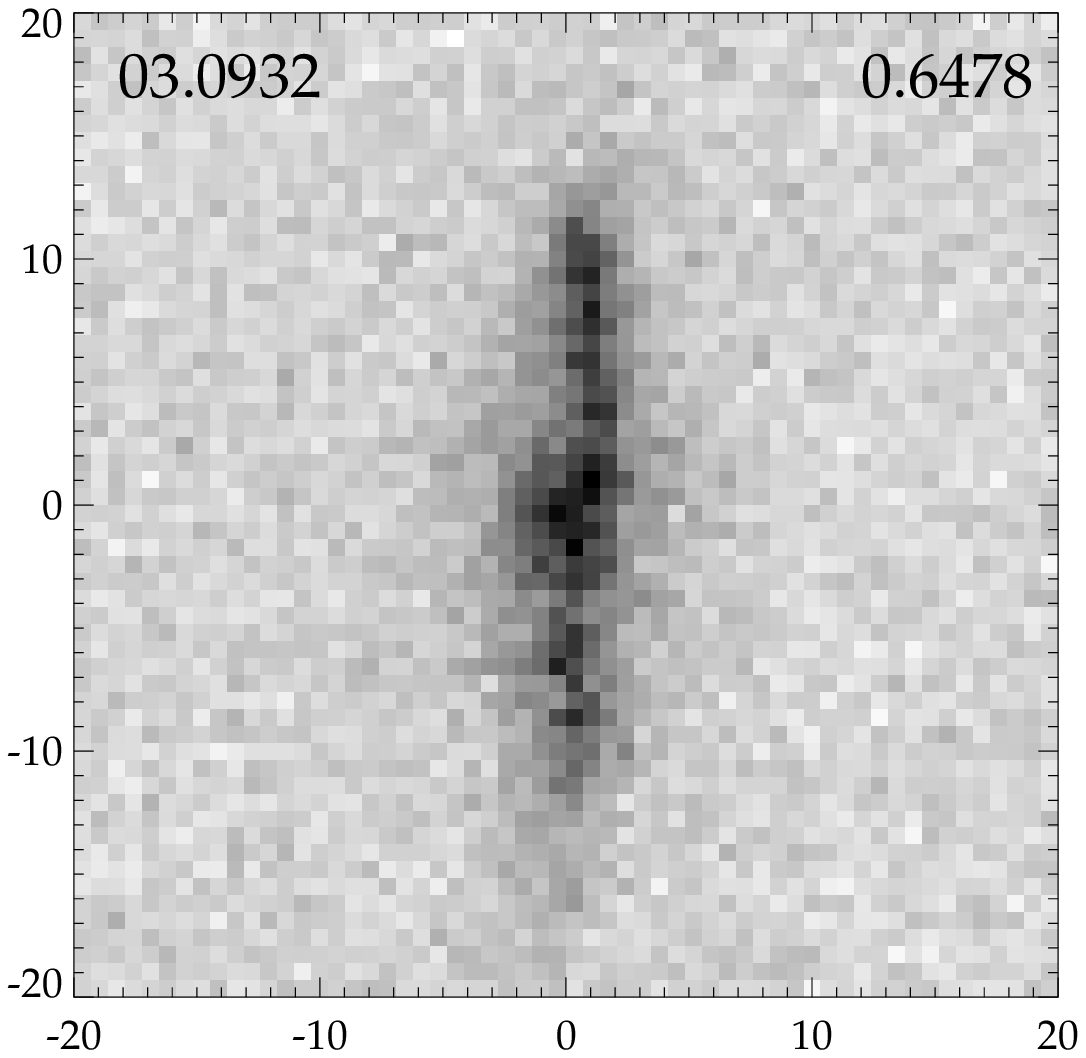}    \includegraphics[height=0.21\textwidth,clip]{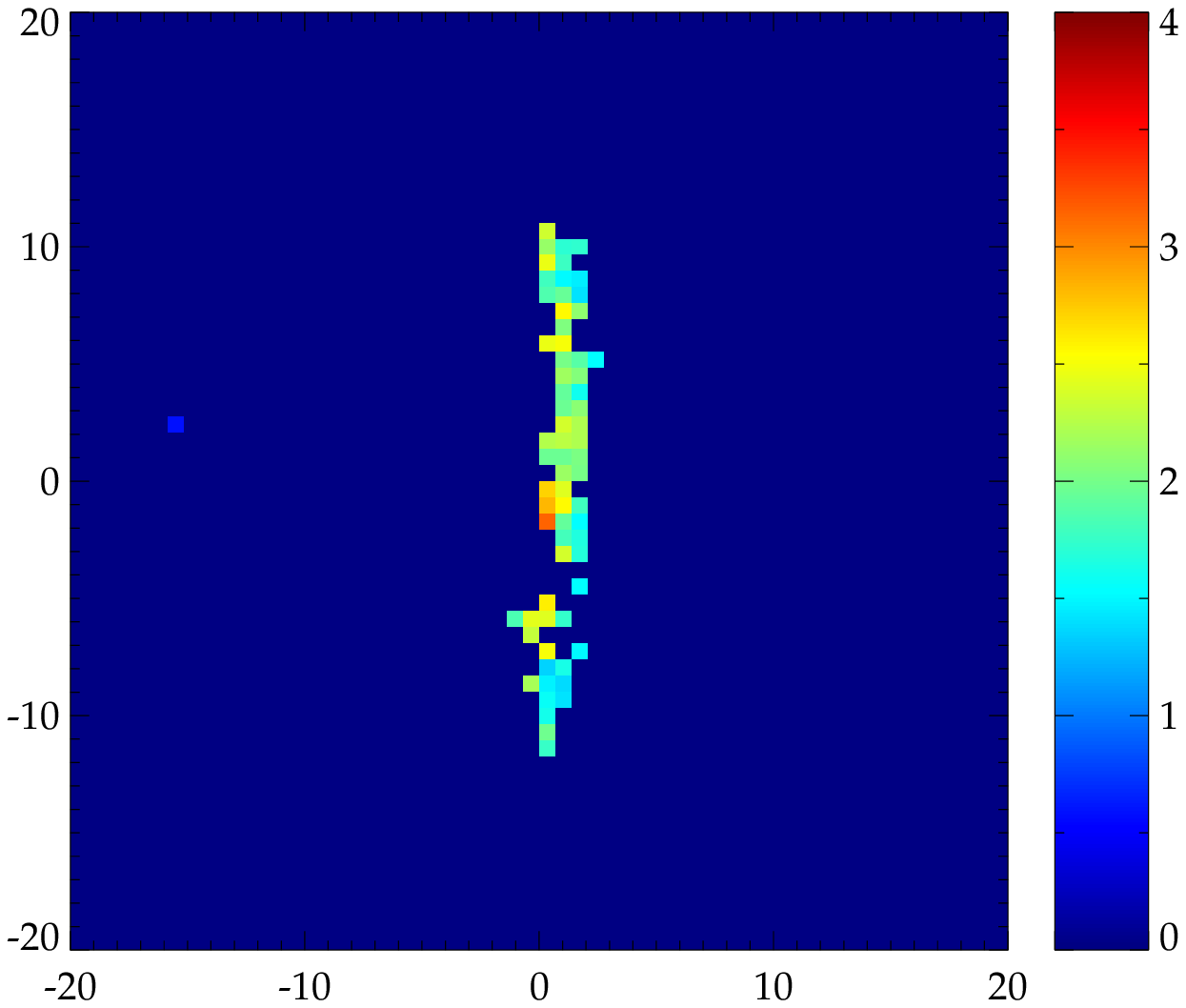}
\includegraphics[height=0.21\textwidth,clip]{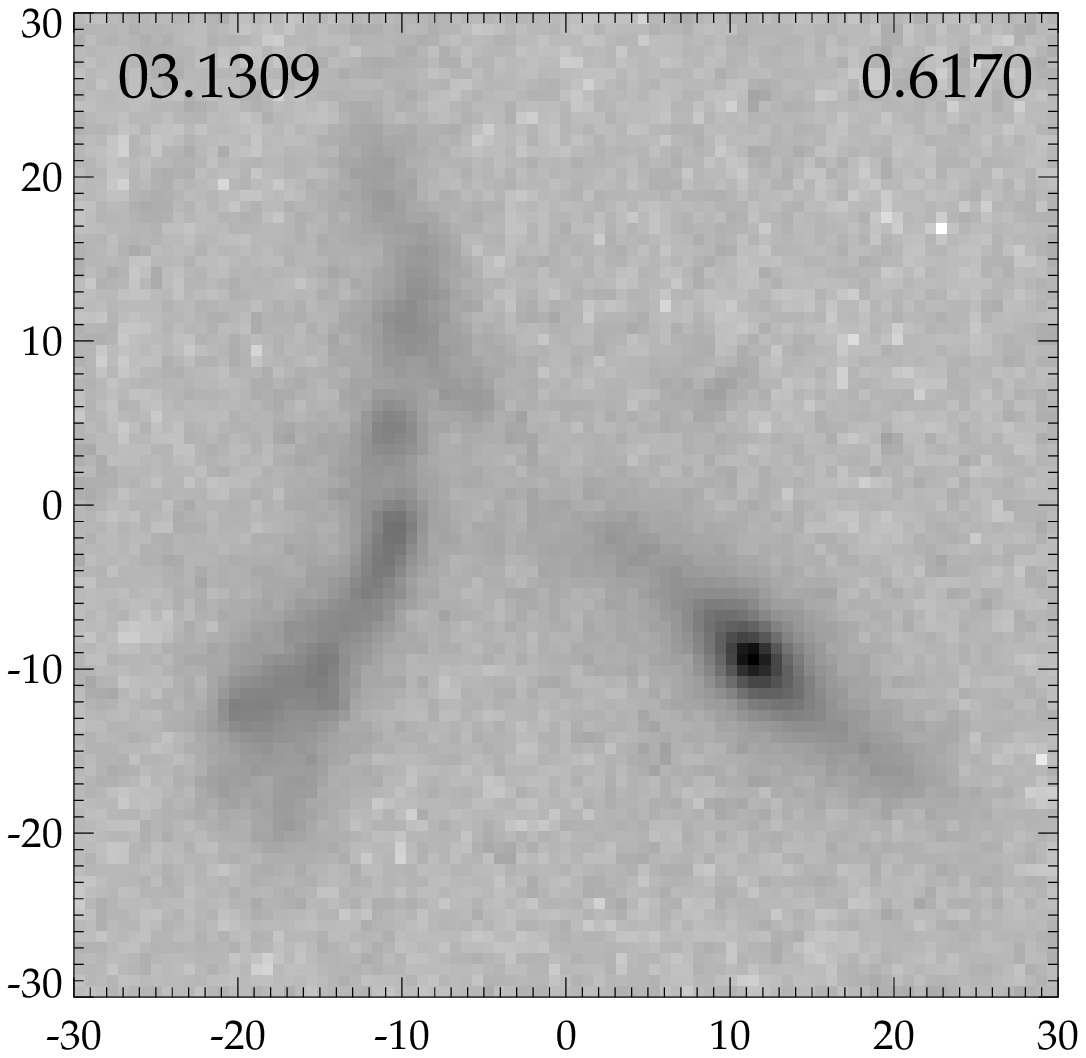}   \includegraphics[height=0.21\textwidth,clip]{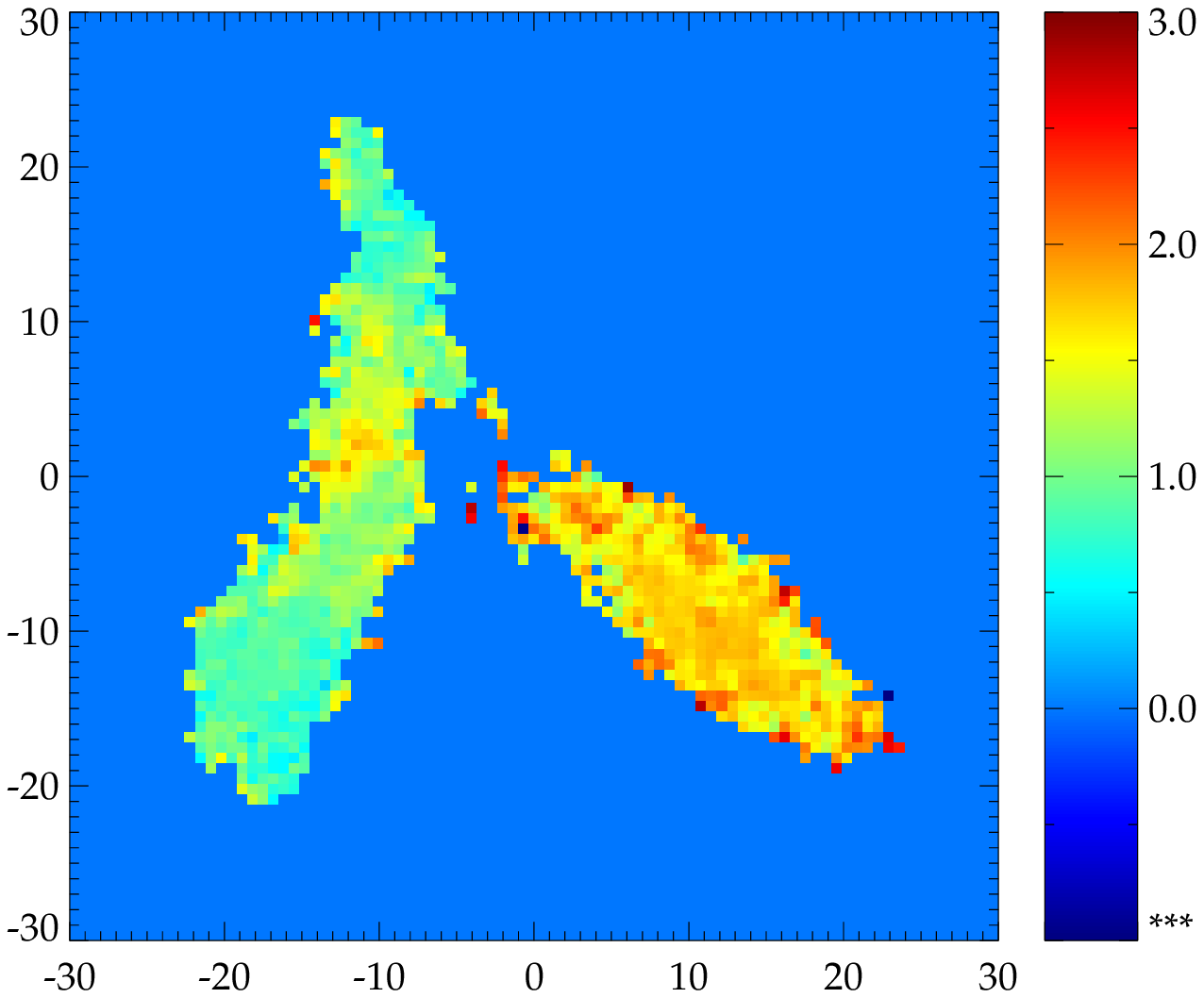}
\includegraphics[height=0.21\textwidth,clip]{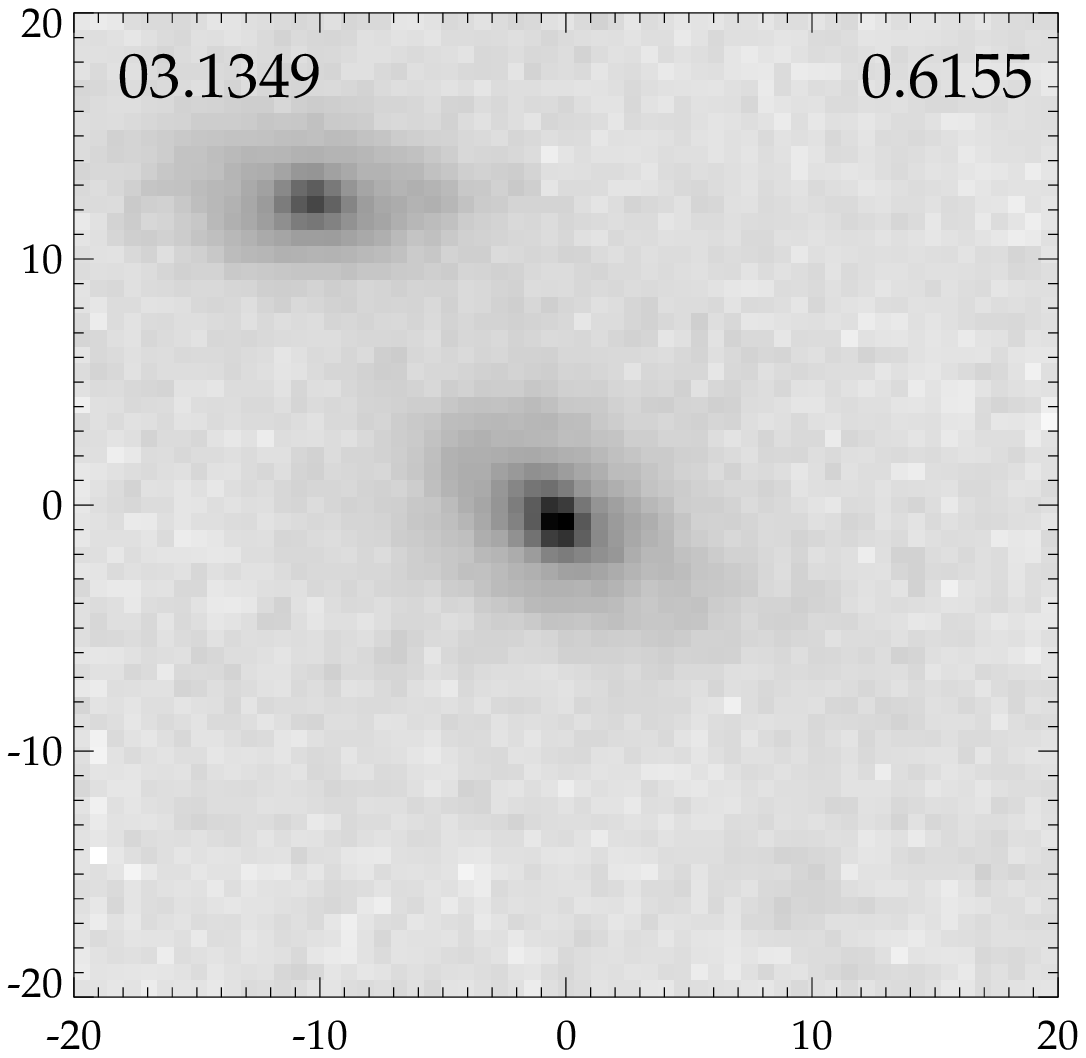}    \includegraphics[height=0.21\textwidth,clip]{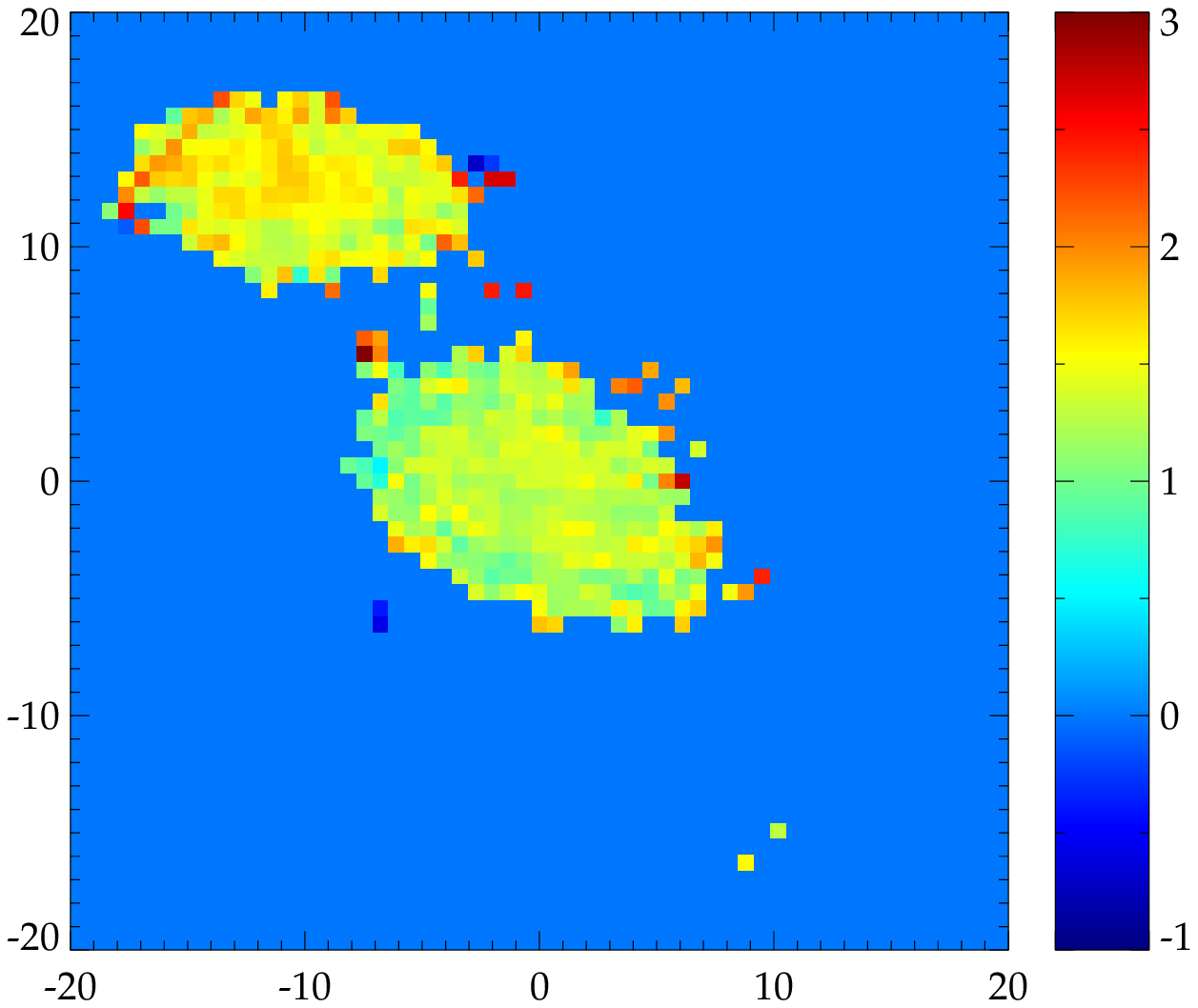}
\includegraphics[height=0.21\textwidth,clip]{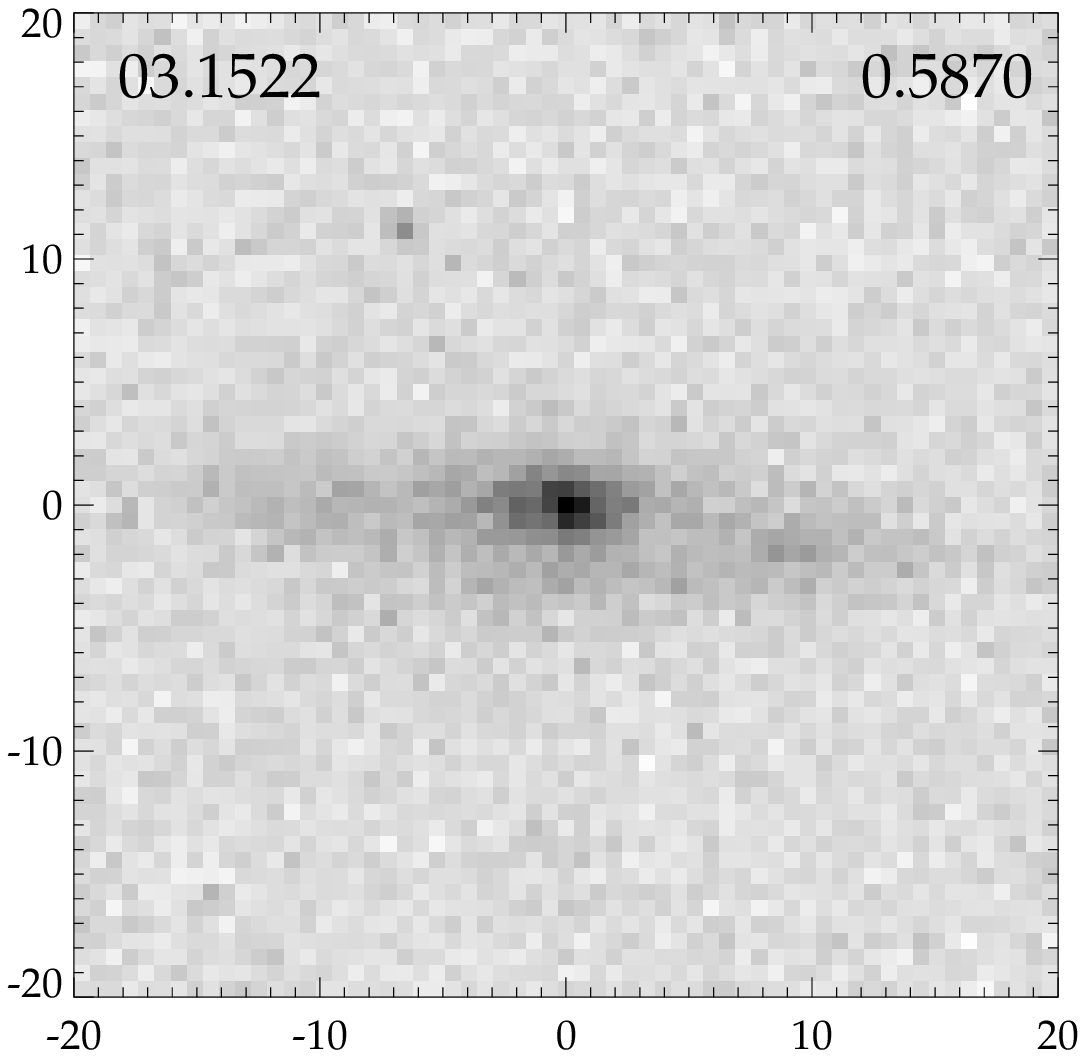}    \includegraphics[height=0.21\textwidth,clip]{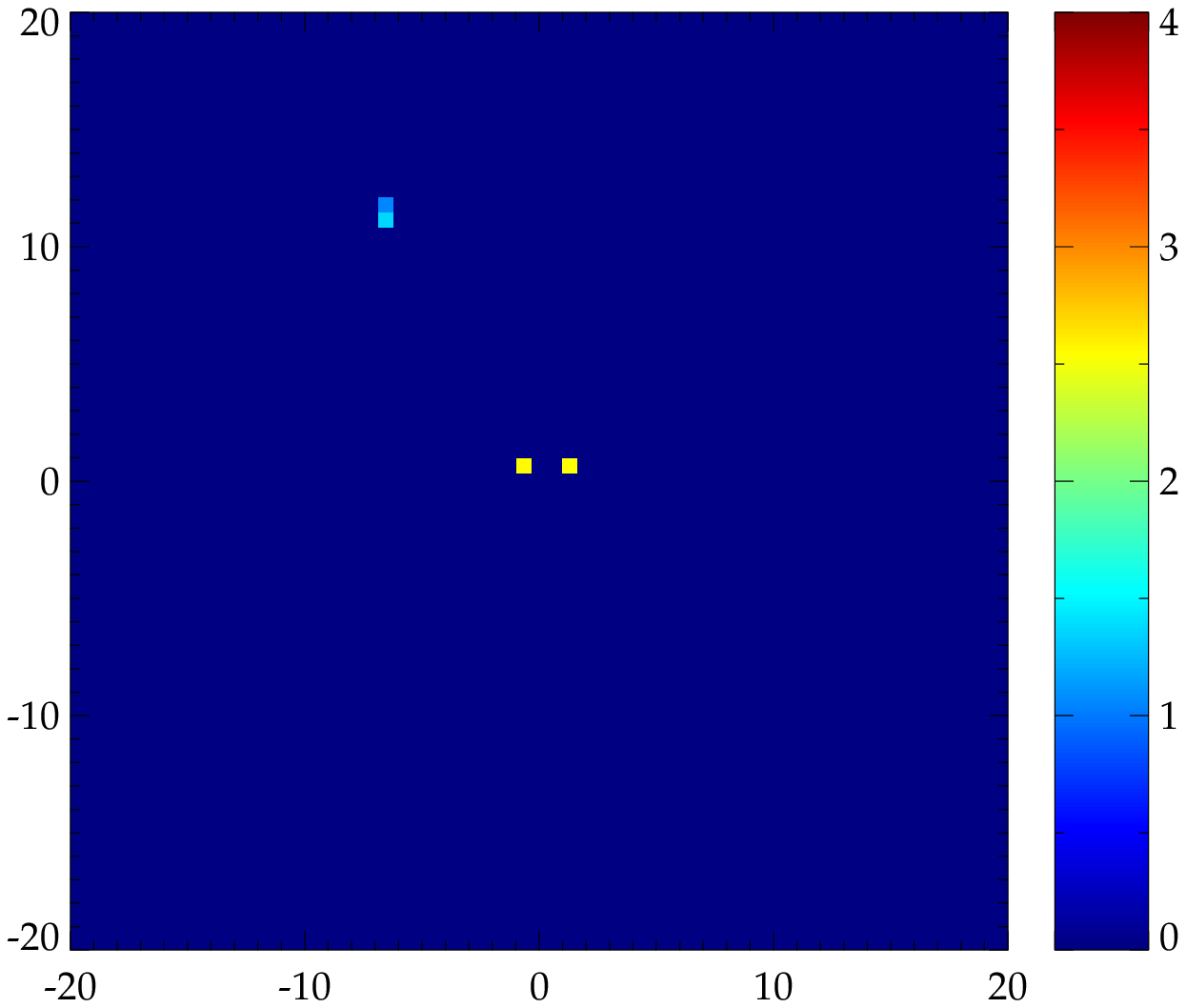}
\includegraphics[height=0.21\textwidth,clip]{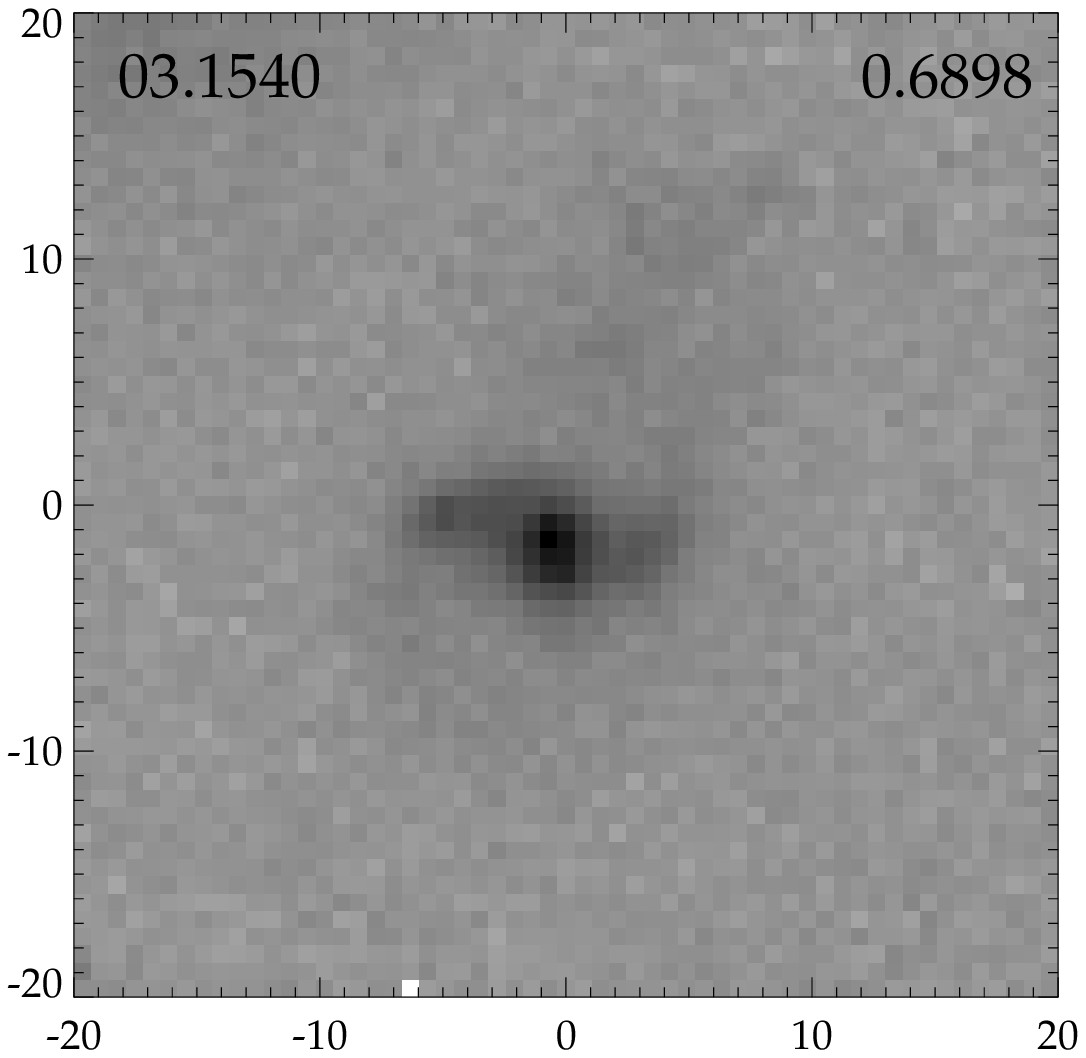}    \includegraphics[height=0.21\textwidth,clip]{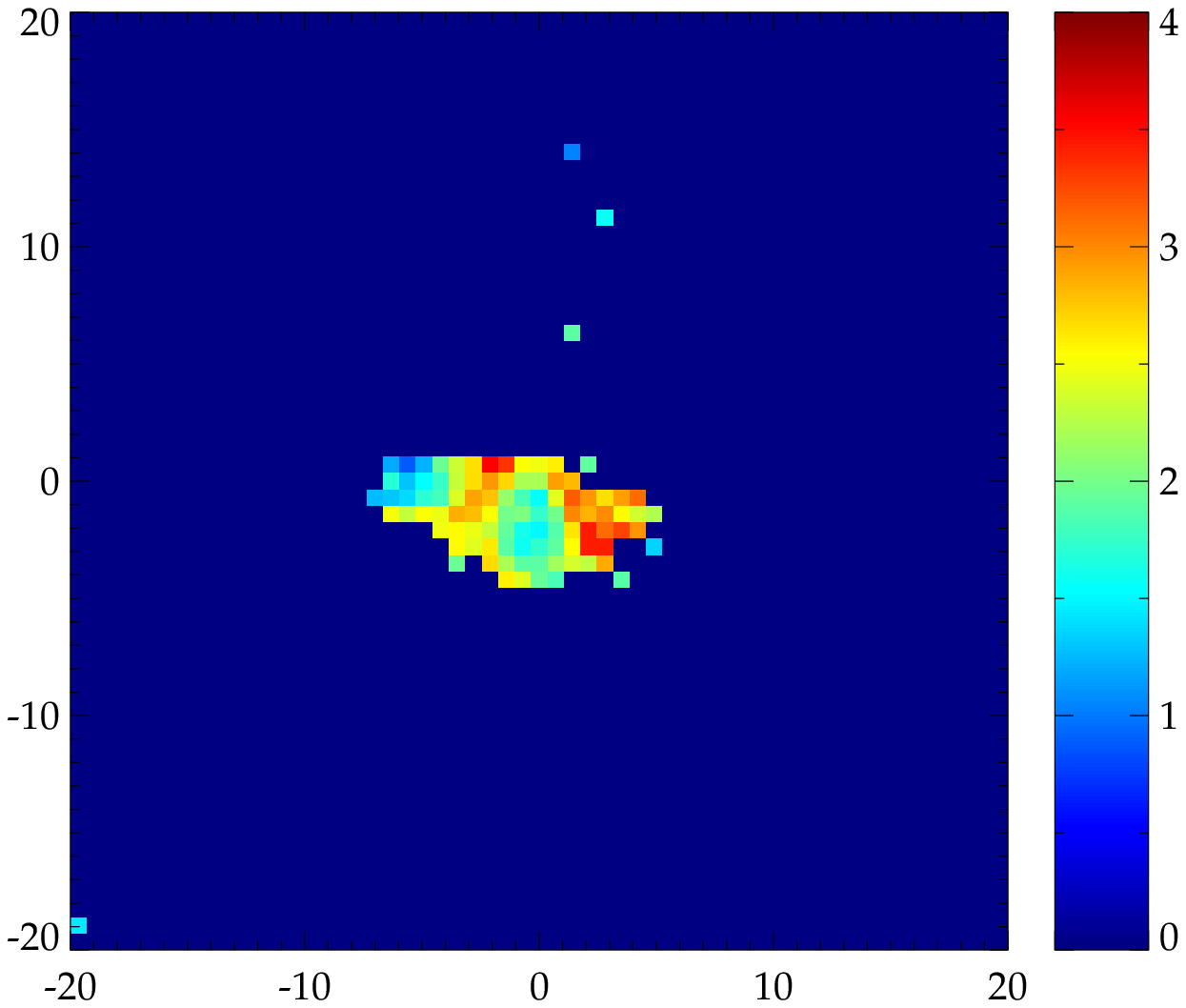}
\includegraphics[height=0.21\textwidth,clip]{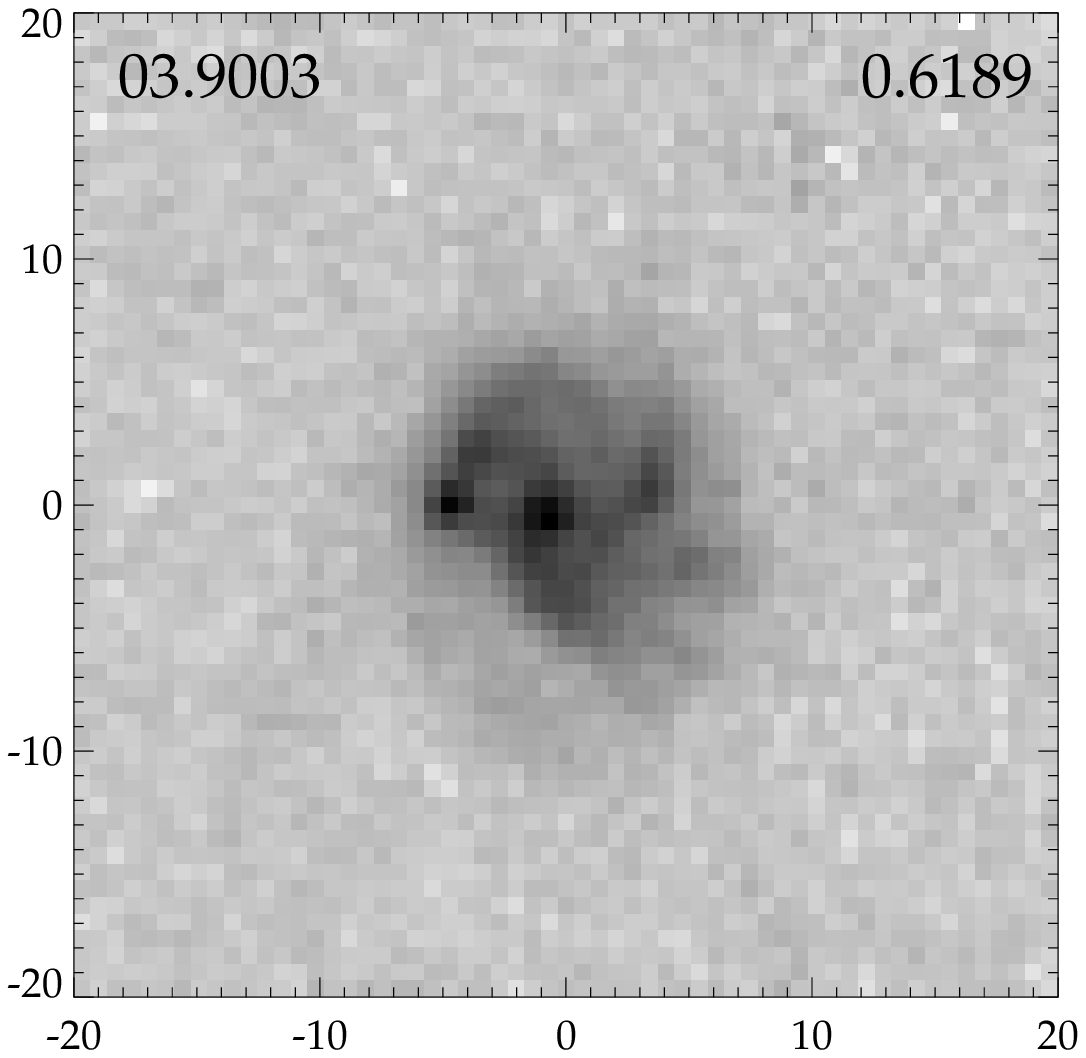}    \includegraphics[height=0.21\textwidth,clip]{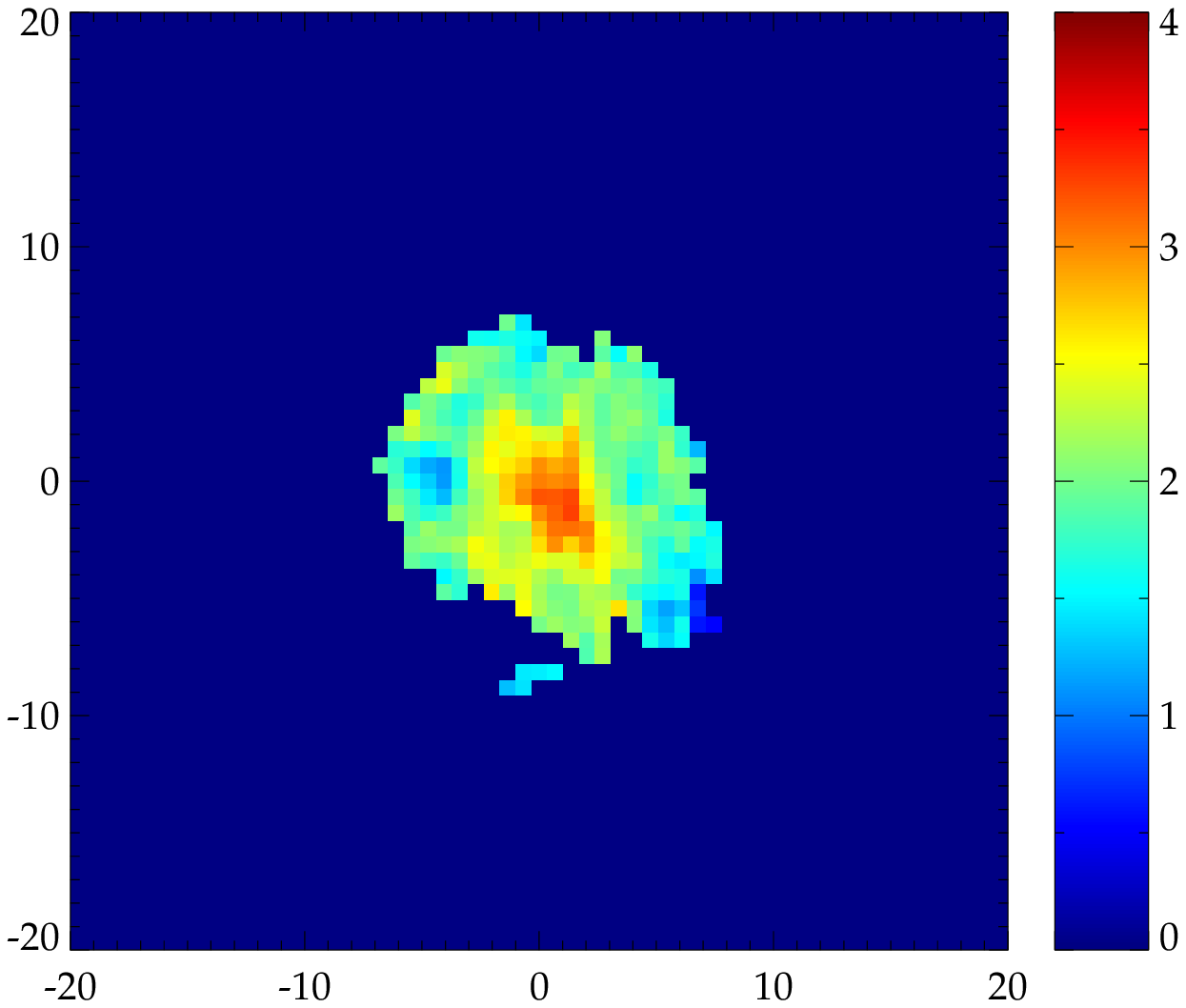}
\includegraphics[height=0.21\textwidth,clip]{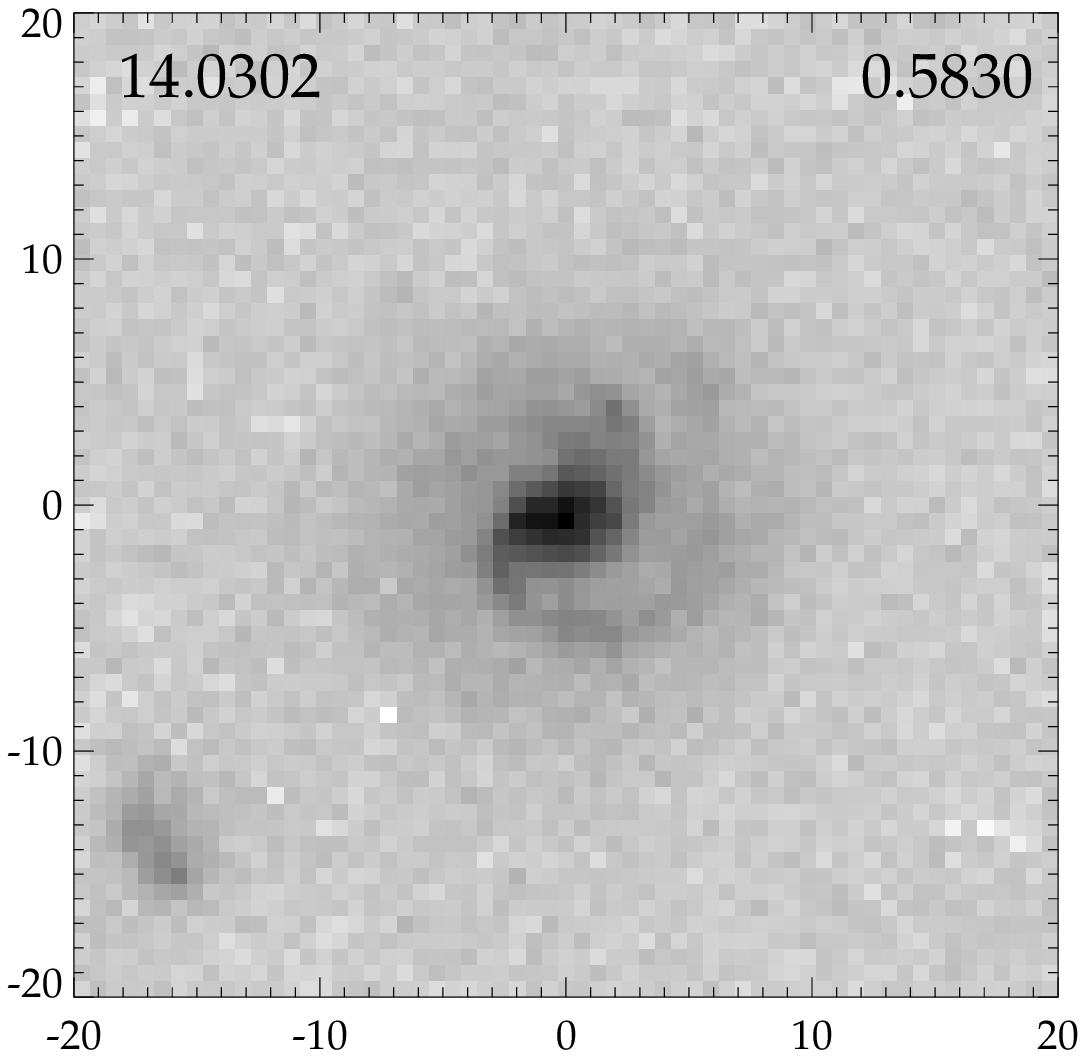}    \includegraphics[height=0.21\textwidth,clip]{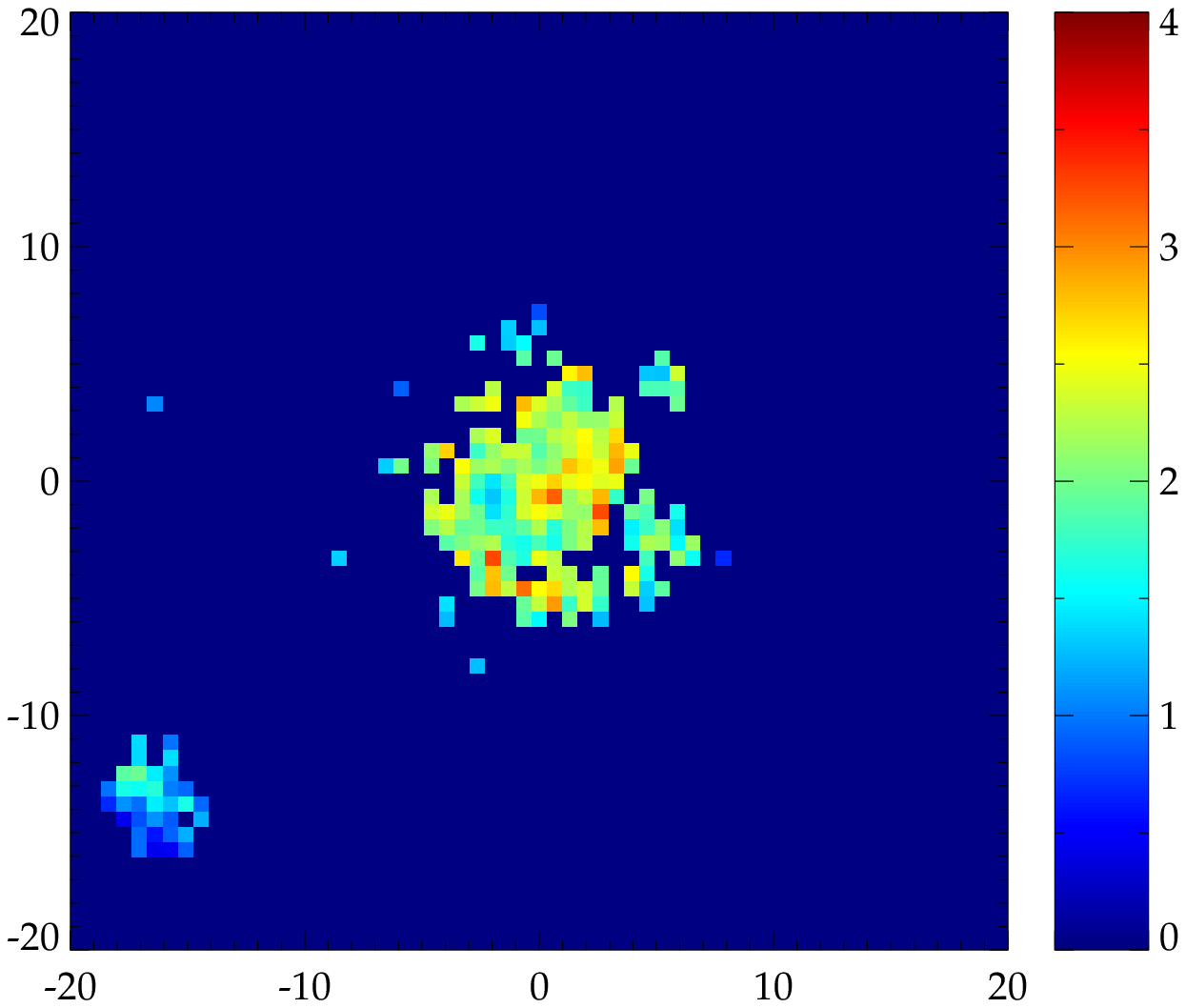}
\includegraphics[height=0.21\textwidth,clip]{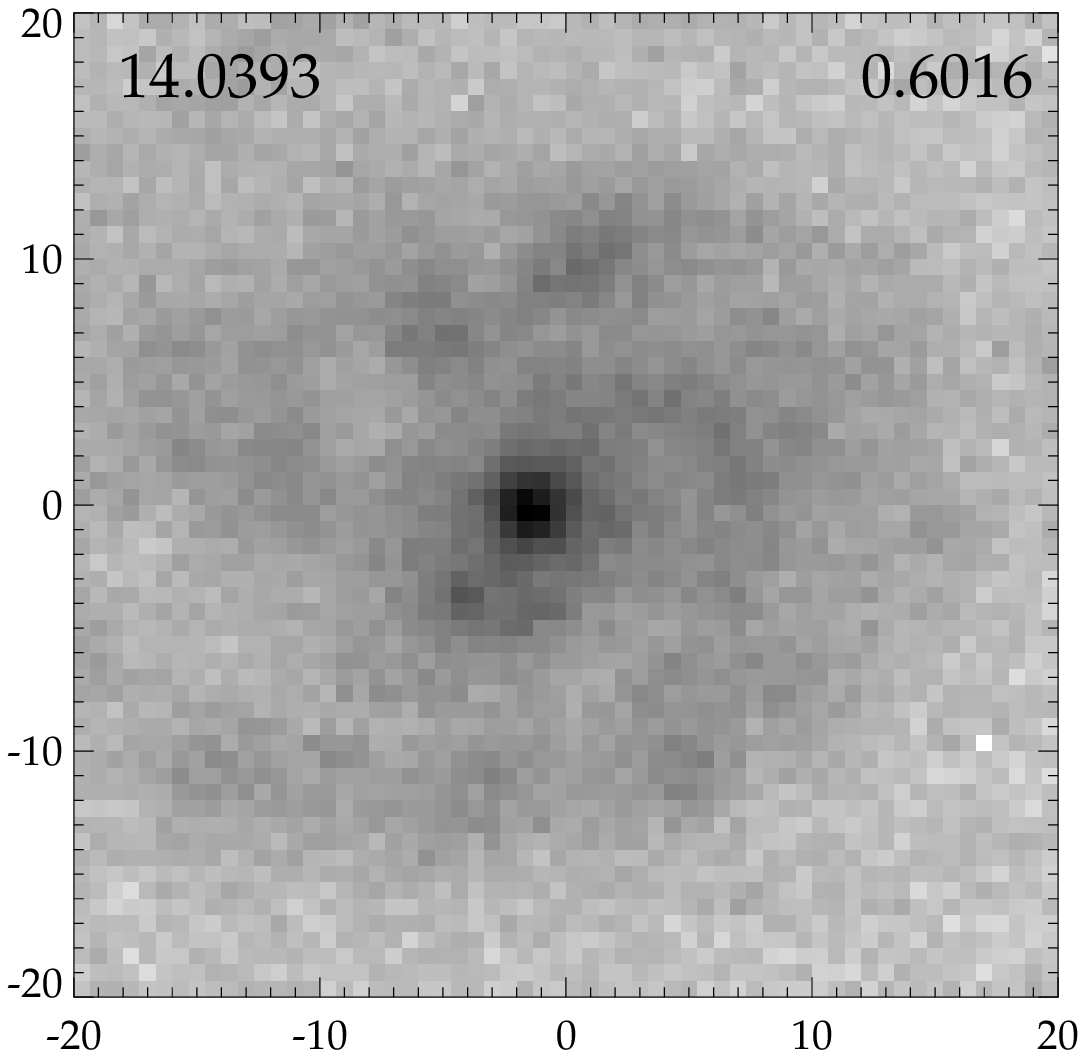}    \includegraphics[height=0.21\textwidth,clip]{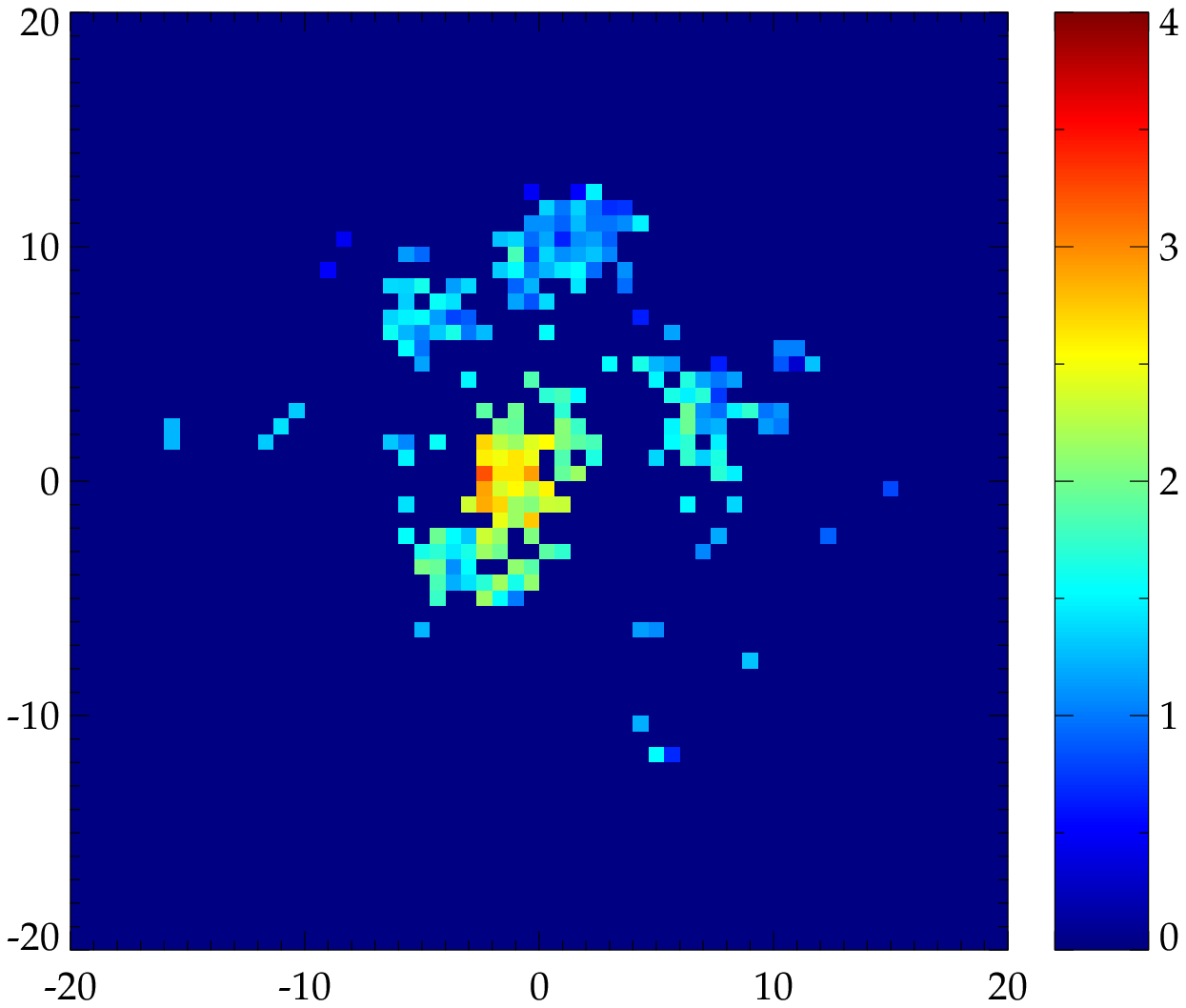}
\includegraphics[height=0.21\textwidth,clip]{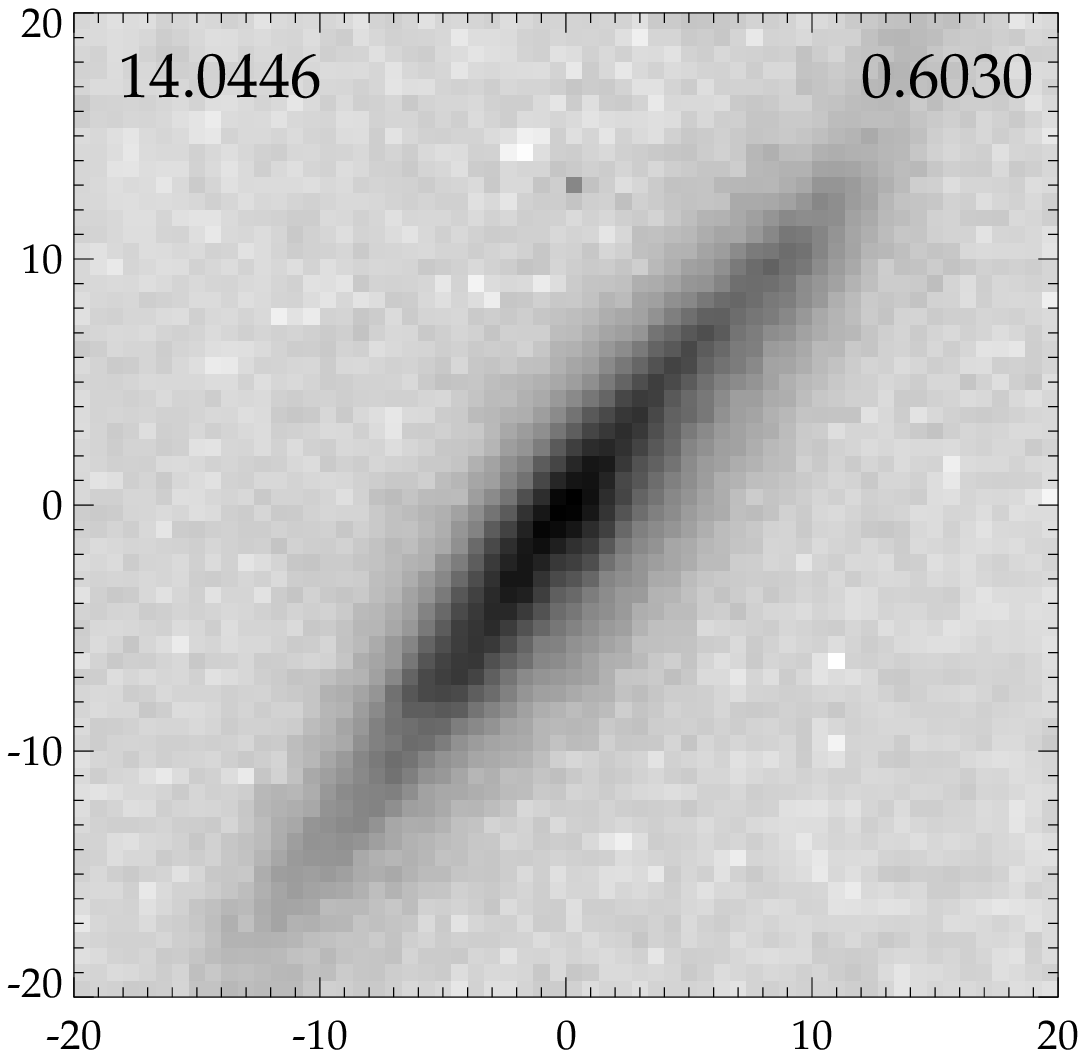}    \includegraphics[height=0.21\textwidth,clip]{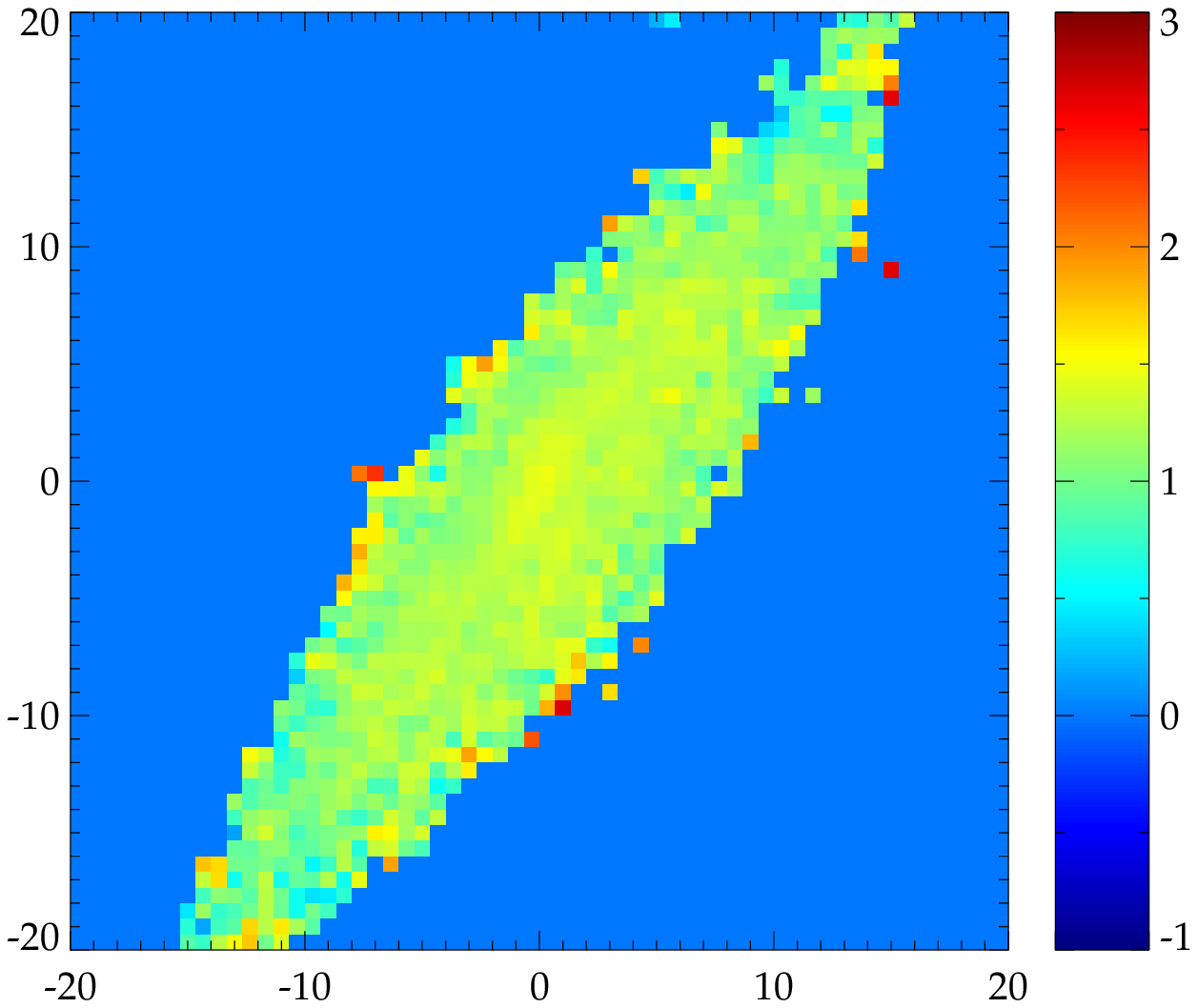}
\includegraphics[height=0.21\textwidth,clip]{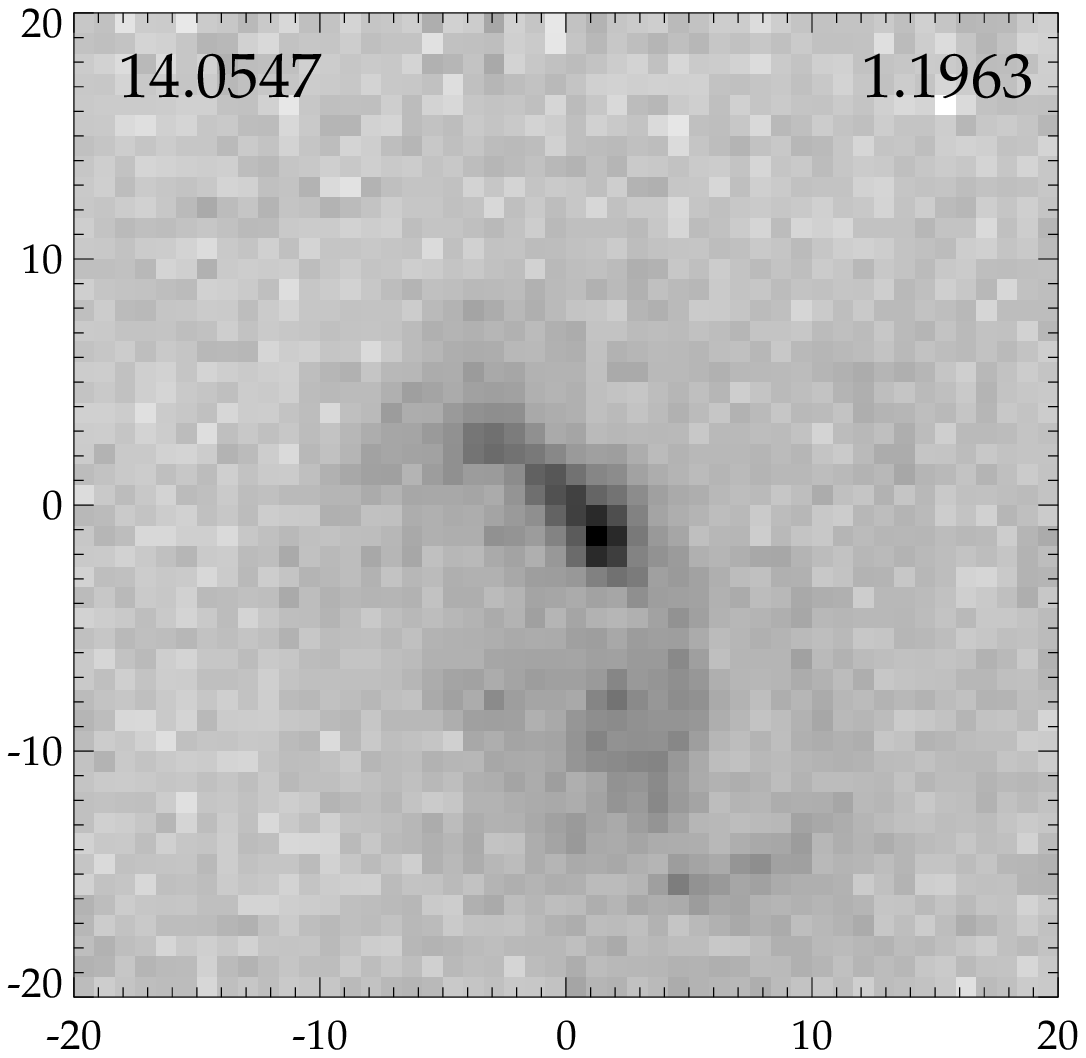}    \includegraphics[height=0.21\textwidth,clip]{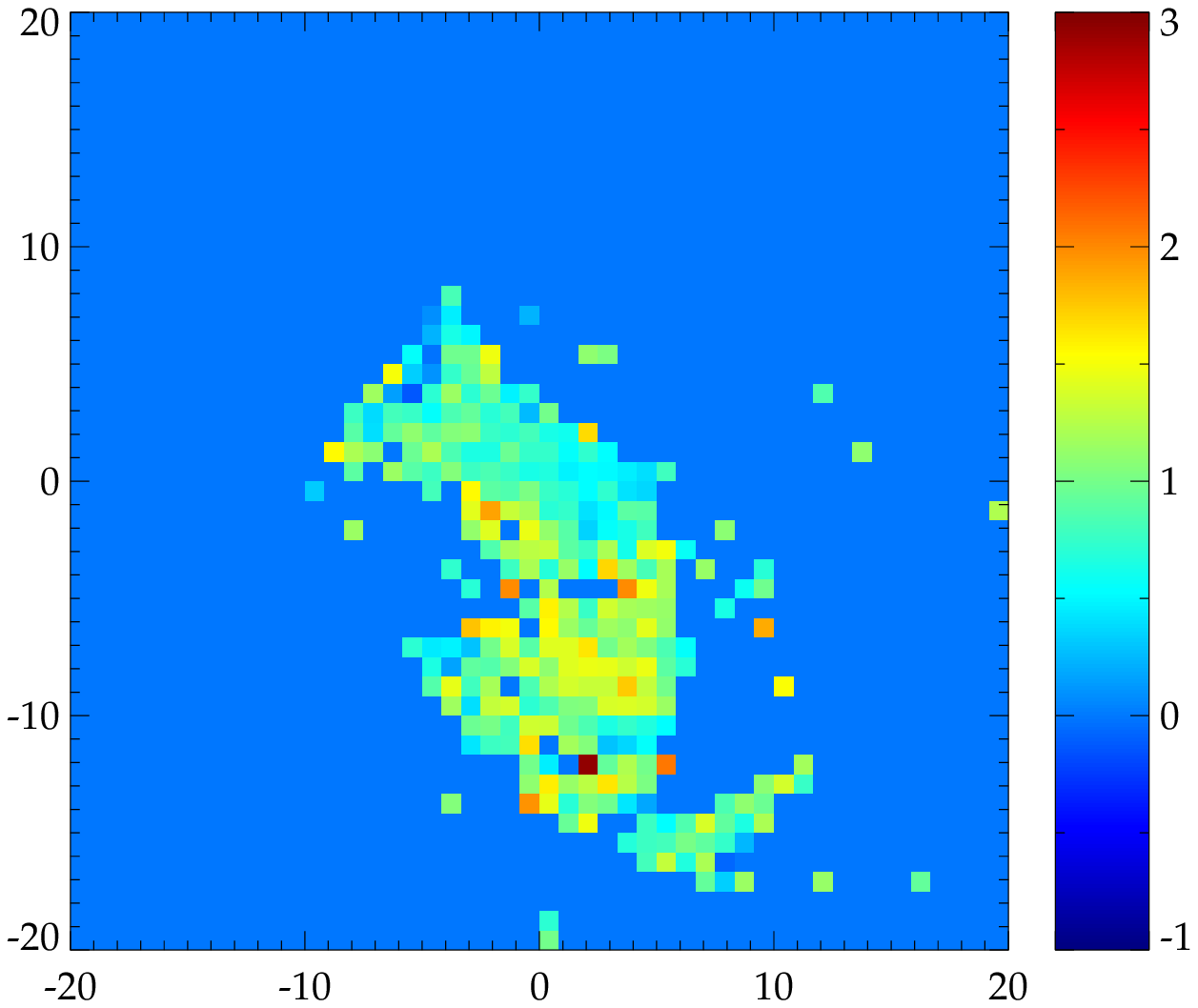}
\includegraphics[height=0.21\textwidth,clip]{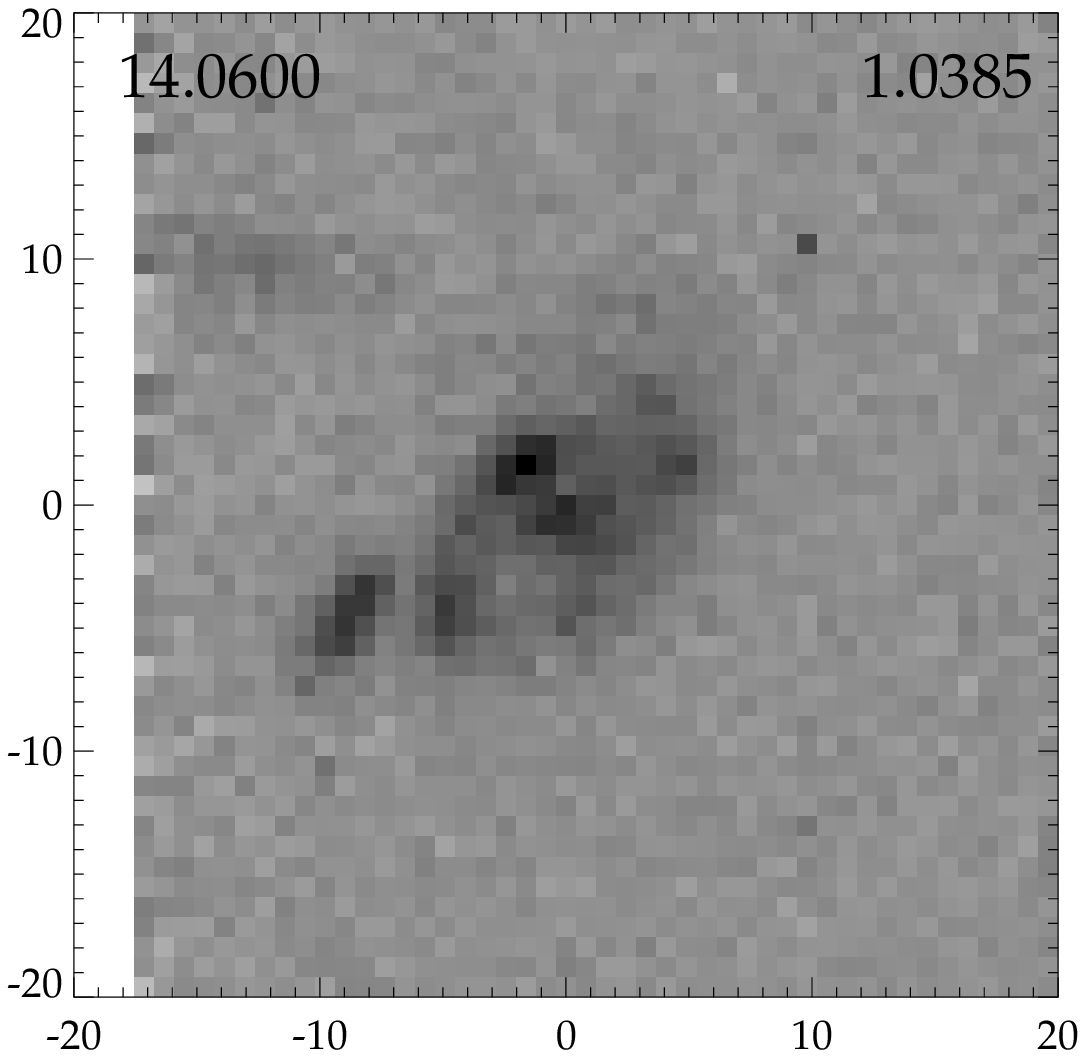}    \includegraphics[height=0.21\textwidth,clip]{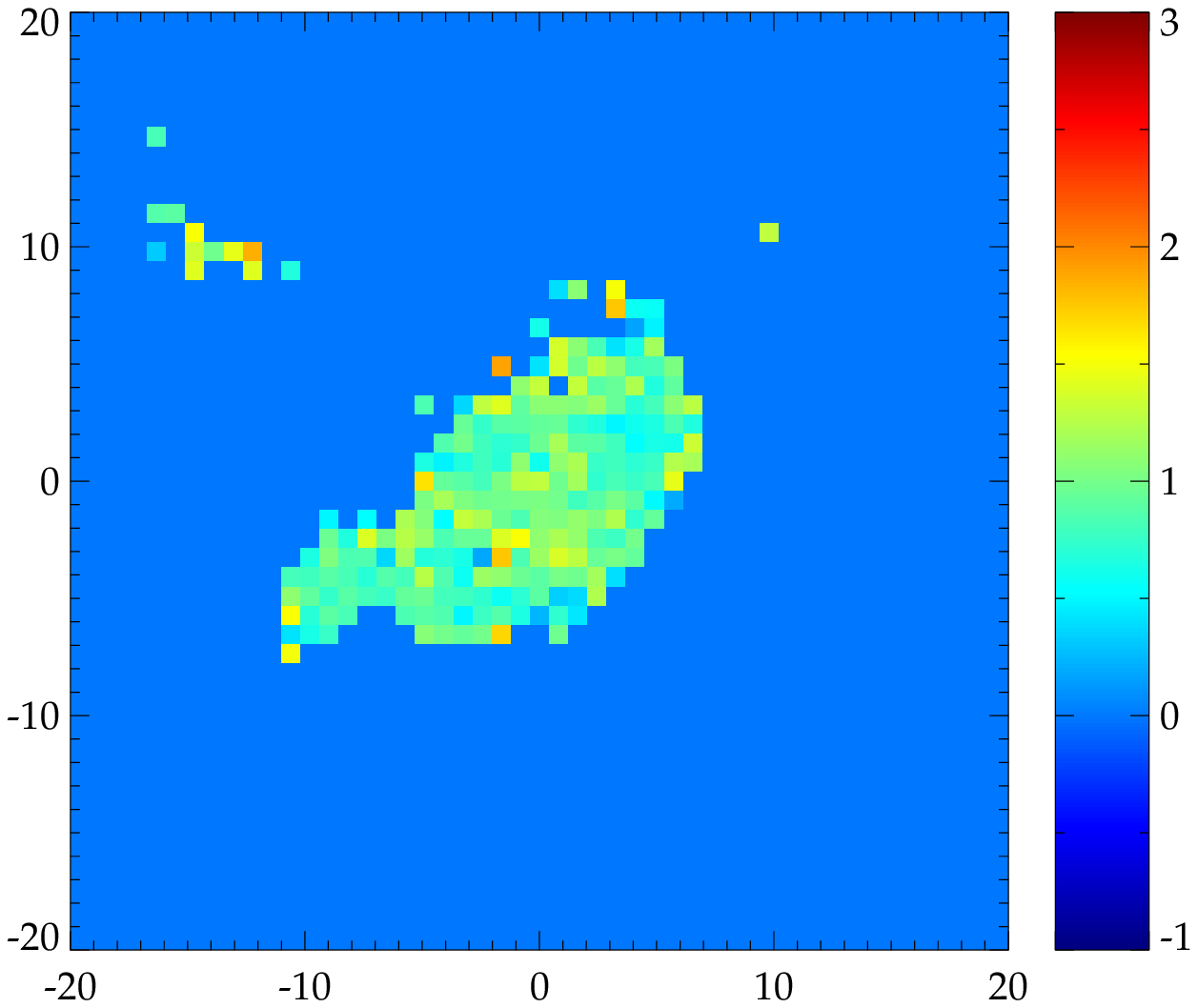}
\includegraphics[height=0.21\textwidth,clip]{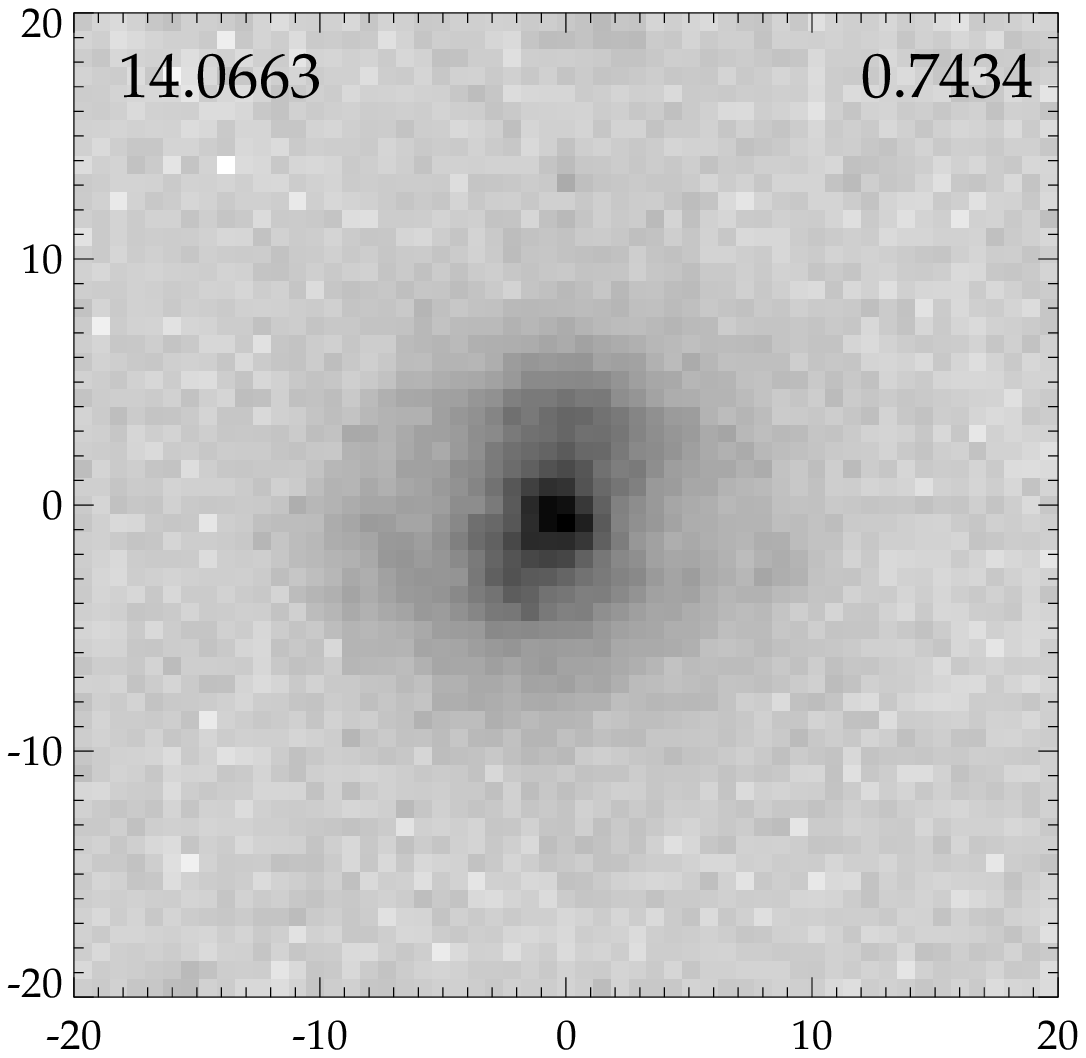}    \includegraphics[height=0.21\textwidth,clip]{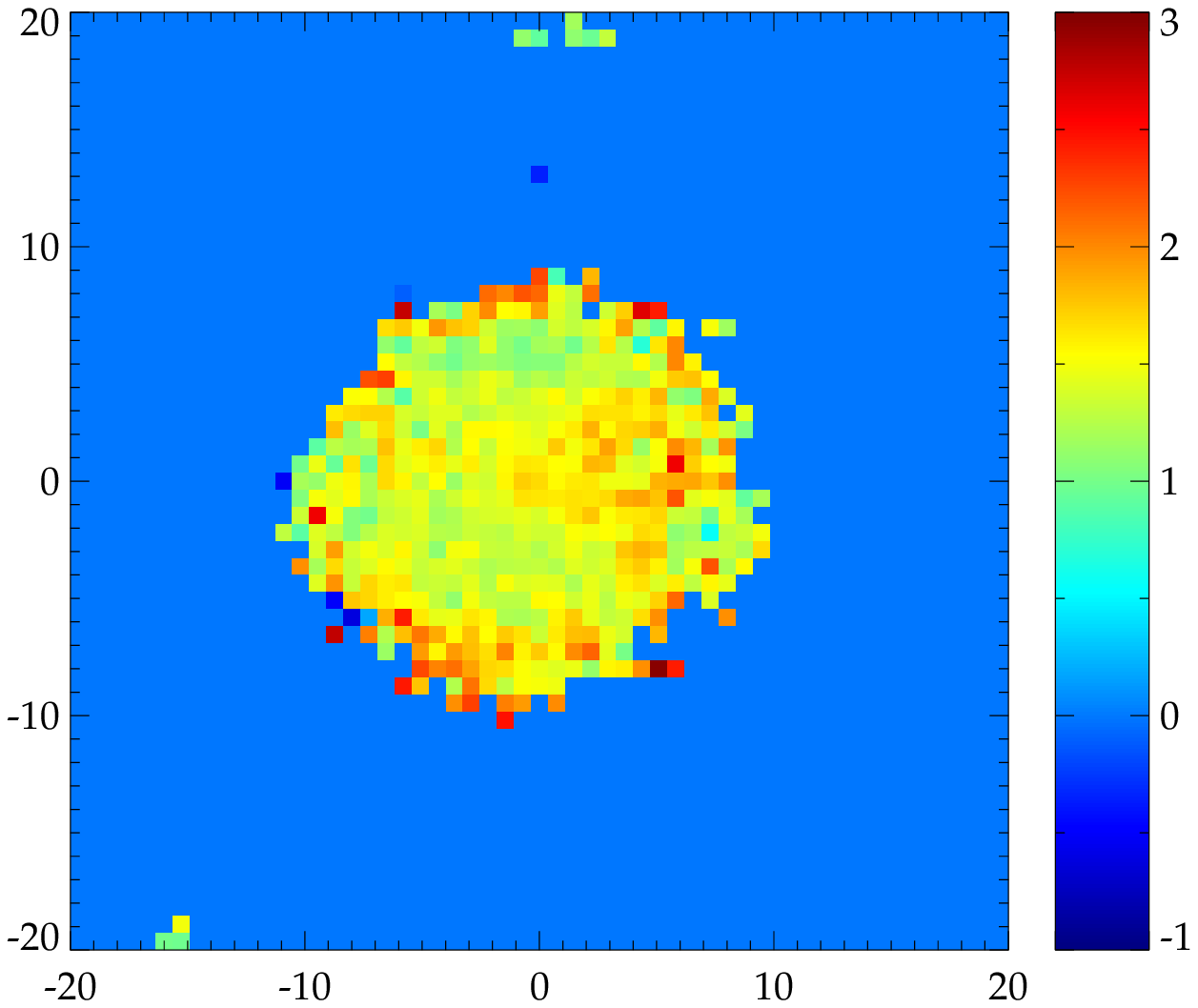}

\caption{Continued.} \end{figure*}

\addtocounter{figure}{-1}
\begin{figure*}[] \centering

\includegraphics[height=0.21\textwidth,clip]{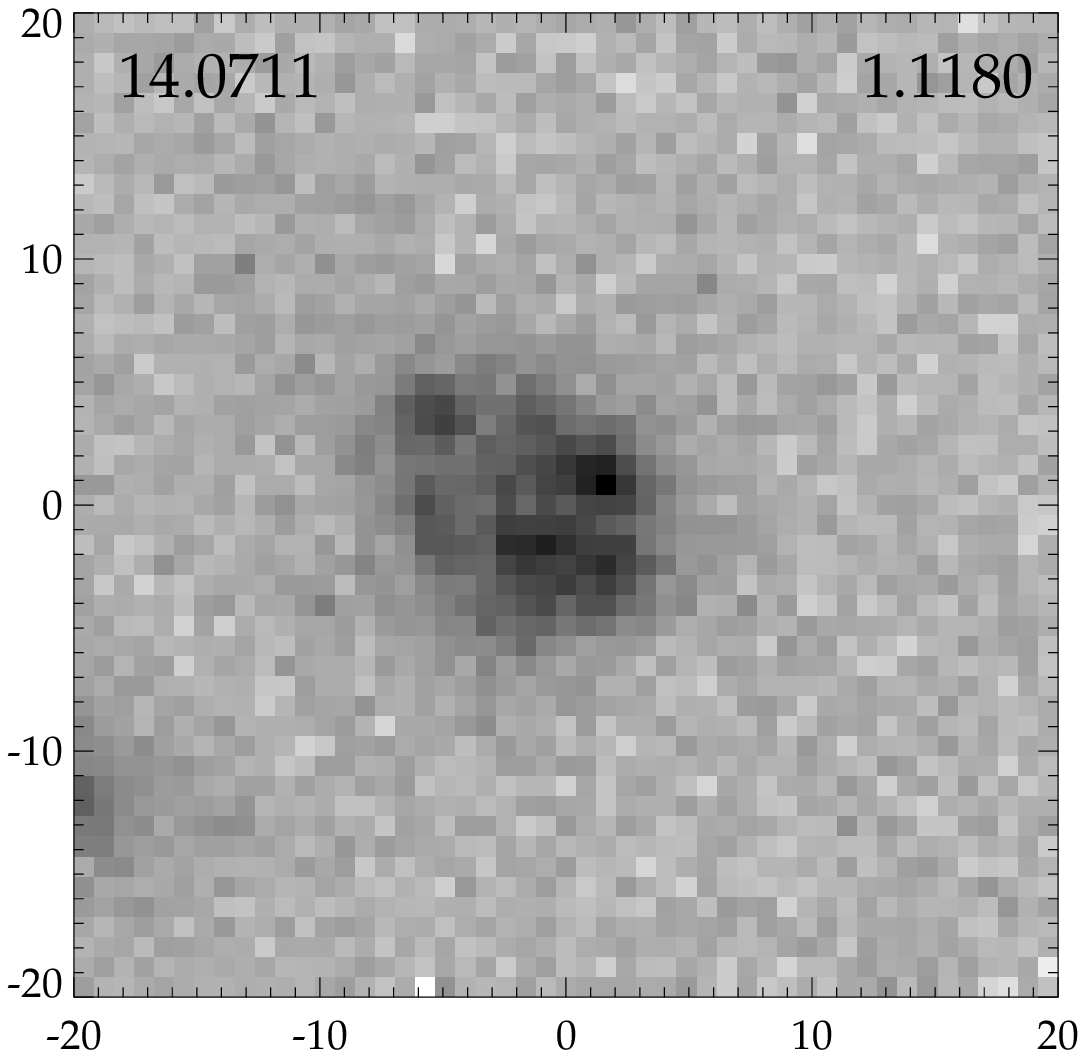} \includegraphics[height=0.21\textwidth,clip]{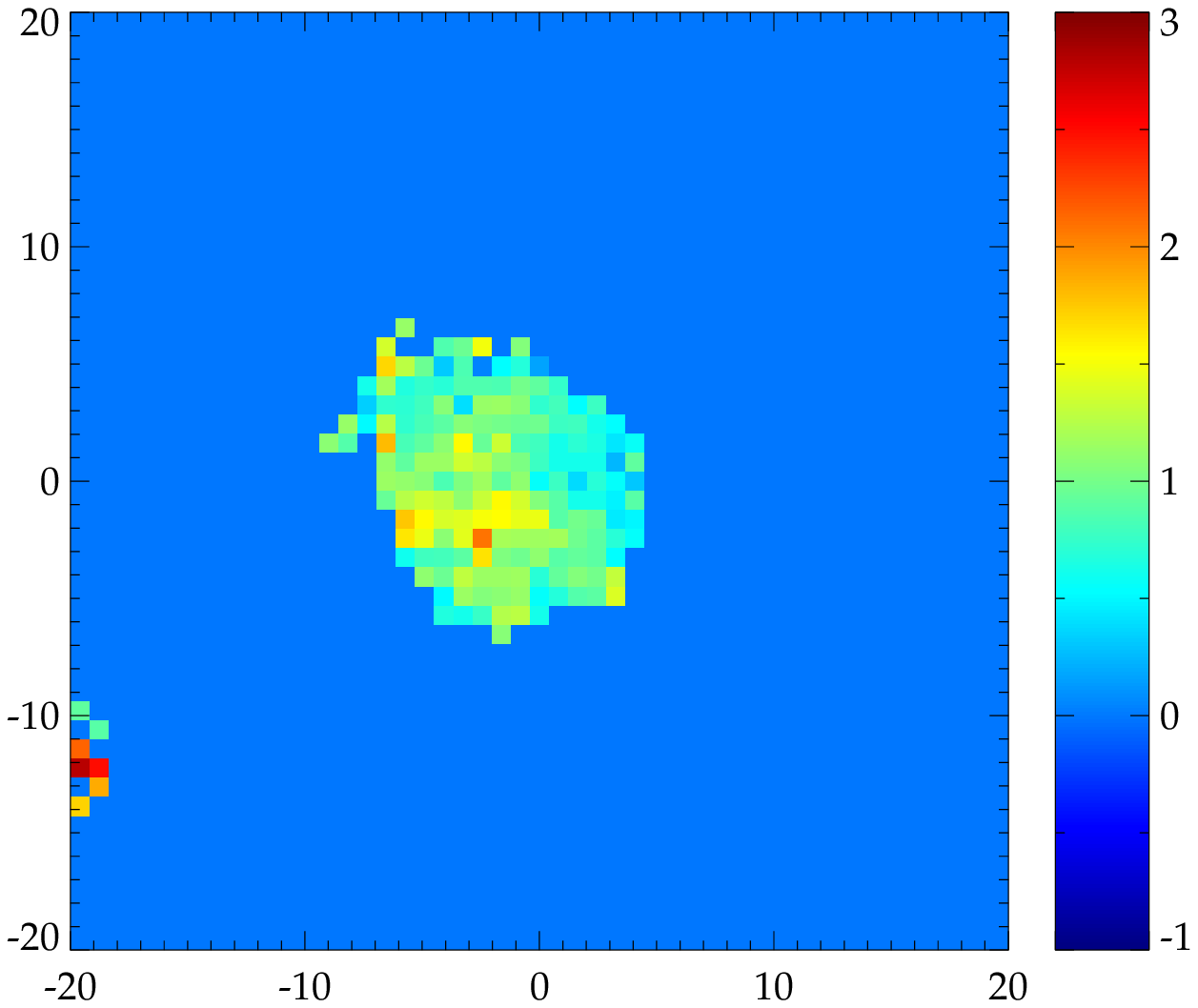}
\includegraphics[height=0.21\textwidth,clip]{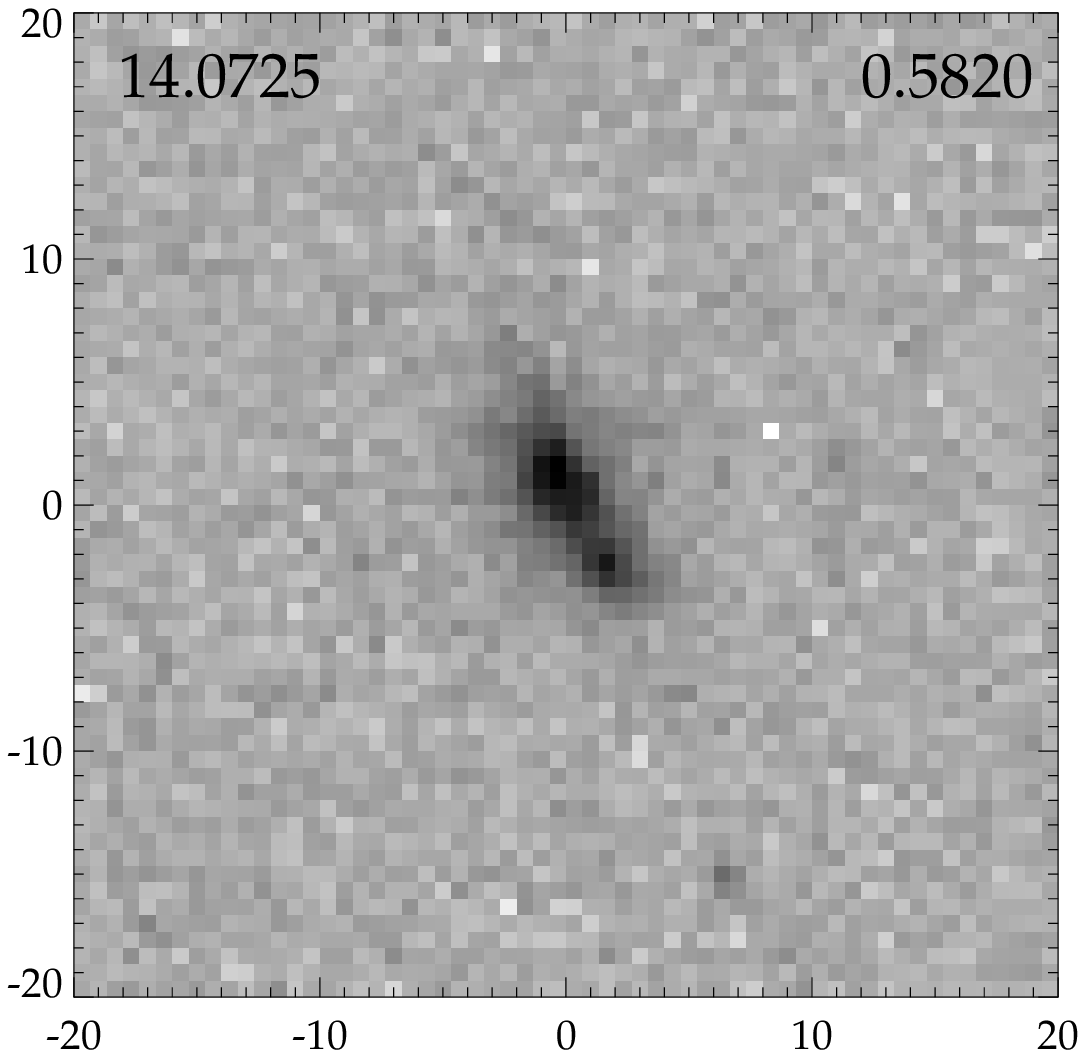}    \includegraphics[height=0.21\textwidth,clip]{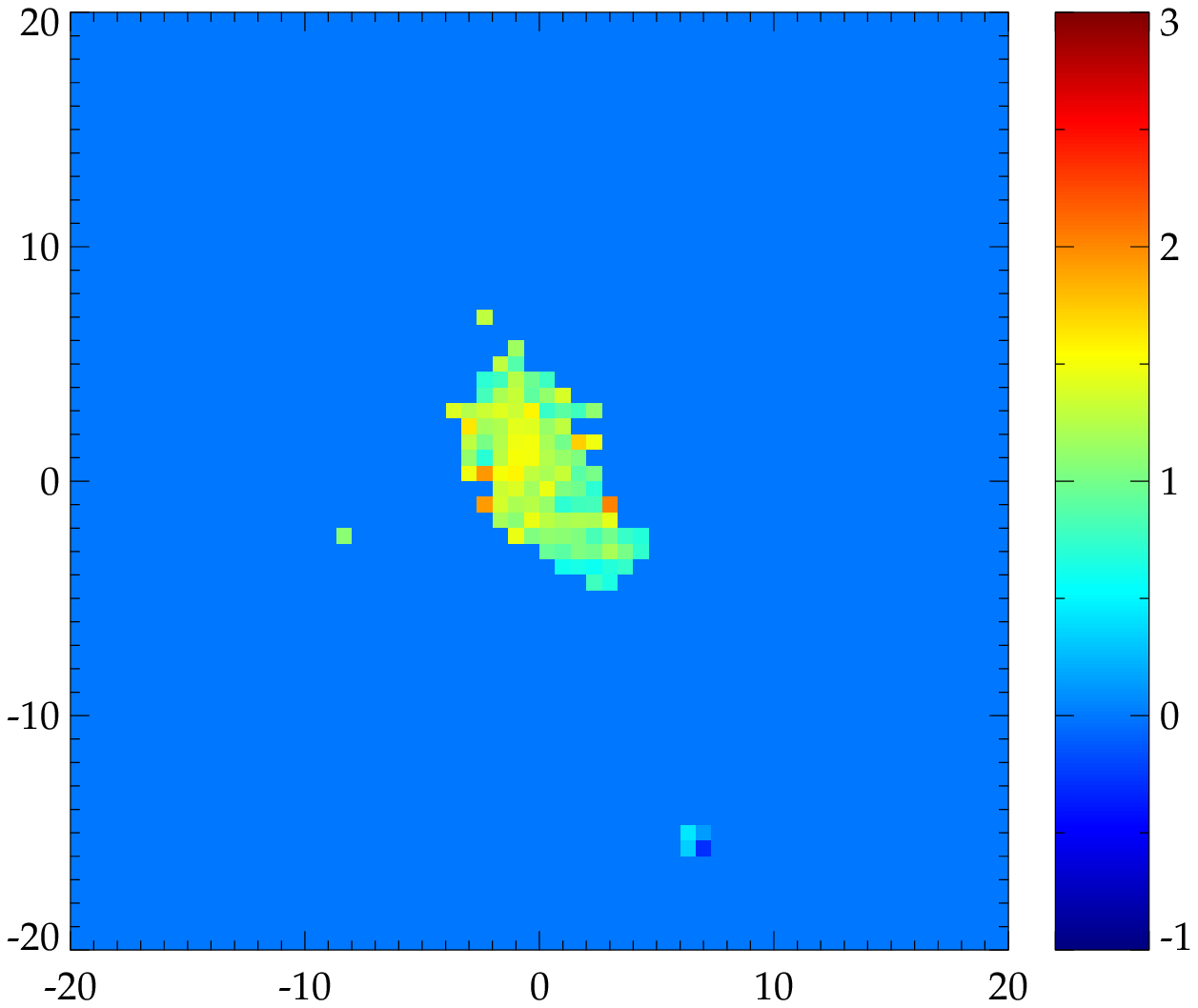}
\includegraphics[height=0.21\textwidth,clip]{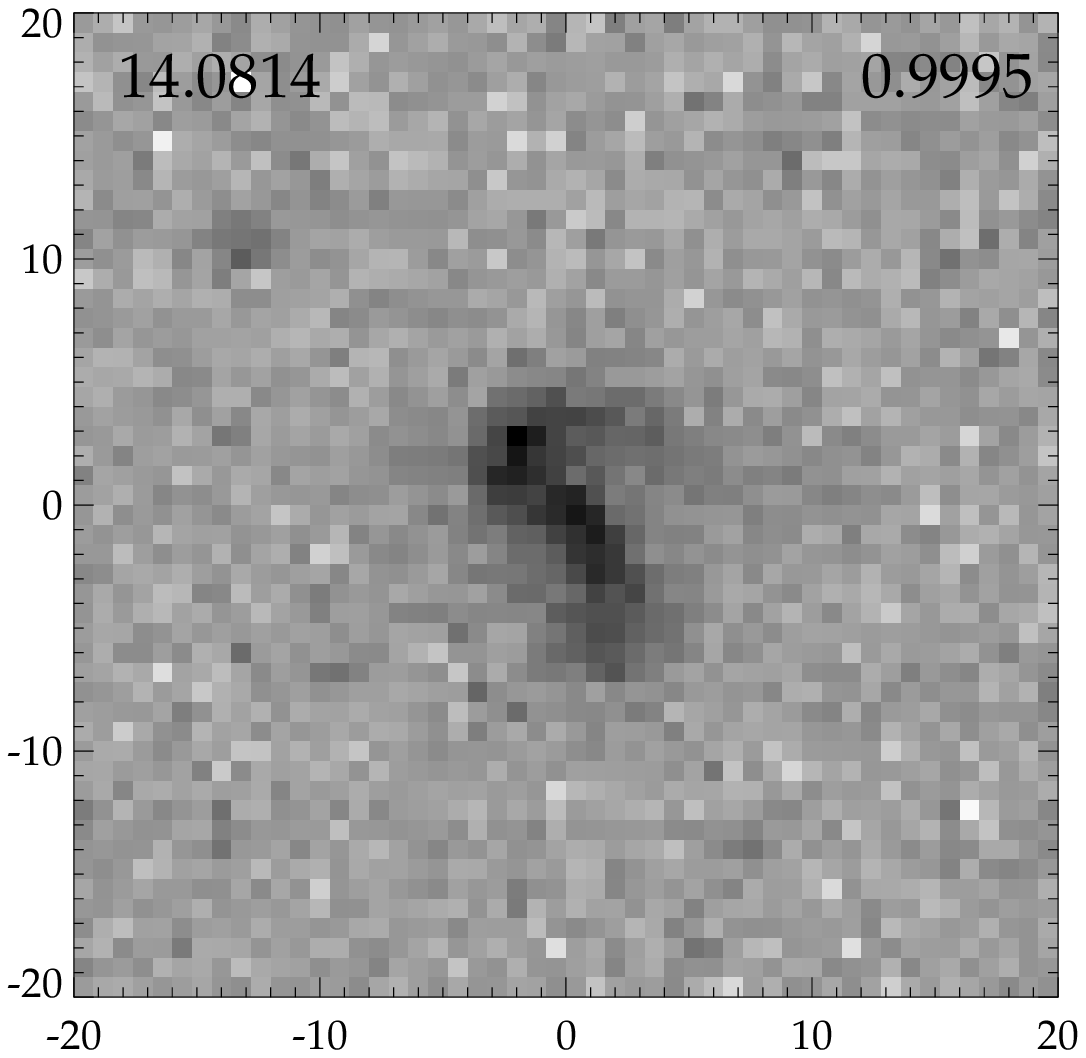}    \includegraphics[height=0.21\textwidth,clip]{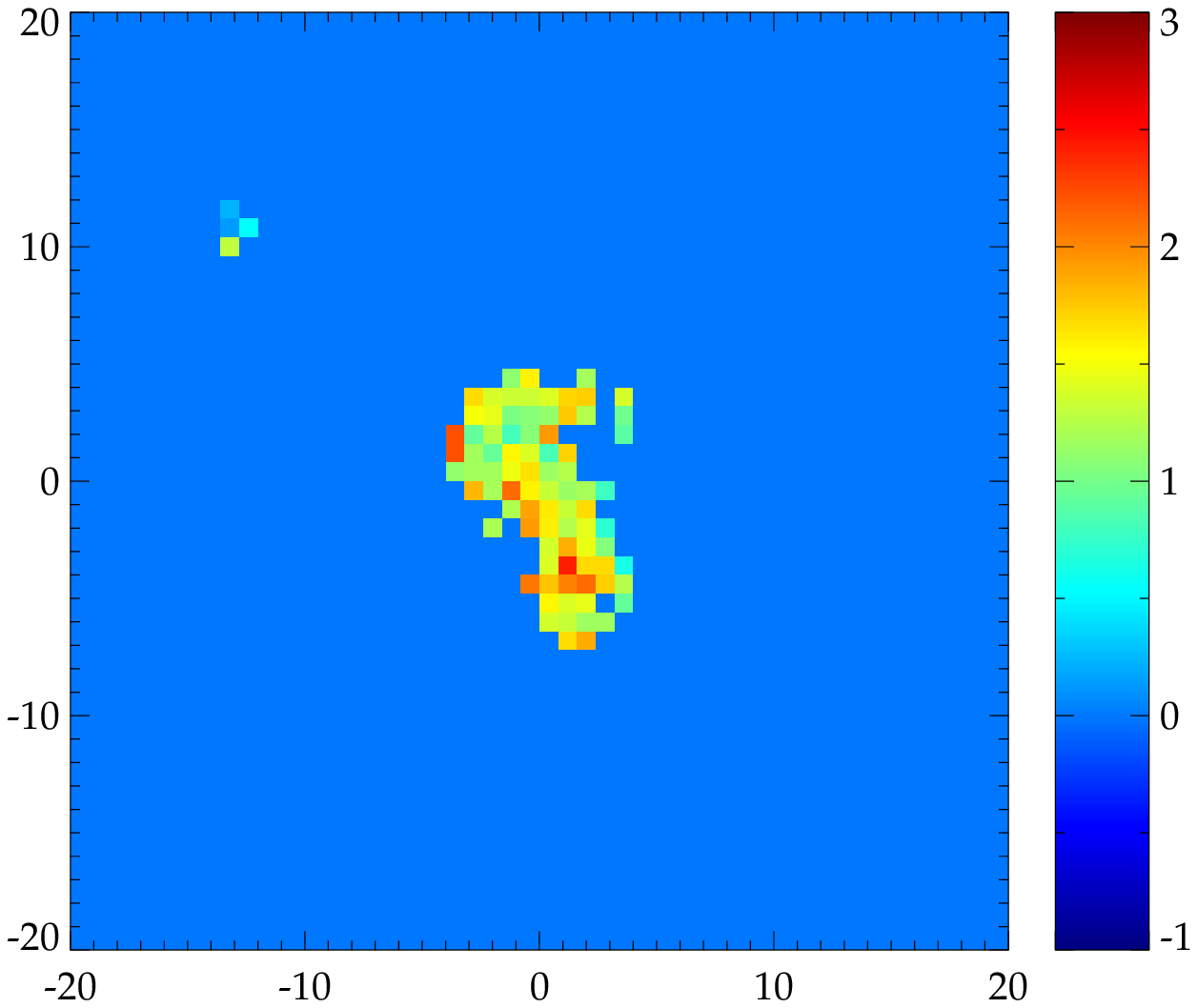}
\includegraphics[height=0.21\textwidth,clip]{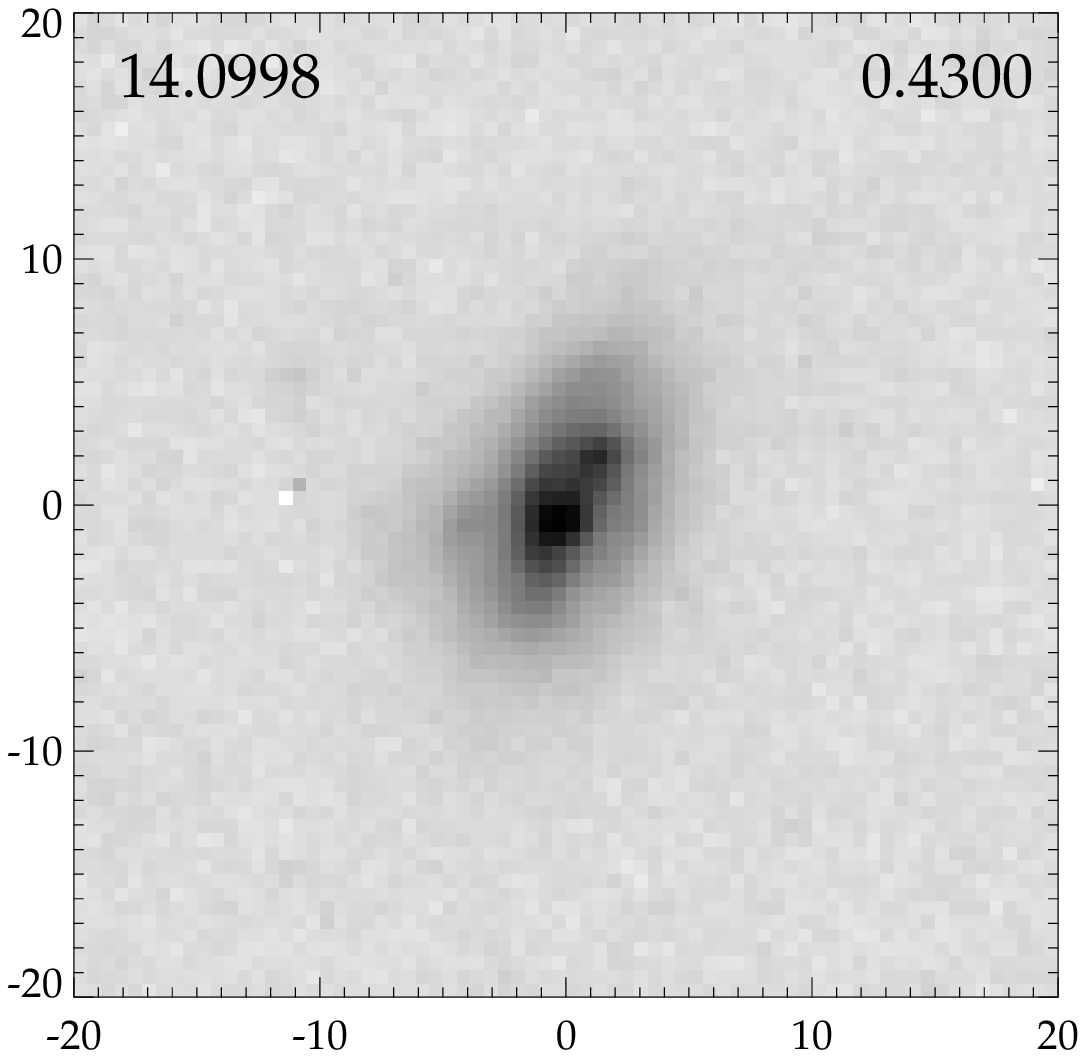}   \includegraphics[height=0.21\textwidth,clip]{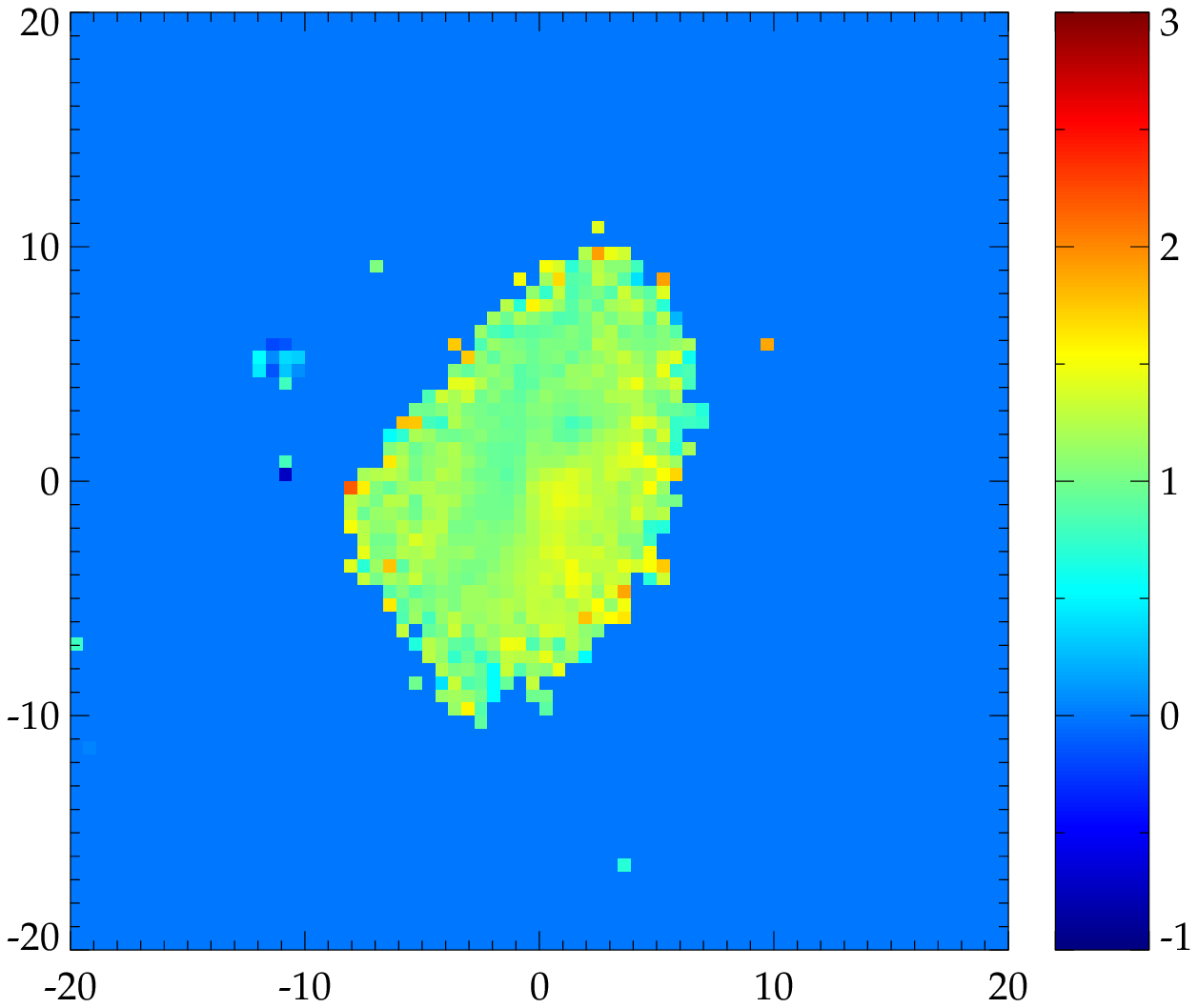}
\includegraphics[height=0.21\textwidth,clip]{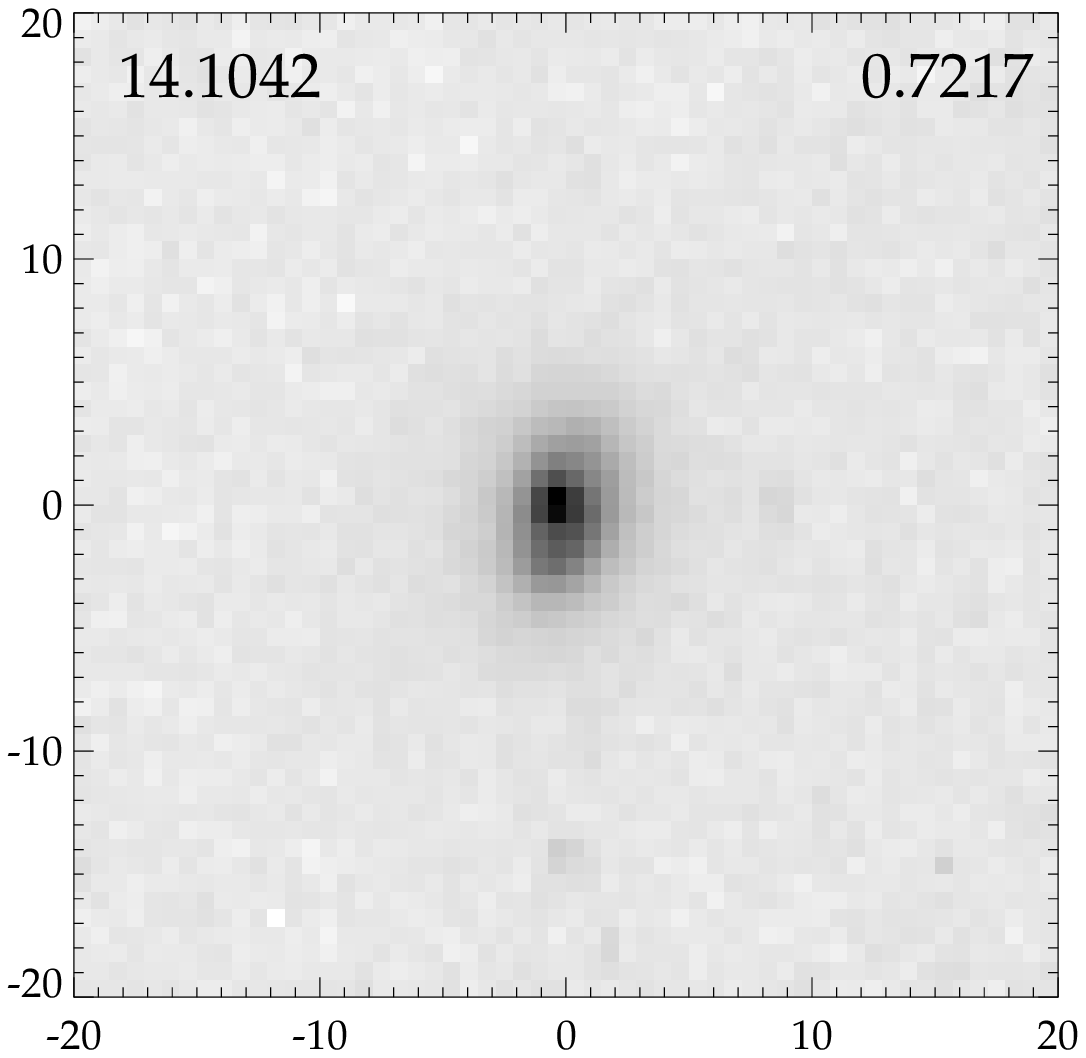}   \includegraphics[height=0.21\textwidth,clip]{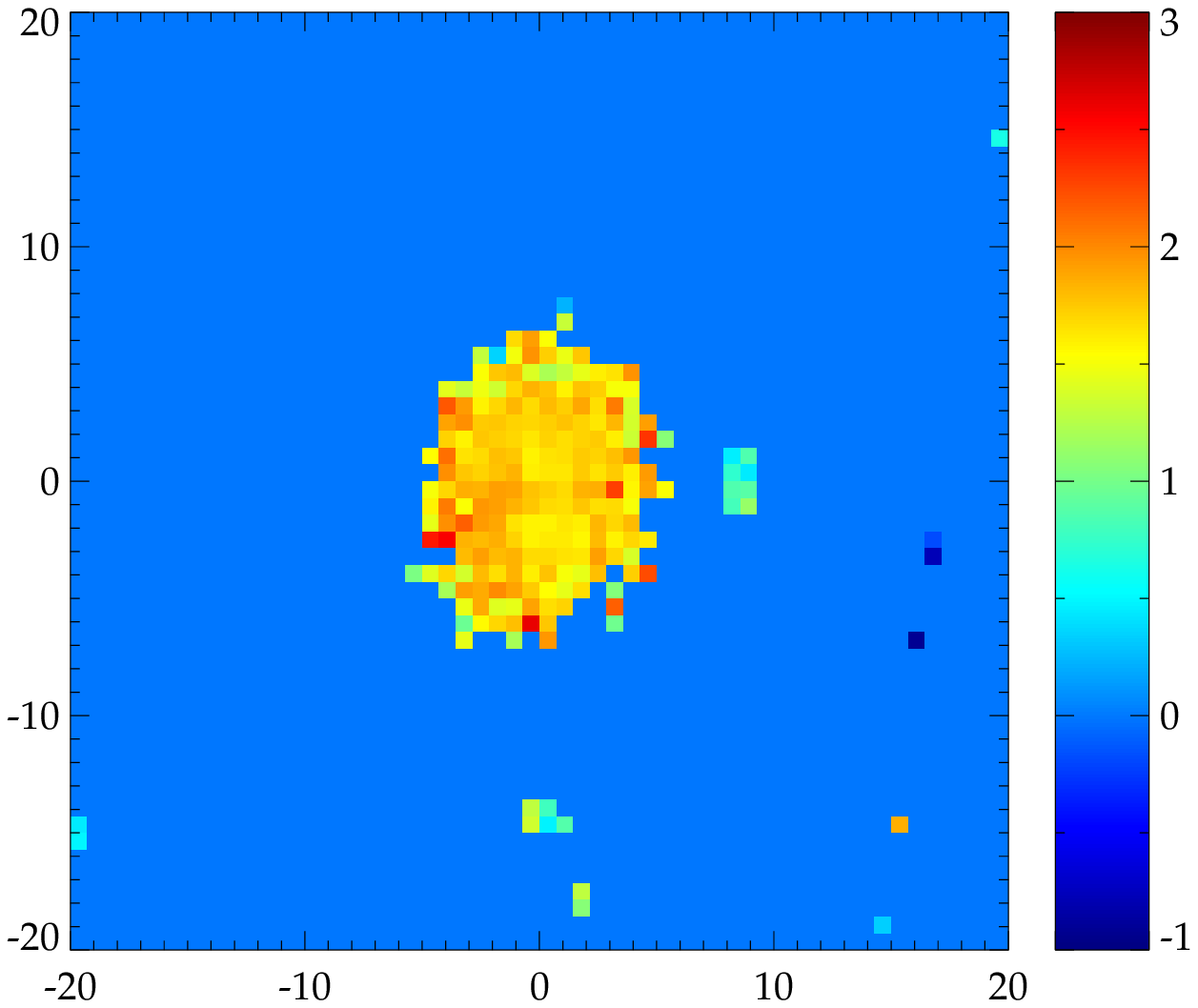}
\includegraphics[height=0.21\textwidth,clip]{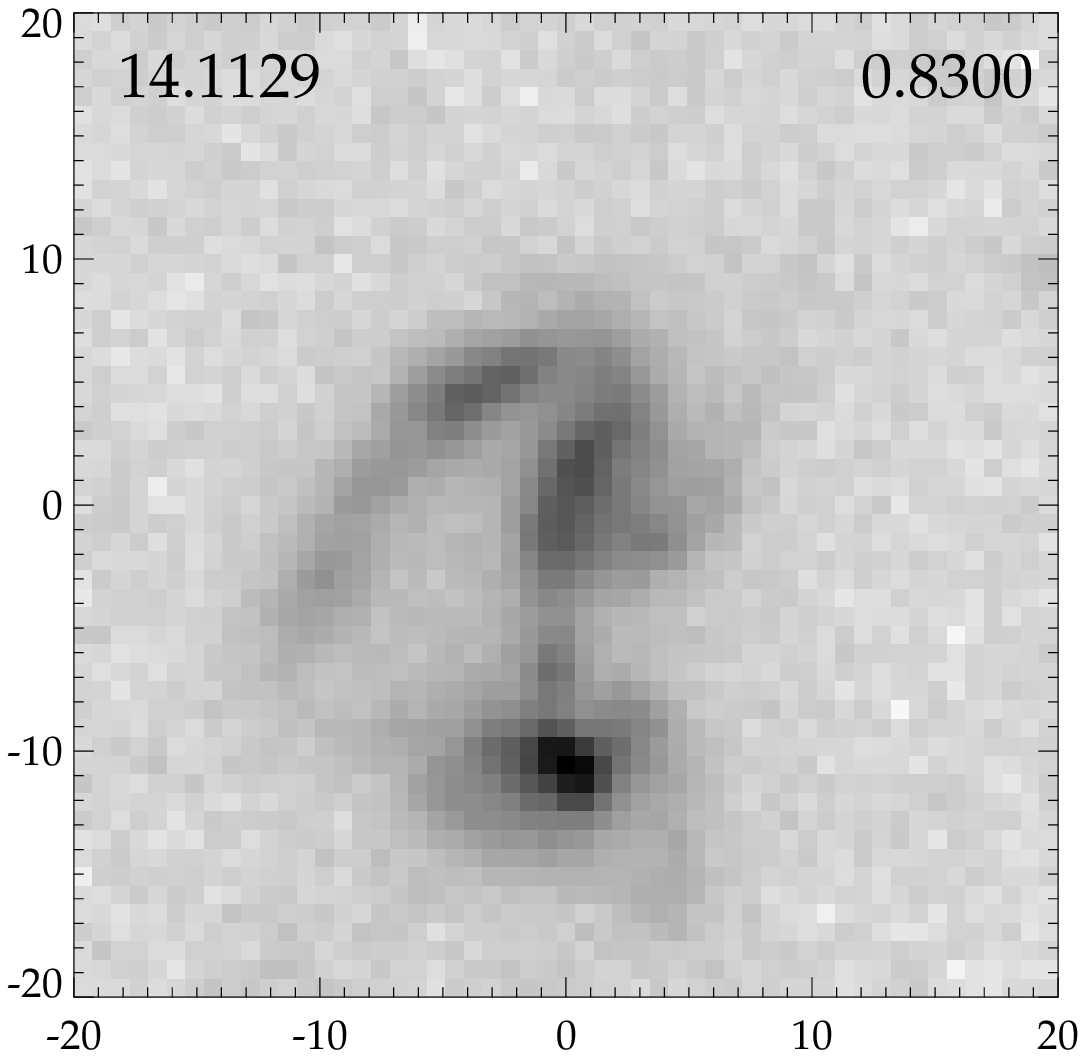}   \includegraphics[height=0.21\textwidth,clip]{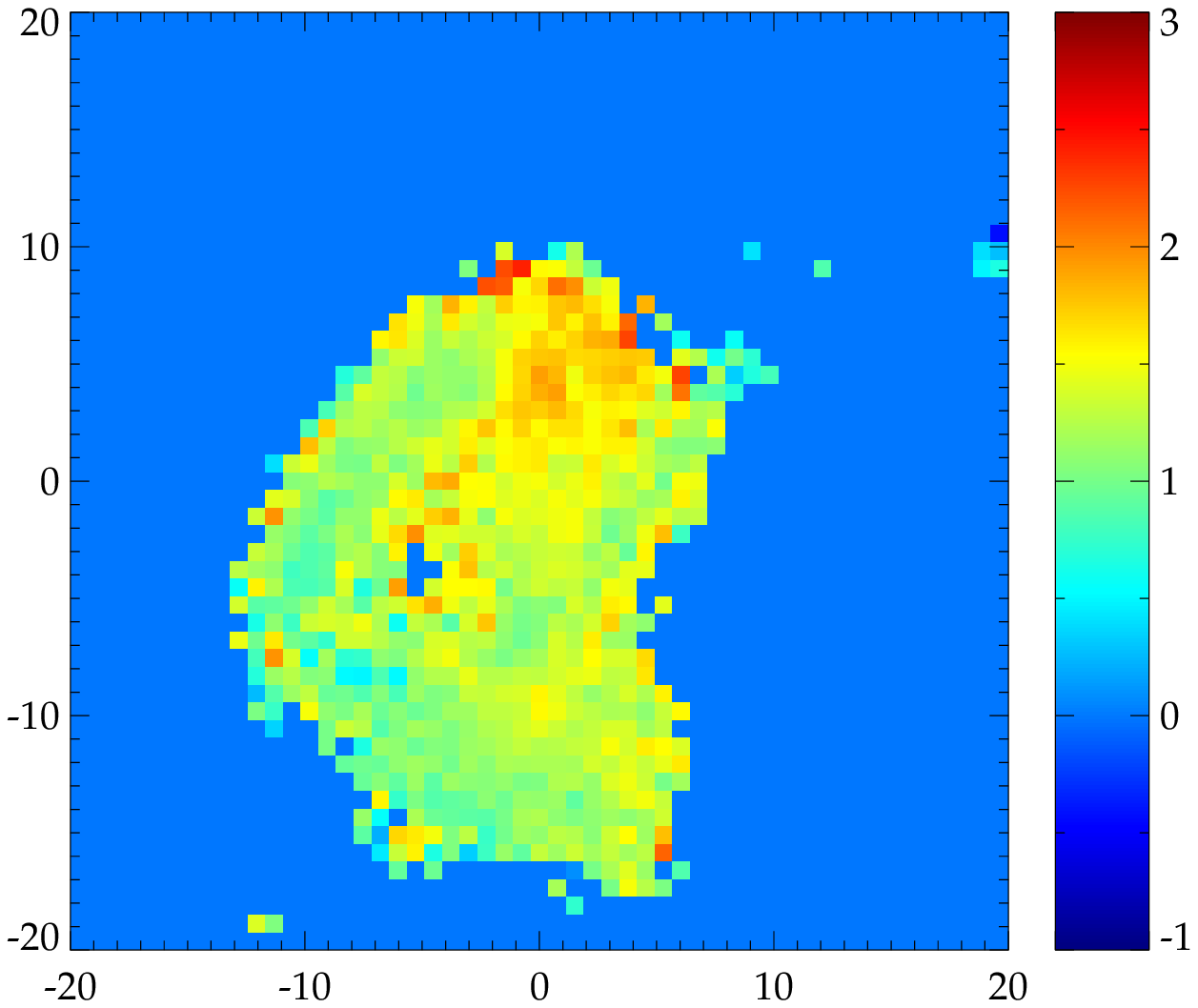}
\includegraphics[height=0.21\textwidth,clip]{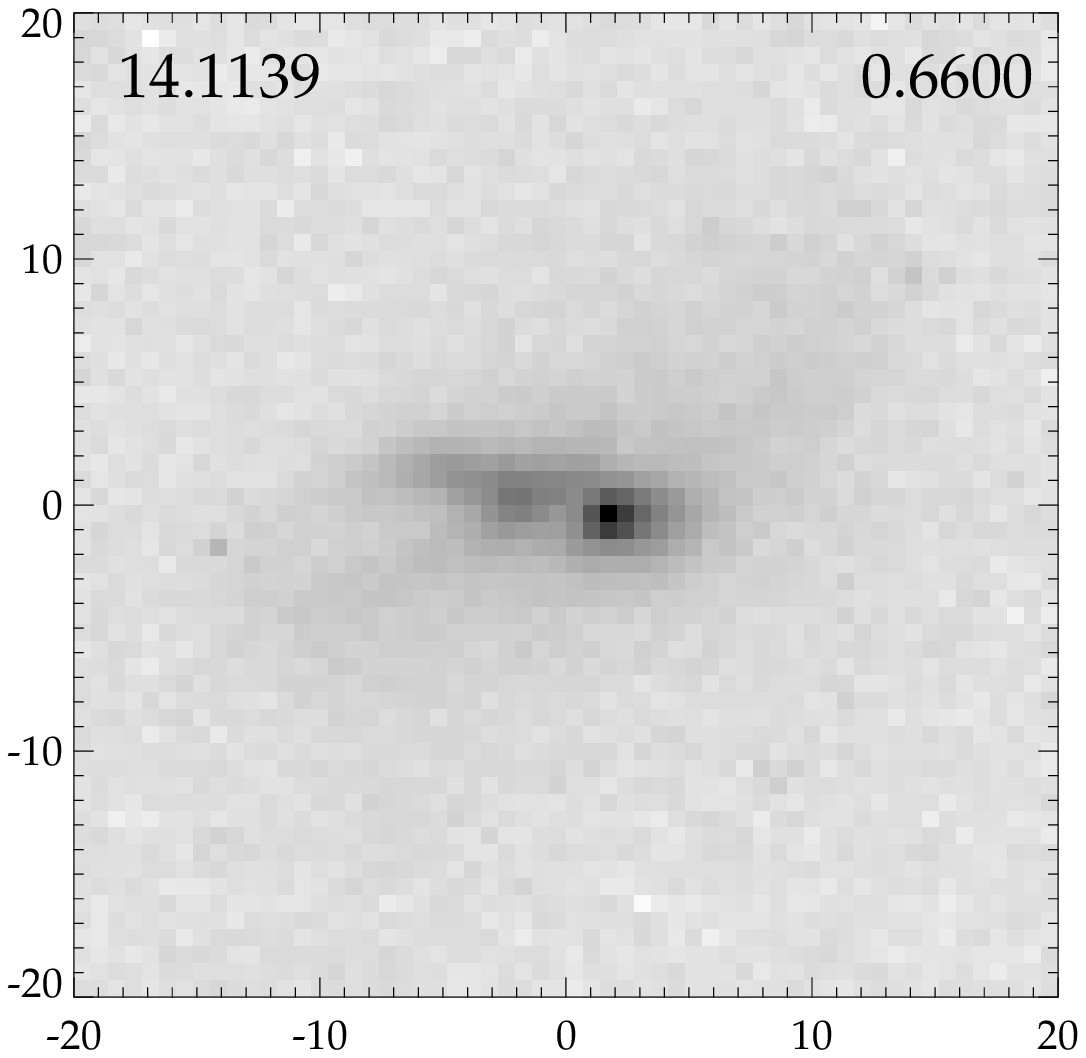} \includegraphics[height=0.21\textwidth,clip]{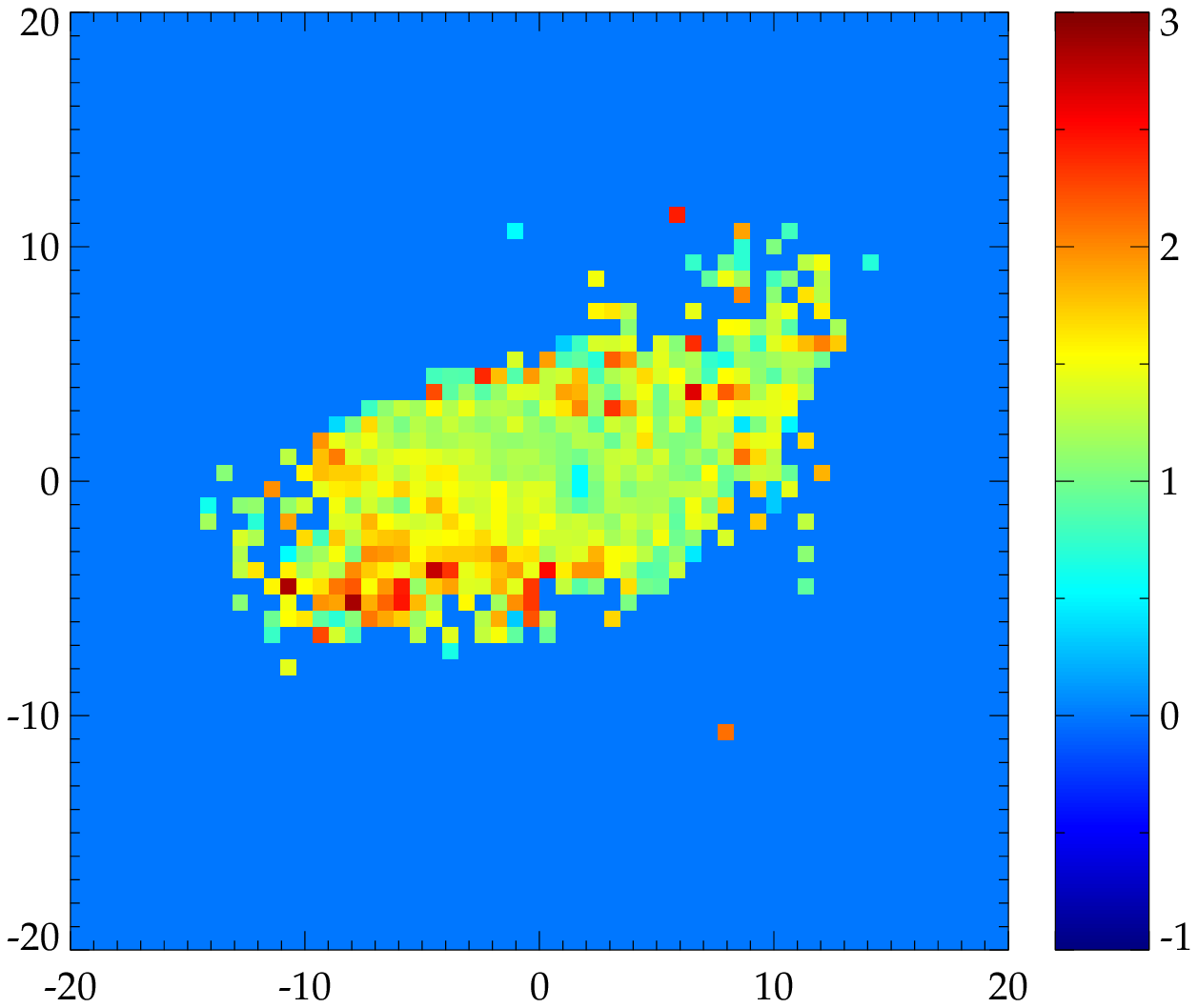}
\includegraphics[height=0.21\textwidth,clip]{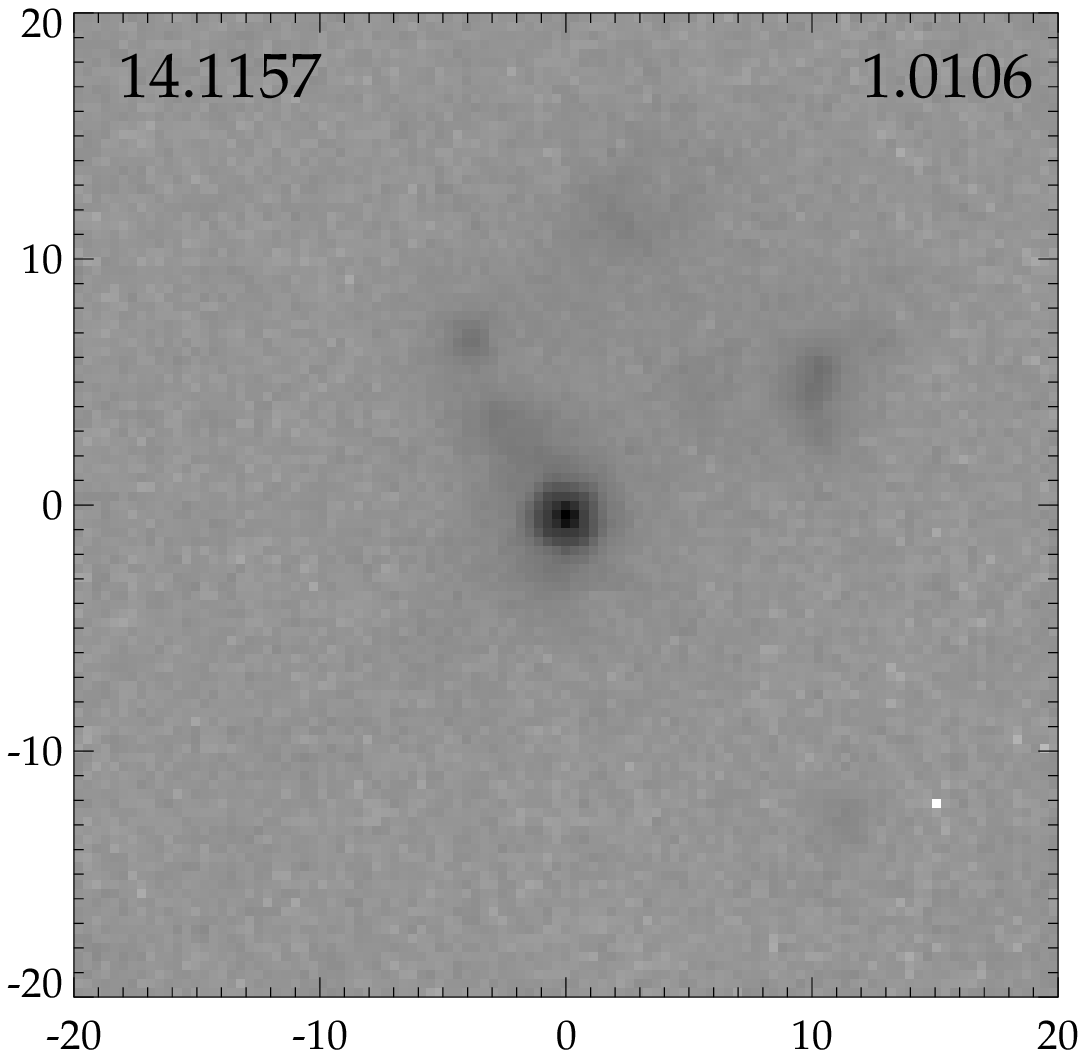} \includegraphics[height=0.21\textwidth,clip]{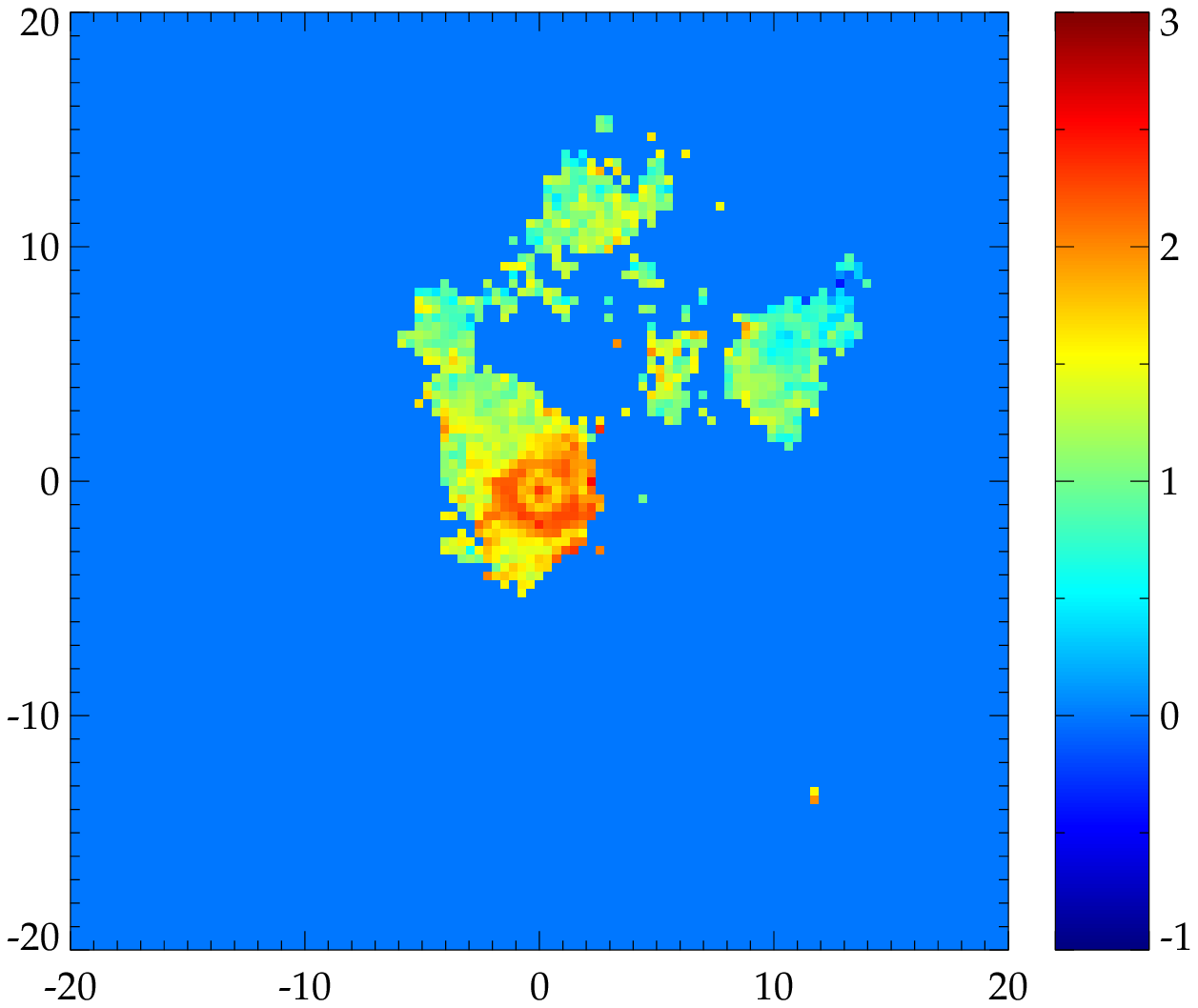}
\includegraphics[height=0.21\textwidth,clip]{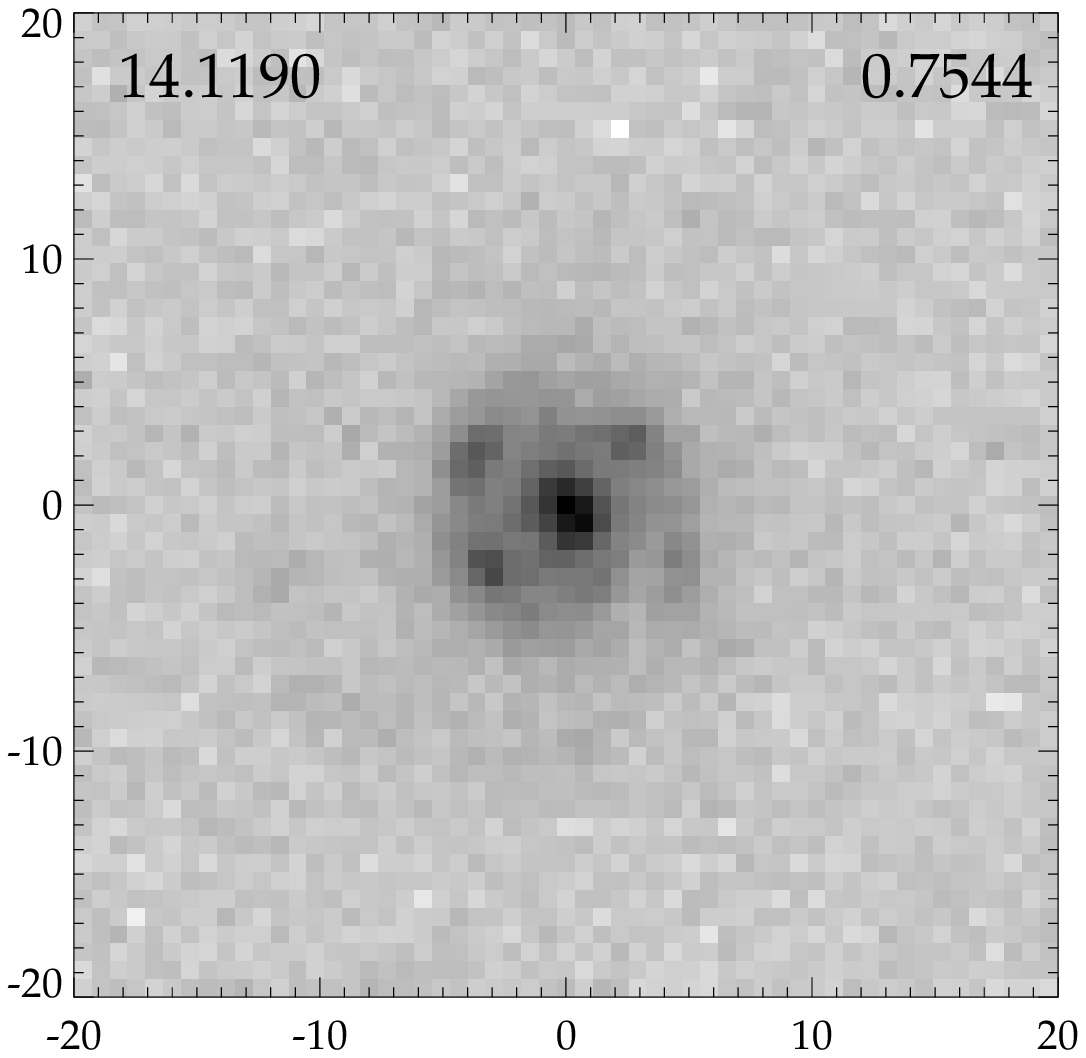}    \includegraphics[height=0.21\textwidth,clip]{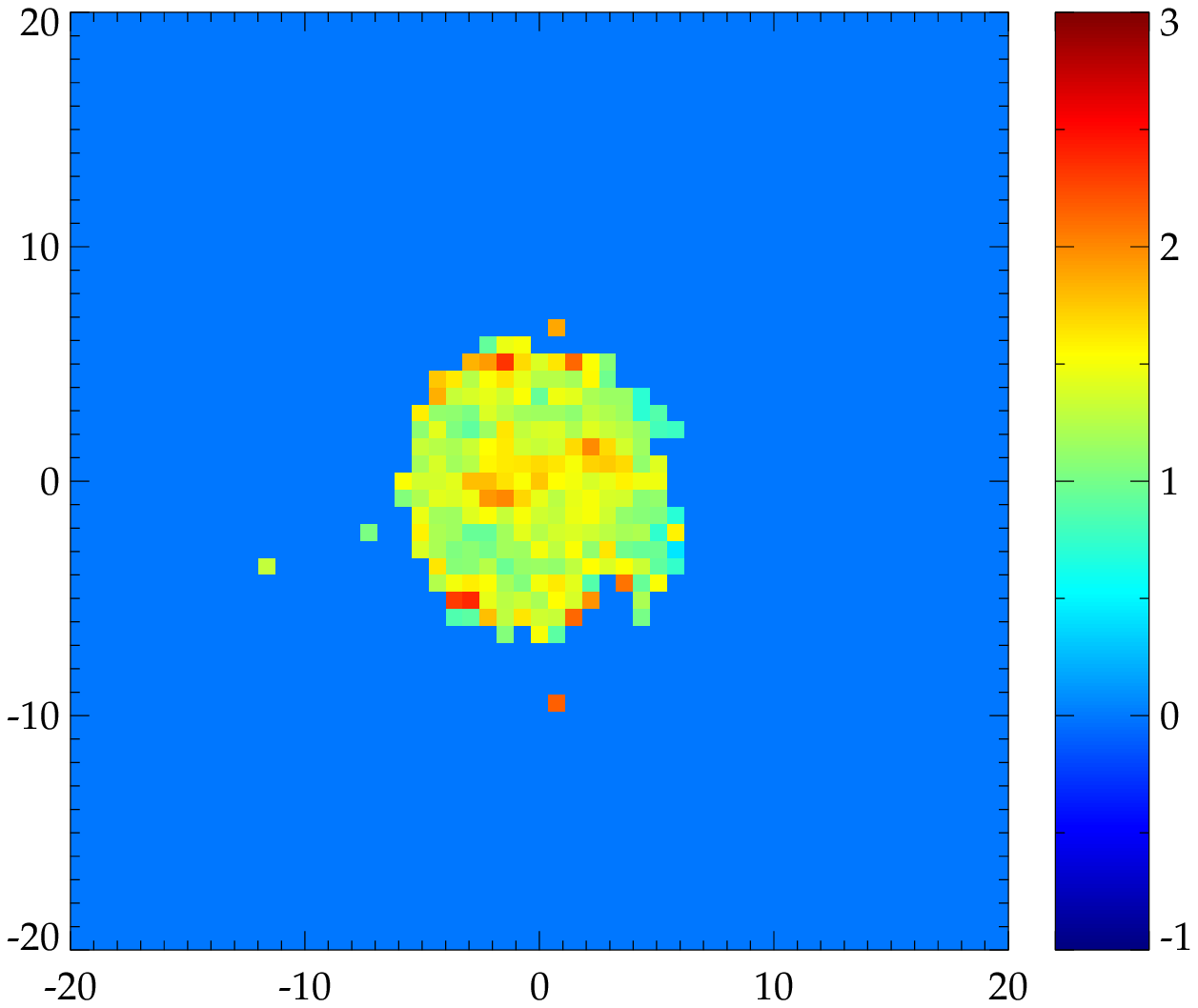}
\includegraphics[height=0.21\textwidth,clip]{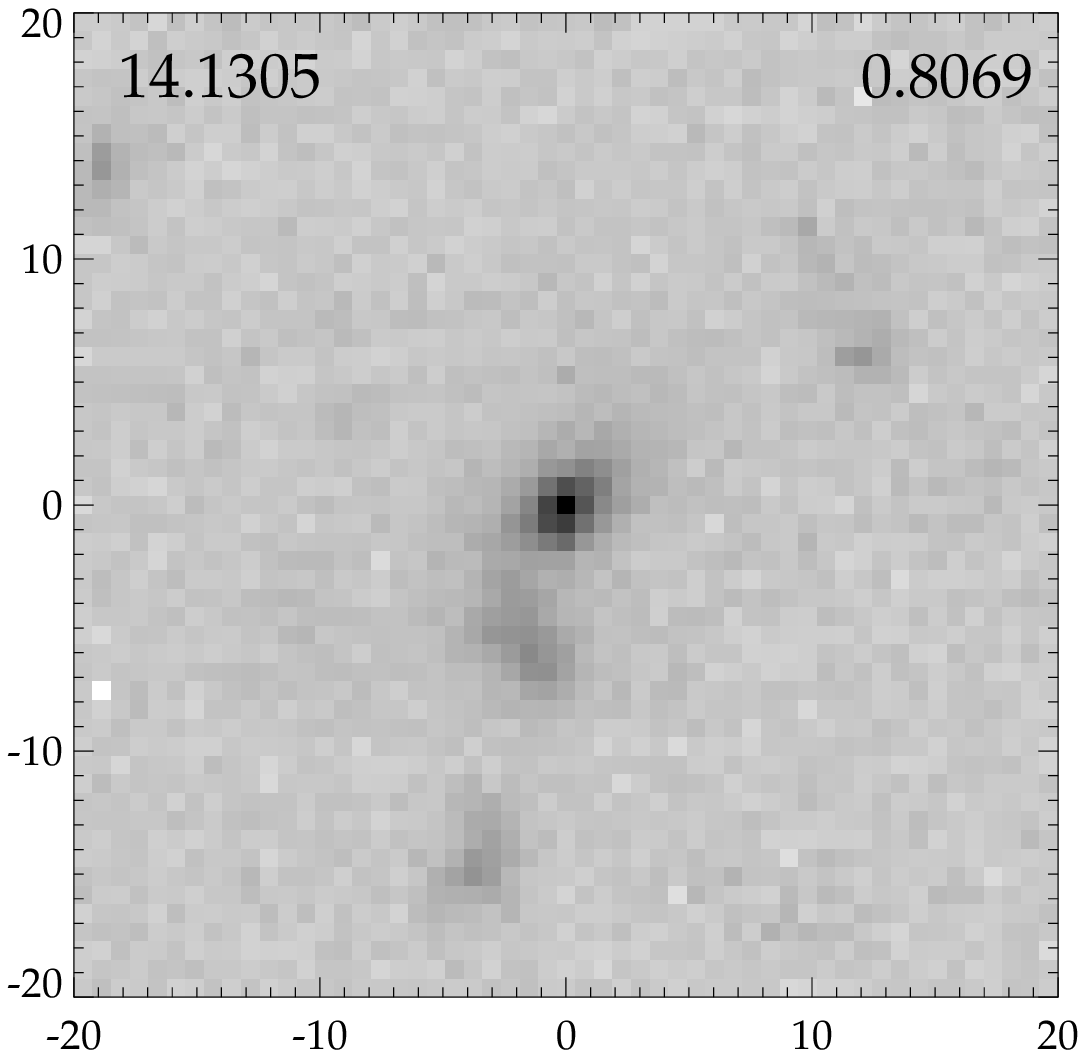}    \includegraphics[height=0.21\textwidth,clip]{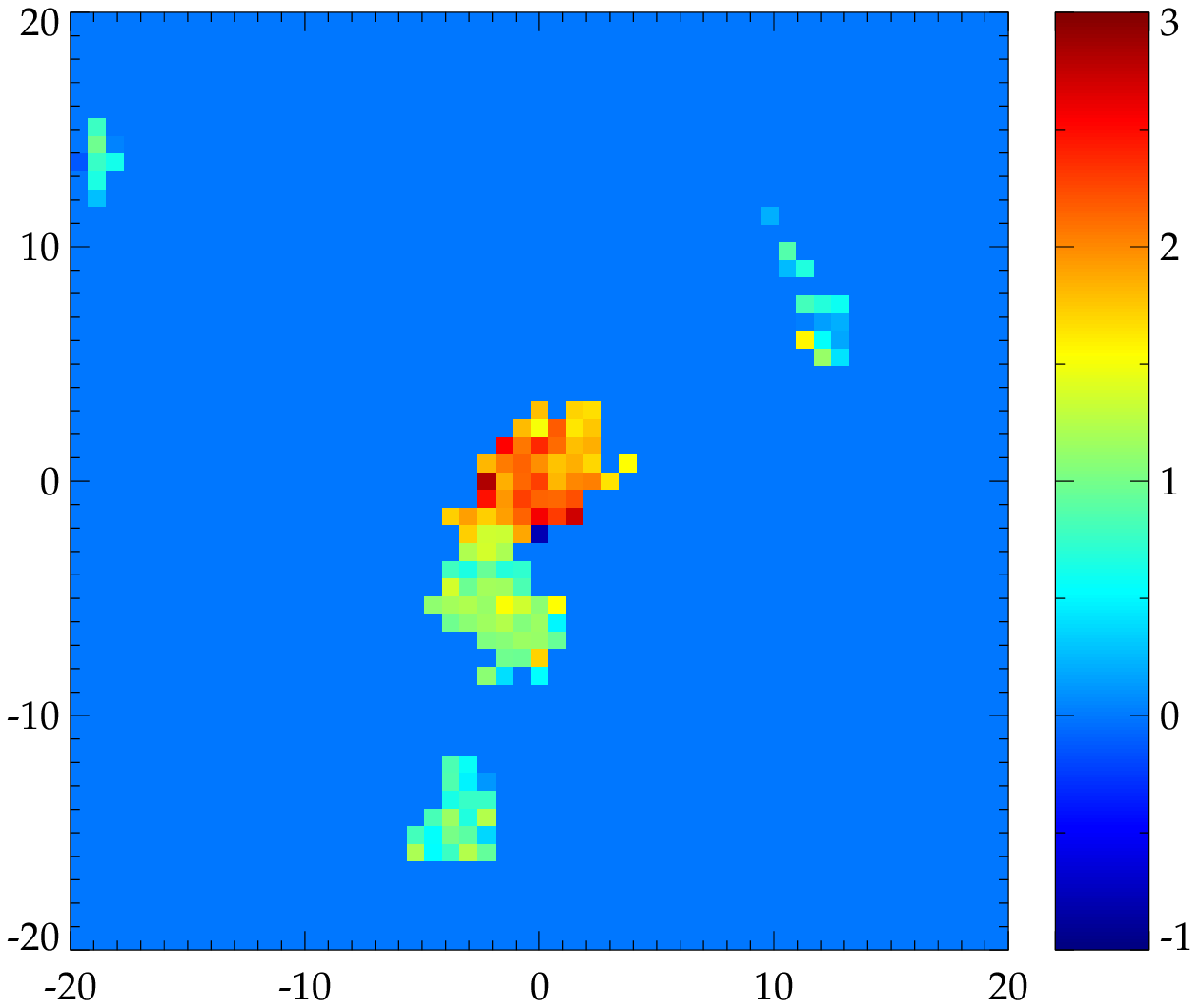}
\includegraphics[height=0.21\textwidth,clip]{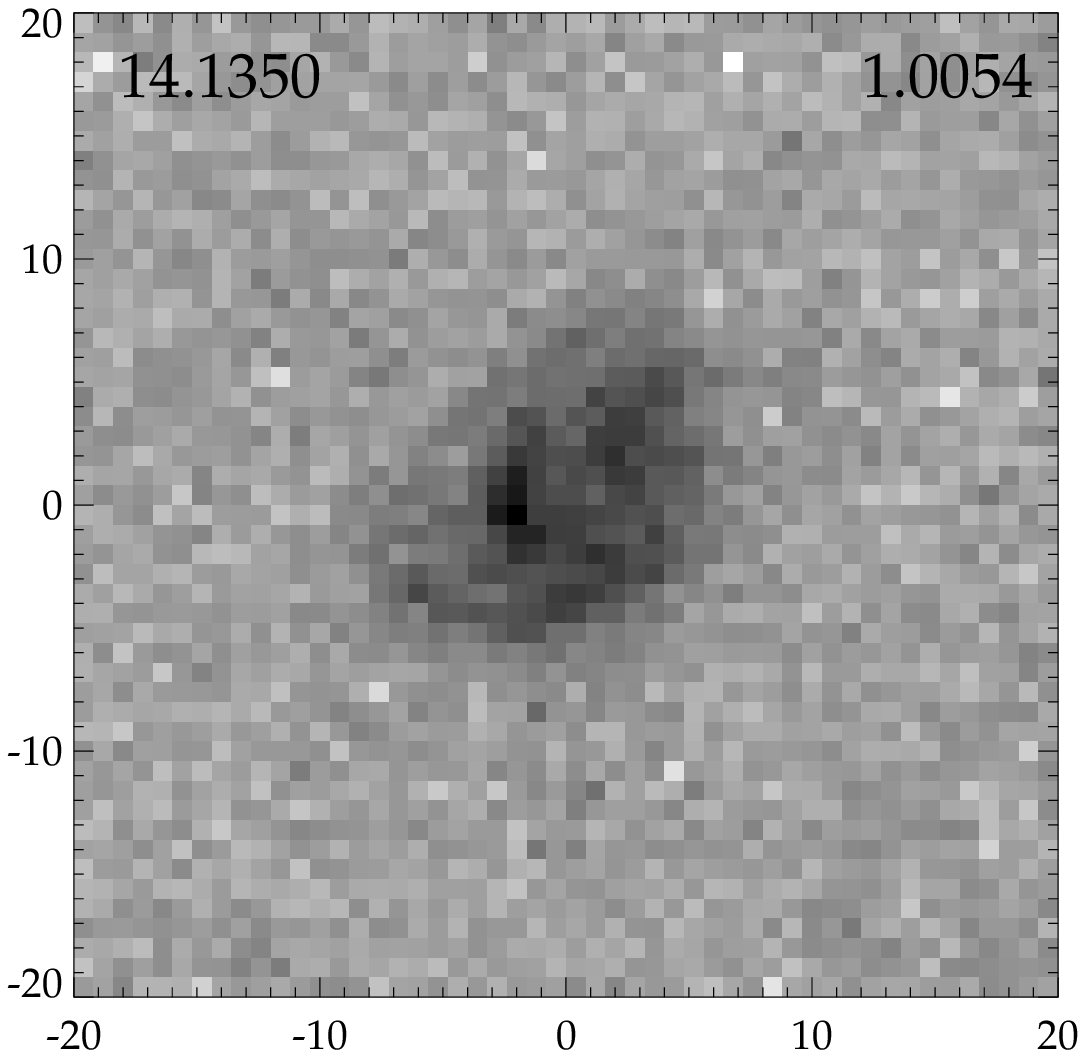} \includegraphics[height=0.21\textwidth,clip]{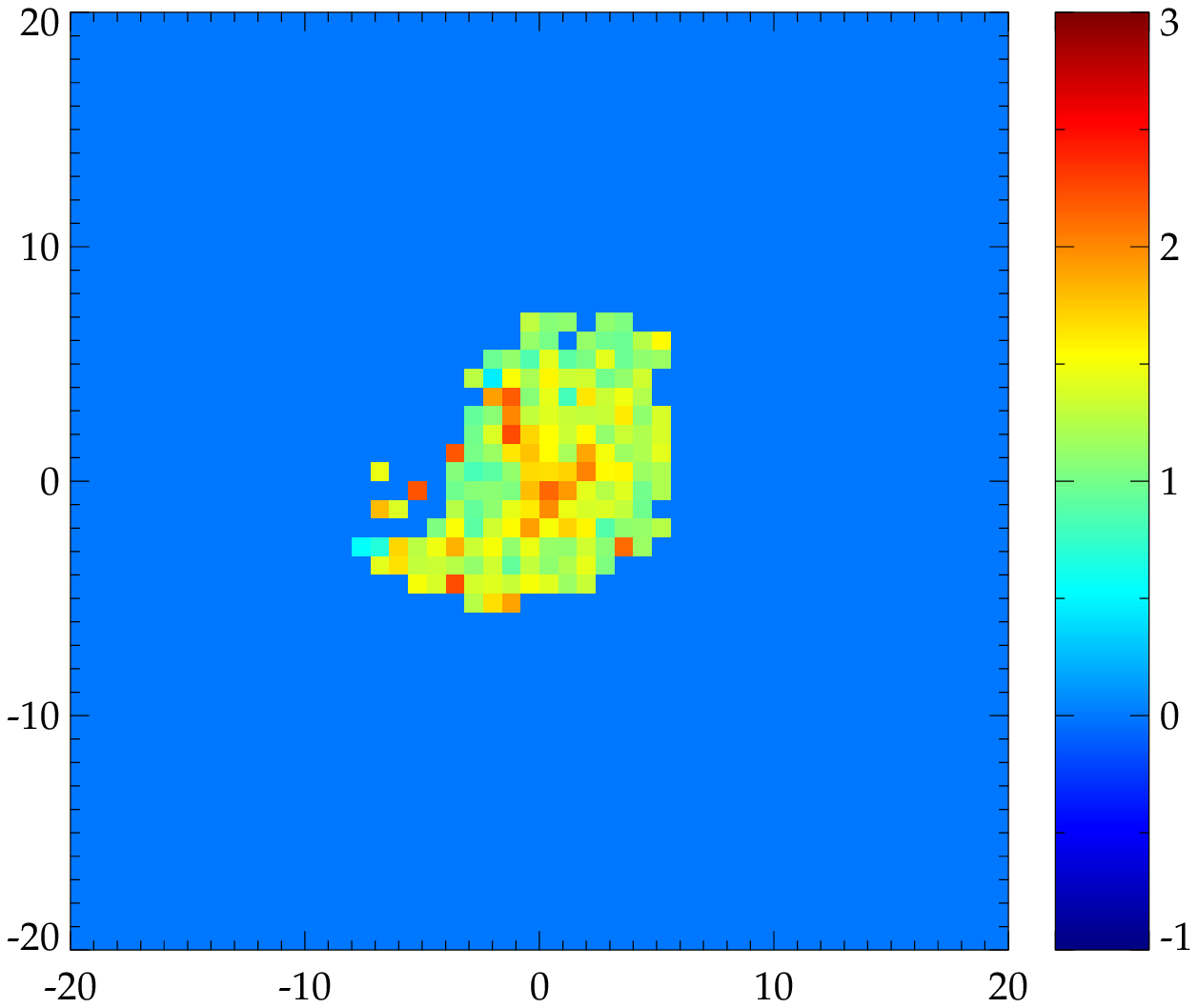}
\includegraphics[height=0.21\textwidth,clip]{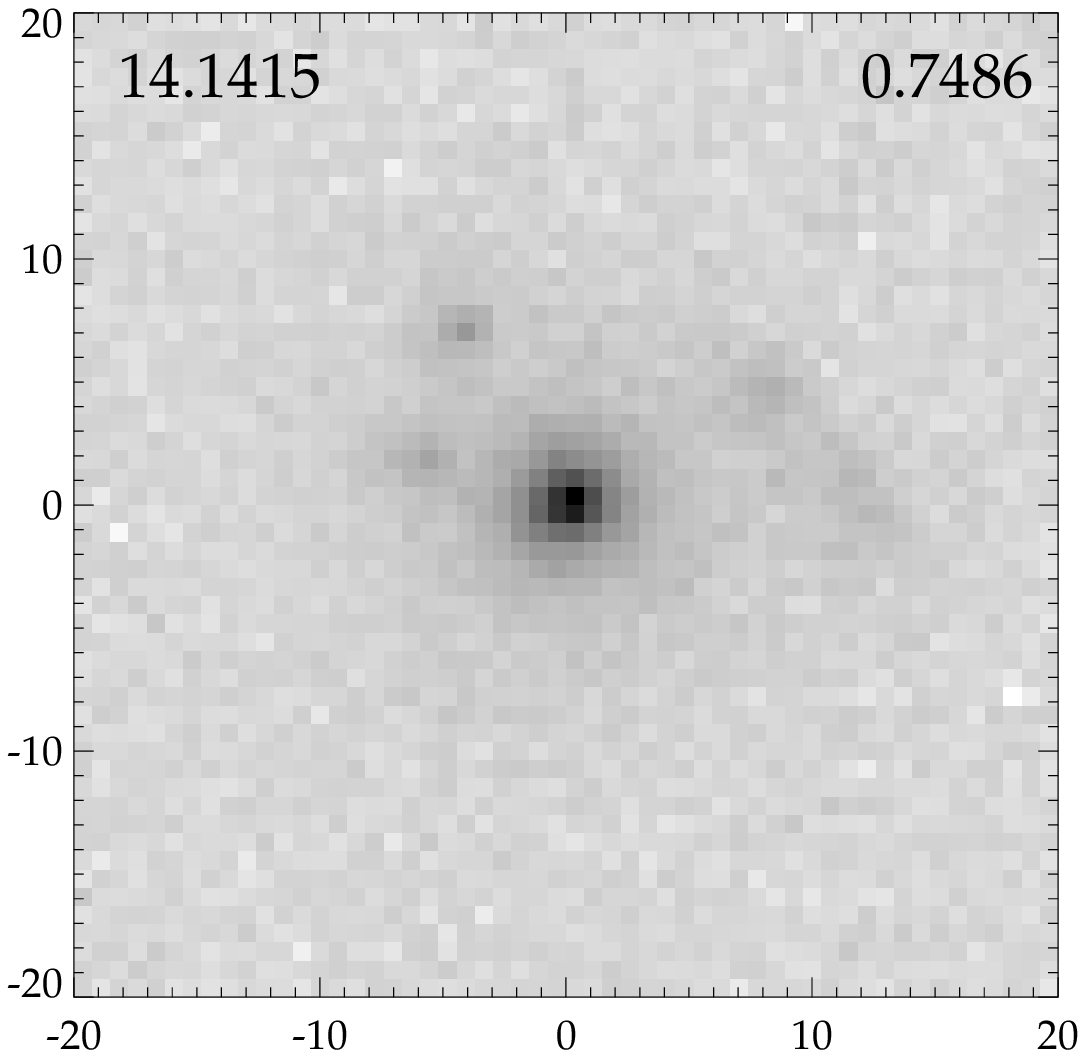}    \includegraphics[height=0.21\textwidth,clip]{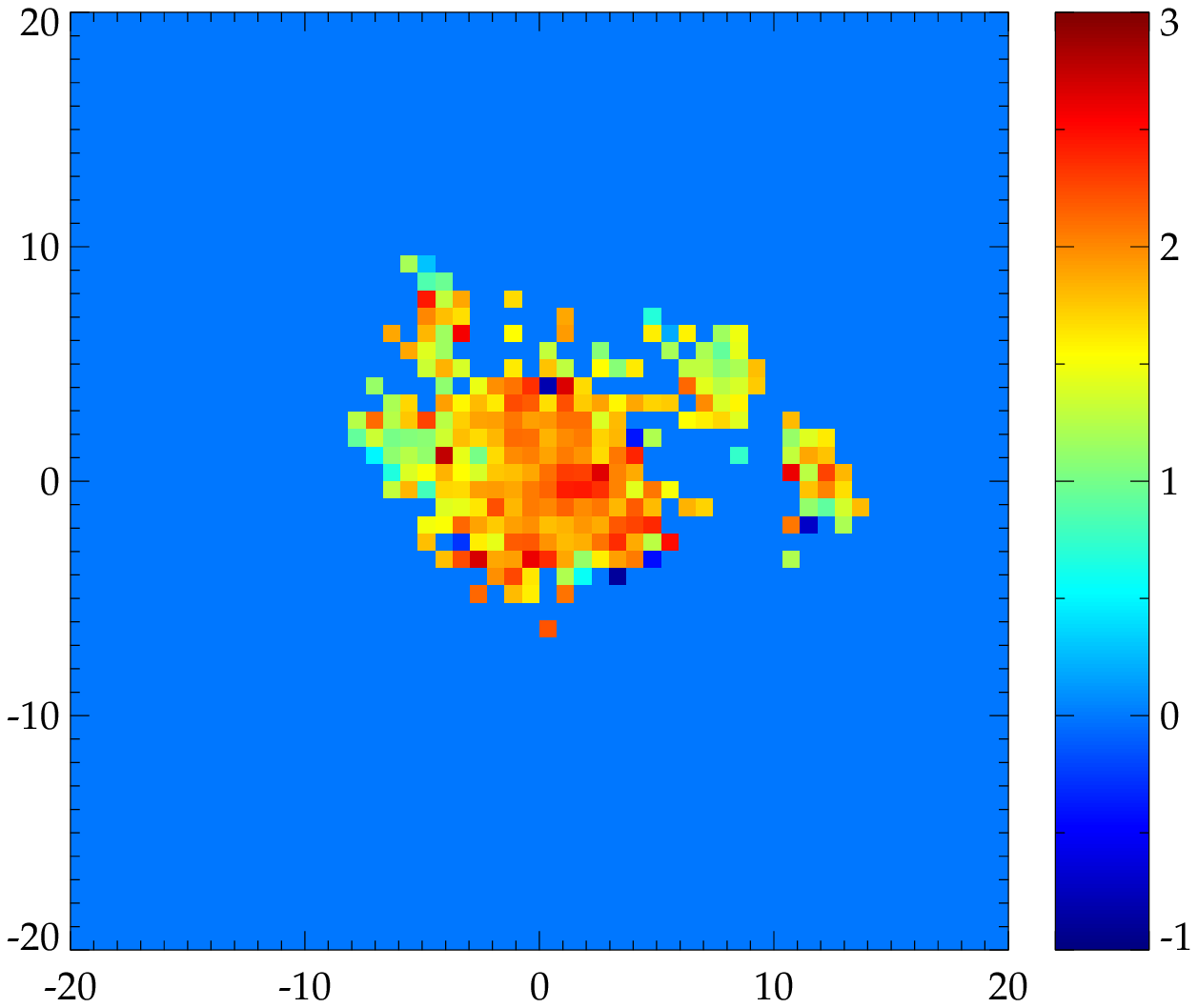}
\caption{Continued.}
\end{figure*}

\section{Discussion}

\subsection{Morphological properties}

From  Table~\ref{morpho},  we  can  derive  the  global  morphological
properties for distant LIRGs. Of  the 36 galaxies,  13 (36\%) are
classified as  disk galaxies  with `Hubble type'  from Sab to  Sd. For
these disk galaxies, their morphological classifications are secure (Q\,$<$\,2).
An extremely  red edge-on disk galaxy is  03.1522. This galaxy is
so red that almost no light is detected in B$_{450}$ band, which leads
to few pixels available in its color map.  We label this galaxy simply
``Spiral''  due to the  large uncertainty  in determining  the bulge
fraction. Of the 13 disk galaxies, 4 of them are very edge-on.
Morphological classification exhibits that 25\% (9 in 36) of the LIRGs
are compact  galaxies as they  show concentrated  light distribution
(see the definition in Hammer et  al.~\cite{Hammer01}).  Such  a  high
fraction is  strikingly similar to  that derived from  optical samples
(Guzm$\acute{\rm a}$n  et al.~\cite{Guzman};  Hammer  et al.~\cite{Hammer01}). Their
two-dimensional  structure fitting suffers  from  large
uncertainties.
Of  the 36  galaxies, 8 (22\%) are  classified as  irregular galaxies
which show complex morphology and clumpy light distribution.
In  the 36  LIRGs,  only 6  cases  (17\%) are  major ongoing  mergers
showing  multiple components  and apparent  tidal tails.   Five (either
compact or  irregular) LIRGs are possibly linked  to merging (labeled
by  M2) and  one  spiral LIRG  is  probably  an interacting  system
(labeled I2). Signs of merging or relics of interactions (possibly) occur in 9 LIRGs (labeled R and R?).
Accounting for them  and for major mergers,  the total
fraction of merging/interacting systems is estimated to be 58\% (21 in
36) of the  LIRGs.  

For  galaxies at  redshift  $\sim$1, the  cosmological dimming  effect
becomes significant  and the morphological  classifications might have
some  uncertainties since faint  features are  hardly detected  at the
detection limits  of our WFPC2  imaging data. Some  irregular galaxies
could  be indeed spiral  galaxies.  The  objects classified  as spiral
galaxies and  major mergers are free from  those uncertainties because
their main  structure properties have already  been recognized. Hence,
in the present  LIRG sample, the fraction for  spiral galaxies is well
estimated   or   slightly   underestimated   if  there   are   spirals
misclassified as  irregular galaxies.   HST I$_{814}$ band  imaging is
available for  all LIRGs. Thus the  compactness of each  galaxy can be
well determined and  the fraction of the compact  galaxies is reliable
(see e.g. Hammer et al.~\cite{Hammer01}).

\subsection{LIRGs: massive systems related to large disks?}

\begin{figure}[]
   \centering \includegraphics[width=0.40\textwidth]{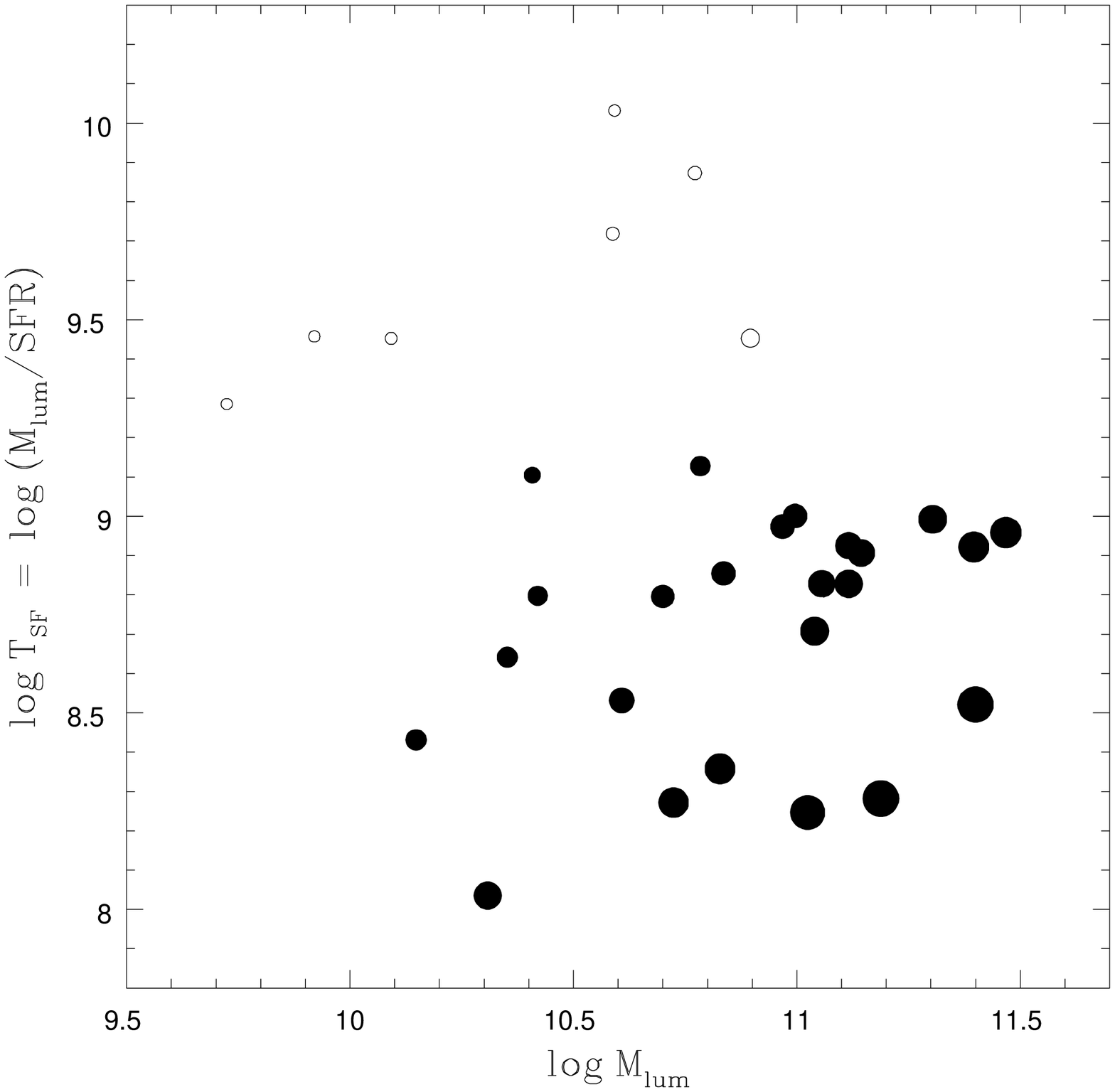}
   \caption{Stellar mass  versus time scale  to duplicate the  bulk of
stars at  a rate of  the observed value.  The stellar mass  is derived
from  K  band  luminosity.   The  size  of  the  point  is  scaled  by
SFR.  Nearby  galaxies  listed   in  Table~\ref{morpho}  with 
available K band luminosity are shown in open circles for a comparison.}
   \label{Mlum}%
    \end{figure}

K band  luminosity is widely used  to estimate the  stellar mass.  For
our 36  sample LIRGs, K band  luminosity is available for  24 of them.
Following  Hammer  et  al.    (\cite{Hammer01}),  we  assume  a  unity
mass-to-luminosity ratio in K band and estimate the stellar masses for
the   24    LIRGs.    The   derived   stellar    masses   range   from
1.4$\times$10$^{10}$ to 2.9$\times$10$^{11}$\,M$_\odot$, % for H0=70
% 1.8$\times$10$^{10}$ to 3.2$\times$10$^{11}$\,M$_\odot$, % for H0=50
compared to 1.8$\times$10$^{11}$\,M$_\odot$ of the Milk Way mass. Note
that extinction  and age of  stellar population are not  considered in
estimating  the   stellar  mass   due to  the   numerous  related
uncertainties  (e.g.   those   associated  with  the  assumed  stellar
population mixing and the  dust extinction modeling).  We notice that
the stellar masses we obtained are, systematically, from 50\% to 100\% of the values
derived  by  Franceschini  et  al.  (\cite{Franceschini03}),  after  a
careful examination of galaxies with similar K band luminosities. Indeed,
Franceschini  et  al.  (\cite{Franceschini03})  have  derived  stellar
masses using  spectral synthesis  modeling of the  overall optical-IR
continuum. Further discussion about uncertainties in the mass estimate can be found in Berta et al. (\cite{Berta}). Adopting M$_{\rm AB}^\ast$(K)\,=\,$-$21.82    for
H$_0$\,=\,70 (Glazebrook et  al. \cite{Glazebrook}), LIRGs are systems
ranging  from   0.3\,L$^\ast$  to  6.7\,L$^\ast$   with  median  value
2.2\,L$^\ast$.  These LIRGs were  undergoing violent star formation at
rates of the order  of $\sim$100\,M$_\odot$\,yr$^{-1}$. They may add a
significant mass contribution during a  short time.  For a galaxy, the
ratio of stellar mass  to SFR is the time scale for the formation of the 
bulk of the
stars.  If  star formation  is sustained at  the observed  rate, LIRGs
could   duplicate    themselves   within   10$^8$    to   10$^9$   yr
(Fig.~\ref{Mlum}).

\begin{figure}[]
   \centering \includegraphics[width=0.40\textwidth]{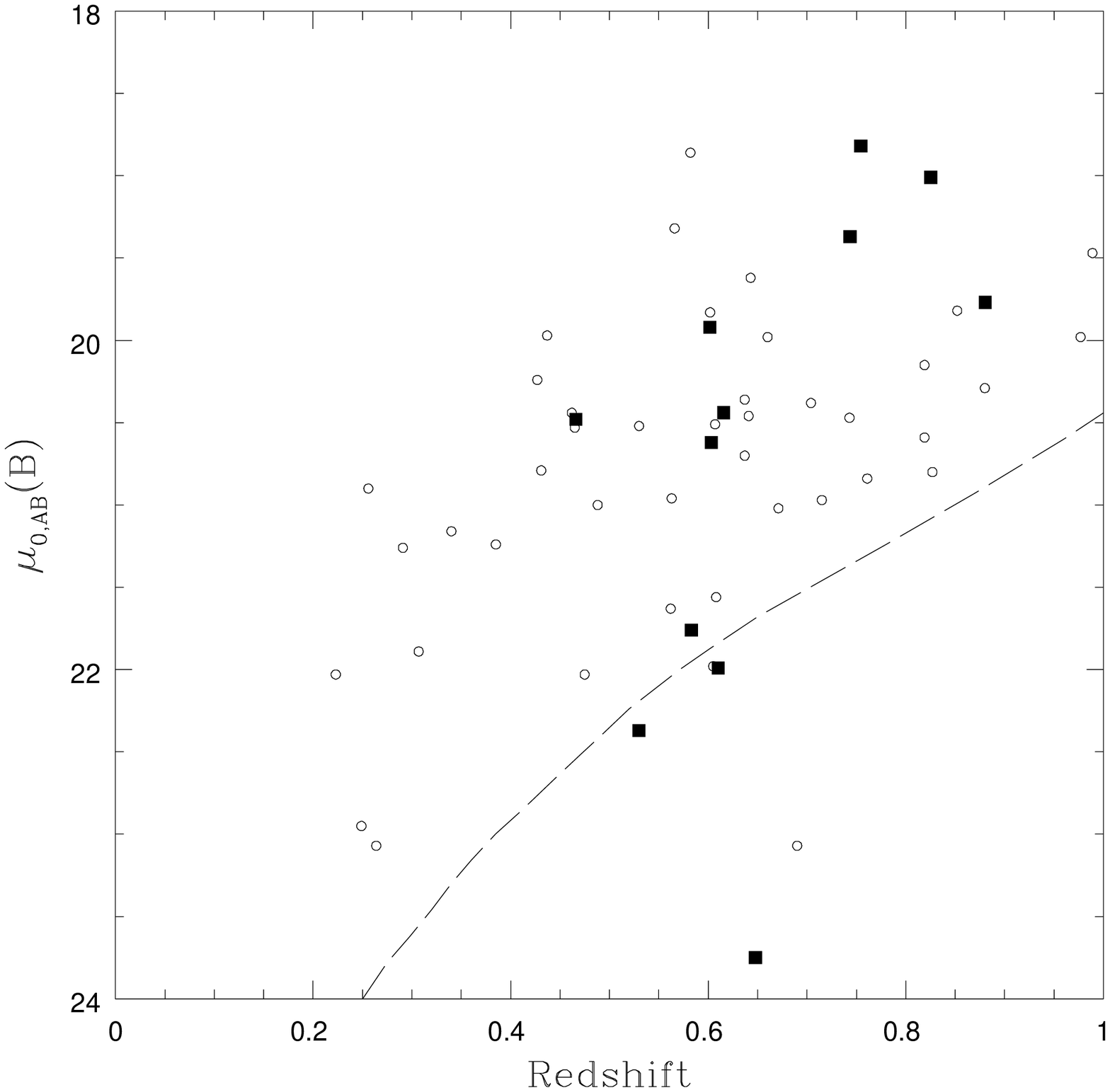}
   \caption{Central  surface brightness  of the  disk components  as a
function  of redshift  for LIRG  disks  in our  sample (solid  square),
compared  with the  large  disk  galaxies in  L98  (open circle).  The
long-dashed   line   curve  is   the   selection  criterion   (I$_{\rm
AB}\,<$\,22.5)   for   a   pure   disk  galaxy   with scale length
4\,h$_{50}^{-1}$\,kpc.}
   \label{diskSB}%
    \end{figure}

A  large  fraction of  LIRGs  are  classified  as disk  galaxies.   In
Table~\ref{morpho}, the  disk scale length R$_{\rm d}$  is also listed
for the LIRGs classified as spiral galaxies. We can see that all the
LIRG  disks are  large galaxies  with  R$_{\rm  d}$\,$\geq$\,2.9\,kpc  
(corresponding to 4\,kpc at H$_0$\,=\,50)  except for the object 14.1190, 
which  is a face-on Sd galaxy from  GSS fields.  We  suspect  that its 
disk scale length has been severely underestimated due to the shallow
detection ($\sim$24.5\,mag\,arcsec$^{-2}$ in  I$_{814}$ band, compared
to $\sim$25.5\,mag\,arcsec$^{-2}$ in other fields).
We compare  the LIRG  disks with size-selected  disk sample  using the
distribution  of  the  disk  central  surface  brightness.   Lilly  et
al. (\cite{Lilly98},  L98 hereafter) presented  a detailed study  of a
large  disk  (R$_{\rm d}$\,$\geq$\,4\,$h_{50}^{-1}$kpc)
sample, which is homogeneous  and essentially complete.  Similar to
L98 (see  their fig. 12), we plot  the LIRG disks in the diagram of
the  central surface  brightness  versus the  redshift.  The  observed
I$_{814}$  band  central surface  brightness  of  the disk  components
derived from the  two-dimensional structure fit is used  to obtain the
rest frame B$_{\rm AB}$ band central surface brightness.  Cosmological
dimming  and K-correction  are corrected  following L98.   We  use the
central    surface    brightness    color   V$_{606}-$I$_{814}$    (or
B$_{450}-$I$_{814}$)  to calculate the K-correction except for LIRG disks
03.0085,  03.0445, 03.0932, and 14.0393, whose integrated HST colors 
are adopted because the central surface brightness is either available 
only in one band or not fairly estimated in B$_{450}$ band. The edge-on 
LIRG disk 03.1522 is not included because of the heavy extinction. 
A  spectral energy distribution is chosen
to match  the observed color  from theoretical ones at  different ages
for a  solar abundance galaxy  with e-folding time  $\tau$\,=\,1\,Gyr (see
Hammer  et  al.  \cite{Hammer01}  for  more  details).   As  shown  in
Fig.~\ref{diskSB}, the  central surface brightness of 8  LIRG disks is
consistent  with that  of the  large disk  galaxies in  L98  at similar
redshift (LIRG  disks 03.0085,  03.0445, 03.0932, 03.1522  and 14.0393
are absent because of no secure structural parameters available in one
band).  This confirms that
LIRG disks belong to the large disk galaxy population.

We  also investigated  if  the size-selected  sample  of distant  disk
galaxies includes  LIRGs.  In  L98, 5, 14  and 5 objects  are selected
from the CFRS 0300+00 and 1415+52 fields in redshift bins 0.2\,$<\,z\,<$\,0.5,
0.5\,$<\,z\,<$\,0.75 and 0.75\,$<\,z\,<$\,1.0, respectively. After a
cross-identification with  ISOCAM observations, 6 of  19 disk galaxies
in  0.5\,$<\,z\,<$\,1.0 are found luminous in the IR band with IR luminosities
ranging from  1.6 $\times$ to  12$\times 10^{11}$\,L$_{\odot}$ (to be the median
value is 5.3$\times 10^{11}$\,L$_{\odot}$).  No  galaxy in  the redshift
bin  0.2\,$<\,z\,<$\,0.5 is  identified  as nearby  IR  luminous or  starburst
galaxies.  Of the 6 IR luminous large disk galaxies, 5 are included in
our HST sample, including 3 morphologically classified as spirals, one
as bar-dominated compact galaxy  (03.1540), although it shows evidence
for an extended faint disk, and one as merger (14.1139). This indicates
that in the large disk galaxy population at redshifts ranging from 0.5
to 1.0, about  30$\pm$ 12\% are infrared luminous.  This confirms that
LIRG disks are large (massive?)  disks.
LIRGs are  systems with  stellar content averaged  to 1.4\,M$^\ast$,
and they are intimately linked to large disks from their morphologies,
as well as the  fact that  size  selected disks  include a  significant
fraction of  them. We believe that  the star formation  in large disks
(derived   from   UV   measurements)   by  L98   had   been   severely
underestimated.
Indeed from their IR luminosities,  LIRG disks were forming stars at a
high rate, ranging from  18 to 210\,M$_\odot$\,yr$^{-1}$ and averaged
to  110\,M$_\odot$\,yr$^{-1}$. This is contradictory to the modest  star
formation rate of about 3 -- 10\,M$_\odot$\,yr$^{-1}$ reported by L98.
This  indicates   that   UV  and   [OII]
luminosities are poor tracers of the star formation rate.

The  number  density  of  LIRGs   is  much  larger  at  $z\sim$1  than
the present day by a factor of more than 40 (Elbaz et al. \cite{Elbaz}). 
From the above discussions, we believe that they significantly
contribute to the large and  massive disks population at z\,=\,0.5 -- 1. 
Our result is contradictory to that of Brinchmann \& Ellis 
(\cite{BrinchmannEllis}), which claimed that dwarf galaxies, rather than
massive systems were responsible for the star formation activity since
$z\sim$1. As  an example, we assume that the L98 SFR of 
$\sim$7\,M$_\odot$\,yr$^{-1}$ apply to the  large disks which have not been
detected  by ISO (which  is somewhat  unrealistic since  ISO detection
limit at  $z$\,=\,0.75  is $\sim$ 40\,M$_\odot$\,yr$^{-1}$). Assuming the SFR
derived  from IR luminosity  for the  ISO detected  disks in  L98, the
averaged   SFR   at   redshift   $\sim$0.75  would   be   on   average
$\sim$40\,M$_\odot$\,yr$^{-1}$, or 6 times
larger than the L98 average value.  Further investigation will address
the  contribution of  the  massive  galaxies to  the  CSFD (Hammer  et
al. \cite{Hammer04}).

\subsection{Central color versus concentration index}

   \begin{figure}[]
   \centering
   \includegraphics[width=0.40\textwidth]{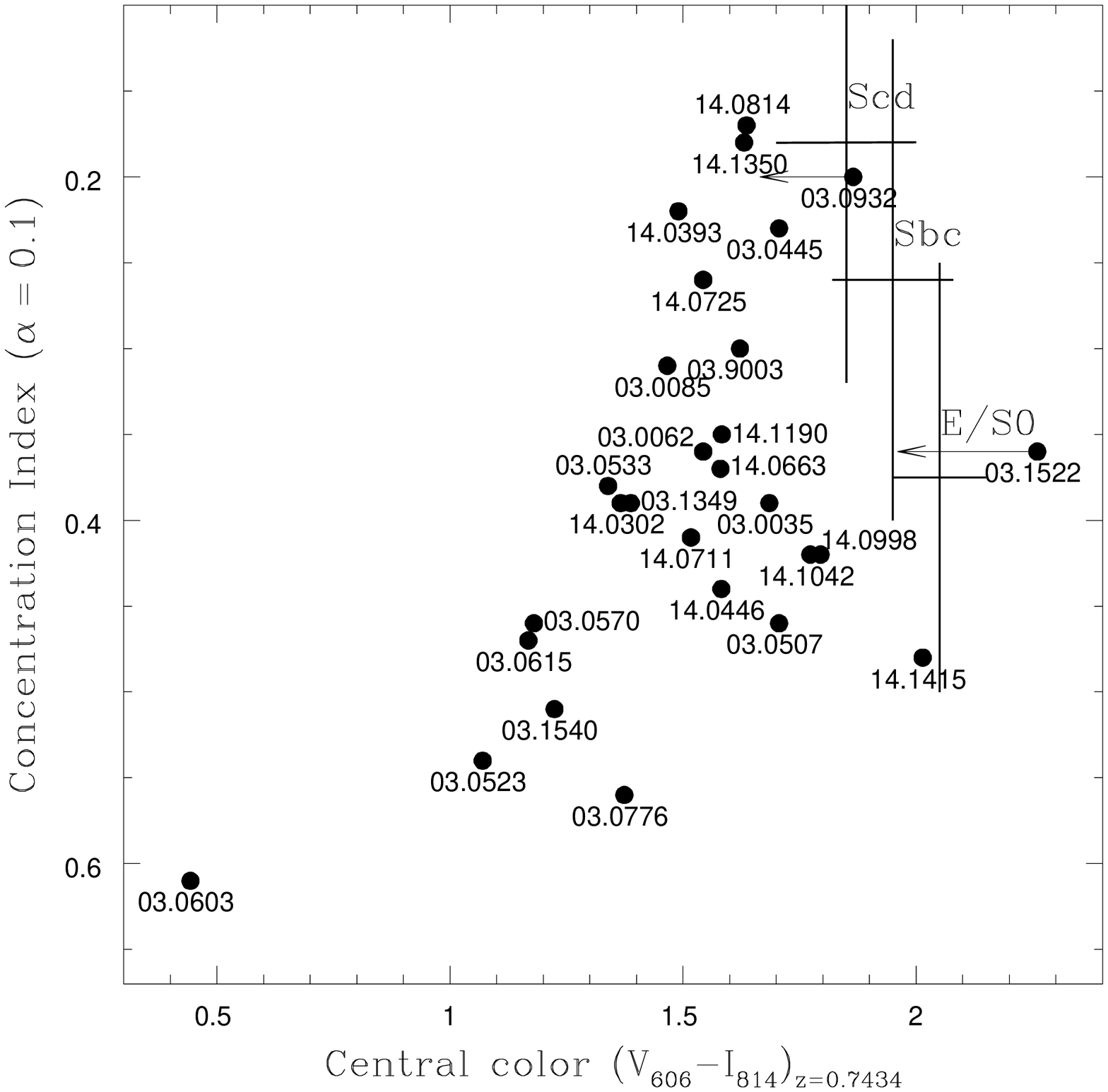}
   \caption{Central   color    versus   concentration   index.    
The distributions (large  crosses)  of normal  elliptical,  Sab  and  Sbc 
galaxies are derived from a nearby galaxy sample (Frei et al. \cite{Frei}).
The modeling color for the galaxies E, S0, Sbc and  Irr at this redshift 
are 2.05, 1.88, 1.10 and 0.38 respectively (see Sect.  5.2.2).  
We assume that Sbc galaxies
are bluer  0.1 mag  than E/S0  galaxies and Scd  galaxies are  0.1 mag
bluer than Sbc galaxies in the bulge color (de Jong \cite{deJong}).  A
central  color of  2.05  is adopted  for  E/S0 galaxies.   CFRS ID  is
labeled for  each data  point. Spearman rank-order  (S-R) correlation
analysis  reveals   a  correlation  coefficient  of   $-$0.42  with  a
probability of  0.026 that  the null hypothesis  of no  correlation is
true.   
}
   \label{colorconcen}%
    \end{figure}

We have  noticed that for  most compact LIRGs (e.g.  03.0523, 03.0603,
03.0615 and  03.1540), the  color map have  revealed a  central region
strikingly bluer  than the outer  regions. These blue  central regions
have a size similar to that of bulges and a color comparable to that of 
star-forming regions.  Since the bulge/central region in local spirals 
is relatively red, such blue-core structure could imply that the galaxy was
(partially) forming  the bulge. This  is consistent with  the scenario
proposed  by Hammer  et al.   (\cite{Hammer01}) that  luminous compact
galaxies are  counterparts of the  bulges in local spiral  galaxies at
intermediate redshifts.  From color maps, the central regions (bulges?) 
in formation are revealed directly.

Let us  assume that luminous  compact galaxies are progenitors  of the
spiral cores/bulges and that disk  components are formed later. Then a
correlation is anticipated among star forming systems (LIRGs: galaxies
forming disk and/or  bulge) between the color of  their central region
and the  galaxy compactness. Star  formation would first occur  in the
center  (bulge) and  hence would  gradually migrate  to  the outskirts
(disk), leading  to redder colors of  the central regions  as the disk
stars  were forming.  We  have investigated  the relation  between the
central color and the compactness for the LIRGs excluding the six major
ongoing mergers,  for which no central color is  available due to their
two separated components, and two objects whose structural 
parameters are not available.  To reduce the  contamination  from  the
surrounding disk, a  circle aperture with radius 1\,kpc centered at 
I$_{814}$ band brightness peak is adopted to
determine   the  central  color.    PHOT  in   IRAF$\footnote{IRAF  is
distributed by the National Optical Astronomy Observatories, which are
operated by the Association of Universities for Research in Astronomy,
Inc.   under   cooperative  agreement   with   the  National   Science
Foundation.}$  is  used  to  do aperture  photometry.   This  aperture
includes 5 pixels for objects at redshift 1 and the random fluctuation
is marginal due to the high S/N in the central regions. We convert the
observed central  color V$_{606}-$I$_{814}$ or  B$_{450}-$I$_{814}$ to
the observed color V$_{606}-$I$_{814}$  at median redshift 0.7434 (see
Hammer  et  al.~\cite{Hammer01}). At this reference redshift, The centroid 
wavelengths of the HST filters I$_{814}$ and V$_{606}$ correspond roughly
to  B  band (4596\,\AA)  and  U  band  (3440\,\AA).  The  same  method
described in the previous subsection is used to estimate the K-correction.  
We use the  concentration index ($\alpha$\,=\,0.1) defined  in Abraham et
al. (\cite{Abraham94}) to measure  the compactness.  It is provided as
an output  of GIM2D  in the two-dimensional  structure fitting.  
We  also compute  the same parameters for  a population  of local
galaxies  from Frei  et  al. (\cite{Frei}). The central color  of the local
galaxies have been assumed to be  that of local ellipticals (for E and
S0) or  local bulge (for Sab to  Sbc), and have been  transformed to B
(4596\,\AA)  $-$ U  (3440\,\AA)  color system  using  GISSEL98 model  with
$\tau$\,=\,1\,Gyr.   Fig.~\ref{colorconcen} shows the  investigation of the
concentration index from I$_{814}$ band as a function of the central 
color. In  this
diagram, the LIRGs are distributed along a sequence from galaxies with
blue central color and  compact morphology to galaxies with relatively
red color  and extended light distribution.  Almost  all distant LIRGs
are  discrepant from  the  sequence delineated  by  local spirals  and
ellipticals.   There are two  extreme cases  03.0603 and  03.1522. The
object 03.0603 could  be contaminated by AGN because  of the very blue
point-like core.  As  mentioned in Sect. 6.1, the  object 03.1522 is a
very  dusty edge-on  spiral  galaxy.  Its very  red  central color  is
related to  a (dust-screen) extinction.   It is worth noting  that the
sequence still exist when we  replace the central color with the color
contrast, which is defined as the difference between the central color
and the integrated color (3$\arcsec$ aperture).

The compact galaxies  with blue central color are  suggested to be the
galaxies  forming their  cores (bulges). Fig.~\ref{colorconcen}
clearly shows  that these  compact galaxies are located at one end of the
sequence.  The other  end is  occupied by  extended galaxies  with the
concentration  index  and  central  color  closer to  those  of  local
galaxies. 
It appears that  such a sequence linking  compact LIRGs to extended
ones  hints at a formation/assembly  scenario  for  at  least  part  of
the present day spiral galaxies: the  bulge formed first and dominated
a  galaxy with  a compact  morphology and  a relatively blue central 
color, and later  the disk was assembled around  the bulge, resulting to
an  extended light  distribution which  ultimately resembles  those of
local galaxies.

In trying  to make sense  of the sequence,  we point out  that intense
star formation  could happen  in processes  of both  bulge and disk formation.
In our color maps we see evidence that for LIRG spirals,
star formation spreads over all their disks. In fact, a star formation
history with several episodes  of massive starbursts are suggested for
LIRGs according  to their composite  stellar populations (Franceschini
et al. \cite{Franceschini03}).
The star  formation history with  multiple starburst episodes  is also
implied  by the  morphological classification  of LIRGs  that massive
starburst  (luminous infrared phase)  can happen  in spirals/irregular
phase, major merger  phase or compact phase.  
The major mergers in LIRGs  would most likely
result  in  spheroidal systems  with  violent  starbursts, and  become
similar  to  the compact  LIRGs  with  blue  central colors  as  those
described in this paper. 
The spiral LIRGs could have undergone a compact stage before the observed epoch.

However, a simple evolutionary sequence might not account for
all LIRGs. For example, among 12 LIRG disks with known bulge fraction,
3 are classified as Sd, i.e. with  a very small bulge, and 8 LIRGs
are classified as irregulars. Examination of these 11 Sd and Irr, generally
reveals the  presence of a well  defined central component  in most of
them  (03.0085, 03.0533, 03.9003,  14.0711, 14.0725,  14.0914, 14.0998
and 14.1190).  On the other hand, such component might not be a bulge
(or a  forming bulge), because they  can be elongated  or with peculiar
shapes (14.0814  has an S-shape and  it could be  a giant bar).  It is
beyond the scope of this  paper to evaluate the relative importance of
each  physical   processes  (major  and  minor   merging,  bars,  disk
formation) which are driving the galaxy formation.

The scenario  suggested by the sequence  in Fig.~\ref{colorconcen} for
disk galaxy  formation is also supported  by metallicity investigation
of LIRGs by  Liang et al. (\cite{Liang}). They found that, on average,
the metal abundance of LIRGs is less than half that of the local disks with
comparable brightness. They suggested  that LIRGs would form nearly half
of their metals and stars since $z\sim1$, assuming that they eventually
evolve into the local massive disk galaxies.

It  should be  borne in  mind that  contribution from  active galactic
nuclei (AGNs) might strengthen this sequence since AGNs usually appear
in blue color with respect to stellar populations (the situation could
be  opposite  if the  AGNs  are obscured  severely  by  dust) and  the
existence  of  bright AGNs  will  bias  the  galaxies to  the  compact
ones. However this concerns only a small fraction of our compact LIRGS
(2 among 9).   On the other hand,  dust extinction
will redden the color and smear the sequence.

\section{Conclusions}

Specific efforts were made to  obtain the color maps for galaxies with
complex morphology,  including the accurate alignment of  the blue and
red  band images  and a  new  method to  quantitatively determine  the
reliability  of each pixel  in the color  map. These  efforts allow  us to
access the spatially resolved color distribution of the distant LIRGs,
which    often     have    complex    morphologies,     relating    to
interactions/mergers.   In   two  10$\arcmin$$\times$10$\arcmin$  CFRS
fields 0300+00 and 1415+52, HST  WFPC2 imaging in F606W (or F450W) and
F814W filters is available  for 87 square arcminute area. From these fields
, we select a representative  sample of 36 distant ($0.4\,<\,z\,<\,1.2$) LIRGs
detected  in  deep  ISOCAM  observations.   Two-dimensional  structure
analysis  is  carried  out   using  GIM2D  software.   With  structure
parameters   and   color   distribution,   a   careful   morphological
classification was performed for the distant LIRGs.  We find that
about 36\% LIRGs are spiral galaxies and about 25\% LIRGs show compact
morphology.  About 22\%  LIRGs are  classified as  irregular galaxies,
showing complex and  clumpy structures. Among 36 LIRGs,  only 6 (17\%)
of  them   were  undergoing  a  major  merger   episode,  revealed  by
distinctive close  galaxy pair with distorted  morphology and apparent
tidal tails. The  fraction of mergers could reach  58\% if all of the 
possible post-mergers/pre-mergers are included.

Inspection  of  their stellar  masses  derived  from  K band  absolute
magnitude  evidences  that  LIRGs  are  massive  systems.   The  LIRGs
classified as  disk galaxies  indeed belong to  the large  disk galaxy
population, and  become a significant fraction of  large distant disks
selected by their sizes.

We find  that LIRGs  are distributed along  a sequence in  the central
color versus compactness diagram. The sequence links the compact LIRGs
with  relatively blue  central color  to that  of extended  LIRGs with
central  color and  compactness close  to  those of  the local  normal
galaxies.  The compact LIRGs  showing blue central color are suggested
to  be  the  systems  forming  their bulges,  in  agreement  with  the
suggestion of Hammer et al. (\cite{Hammer01}). We argue that the sequence
suggests that distant compact LIRGs  would eventually evolve  into the
spiral galaxies in the local universe.

\begin{acknowledgements}
We are grateful to Francoise Combes for helpful discussions and help in 
our morphological classification. We thank Jun Cui for his help on improving 
this manuscript. We thank the referee Dr. B. Mobasher for his helpful comments.
This work is based  on observations  made with  the NASA/ESA  Hubble Space
Telescope,  obtained from  the  data archive  at  the Space  Telescope
Institute. STScI  is operated by  the association of  Universities for
Research in Astronomy, Inc. under the NASA contract NAS 5-26555.

\end{acknowledgements}

\begin{center}{\bf\large Appendix: Signal-to-noise ratio for color map} \end{center}

\begin{figure}[]
   \centering
   \includegraphics[angle=-90,width=0.40\textwidth]{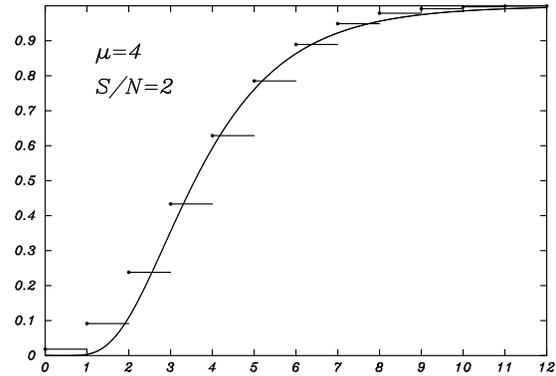}
   \caption{The distribution function of a Poisson
noise at S/N\,=\,2 and its Log-normal approximation (continuous line) for the
same S/N ratio. Even at this low S/N the two functions differ by a value of 
0.15 at most.  }
   \label{lognormal}%
    \end{figure}

Color image is a difference of two images in a logarithm function
(here we ignore the scaling constant which will not affect the final results). 
Given one HST image in blue band with flux $F_\mathrm{B}$ and another one
in red band with flux$F_\mathrm{R}$, the color image is defined as
\begin{equation}
Y\,=\,\log F_\mathrm{B} - \log F_\mathrm{R}.
\end{equation}
Here, the flux is a function of 
the position in an image, including signals of background and object.
The HST images have Poisson noise. 
We know the signals $\mu_{F_\mathrm{B}}$, $\mu_{F_\mathrm{R}}$ and the noises
$\sigma^2_{F_\mathrm{B}}$, $\sigma^2_{F_\mathrm{R}}$ in the two images, 
respectively.  We intend to derive signal $\mu$ and noise $\sigma^2$ 
of the color image and then obtain its S/N ratio.
The issue with color representation of an image is to avoid the divergence 
at zero value implied by the logarithm function. A way to do this is to replace
the Poisson noise by a random variable  possessing zero probability at zero 
value. This approximation must be valid at low S/N ratio, say 
S/N\,$\geq$\,2. It is clear that such an approximation must break down at 
very low S/N, say S/N\,$\leq$\,1 where a Poisson random variable is zero most 
of the time. 
Here we adopt an approximation that the Poisson noise distribution function 
in the HST image is close to a Log-normal law. A similar approach was tested
in a spectroscopic context by Rola \& Pelat (\cite{Rola}).
It is shown in Fig.~\ref{lognormal} that even at
S/N\,=\,2 a Log-normal variable is an acceptable approximation of a Poisson one.

In logarithm, the blue image $B$\,=\,$\log F_\mathrm{B}$ and the red image
$R$\,=\,$\log  F_\mathrm{R}$ satisfy a Normal law.
A difference between two Normal law images is also a Normal law image. 
This property is the main advantage of using the Log-normal law to approximate 
the Poisson noise.
The signal $\mu$ and noise $\sigma^2$ of the color image are given by
\begin{equation}
\mu = \mu_\mathrm{B-R} = \mu_\mathrm{B} - \mu_\mathrm{R} 
\end{equation}
and
\begin{equation}
\sigma^2 = \sigma_\mathrm{B-R}^2 = \sigma_\mathrm{B}^2 + \sigma_\mathrm{R}^2,
\end{equation}
where $\mu_\mathrm{B}$ and $\sigma_\mathrm{B}^2$, $\mu_\mathrm{R}$ and 
$\sigma_\mathrm{R}^2$ are the signal and the noise of the Normal law images
B and R respectively.
For Log-normal function $\exp Y$, its signal $m$ and noise $s^2$ 
can be written as
\begin{equation}
m\,=\,\exp (\mu\,+\,{1 \over 2}\sigma^2)
\end{equation}
and
\begin{equation}
s^2\,=\,(\exp \sigma^2\,-\,1)\,\exp (2\mu\,+\,\sigma^2).
\end{equation}
Expressing $\mu$ and $\sigma^2$ in $m$ and $s^2$, we obtain
\begin{equation}
\mu\,=\,-{1 \over 2}\log ({s^2 \over m^2}\,+\,1)\,+\,\log m
\end{equation}
and
\begin{equation}
\sigma^2\,=\,\log ({s^2 \over m^2}\,+\,1).
\end{equation}
With $\exp Y$\,=\,$F_\mathrm{B} \over F_\mathrm{B}$,
 we obtain the signal and noise for the color image as
\begin{equation}\label{signal}
\mu\,=\,-{1 \over 2}\log ({\sigma_{F_\mathrm{B}}^2 \over \mu_{F_\mathrm{B}}}\,+\,1)\,+\,\log \mu_{F_\mathrm{B}}\,+\,{1 \over 2}\log ({\sigma_{F_\mathrm{R}}^2 \over \mu_{F_\mathrm{R}}}\,+\,1)\,-\,\log \mu_{F_\mathrm{R}}
\end{equation}
and
\begin{equation}
  \sigma^2 = \log ({\sigma^2_{F_\mathrm{B}} \over \mu^2_{F_\mathrm{B}}} + 1) + \log ({\sigma^2_{F_\mathrm{R}} \over \mu^2_{F_\mathrm{R}}} + 1), 
\end{equation}
respectively.

\begin{landscape}
\centering
\begin{table}
  \caption[]{Catalog of the deep ISOCAM detected sources in CFRS fields 1300+00 and 1415+52}
  \label{morpho}
  \begin{tabular}{cccccccccccccclll}
  \hline
  \noalign{\smallskip}
         & & & & & &    &   &  \multicolumn{2}{c}{\hrulefill B$_{450}$\hrulefill} & \multicolumn{2}{c}{\hrulefill V$_{606}$ \hrulefill} & \multicolumn{2}{c}{\hrulefill I$_{814}$ \hrulefill}  &  & & \\
  CFRS ID & $z$ & m$_{450}$ & m$_{606}$ & m$_{814}$ & M$_{\rm AB}$(B) & M$_{\rm AB}$(K) & L$_{\rm IR}(10^{10}$L$_\odot$) & B/T & $\chi^2$ & B/T &  $\chi^2$ &  B/T &  $\chi^2$ & Type$^{\mathrm{a}}$ & Q$^{\mathrm{b}}$ & Int/M$^{\mathrm{c}}$ \\
(1) & (2) & (3) & (4) & (5) & (6) & (7) & (8) & (9) & (10) & (11) & (12) & (13) & (14) & (15) & (16) & (17) \\
  \noalign{\smallskip}
  \hline
  \noalign{\smallskip}
  \multicolumn{17}{c}{No redshift identification}  \\
  \noalign{\smallskip}
  03.0028 &   --   &  --   & 21.16 & 19.30 &    --    &    --    &   --              &      --                &   --   & $0.71_{-0.02}^{+0.01}$ &  1.078 & $0.73_{-0.01}^{+0.01}$  &  1.047 &     &   &     \\  
  03.0115 &   --   &  --   & 21.15 & 19.53 &    --    &    --    &   --              &      --                &   --   & $0.13_{-0.01}^{+0.01}$ &  1.588 & $0.22_{-0.01}^{+0.01}$  &  1.474 &     &   &     \\  
  03.0121 &   --   &  --   & 21.40 & 19.76 &    --    &    --    &   --              &      --                &   --   & $0.20_{-0.07}^{+0.02}$ &  1.947 & $0.18_{-0.02}^{+0.01}$  &  1.413 &     &   &     \\  
  03.0346 &   --   & 23.77 &  --   & 21.52 &    --    &    --    &   --              & $0.72_{-0.06}^{+0.10}$ &  1.089 &       --               &   --   & $0.43_{-0.01}^{+0.01}$  &  2.375 &     &   &     \\  
  14.0324 &   --   &  --   & 22.85 & 21.47 &    --    &    --    &   --              &    & -- & $0.27_{-0.05}^{+0.13}$ &  1.0156 &  $0.38_{-0.05}^{+0.05}$ &  1.0075  &     &   &    \\
  14.0405 &   --   &  --   & 22.26 & 20.38 &    --    &    --    &   --              &    & -- & $0.36_{-0.04}^{+0.06}$ &  1.0004 &  $0.19_{-0.02}^{+0.02}$ &  1.0183  &     &   &    \\
  \noalign{\smallskip}
  \hline
  \noalign{\smallskip}
  \multicolumn{17}{c}{ $z \leq 0.4$}  \\
  \noalign{\smallskip}
03.0355 & 0.0870 & 19.06 &  --   & 16.91 & $-$19.31 &   --     &   1.52$\pm$0.32   & $0.34^{-0.02}_{+0.02}$ & 1.183 &       --       &       --      & $0.70^{-0.01}_{+0.00}$ & 1.383 &    &   &    \\
03.0364 & 0.2513 & 20.97 &  --   & 18.95 & $-$20.30 & $-$22.14 &   4.56$\pm$0.81   & $0.00^{-0.00}_{+0.01}$ & 1.520 &       --       &       --      & $0.03^{-0.00}_{+0.00}$ & 1.787 &    &   &    \\
03.0365 & 0.2187 & 21.10 &  --   & 18.89 & $-$19.57 & $-$21.68 &   4.26$\pm$0.87   & $0.18^{-0.02}_{+0.02}$ & 1.374 &       --       &       --      & $0.00^{-0.00}_{+0.00}$ & 1.420 &    &   &    \\
03.0443 & 0.1178 & 20.37 &  --   & 19.12 & $-$19.15 &   --     &   0.78$\pm$0.16   & $0.72^{-0.07}_{+0.05}$ & 0.952 &       --       &       --      & $0.70^{-0.02}_{+0.02}$ & 1.093 &    &   &    \\
03.0495 & 0.2614 &  --   & 20.38 & 19.25 & $-$19.82 &   --     &   5.38$\pm$1.22   &       --       &       --      & $0.32^{-0.01}_{+0.01}$ & 2.174 & $0.40^{-0.06}_{+0.07}$ & 1.906 &    &   &    \\
03.0569 & 0.1810 &  --   & 22.41 & 21.27 & $-$16.73 & $-$20.01 &   1.67$\pm$0.37   &       --       &       --      &       --       &       --      &       --       &       --      &    &   &    \\
03.0578 & 0.2200 &  --   & 21.59 & 20.93 & $-$18.64 & $-$19.52 &   1.58$\pm$0.36   &       --       &       --      & $0.13^{-0.01}_{+0.02}$ & 1.084 & $0.21^{-0.05}_{+0.05}$ & 0.989 &    &   &    \\
03.0711 & 0.2620 &  --   & 21.34 & 20.51 & $-$18.55 & $-$20.44 &   2.51$\pm$0.58   &       --       &       --      & $0.00^{-0.00}_{+0.00}$ & 1.298 & $0.00^{-0.00}_{+0.00}$ & 1.136 &    &   &    \\
03.0949 & 0.0330 & 19.40 &  --   & 17.56 & $-$16.86 &   --     &   0.18$\pm$0.02   & $0.24^{-0.08}_{+0.03}$ & 0.897 &       --       &       --      & $0.13^{-0.07}_{+0.12}$ & 1.225 &    &   &    \\
03.1299 & 0.1760 &  --   & 19.66 & 18.76 & $-$19.94 &   --     &   5.02$\pm$1.23   &       --       &       --      & $0.13^{-0.02}_{+0.02}$ & 3.688 & $0.10^{-0.01}_{+0.07}$ & 1.005 &    &   &    \\
03.1311 & 0.1760 &  --   & 20.30 & 19.36 & $-$18.86 & $-$21.69 &   2.09$\pm$0.49   &       --       &       --      & $0.00^{-0.00}_{+0.00}$ & 1.562 & $0.04^{-0.01}_{+0.01}$ & 1.327 &    &   &    \\
14.0435 & 0.0684 &  --   & 18.71 & 17.91 & $-$18.30 &   --     &   0.14$\pm$0.02   &       --       &       --      & $0.00^{-0.00}_{+0.00}$ & 1.481 & $0.00^{-0.00}_{+0.00}$ & 1.395 &    &   &    \\
14.1257 & 0.2927 &  --   & 21.24 & 20.16 & $-$19.44 & $-$22.45 &  15.97$\pm$2.15   &       --       &       --      & $0.00^{-0.00}_{+0.00}$ & 1.859 & $0.00^{-0.00}_{+0.00}$ & 1.490 &    &   &    \\
14.1329 & 0.3750 &  --   & 20.21 & 19.19 & $-$21.00 &   --     &  13.03$\pm$2.47   &       --       &       --      & $0.43^{-0.01}_{+0.02}$ & 1.976 & $0.48^{-0.02}_{+0.01}$ & 1.816 &    &   &    \\
14.9025 & 0.1550 &  --   & 19.57 & 18.69 & $-$19.77 &   --     &   1.69$\pm$0.39   &       --       &       --      & $0.00^{-0.00}_{+0.00}$ & 3.280 & $0.07^{-0.00}_{+0.00}$ & 2.186 &    &   &    \\
  \noalign{\smallskip}
  \hline
  \noalign{\smallskip}
  \multicolumn{17}{c}{ $0.4\,<\,z\,<\,1.2$}  \\
  \noalign{\smallskip}
03.0035 & 0.8804 & 24.06 & 22.69 & 21.12 & $-$21.87 & $-$23.70 & 171.47$\pm$23.41  &       --       &       --      & $0.28^{-0.08}_{+0.06}$ & 1.349 & $0.31^{-0.06}_{+0.04}$ & 0.993 & Sab (3.6) & 2 & R?  \\
03.0062 & 0.8252 &  --   & 22.26 & 20.90 & $-$21.84 & $-$23.00 & 111.73$\pm$13.74  &       --       &       --      & $0.00^{-0.00}_{+0.00}$ & 1.545 & $0.01^{-0.01}_{+0.01}$ & 1.439 & Sbc (3.0) & 1 &   \\
03.0085 & 0.6100 &  --   & 22.58 & 21.51 & $-$19.98 & $-$21.09 &  29.55$\pm$2.67   &       --       &       --      & $0.00^{-0.00}_{+0.01}$ & 1.274 &       --       &       --      & Sd (5.1) & 2 &   \\
03.0445 & 0.5300 & 22.84 &  --   & 20.40 & $-$21.25 &    --    &  21.28$\pm$4.27   & $1.00^{-0.03}_{+0.00}$ & 0.992 &       --       &       --      & $0.01^{-0.00}_{+0.00}$ & 1.038 & Sbc (3.9) & 2 & R? \\
03.0507 & 0.4660 &  --   & 21.33 & 20.48 & $-$20.47 & $-$21.23 &  11.57$\pm$2.71   &       --       &       --      & $0.07^{-0.01}_{+0.12}$ & 1.325 & $0.24^{-0.04}_{+0.08}$ & 1.227 & Sab (3.2) & 2 &  \\
03.0523 & 0.6508 & 23.03 & 22.00 & 20.99 & $-$20.93 & $-$21.73 &  68.65$\pm$4.36   & $0.34^{-0.29}_{+0.55}$ & 0.931 & $0.53^{-0.03}_{+0.04}$ & 1.789 & $0.42^{-0.05}_{+0.05}$ & 1.292 & C/T & 1 & M2  \\
03.0533 & 0.8290 &  --   & 22.30 & 21.02 & $-$21.38 & $-$22.28 & 170.04$\pm$23.67  &       --       &       --      & $0.00^{-0.00}_{+0.00}$ & 1.096 & $0.00^{-0.00}_{+0.00}$ & 1.170 & Irr & 1 & R   \\
03.0570 & 0.6480 &  --   & 22.67 & 21.69 & $-$20.12 & $-$20.58 &  29.97$\pm$2.75   &       --       &       --      & $0.03^{-0.03}_{+0.05}$ & 1.242 & $0.42^{-0.08}_{+0.06}$ & 1.111 & C   & 1 &   \\
03.0603 & 1.0480 &  --   & 21.26 & 20.37 & $-$23.01 & $-$23.71 & 435.58$\pm$83.86  &       --       &       --      & $0.43^{-0.01}_{+0.01}$ & 1.464 & $0.73^{-0.01}_{+0.01}$ & 1.290 & C   & 1 & R   \\
03.0615 & 1.0480 &  --   & 22.52 & 21.51 & $-$21.66 & $-$22.77 & 344.69$\pm$62.39  &       --       &       --      & $0.35^{-0.03}_{+0.04}$ & 1.390 & $0.33^{-0.03}_{+0.02}$ & 1.306 & C   & 3 &     \\
03.0776 & 0.8830 &  --   & 23.28 & 22.05 & $-$20.69 & $-$20.98 & 107.84$\pm$14.25  &       --       &       --      & $0.00^{-0.00}_{+0.08}$ & 0.992 & $0.04^{-0.04}_{+0.09}$ & 1.061 & C   & 1 &   \\
03.0916$^d$ & 1.0300 & 22.64 &  --   & 21.17 & $-$22.23 &    --    & 503.26$\pm$68.64  &       --       &       --      &       --       &       --      &       --       &       --      & C   & 1 &   \\
03.0932 & 0.6478 & 23.54 &  --   & 21.18 & $-$20.73 & $-$22.85 &  97.18$\pm$6.95   &       --       &       --      &       --       &       --      & $0.00^{-0.00}_{+0.01}$ & 1.046 & Sbc (7.0) & 1 &     \\
03.1309 & 0.6170 &  --   & 21.65 & 20.77 & $-$21.49 & $-$23.00 &  89.28$\pm$6.32   &       --       &       --      &       --       &       --      &       --       &       --      &     & 4 & M1  \\
  \noalign{\smallskip}
  \hline
  \end{tabular}
\end{table}
\newpage
\end{landscape}

\addtocounter{table}{-1}
\begin{landscape}
\begin{table}
  \caption[]{Catalog of the deep ISOCAM detected sources in CFRS fields 1300+00 and 1415+52  --- continued}
  \begin{tabular}{cccccccccccccclll}
  \hline
  \noalign{\smallskip}
         & & & & & &    &   &  \multicolumn{2}{c}{\hrulefill B$_{450}$\hrulefill} & \multicolumn{2}{c}{\hrulefill V$_{606}$ \hrulefill} & \multicolumn{2}{c}{\hrulefill I$_{814}$ \hrulefill}  &  & & \\
  CFRS ID & $z$ & m$_{450}$ & m$_{606}$ & m$_{814}$ & M$_{\rm AB}$(B) & M$_{\rm AB}$(K) & L$_{\rm IR}(10^{10}$L$_\odot$) & B/T & $\chi^2$ & B/T &  $\chi^2$ &  B/T &  $\chi^2$ & Type$^{\mathrm{a}}$ & Q$^{\mathrm{b}}$ & Int/M$^{\mathrm{c}}$ \\
(1) & (2) & (3) & (4) & (5) & (6) & (7) & (8) & (9) & (10) & (11) & (12) & (13) & (14) & (15) & (16) & (17) \\
  \noalign{\smallskip}
  \hline
  \noalign{\smallskip}
03.1349 & 0.6155 &  --   & 21.67 & 20.38 & $-$21.15 & $-$22.63 &  56.77$\pm$3.38    &       --       &       --      & $0.28^{-0.01}_{+0.01}$ & 1.179 & $0.39^{-0.09}_{+0.08}$ & 0.873 & Sab (4.5) & 1 & I2  \\
03.1522 & 0.5870 & 25.73 &  --   & 21.74 & $-$19.59 & $-$21.96 &  46.13$\pm$3.02    &       --       &       --      &       --       &       --      & $0.22^{-0.16}_{+0.28}$ & 0.994 & S (9.0)  & 1 &     \\
03.1540 & 0.6898 & 22.95 &  --   & 20.61 & $-$21.27 &    --    &  90.23$\pm$9.19    & $0.64^{-0.64}_{+0.36}$ & 0.932 &       --       &       --      & $0.52^{-0.04}_{+0.06}$ & 1.223 & C   & 1 & R?  \\
03.9003 & 0.6189 & 22.58 &  --   & 20.47 & $-$21.24 &    --    &  54.74$\pm$3.73    & $0.13^{-0.02}_{+0.04}$ & 1.863 &       --       &       --      & $0.00^{-0.00}_{+0.00}$ & 2.017 & Irr & 1 & M2  \\
14.0302 & 0.5830 & 22.78 &  --   & 20.44 & $-$20.98 & $-$22.70 &  56.94$\pm$3.17    & $0.28^{-0.18}_{+0.16}$ & 1.005 &       --       &       --      & $0.22^{-0.03}_{+0.04}$ & 1.069 & Sab (3.0) & 2 & R?  \\
14.0393 & 0.6016 & 23.07 &  --   & 20.94 & $-$21.64 & $-$22.30 &  55.15$\pm$3.57    &       --       &       --      &       --       &       --      & $0.04^{-0.01}_{+0.01}$ & 1.157 & Sbc (7.8) & 1 &   \\
14.0446 & 0.6030 &  --   & 21.15 & 19.88 & $-$21.95 &    --    &  56.87$\pm$3.29    &       --       &       --      & $0.00^{-0.00}_{+0.00}$ & 1.243 & $0.00^{-0.00}_{+0.00}$ & 1.363 & Sd (6.8) & 1 &     \\
14.0547 & 1.1963 &  --   & 22.02 & 21.17 & $-$23.12 & $-$23.88 & 186.03$\pm$43.71   &       --       &       --      &       --       &       --      &       --       &       --      &     & 4 & M1  \\
14.0600$^e$ & 1.0385 &  --   & 22.03 & 21.13 & $-$22.04 &    --    & 369.49$\pm$61.12   &       --       &       --      &       --       &       --      &       --       &       --      & Irr & 1 &    \\
14.0663 & 0.7434 &  --   & 21.98 & 20.51 & $-$21.65 &    --    & 148.43$\pm$18.98   &       --       &       --      & $0.02^{-0.01}_{+0.01}$ & 1.241 & $0.15^{-0.01}_{+0.01}$ & 1.155 & Sbc (3.1) & 2 &    \\
14.0711 & 1.1180 &  --   & 22.30 & 21.24 & $-$22.50 & $-$23.18 & 463.42$\pm$96.59   &       --       &       --      & $0.42^{-0.16}_{+0.06}$ & 1.166 & $0.03^{-0.03}_{+0.05}$ & 1.228 & Irr & 1 & M2 \\
14.0725 & 0.5820 &  --   & 22.86 & 21.66 & $-$19.47 & $-$21.26 &  24.12$\pm$3.27    &       --       &       --      & $0.00^{-0.00}_{+0.01}$ & 1.027 & $0.00^{-0.00}_{+0.00}$ & 1.091 & Irr & 3 &    \\
14.0814 & 0.9995 &  --   & 23.56 & 22.04 & $-$20.73 & $-$22.02 & 163.02$\pm$84.25   &       --       &       --      & $0.01^{-0.01}_{+0.02}$ & 1.076 & $0.00^{-0.00}_{+0.01}$ & 1.159 & Irr & 3 & R? \\
14.0998 & 0.4300 & 23.02 & 21.27 & 20.09 & $-$20.32 & $-$22.17 &  26.08$\pm$4.24    & $0.08^{-0.08}_{+0.32}$ & 0.867 & $0.00^{-0.00}_{+0.00}$ & 1.891 & $0.00^{-0.00}_{+0.00}$ & 1.582 & Irr & 3 & M2 \\
14.1042 & 0.8916 &  --   & 22.75 & 20.98 & $-$21.60 & $-$23.07 &  99.37$\pm$10.51   & $0.21^{-0.21}_{+0.28}$ & 1.008 & $0.01^{-0.01}_{+0.00}$ & 1.189 & $0.31^{-0.01}_{+0.01}$ & 1.352 & C   & 1 & M2 \\
14.1129 & 0.8443 &  --   & 22.59 & 21.36 & $-$21.86 & $-$22.81 & 123.68$\pm$15.89   &       --       &       --      &       --       &       --      &       --       &       --      &     & 4 & M1 \\
14.1139 & 0.6600 & 22.40 & 21.20 & 20.24 & $-$22.02 & $-$23.47 & 118.00$\pm$11.36   &       --       &       --      &       --       &       --      &       --       &       --      &     & 4 & M1 \\
14.1157 & 1.0106 &  --   & 22.44 & 19.95 & $-$23.11 &    --    &1786$^f$        &       --       &       --      &       --       &       --      &       --       &       --      &     & 4 & M1 \\
14.1190 & 0.7544 &  --   & 21.97 & 20.57 & $-$21.58 &    --    &  92.43$\pm$9.42    &       --       &       --      & $0.02^{-0.01}_{+0.01}$ & 1.309 & $0.03^{-0.00}_{+0.00}$ & 1.373 & Sd (2.6) & 2 & R? \\
14.1305 & 0.8069 &  --   & 23.29 & 21.59 & $-$20.73 &    --    &  67.30$\pm$7.23    &       --       &       --      &       --       &       --      &       --       &       --      &     & 4 & M1 \\
14.1350 & 1.0054 &  --   & 22.70 & 21.43 & $-$21.44 &    --    & 297.35$\pm$139.92  &       --       &       --      & $0.01^{-0.01}_{+0.01}$ & 1.085 & $0.00^{-0.00}_{+0.00}$ & 1.171 & Irr & 2 &   \\
14.1415 & 0.7486 &  --   & 22.47 & 20.67 & $-$21.48 &    --    &  87.44$\pm$9.98    &       --       &       --      & $0.23^{-0.02}_{+0.03}$ & 1.076 & $0.42^{-0.02}_{+0.01}$ & 1.142 & C   & 2 & R  \\

  \noalign{\smallskip}
  \hline
  \end{tabular}
  \begin{list}{}{}
  \item[$^{\mathrm{a}}$]
Galaxy type --- E/S0: 0.8\,$<$\,B/T\,$\leq$\,1, S0: 0.5\,$<$\,B/T\,$\leq$\,0.8, Sab: 0.15\,$<$\,B/T\,$\leq$\,0.5, Sbc: 0$<$\,B/T\,$\leq$\,0.15, Sd: B/T\,=\,0, C: compact, T: tadpole, Irr: irregular. Note that disk scale length R$_{\rm d}$ is given in parentheses in unit of kiloparsec for spiral type galaxies.
  \item[$^{\mathrm{b}}$]
Q quality factor --- 1: secure, 2: merely secure, 3: insecure, 4: undetermined.
  \item[$^{\mathrm{c}}$]
Interaction/Merging --- M1: obvious merging, M2: possible merging, I1: obvious interaction, I2: possible interaction, R: relics of merger/interaction.
  \item[$^{\mathrm{d}}$] This object is too compact to obtain structural parameters.
  \item[$^{\mathrm{e}}$] Structural parameters are not successfully derived for this object from the two-dimensional structure fitting because of its clumpy light distribution and imagery close to the CCD chip border.
  \item[$^{\mathrm{f}}$] Infrared luminosity is estimated according to an obscured AGN model.
  \item[Notes: Structural parameters are not provided for the merging systems with multiple distinctly separated components.]
  \end{list}
\end{table}
\end{landscape}

\end{document}